\documentclass[a4paper,11pt]{report}

\usepackage{epsfig}

\usepackage[english]{babel}

\usepackage{amsmath}

\usepackage{subfigure}

 \usepackage{amssymb}
 
 \usepackage{latexsym}

\usepackage{USCthesis2004}

          \renewcommand{\bibname}{BIBLIOGRAPHY}
        \renewcommand\contentsname{\centerline{\bf Table of Contents}}

\renewcommand{\thechapter}{\arabic{chapter}}

\title{ASPECTS OF THE HOLOGRAPHIC STUDY OF FLAVOR DYNAMICS}

\author{Veselin Filev}

\date{August 2008}

\renewcommand{\copyrightyear}{2008}

\renewcommand{\copyrightname}{Veselin Filev}

\DeclareGraphicsExtensions{eps,.pdf}

\newcommand{\be}{\begin{equation}}
\newcommand{\ee}{\end{equation}}
\newcommand{\ber}{\begin{eqnarray}}
\newcommand{\eer}{\end{eqnarray}}

\def\lsim{\mathrel{\rlap{\lower4pt\hbox{\hskip1pt$\sim$}}
    \raise1pt\hbox{$<$}}}                
\def\gsim{\mathrel{\rlap{\lower4pt\hbox{\hskip1pt$\sim$}}
    \raise1pt\hbox{$>$}}}                
\DeclareSymbolFont{AMSb}{U}{msb}{m}{n}
\DeclareMathSymbol{\IN}{\mathbin}{AMSb}{"4E}
\DeclareMathSymbol{\IZ}{\mathbin}{AMSb}{"5A}
\DeclareMathSymbol{\IR}{\mathbin}{AMSb}{"52}
\DeclareMathSymbol{\Q}{\mathbin}{AMSb}{"51}
\DeclareMathSymbol{\II}{\mathbin}{AMSb}{"49}
\DeclareMathSymbol{\IC}{\mathbin}{AMSb}{"43}
\DeclareMathSymbol{\IP}{\mathbin}{AMSb}{"50}
\DeclareMathSymbol{\IH}{\mathbin}{AMSb}{"48}
\DeclareMathSymbol\IA{\mathalpha}{AMSb}{"41}
\DeclareMathSymbol\IS{\mathalpha}{AMSb}{"53}

\makeatletter
\let\l@figureOLD \l@figure
\renewcommand{\l@figure}{\vspace{\baselineskip}\l@figureOLD}
\makeatother

\makeatletter
\let\l@tableOLD \l@table
\renewcommand{\l@table}{\vspace{\baselineskip}\l@tableOLD}
\makeatother

\begin{document}

\renewcommand\contentsname{Table of Contents}

\maketitle

\begin{preface}

\pagebreak   

\chapter*{Dedication}

\begin{center}

{ \large To my wife Denitsa, the love of my life.}

\end{center}


\addcontentsline{toc}{chapter}{Dedication}

\chapter*{Acknowledgements}

\addcontentsline{toc}{chapter}{Acknowledgements}

First, I would like to thank and express gratitude to my advisor Clifford V. Johnson for introducing me to the subject of my research. I am grateful for his constant encouragement of independence of thought and creativity. I have always been impressed by his readiness to embrace new research endeavors and by his open-minded approach to science. I highly appreciate the fact that he never exerted pressure on my studies and was available to provide me with direction, assistance, and guidance through my research. This created an environment extremely fruitful for my development as a  researcher and stimulated me towards a successful completion of my studies. I will always be grateful for that.

Next, I would like to thank Nicholas Warner for his willingness to clarify and discuss important physical concepts. I have benefited immensely  from discussions during group meetings and from attending his excellent courses in general relativity and superstring theory. I am also grateful for his constant encouragement during my studies, for his friendly attitude and for his readiness to discuss issues on a broad variety of topics.  

I also wish to thank Krzysztof Pilch for his insightful comments and suggestions during group meetings. I have learned a lot from interacting with him. I am also grateful for his supporting and encouraging attitude and for his down to earth advises on manners important for  the successful development of a young scientist. 

I am also greatly indebted to Itzhak Bars for particularly useful discussions on a variety of scientific themes. He has always been available to answer my questions, specifically on Standard Model of Interactions and Quantum Field Theory. I have benefited exceedingly from attending his wonderful courses in Quantum Mechanics and Standard Model of Interactions. 

I am also thankful to Stephan Haas for his readiness and willingness to assist me in any difficulties of both scientific and general character that I have endured throughout my Ph. D. studies. He had always been extremely conscientious and responsible as a graduate advisor.

I wish to thank Elena Pierpaoli and Todd Brun for serving in my dissertation committee. 

Thanks are due to Tameem Albash and Arnab Kundu for the extremely fruitful collaboration during my studies. The creative and stimulating environment that they have created was crucial for the successful completion of virtually every research projects discussed in this work.

Special thanks to all my fellow graduate students and everybody in the department, who I had the privilege to interact with but did not mention by name, for creating a friendly and lively environment which made my experience at USC enjoyable and a rewarding one. 

Finally, I would like to thank my family: my mother Gergana for nurturing my early interest in exact science and for her constant encouragement and support  through every step of my life. I would like to thank my deceased father Georgi for he has always been an inspiration for me. I would also like to thank my uncle Dimitar for supporting my interest in science even in the most difficult periods of my studies and for being a role model for me. And last but not least I would like to thank my wife Denitsa for never loosing her belief in me and supporting me through this long journey. None of this work would be possible if it weren't for her love and devotion.


\begin{singlespace}

\newcommand*\oldhss{}
\let\oldhss\hss
\renewcommand*\hss{\oldhss\normalfont}
\tableofcontents
\let\hss\oldhss

\newpage

\addcontentsline{toc}{chapter}{List of Tables}

\addtocontents{lot}{\vspace*{-\baselineskip}}

\listoftables


\newpage

\addcontentsline{toc}{chapter}{List of Figures}

\addtocontents{lof}{\vspace*{-\baselineskip}}

\listoffigures


\end{singlespace}

\chapter*{Abstract}
\addcontentsline{toc}{chapter}{Abstract}

 This thesis is dedicated to the holographic study of flavor dynamics. The technique employed is a D7--brane probing of various D3--brane backgrounds. The first topic covered studies the influence of an external magnetic field on a flavored large N Yang--Mills theory. The theory exhibits spontaneous chiral symmetry breaking. A discrete self--similar structure of the spontaneous symmetry breaking mechanism is observed, this structure is reflected in the meson spectrum. The meson spectrum exhibits Zeeman splitting and characteristic GMOR relation. The second topic examines thermal properties of the dual gauge theory. The study reveals a first order phase transition associated to the melting of mesons. The critical ratio of the bare quark mass and temperature at which the transition happens is computed. The third topic studies the phase structure of the finite temperature dual gauge theory in the presence of magnetic field. A phase diagram of the theory is obtained. The temperature restores the chiral symmetry while the magnetic field has a freezing effect on the meson melting. The meson spectrum exhibits Zeeman splitting and characteristic GMOR relation. Thermodynamic quantities such as free energy, entropy, and magnetization are computed. The fourth topic studies the addition of an external electric field. The observed effect is dissociation of the bound quarks, favoring the meson melting. For sufficiently strong electric fields a global electric current is induced. Thus, the dissociation of mesons corresponds to an insulator/conductor phase transition. This transition persists at vanishing temperature. The fifth topic studies the addition of an R--charge chemical potential via brane probing of the spinning D3--brane geometry. The corresponding phase diagram is obtained. The chemical potential favors the dissociation of mesons. For high chemical potential, a finite phase difference between the bare quark mass and the quark condensate is induced. The last topic explores universal properties of gauge theories dual to the D$p$/D$q$ system. A universal discrete self--similar behavior associated to the insulator/conductor phase transition is observed and the corresponding scaling exponents are computed. A similarity between the electric field and R--charge chemical potential cases is discussed. 

\end{preface}

\chapter*{Chapter 1:  \hspace{1pt} Introduction}

\addcontentsline{toc}{chapter}{Chapter 1:\hspace{0.15cm}
Introduction}

\section*{1.1 \hspace{2pt} Introduction to the AdS/CFT correspondence}
\addcontentsline{toc}{section}{1.1 \hspace{0.15cm} Introduction to the AdS/CFT correspondence}

 The goal of theoretical physics is to provide an economical and self--consistent description of physical reality, by means of physical laws and first principles. This implies borrowing the structure of mathematics, while using the concepts of philosophy. However, the interplay between these subjects has always been mutual, as the discovery of new physical phenomena required the development of both novel philosophical concepts and mathematical tools. Indeed, it is now difficult to draw a solid line between mathematical and theoretical physics, while the philosophical aspects of general relativity and quantum mechanics remain challenging.

Despite the broad meaning of the term theoretical physics, a considerable number of theoretical physicists are interested in the study of the fundamental interactions of matter. It is well established that there are four basic interactions of matter, namely, the electromagnetic, weak nuclear, strong nuclear, and the gravitational interactions. It is somewhat ironic that albeit the concept of gravitational interaction was the first to be developed, it remains a challenge to come up with a consistent quantum description of gravity. It is believed that string theory, which has the graviton present in the spectrum of the fundamental string, is the best candidate of quantum theory of gravity. Furthermore, string theory provides a natural framework for a unified theory of fundamental interactions \cite{{Green:1987sp},{Johnson:2003gi},{Polchinski:1998rq}}.

Historically, string theory emerged as an attempt to delineate the strong interactions by what was called dual resonance models. However, shortly after its discovery Quantum Chromo Dynamics (QCD) which is a $SU(3)$ Yang--Mills gauge theory, superseded it. The matter degrees of QCD consist of quarks which are in the fundamental representation of the gauge group, while the interaction between the fundamental fields is being mediated by the gluons which are the gauge fields of the theory thus transforming in the adjoint representation of $SU(3)$. A remarkable property of QCD is the fact that it is asymptotically free, meaning that at large energy scales, or equivalently at short distances, it has a vanishing coupling constant. This makes QCD perturbatively accessible at ultraviolet. However, the low energy regime of the theory is quite different. At low energy QCD is strongly coupled, the interaction force between the quarks grows immensely and they are bound together, they form hadrons. This phenomenon is called confinement. Additional property of the low energy dynamics of QCD is the formation of a quark condensate which mixes the left and right degrees of the fundamental matter and leads to a breaking of their chiral symmetry. It is extremely hard to examine the properties of the strongly coupled low energy regime of QCD, since the usual perturbative techniques are not applicable. We need to coin new non--perturbative tools. One such approach is lattice QCD which has had a tremendous success in describing the static properties of the theory, with the use of large scale computer simulations. However, some of the most interesting and important problems of QCD, such as the description of the low energy mechanisms for confinement and spontaneous chiral symmetry breaking remain unsolved. This is why it is of extreme domination to develop novel analytic techniques applicable in the strongly coupled regime of Yang--Mills theories.

The AdS/CFT correspondence, as we shall demonstrate in this section, is a powerful analytic tool providing a non--perturbative dual description of non--abelian gauge theories, in terms of string theory defined on a proper gravitational background. Let us proceed by tracing the sequence of the ideas that lead to the development of this gauge/string correspondence.
 
A crucial milestone was the large $N$ limit proposed by t'Hooft \cite{ 't Hooft:1973jz}. 
Instead of using $SU(3)$ as a gauge group, t'Hooft proposed to consider $SU(N)$ Yang--Mills theory and take the limit $N\to\infty$, while keeping the so--called t'Hooft coupling fixed $\lambda=Ng_{YM}^2$. t'Hooft proved that in this limit only planar diagrams contribute to the partition function which makes the theory more tractable. On the other side, the expansion in $1/N$ corrections of the QCD partition function and  the genus expansion of the string partition function exhibit the same qualitative behavior, suggesting that perhaps a dual description of the large $N$ limit of non--abelian gauge theories might be attainable in the frame work of string theory.

These early hints about a possible gauge/string duality came very close to reality with the development of the concept of D$p$--branes and their identification as the sources of the well--known black $p$--brane solutions of type IIB supergravity. The key observation was that the low energy dynamics of a stack of $N$ coincident D$p$--branes can be equally well described by a $SU(N)$ supersymmetric Yang--Mills theory in $p+1$ dimensions and an appropriate limit of a $p$--brane gravitational background. The first gauge theory studied in this context is the large $N$ ${\cal N}=4$ $SU(N)$ supersymmetric Yang--Mills theory in $1+3$ dimensions which is a maximally supersymmetric conformal field theory. The corresponding gravitational background is, as proposed by Maldacena \cite{Maldacena:1997re}, the near horizon limit of the extremal 3--brane solution of type IIB supergravity. This was the original formulation of the standard (by now) AdS/CFT correspondence. 

On the other side, by the time when the AdS/CFT correspondence was proposed, there was enough evidence that a consistent theory of quantum gravity should exhibit holographic properties \cite{Susskind:1994vu}, more precisely the dynamics of the physical degrees of freedom contained in a certain volume of space-time should be captured by the physics of its boundary. As we shall demonstrate, the framework of the AdS/CFT correspondence provides a natural description of string theory on certain gravitational background, in terms of a dual field theory defined at a proper asymptotic boundary. This is how the AdS/CFT correspondence satisfies the holographic principle.

\subsection*{1.1.1 \hspace{2pt} Low energy dynamics of D3--branes}
\addcontentsline{toc}{subsection}{1.1.1 \hspace{0.15cm} Low energy dynamics of D3--branes}

Let us consider a stack of $N$ coincident D3--branes. This system has two different kinds of perturbative type IIB string theory excitations, namely open strings that begin and end on the stack of branes and closed strings which are the excitations of empty space. Let us focus on the low energy massless sector of the theory.

The first type of excitations corresponds to zero length strings that begin and end on the D3--branes. The orientation of these strings is determined by the D3--brane that they start from and the D3--brane that they end on. Thus, the states describing the spectrum of such strings are labeled by 
$\lambda_{ij}$, where $i,j=1,\dots,N$. It can be shown that in the case of oriented strings \cite{Johnson:2003gi}  $\lambda_{ij}$ transform in the adjoint representation of $U(N)$. On the other side, the massless spectrum of the theory should form a ${\cal N}=4$ supermultiplet in $1+3$ dimensions. The possible form of the interacting theory (if we take into account only interactions among the open strings) is thus completely fixed by the large amount of supersymmetry that we have and is the ${\cal N}=4$ $U(N)$ supersymmetric Yang--Mills theory. Note that $U(N)$ came from the transformation properties of $\lambda_{ij}$. On the other side, $U(N)$ can be thought of as a direct product of $U(1)$ and $SU(N)$, geometrically the $U(1)$ corresponds to the collective coordinates of the stack of D3--branes. We will restrict ourselves to the case when those modes were not excited, we refer the reader to ref.~\cite{Aharony:1999ti} for further discussion on this point.

The second kind of excitations is that of type IIB closed strings in flat space. The low energy massless sector is thus a type IIB supergravity in $1+9$ dimensions. 

The complete action of the system is a sum of the actions of those two different sectors plus an additional interaction term. This term can be arrived at by covariantizing the brane action after introducing the background metric for the brane~\cite{Leigh:1989jq}. It can be shown \cite{Aharony:1999ti} that in the $\alpha'\to 0$ limit the interaction term vanishes and the two sectors of the theory decouple. 

To summarize: the low energy massless perturbative excitations of the stack of D3--branes are given by two decoupled sectors, namely ${\cal N}=4$ $SU(N)$ supersymmetric Yang--Mills theory and supergravity in flat $1+9$ space-time. Our next step is to consider an equivalent description of this system in terms of effective supergravity solution.

\subsection*{1.1.2 \hspace{2pt} The decoupling limit}

\addcontentsline{toc}{subsection}{1.1.2 \hspace{0.15cm} The decoupling limit}

Let us consider the extremal black 3--brane solution of type IIB supergravity. The appropriate gravitational background is given by \cite{Johnson:2003gi}:
\begin{eqnarray}
ds^2&=&H_3^{-1/2}\eta_{\mu\nu}dx^{\mu}dx^{\nu}+H_3^{1/2}dx^idx^i \ ,\label{D3--branes}\\
e^{2\Phi}&=&g_s^2 \ , \nonumber\\
C_{(4)}&=&H_3^{-1}g_s^{-1}dx^0\wedge\dots\wedge dx^3\ , \nonumber
\end{eqnarray}
where $\mu=0,\dots, 3$, $i=4,\dots,9$, and the harmonic (in six dimensions) function $H_3$ is given by:
\begin{equation}
H_3=1+\frac{4\pi g_3 N\alpha'^2}{r^4}\ .
\end{equation}
The integer number $N$ quantizes the flux of the five-form field strength $dC_{(4)}
$. It can also be interpreted as the number of D3--branes sourcing the geometry.
Taking the near horizon limit of the geometry corresponds to sending $\alpha'\to 0$ while keeping the quantity $u=r/\alpha'$ fixed. Such a limit serves two goals: first it enables one to zoom in the geometry near the extremal horizon and second it corresponds to a low energy limit in the string theory defined on this background. After leaving only the leading terms in $\alpha'$, one can obtain the following metric \cite{Maldacena:1997re}:
\begin{eqnarray}
ds^2&=&\frac{u^2}{R^2}(-dx_{0}^2+dx_1^2+dx_2^2+dx_3^2)+R^2\frac{du^2}{u^2}+R^2d\Omega_{5}^2\ ,\label{AdS-intro}\\
C_{(4)}&=&\frac{1}{g_s}\frac{u^4}{R^4}dx^0\wedge dx^1\wedge dx^2 \wedge dx^3\ ,\nonumber\\\
e^\Phi&=&g_s\ ,\nonumber\\
R^4&=&4\pi g_{s}N_{c}\alpha'^2\ ,\nonumber
\end{eqnarray}
The background in equation (\ref{AdS-intro}) is that of an AdS$_5\times S^5$ space-time of radius $R$. Note that from a point of view of an observer at $r\to\infty$, the type IIB string theory excitations living in the near horizon area, namely superstring theory on the background (\ref{AdS-intro}), will be redshifted by an infinite factor of $\sqrt{g_{tt}}=H_3^{-1/4}$. Therefore, we conclude that type IIB superstring theory on the background of AdS$_5\times S^5$ should contribute to the low energy massless spectrum of the theory seen by an observer at infinity. However, an observer at infinity has another type of low energy massless excitations of type IIB string theory, namely type IIB supergravity on flat $1+9$ space or gravitational waves. Those two different types of excitations can be shown to decouple form each other. To verify this one can consider the scattering amplitudes of incident gravitons of the core of the geometry (the near horizon area). It can be shown that at low energies ($\omega\ll 1/R$) the absorption cross-section of such a scattering $\sigma_{\rm{abs}}$ goes like \cite{Gubser:1997yh,Klebanov:1997kc} $\sigma_{\rm{abs}}\propto \omega^3R^8$. Therefore, one has that $\sigma_{\rm{abs}}\to 0$ and those two types of excitations decouple in the low energy limit $\omega\to 0$.

Similar to the description from the previous subsection, we learned that the massless low energy spectrum of the theory contains two types of perturbative excitations, namely type IIB string theory on the background of AdS$_5\times S^5$, and type IIB supergravity in flat $1+9$ space-time. Now we are ready to conjecture the AdS/CFT correspondence.
\subsection*{1.1.3 \hspace{2pt} The AdS/CFT correspondence}
\addcontentsline{toc}{subsection}{1.1.3 \hspace{0.15cm} The AdS/CFT correspondence} 

\begin{figure}[h] 
   \centering
   \includegraphics[ width=13cm]{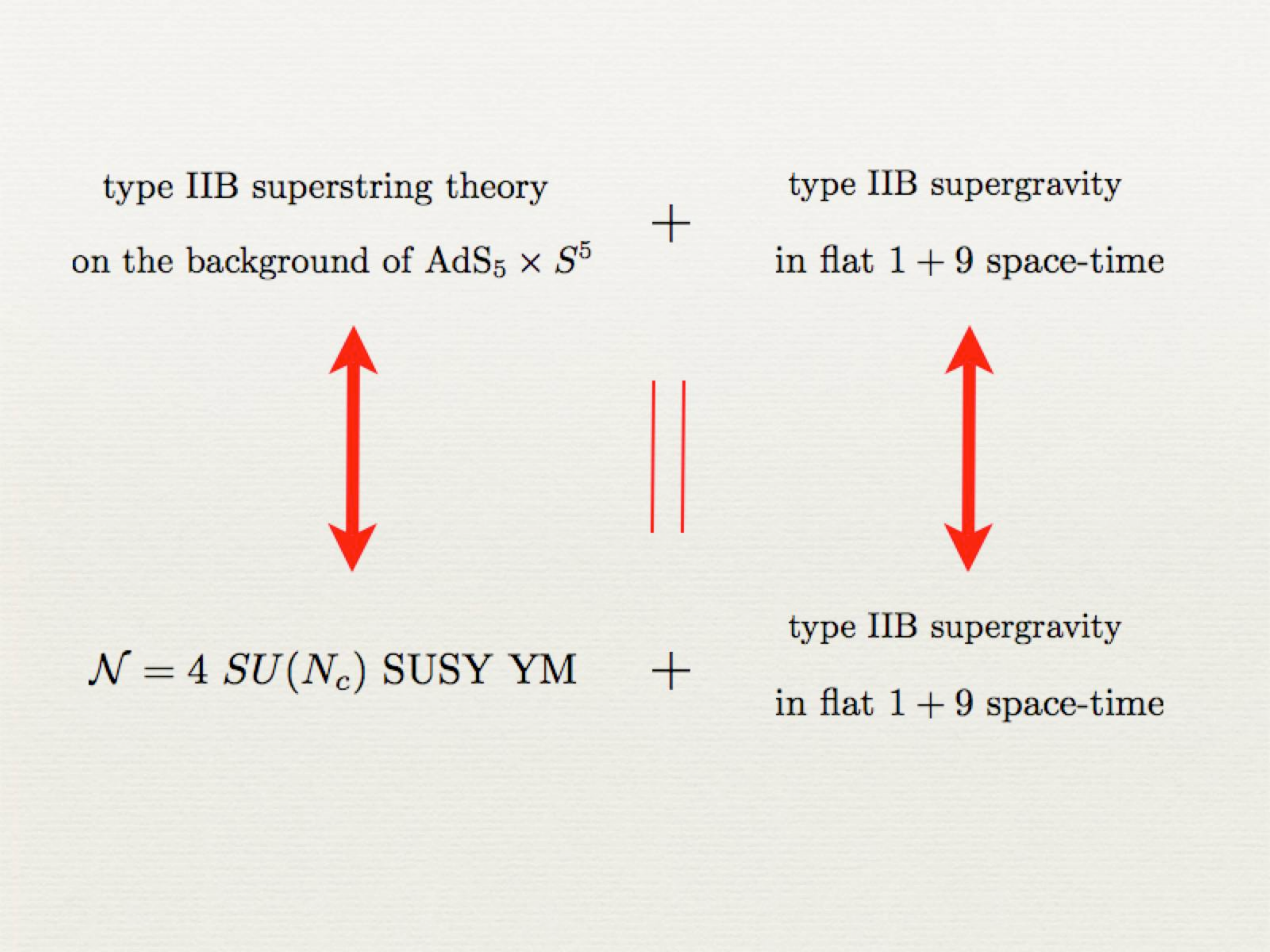}
   \caption{A diagrammatic statement of the AdS/CFT correspondence. }
   \label{fig:correspondence}
\end{figure}

As we learned from the previous subsections, the massless sector of the low energy dynamics of $N$ coincident D3--branes allows two possible descriptions. Conjecturing that these descriptions are equivalent is the core of Maldacena's AdS/CFT correspondence \cite{Maldacena:1997re}. Notice that in both descriptions one part of the decoupled sectors is a type IIB supergravity in flat space. Thus, we are naturally led to the conclusion that 
{\it type IIB superstring theory on the background of AdS$_5\times S^5$ background is dual to  ${\cal N}=4$ $SU(N)$ supersymmetric Yang--Mills theory in $1+3$ dimensions.}
We have presented this statement in a diagrammatic way in Figure~\ref{fig:correspondence}.

A further hint supporting the AdS/CFT correspondence is that the global symmetries of the proposed dual theories match. Indeed, an AdS$_5$ space-time of radius $R$ can be embedded in a $\IR^{2,6}$ flat space. It can then be naturally described as a hyperboloid of radius $R$ and thus has a group of isometry $SO(2,6)$ which is also the group of rotations in $2+4$ dimensions. On the other side, the $S^5$ part of the geometry has a group of isometry $SO(6)$ (the group of rotations in $6$ dimensions). This leads us to the conclusion that the total global symmetry of string theory on AdS$_5\times S^5$ gravitational background is $SO(2,4)\times SO(6)$. It is satisfying that the corresponding gauge theory has  the same global symmetry. Indeed, it is a well--known fact that the ${\cal N}=4$ supersymmetric Yang--Mills theory in $1+3$ dimensions is a conformal field theory. As such it should has the global symmetry of the conformal group in $1+3$ dimensions, but this is precisely $SO(2,4)$. Actually, since the theory is supersymmetric,  the full global symmetry group is the superconformal group which includes an $SU(4)$ global R--symmetry. In particular this group rotates the gauginos of the super Yang--Mills theory. However, it is well known that $SU(4)\cong SO(6)$, and therefore the global symmetry of the gauge theory is indeed $SO(2,4)\times SO(6)$!

An important aspect of the correspondence is the regime of the validity of the dual description. Depending on the precise way in which we are taking the $\alpha'\to 0$ limit, there are two basic forms of the AdS/CFT correspondence. The strongest form is that the string/gauge correspondence holds for any $N$. Unfortunately, this strong form of the conjecture cannot be tested directly since it is not known how to quantize superstring theory on a curved background in the presence of Ramond-Ramond fluxes \cite{Erdmenger:2007cm}. The second form of the conjecture holds in the t'Hooft limit when $N\to\infty$ and the t'Hooft coupling $\lambda\propto g_sN$ is kept fixed. In this way on gauge side of the correspondence only planar diagrams contribute to the partition function while on string side the required $g_s\to 0$ limit suggests semiclassical limit of the superstring theory on AdS$_5\times S^5$.  An important observation is that large $\lambda \gg1 $ suggests large AdS radius $R\propto \lambda^{1/4}$ and hence small curvature of the AdS background. This implies that the supergravity description is perturbative and thus provides an analytic tool for perturbative studies of non--perturbative field theory phenomena. On the other side, if we are at weak t'Hooft coupling ($\lambda \ll 1$) we can perform perturbative studies on gauge side of the correspondence and transfer the result to the perturbatively inaccessible regime of the supergravity description. Therefore, we conclude that the AdS/CFT correspondence is a strong/weak duality. In this work we will concentrate solely on the study of strongly coupled $\lambda \gg 1$ Yang--Mills theories. Hence, we will perform the analysis on the supergravity side of the AdS/CFT correspondence.

\subsection*{1.1.4 \hspace{2pt} The AdS/CFT dictionary}
\addcontentsline{toc}{subsection}{1.1.4 \hspace{0.15cm} The AdS/CFT dictionary}
  Let us focus on the precise way that the AdS/CFT correspondence is implemented. After a closer look at the geometry of the AdS$_5\times S^5$ background, we conclude that it has five non--compact directions. Four of them are parallel to the D3--branes world volume and correspond to the $1+3$ directions of the dual gauge theory. The fifth non--compact direction is the radial direction $u$ (radial in the transverse, to the D3--branes, $\IR^6$ space) and its interpretation in the dual gauge theory is not obvious. To shed more light on it, let us consider the action of a free massless scalar field in $1+3$ dimensions~\cite{Erdmenger:2007cm}:
\begin{equation} 
S=\int d^4x(\partial\phi)^2\ .
\label{free}
\end{equation}
The corresponding field theory is conformal and thus has a global symmetry $SO(2,4)$ which is the conformal group in $1+3$ dimensions. Therefore, we can consider the transformation properties of the scalar field $\phi$ under the action of the dilatation operator. One can verify that the transformation:
\begin{equation} 
x\to e^{\alpha}x;~~~\phi\to e^{-\alpha}\phi;
\end{equation}
leaves the action (\ref{free}) invariant. Furthermore, we learn that the scalar field $\phi$ has a scaling dimension one. On the other side, the $SO(2,4)$ group is the group of isometry of AdS$_5$ and one can verify from equation (\ref{AdS-intro}) that the transformation $x\to e^{\alpha}x$, suggests:
\begin{equation} 
u\to e^{-\alpha}u\ .
\end{equation}
Therefore, we learn that the coordinate $u$ scales as energy under dilations and thus has a natural interpretation as an energy scale of the dual gauge theory. 

The further development of the AdS/CFT correspondence resulted in a map between gauge invariant operators in ${\cal N}=4$ super Yang -Mills in a particular irreducible representation of the $SU(4)$ R-symmetry group and supergravity fields in the isomorphic representation of the $SO(6)$ global symmetry. These representations are obtained after Kaluza-Klein reduction of the supergravity fields on the internal $S^5$ sphere. Let us consider for simplicity the case of a scalar field of mass $m$, propagating on the AdS$_{d+1}$ background.
The relevant action is \cite{Erdmenger:2007cm}:
\begin{equation}
S=\int d^dxdu\sqrt{-g}(g^{ab}\partial_a\phi\partial_b\phi-m^2\phi^2)\ .
\end{equation}
The solution of the corresponding equation of motion have the following asymptotic behavior at large $u$:
\begin{equation}
\phi(u,x)=\left(\frac{1}{u}\right)^{4-\Delta}\phi_0(x)+\left(\frac{1}{u}\right)^{\Delta}\langle{\cal O}(x)\rangle\ ,
\label{asdict}
\end{equation}
where
\begin{equation}
\Delta=\frac{d}{2}+\sqrt{\frac{d^2}{4}+R^2m^2}\ .
\end{equation}
Note that the supergravity field $\phi(u)$ is a scalar field and is thus invariant under the action of the dilatation operator because the latter is one of the generators of the global symmetry $SO(2,4)$. Therefore, we conclude that $\phi_0$ and $\langle{\cal O}(x)\rangle\ $ carry scaling dimensions $4-\Delta$ and $\Delta$, respectively. In \cite{Gubser:1998bc} it was suggested that $\phi_0$ and $\langle{\cal O}(x)\rangle$ correspond to the source and the vacuum expectation value of the gauge invariant operator ${\cal O}(x)$. It is also worth noting that the expression:
\begin{equation}
\int d^dx\phi_0(x)\langle{\cal O}(x)\rangle=\mathrm{inv.}
\end{equation}
is invariant under the $SO(2,4)$ global symmetry. It was suggested that the exact form of the map is given by the relation \cite{Gubser:1998bc,Witten:1998qj}:
\begin{equation}
\langle e^{\int d^dx\phi_0(x)\langle{\cal O}(x)\rangle}\rangle_{\mathrm{CFT}}={\cal Z}_{\mathrm{Sugra}}[\phi_0(x)]\ ,
\label{dictwit}
\end{equation}
where
\begin{equation}
{\cal Z}_{\mathrm{Sugra}}[\phi_0(x)]=\lim_{\epsilon\to 0}{\cal Z}_{\mathrm{Sugra}}[{\phi_0(1/\epsilon,x)=\epsilon^{d-\Delta}\phi_0(x)}]\ .\label{regeps}
\end{equation}
{\it i.e.} the generating functional of the conformal field theory coincide with the generating functional for tree level diagrams in supergravity. We refer the reader to the extensive review ref.~\cite{Aharony:1999ti} for more subtleties on the precise way of taking the $\epsilon\to 0$  limit in equation (\ref{regeps}). 

Formula (\ref{dictwit}) has been tested by comparing correlation functions of the ${\cal N}=4$ quantum field theory with classical correlation functions in AdS$_d$. Note that the tree level approximation on supergravity side is valid only at strong t'Hooft coupling and therefore the corresponding conformal field theory is strongly coupled. This is why the correspondence was tested in this way only for correlation functions which satisfy non--renormalization theorems and hence are independent on the coupling \cite{Erdmenger:2007cm}. In particular it applies for the two- and three- point functions of $1/2$ BPS operators \cite{Freedman:1998tz, Lee:1998bxa}.

Further checks of the correspondence beyond the $1/2$ BPS sector was started with the so--called  plane--wave string/gauge theory duality, where one takes appropriate plane--wave limit of the AdS$_5\times S^5$ background \cite{Berenstein:2002jq}. Key point of this limit is that superstring theory on this background can be exactly quantized. Recently a significant progress towards quantizing superstring theory on AdS$_5\times S^5$ has been achieved using integrable spin chain models. We refer the reader to refs.~\cite{{Beisert:2004ry},{Plefka:2003nb}} for more details on these vast subjects.

\section*{1.2 \hspace{2pt} Introducing flavor to the correspondence}

\addcontentsline{toc}{section}{1.2 \hspace{0.15cm} Introducing flavor to the correspondence}

Direct consequence of the confining property of QCD is the fact that the low energy dymanics of the theory is governed by color singlets, such as mesons, baryons and glueballs. Mesons and baryons are bound states of quarks, the latter transform in the fundamental representation of $SU(3)$. The fact that at low energy QCD is strongly coupled suggests that it is not accessible for perturbative studies. This is why it is important to come up with an alternative non--perturbative techniques describing the strongly coupled regime of Yang--Mills theories and in particular Yang--Mills theories containing matter in the fundamental representation of the gauge group, such as QCD.
 
Further need of alternative non--perturbative techniques applicable to the properties of the fundamental fields in the strongly coupled regime of non--abelian gauge theories is required by the very recent discoveries of the properties of matter obtained in heavy ion collision experiments. More precisely the fact that the quark--gluon plasma which is the phase of matter of the fireballs obtained in such experiments, is not the expected weakly coupled quark--gluon plasma predicted by the standard perturbative QCD but is classified as a strongly coupled quark-gluon plasma. A novel phase of matter that provides challenge for the society of theoretical physicists.

One of the purpose of the study of the AdS/CFT correspondence is to develop the above mentioned analytic tools for the study of strongly coupled Yang--Mills theories. The original form of the conjecture that we described in the previous section, focuses on a gauge theory with a huge amount of symmetry, namely the ${\cal N}=4$ super Yang--Mills theory. On way to make the correspondence more applicable to realistic gauge theories, such as QCD, is to reduce the amount of the supersymmetry of the theory by introducing additional gauge invariant operators. This approach though fruitful still has the weakness that the matter content of the dual gauge theory, more precisely the fermionic degrees of freedom, transform in the adjoint representation of the gauge group. In other words there are no fundamental fields in the theory. The reason is that both ends of the strings, producing the field content of the gauge theory, are attached to the same stack of D3--branes and the corresponding states transform in the adjoined representation of the gauge group.  In order to introduce fundamental matter, one needs to consider separate stack of D--branes. 

The easiest way to introduce fundamental fields in the context of the AdS/CFT correspondence is to consider an additional stack of $N_f$ D7--branes \cite{Karch:2002sh}. (From now on we will use $N_c$ as a notation for the number of the D3--branes sourcing the gravitational background.) Since the D7--branes' world volume is higher dimensional and non--compact in the transverse to the D3--branes dimensions, the D7--branes have infinite ``internal" volume and thus their gauge coupling vanish making their gauge symmetry group a global symmetry. In this way we introduce family of fundamental matter with global flavor symmetry $SU(N_f)$. To be more precise let us consider two stacks of parallel $N_{c}$ D3--branes and $N_f$ D7--branes embedded in the following way: 
      \begin{table}[h]
\begin{center}
\begin{tabular}{|c|c|c|c|c|c|c|c|c|c|c|}
\hline
 &0&1&2&3&4&5&6&7&8&9\\\hline
 D3&-&-&-&-&$\cdot$&$\cdot$&$\cdot$&$\cdot$&$\cdot$&$\cdot$\\\hline
 D7&-&-&-&-&-&-&-&-&$\cdot$&$\cdot$\\\hline
 
\end{tabular}
\end{center}
\caption{ Embedding of the flavor D7--branes. }
\label{default}
\end{table}%
 
 The low energy spectrum of the strings stretched between the D3-- and D7--branes directions give rise to the ${\cal N}=2$ hypermultiplet containing two Weyl fermions of opposite chirality coming from the light-cone modes of strings stretched along the NN and DD directions (2,3,8,9) and two complex scalars coming form strings stretched along the ND directions, namely 4,5,6,7. Now if we consider $N_f \ll N_c$ and take the large $N_c$ limit. We can substitute the stack of D3--branes with a $AdS_5\times S^5$ space and study the $N_f$ D--branes in the probe limit using their Dirac--Born--Infeld action. On gauge side this corresponds to working in the quenched approximation ($N_f \ll N_c$) and taking the large $N_c$ t'Hooft limit.
 If the D3-- and D7--branes are separated in their transverse (8,9)-plane, then the strings stretched between them has a final length and hence final energy. It can be shown that \cite{Johnson:2003gi} the mass of the hypermultiplet is given by the energy of the string or the distance of separation $L$ multiplied by the string tension $m_q=L/{(2\pi\alpha')}$.
 
 Let us study closer the symmetry of the set up. If the D3-- and the D7--branes overlap the hypermultiplet is massless ($m_q=0$). In this case the $SO(6)$ rotational symmetry of the transverse $\IR^6$ space is broken to the product $SO(4)\times SO(2)$, corresponding to rotations along the ND directions (4,5,6,7) and the DD directions (8,9), respectively. This is equivalent to a $SU(2)_L\times SU(2)_R\times U(1)_R$ global symmetry, and suggests that the gauge theory has a R--symmetry group $SO(2)_R\times U(1)_R$ \cite{Kruczenski:2003be}, which is indeed the case, when the hypermultiplet is massless. If the D3-- and D7--branes are separated it is known that the R--symmetry is just $SU(2)_R$, which again fits that fact that the $SO(2)$ rotational symmetry in the (8,9)-plane is broken.
 
\subsection*{1.2.1 \hspace{2pt} The dictionary of the probe brane}
\addcontentsline{toc}{subsection}{1.2.1 \hspace{0.15cm} The dictionary of the probe brane}
 Let us now focus on the precise way that the AdS/CFT dictionary is implemented. The dynamics of the D7--brane probe is described by the Dirac--Born--Infeld action including the Chern-Simons term~\cite{Johnson:2003gi}:
\begin{equation}
\frac{S}{N_f}=-\mu_7\int\limits_{{\cal M}_8}e^{-\Phi}d^8\xi\sqrt{-det(P[G_{ab}]+(2\pi\alpha')^2{\cal F}_{ab})}+\frac{(2\pi\alpha')^2}{2}\mu_7\int\limits_{{\cal M}_8}P[C_{(4)}]\wedge{\cal F}\wedge{\cal F}\ ,
\end{equation}
where $(2\pi\alpha'){\cal F}_{ab}=P[B_{ab}]+(2\pi\alpha')F_{ab}$ and $\mu_7=[(2\pi)^7\alpha'^4]^{-1}$.
It is convenient to introduce the following coordinates:
\begin{equation}
\rho=u\cos\theta;~~~L=u\sin\theta;
\end{equation}
and consider the ansatz: 
\begin{equation}
L=L(\rho);~~~\phi=const \ .
\end{equation}
Then the lagrangian describing the D7--brane embedding is:
\begin{equation}
{\cal L}\propto \rho^3\sqrt{1+L'(\rho)^2}
\end{equation}
leading to the equation of motion:
\begin{equation}
L'(\rho)=-\frac{2c}{\sqrt{\rho^6-4c^2}}\ .
\label{adssolint}
\end{equation}
At large $\rho$ the solution has the behavior:
\begin{equation}
L(\rho)=m+\frac{c}{\rho^2}+\dots\ .
\end{equation}
Now if we introduce the field: 
\begin{equation}
\chi(u)=\frac{L(\rho)}{\rho^2+L(\rho)^2}=\frac{1}{u}m+\frac{1}{u^3}c+\dots;~~~u^2=\rho^2+L(\rho)^2 \ ,
\end{equation}
we can see that $\chi(u)$ has the same behavior as the field $\phi(u,x)$ from equation (\ref{asdict}). This is quite suggesting. The asymptotic value of $L(\infty)=m$ is exactly the separation of the D3-- and D7--branes and is thus related to the mass of the hypermultiplet via $m_q=m/(2\pi\alpha')$. Since the hypermultiplet chiral fields are our quarks we will call $m_q$ the bare quark mass. Therefore, the coefficient $c$ should be proportional to the vev of the operator that couples to the bare quark mass but this is the quark condensate! This is an example of the how the generalized AdS/CFT dictionary works at the level of a D7--brane probing. Let us provide the exact relation between the quark condensate $\langle{\cal O}_q\rangle$ and the coefficient $c$:
\begin{equation}
\langle{\cal O}_q\rangle=-\frac{N_f}{(2\pi\alpha')^3g_{YM}^2}c\ .
\label{cond-intro}
\end{equation}
We refer the reader to the appendix of Chapter~2 for more details on the last calculation and to ref.~\cite{Karch:2005ms} for an elegant presentation of the holographic renormalization of probe D--branes in AdS/CFT.

Now let us go back to equation (\ref{adssolint}) and note that in order for the D7--brane to close smoothly in the bulk of the geometry, we need to impose $L'(0)=0$. This is possible only for $c=0$ and thus we conclude that the condensate of the theory vanish. But the dual gauge theory is supersymmetric this is why it is not surprising that the quark condensate is zero. Furthermore, since there is no potential between the D3-- and D7--branes (because of the unbroken supersymmetry), the D7--brane should not bend at infinity and this is why the solution for the D7--brane embedding should be simply $L\equiv m$, as it is.

Note that the analysis that led to equation (\ref{cond-intro}) requires that the gravitational background be only asymptotically AdS$_5\times S^5$. In fact, in all cases that we are going to consider in this work, there will be some sort of the deformation of the bulk physics, coming either from the gravitational background or from the introduction of external fields. This will break the supersymmetry and will capacitate the dual gauge theory to develop a quark condensate. In Chapter~2, we will use this approach to provide a holographic description of magnetic catalysis of chiral symmetry breaking. Through the rest of the thesis, the study of the quark condensate as a function of the bare quark mass will enable us to explore the phase structure of the dual gauge theory and uncover a first order phase transition associated to the melting of the light mesons of the theory. Let us conclude by redirecting the reader to the very recent extensive review on the subject ref.~\cite{Erdmenger:2007cm}.

\section*{1.3 \hspace{2pt} Outline}

\addcontentsline{toc}{section}{1.3 \hspace{0.15cm} Outline}

 This work is dedicated to the holographic study of fundamental flavor. We focus on the D7--brane probing of asymptotically AdS$_5\times S^5$ gravitational backgrounds. In particular we concentrate on meson spectroscopy, description of non--perturbative phenomena such as chiral symmetry breaking, and study of the thermal properties of the theory.

In Chapter~2, we consider a D7--brane probe of AdS$_{5}\times S^5$ in the presence of a pure gauge $B$--field. In the dual gauge theory, the $B$--field couples to the fundamental matter introduced by the D7--brane and acts as an external magnetic field. The $B$--field supports a 6-form Ramond-Ramond potential on the D7--branes world volume that breaks the supersymmetry and enables the dual gauge theory to develop a non--zero quark condensate. We explore the dependence of the quark condensate on the bare quark mass $m_{q}$ and show that at zero bare quark mass a chiral symmetry is spontaneously broken. We also study the discrete self-similar behavior of the theory near the origin of the parameter space given by the bare quark mass and the condensate of the theory. We calculate the critical exponents of the bare quark mass and the quark condensate. A study of the meson spectrum supports the expectation based on thermodynamic considerations that at zero bare quark mass the stable phase of the theory is a chiral symmetry breaking one. Our study reveals a self-similar structure of the spectrum near the critical phase of the theory, characterized by zero quark condensate. We calculate the corresponding critical exponent of the meson spectrum. A further study of the meson spectrum reveals a coupling between the vector and scalar modes, and in the limit of weak magnetic field we observe a Zeeman splitting of the energy levels. We also observe the characteristic $\sqrt{m_{q}}$ dependence of the ground state corresponding to the Goldstone boson of spontaneously broken chiral symmetry.

Chapter~3 is studying the finite temperature dynamics of flavored large $N_c$, $SU(N_c)$ gauge theory with fundamental quark flavors in the quenched approximation. A quark condensate forms at finite quark mass, and the value of the condensate varies smoothly with the quark mass for generic regions in parameter space. At a particular value of the quark mass, there is a finite discontinuity in the vacuum expectation value of the condensate, corresponding to a first order phase transition. We study the gauge theory via its string dual formulation using the AdS/CFT conjecture, the string dual being the near-horizon geometry of $N_c$ D3--branes at finite temperature, AdS$_5$--Schwarzschild $\times S^5$, probed by a D7--brane. The D7--brane has topology $\IR^4 \times S^3 \times S^1$ and allowed solutions correspond to either the $S^3$ or the $S^1$ shrinking away in the interior of the geometry. The phase transition represents a jump between branches of solutions having these two distinct D--brane topologies. 
 
In Chapter~4, using a ten dimensional dual string background, we study aspects of the physics of finite temperature large $N_c$ four-dimensional $SU(N_c)$ gauge theory. We focus on the dynamics of fundamental quarks in the presence of a background magnetic field. At vanishing temperature and magnetic field, the theory has ${\cal N}=2$ supersymmetry and the quarks are in hypermultiplet representations. In Chapter~2, similar techniques were used to show that the quark dynamics exhibit spontaneous chiral symmetry breaking. In this chapter we begin by establishing the non--trivial phase structure that results from finite temperature. We observe, for example, that above the critical value of the field that generates a chiral condensate spontaneously, the meson melting transition disappears, leaving only a discrete spectrum of mesons at any temperature. We also compute several thermodynamic properties of the plasma such as the free energy, entropy and magnetization. The study of the meson spectrum allows us to examine the stability of the theory. Our study shows that for sufficiently strong magnetic field, bellow the critical value, there is a metastable chiral symmetry--breaking phase that eventually becomes the true stable phase of the theory. We also obtain the meson spectrum of the coupled vector and scalar modes at strong magnetic field and verify the Zeeman splitting of the spectrum.

In Chapter~5 we perform a study analogous to the one in Chapter~4 but for the case of external electric field.  At zero temperature, we observe that the electric field induces a phase transition associated with the dissociation of the mesons into their constituent quarks. This is an analogue of an insulator-metal transition, since the system goes from being an insulator with zero current (in the applied field) to a conductor with free charge carriers (the quarks). At finite temperature this phenomenon persists, with the dissociation transition become subsumed into the more familiar meson melting transition. Here, the dissociation phenomenon reduces the critical melting temperature. We also focus on the geometric aspect of the transition by performing an analogous T--dual study showing that the nature of the instability leading to the observed insulator/conductor phase transition is related to the over spinning of the T--dual D6--branes. We conclude the chapter with the identification of a peculiar class of embeddings that have a conical singularity above the horizon. We propose that those are fixed by stringy corrections. 

Chapter~6 is dedicated to the effect of an R--charge chemical potential on the meson melting phase transition. We begin by introducing the relevant gravitational background which is that of spinning D3--branes. We comment on the deformation of the internal symmetries due to the three angular momentums of the geometry and identify ansatz relevant for the introduction of a probe D7--brane. We also dedicate a special section to the detailed calculation of the temperature and elaborate on the existence of extremal horizons in the case of multiple R--charges that would enable us to study the theory at zero temperature and non--zero R--charge chemical potential. In the second part of the chapter we study the properties of the D7--brane embeddings for two different cases. First we consider the case of only one non--zero R--charge. We calculate the equation of state in the quark condensate versus bare quark mass plane. We use this to study the dependence of the critical mass on the R--charge and develop analytic expression for the bare quark mass at large R--charge. Finally we obtain the phase diagram of the theory using  appropriate dimensionless parameters which at fixed bare quark mass and t'Hooft coupling are proportional to the temperature and the R--charge chemical potential of the theory. After that we consider the more interesting case of three equal charges. Here we perform analysis analogous to the one for the one charge case and again obtain the phase diagram of the theory in coordinates corresponding to the temperature and the chemical potential. Our study shows that the phase diagram is similar to the phase diagram for the Hawking-Page transition of the adjoint fields. In the last section of the chapter we discuss the physical meaning of the constant of integration obtained while regularizing the behavior of the D7--brane probes at the ergosphere of the background. We conclude that this constant is related to the phase difference between the bare quark mass and the quark condensate.

In Chapter~7 we show how two important types of phase transition in large $N_c$ gauge theory with fundamental flavors can be cast into the same classifying framework as the meson-melting phase transition. These are quantum fluctuation induced transitions in the presence of an external electric field, or a chemical potential for R-charge. The classifying framework involves the study of the local geometry of a special D--brane embedding which seeds a self-similar spiral structure in the space of embeddings. The properties of this spiral, characterized by a pair of numbers, capture some key universal features of the transition. Computing these numbers for these non--thermal cases, we find that these transitions are in the same universality class as each other, but have different universal features from the thermal case. The phase transitions that we consider are the thermal studied in Chapter~3, electrically driven studied in Chapter~5 and the R--charge driven phase transition considered in Chapter~6. We present a natural generalization that yields new universality classes that may pertain to other types of transition. 

Chapter~8 is the concluding chapter. It summarizes our results and outlines possible directions for future studies.

\chapter*{Chapter 2: \hspace{1pt} Flavored large $N_c$ gauge theory in an external magnetic field at zero temperature}

\addcontentsline{toc}{chapter}{Chapter 2:\hspace{0.15cm}
Flavored large $N_c$ gauge theory in an external magnetic field at zero temperature}

\section*{2.1 \hspace{2pt} Introductory remarks}
\addcontentsline{toc}{section}{2.1 \hspace{0.15cm} Introductory remarks}

  In  recent years progress has been made towards the study of  matter in fundamental representation in the context of AdS/CFT correspondence. As we discussed in Chapter~1, one way
    to achieve this is by introducing space filling flavor D7--branes in the probe limit \cite{Karch:2002sh} and in order to keep the probe
     limit valid the condition $N_{f} \ll N_{c}$ is imposed. The fundamental strings stretched between the stack of $N_{c}$ D3--branes and
     the flavor $N_{f}$ D7--branes give rise to an $\cal N$=2 hypermultiplet, the separation of the D3- and D7- branes in the transverse
     directions corresponds to the mass of the hypermultiplet, the classical shape of the D7--brane encodes the value of the fermionic
     condensate, and its quantum fluctuations describe the light meson spectrum of the theory \cite{Kruczenski:2003be}. This technique for
      introducing fundamental matter has been widely employed in different backgrounds. Of particular interest was the study of non--supersymmetric backgrounds and phenomena such as spontaneous chiral symmetry breaking. These phenomena were first studied in this context in ref.~\cite{Babington:2003vm}, where the authors developed an appropriate numerical technique. In  recent years 
this approach received further development, and has proven itself as powerful tool for the exploration of confining gauge theories. In 
\vspace{1cm}
particular, for the description of their thermodynamic properties and for the building of phenomenological models relevant to QCD.
The chapter is organized as follows:

In the second section we describe the method of introducing magnetic field to the theory, employed in ref.~\cite{Filev:2007gb}. We describe the basic properties of the D7--brane embedding and the thermodynamic properties of the dual gauge theory, in particular the dependence of the quark condensate on the bare quark mass. We describe the spontaneous chiral symmetry breaking caused  by the external magnetic field and comment on the spiral structure in the condensate versus bare quark mass diagram. We perform analysis similar to the one considered in ref.~\cite{Frolov:2006tc} for the study of merger transitions and calculate the scaling exponents of the bare quark mass and the quark condensate \cite{Filev:2007qu}. We also describe the discrete self-similarity of the spiral and calculate the scaling factor characterizing it.

In the third section we study the light meson spectrum of the dual gauge theory. First we derive the relevant equations of motion for the scalar and vector meson spectrum. The study of the fluctuations along the axial scalar reveals a Zeeman slitting of the energy levels at weak magnetic field and a characteristic Gell-Mann-Oakes-Renner relation \cite{GellMann:1968rz} for the pion of the softly broken chiral symmetry.

Next we consider the meson spectrum of the states corresponding to the spiral. We study the critical embedding corresponding to the center of the spiral and reveal an infinite tower of tachyonic states organized in a decreasing geometrical series. After that we consider the dependence of the meson spectrum on the bare quark mass and confirm the expectations based on thermodynamic considerations that only the lowest branch of the spiral is stable. We observe that at each turn of the spiral there is one new tachyonic state. We comment on the self-similar structure of the spectrum and calculate the scaling exponent of the meson mass. We also consider the spectrum corresponding to the lowest branch of the spiral and for a large bare quark mass reproduce the result for pure ${\cal N}=2$ Supersymmetric Yang Mills Theory obtained in ref.~\cite{Kruczenski:2003be}.

We end with a short discussion of our results and the possible directions of future study.

\section*{2.2 \hspace{2pt} Fundamental matter in an external magnetic field}

\addcontentsline{toc}{section}{2.2 \hspace{0.15cm} Fundamental matter in an external magnetic field}

\subsection*{2.2.1 \hspace{2pt} Basic configuration}
\addcontentsline{toc}{subsection}{2.2.1 \hspace{0.15cm} Basic configuration}

Let us consider the AdS$_{5} \times S^5$ geometry describing the near-horizon physics of a collection of $N_{c}$ extremal D3--branes.
\begin{eqnarray}
ds^2&=&\frac{u^2}{R^2}(-dx_{0}^2+dx_1^2+dx_2^2+dx_3^2)+R^2\frac{du^2}{u^2}+R^2d\Omega_{5}^2\ ,\label{AdS}\\
g_{s}C_{(4)}&=&\frac{u^4}{R^4}dx^0\wedge dx^1\wedge dx^2 \wedge dx^3\ ,\nonumber\\\
e^\Phi&=&g_s\ ,\nonumber\\
R^4&=&4\pi g_{s}N_{c}\alpha'^2\ ,\nonumber
\end{eqnarray}
Where $d\Omega_5^2$ is the unit metric on a round $S^5$. In order to introduce fundamental matter we first rewrite the metric in the following form, with $d\Omega_3^2$ the metric on a unit $S^3$:
\begin{eqnarray}
ds^2&=&\frac{\rho^2+L^2}{R^2}[ - dx_0^2 + dx_1^2 +dx_2^2 + dx_3^2 ]+\frac{R^2}{\rho^2+L^2}[d\rho^2+\rho^2d\Omega_{3}^2+dL^2+L^2d\phi^2],\nonumber\\
d\Omega_{3}^2&=&d\psi^2+\cos^2\psi d\beta^2+\sin^2\psi d\gamma^2, \label{geometry1}
\end{eqnarray}
where $\rho, \psi, \beta,\gamma$ and $L,\phi$ are polar coordinates in the transverse $\mathbb{R}^4$ and $\mathbb{R}^2$ respectively. Note that: $u^2=\rho^2+L^2$.
We use $x_{0},x_{1},x_{2},x_{3},\rho,\psi,\beta,\gamma$ to parameterize the world volume of the D7--brane and consider the following ansatz \cite{Kruczenski:2003be} for its embedding:
\begin{eqnarray}
\phi\equiv {\rm const},\quad L\equiv L(\rho)\nonumber \label{ansatzEmb},
\end{eqnarray}
leading to the following form of the induced metric on its world--volume:
\begin{equation}
d\tilde s=\frac{\rho^2+L(\rho)^2}{R^2}[ - dx_0^2 + dx_1^2 +dx_2^2
+dx_3^2]+\frac{R^2}{\rho^2+L(\rho)^2}[(1+L'(\rho)^2)d\rho^2+\rho^2d\Omega_{3}^2] \  .
\label{inducedMetric}
\end{equation}
Now let us consider the general DBI action (to simplify notations we temporary set the number of flavor branes to one $N_f=1$):
\begin{eqnarray}
S_{DBI}=-\mu_{7}\int\limits_{{\cal M}_{8}}d^{8}\xi e^{-\Phi}[-{\rm det}(G_{ab}+B_{ab}+2\pi\alpha' F_{ab})]^{1/2}\  . \label{DBI}
\end{eqnarray}

Here $\mu_{7}=[(2\pi)^7\alpha'^4]^{-1}$ is the D7--brane tension, $G_{ab}$ and $B_{ab}$ are the induced metric and $B$--field on the D7--brane's world volume, while $F_{ab}$ is its world--volume gauge field. A simple way to
introduce  magnetic field would be to consider a pure gauge $B$--field along parts of the D3--branes' world volume, {\it e.g.}:
\begin{equation}
B^{(2)}= Hdx_{2}\wedge dx_{3} \label{ansatz}\ .
\end{equation}
Since $B_{ab}$ can be mixed with the gauge field strength $F_{ab}$, this is equivalent to a magnetic field on the world--volume~\cite{Filev:2007gb}.
Recently a similar approach was used to study drag force in SYM plasma \cite{Matsuo:2006ws}. Note that since the $B$--field is pure gauge, $dB=0$, the corresponding background is still a solution to the supergravity equations of motion.
On the other hand, the gauge field $F_{ab}$ comes at next order in the $\alpha'$ expansion compared to the metric and the $B$--field components. Therefore
to study the classical embedding of the D--brane one can study only the $(G_{ab}+B_{ab})$ part of the DBI--action. However, because of the
presence of  the $B$--field, there will be terms of first order in $\alpha'$ in the full action linear in the gauge field $F_{ab}$. Hence  integrating out $F_{ab}$ will result in a constraint for the classical embedding of the D7--brane.

Since for our configuration, we have  that:
\begin{eqnarray*}
B^{(2)}\wedge B^{(2)}=0\ ,\quad B^{(2)}\wedge C_{(4)}=0\ ,
\end{eqnarray*}
and at first order in $\alpha'$
the only contribution to the Wess-Zummino is
\begin{eqnarray}
2\pi\alpha'\mu_{7}\int F_{(2)}\wedge C_{(6)}\ .
\label{potentials}
\end{eqnarray}

By using the following  expansion in the DBI action:
\begin{equation}
[-{\rm det}(E_{ab}+2\pi\alpha'F_{ab})]^{1/2}=\sqrt{E}+\pi\alpha'\sqrt{E}E^{ba}F_{ab}+O(F^2);~~~E=-{\rm det}E_{ab}\ ,
\end{equation}
where we have introduced $E_{ab}=G_{ab}+B_{ab}$ as a notation for the generalized induced metric,  we obtain the following action to first order in $\alpha'$:
\begin{equation}
S_{F}=\pi\alpha'\frac{\mu_7}{g_s}\int\limits_{{\cal M}_{8}}d^8\xi\sqrt{E}E^{[ab]}F_{[ab]}+2\pi\alpha'\mu_7\int F_{(2)}\wedge C_{(6)}\ .
\end{equation}
The resulting equation of motion does not contain $A_a$ and sets the following constraint for the $C_{(6)}$ potential induced by the gauge $B$--field.
\begin{equation}
\frac{g_s}{6!}\epsilon^{ab\tilde\mu_{1}\dots\tilde\mu_{6}}\partial_{a}C_{\tilde\mu_{1}\dots\tilde\mu_{6}}=-\partial_{a}(\sqrt{E}E^{[ba]});~~~a,b,\tilde\mu_{1},\dots\tilde\mu_{6}\in{\cal M}_{8}\ .
\label{cosntr}
\end{equation}
Note that  $C_{(6)}$ has a dynamical term proportional to $1/\kappa_{0}^2$  in the supergravity action, and that the D7--brane action is proportional to $\mu_{7}=2\pi/\kappa_{0}^2$. Therefore they are
at the same order in~$\alpha'$ \cite{Johnson:2003gi}. We must solve for $C_{(6)}$ using the action:
\begin{equation}
S_{C_{(6)}}=\mu_{7}\int B_{(2)}\wedge C_{(6)}-\frac{1}{4k_{0}^2}\int d^{10}x\sqrt{-G}|dC_{(6)}|^2\label{dynamics}\ .
\end{equation}
The solution obtained from equation (\ref{dynamics}) has to satisfy the constraint given in equation~(\ref{cosntr}). Our next goal will be to find a consistent
ansatz for $C_{(6)}$. To do this let us consider the classical contribution to the DBI action:
\begin{eqnarray}
S_{NS}=-\frac{\mu_{7}}{g_s}\int d^8\xi\sqrt{E}\label{decopl}\ .
\end{eqnarray}
From equation (\ref{decopl}) one can solve for the classical embedding of the D7--brane, which amounts to second order differential
equation for $L(\rho)$ with some appropriate solution $L_0(\rho)$. After substituting $L_0(\rho)$ in (\ref{decopl}) we can extract the form of the $C_{(6)}$ potential induced by the $B$--field. However, one still has to satisfy the constraint (\ref{cosntr}).
It can be verified that with the choice (\ref{ansatz}) for the $B$--field and the ansatz of equation  (\ref{inducedMetric}) for the induced metric, the right-hand side of equation (\ref{cosntr}) is zero. Then equation (\ref{cosntr}) and the effective action (\ref{dynamics}) boil down to finding a consistent ansatz for $C_{(6)}$ satisfying:
\begin{eqnarray}
&\partial_{\mu}(\sqrt{-G}dC_{6}^{\mu01\rho\psi\alpha\beta})&=-\frac{\mu_{7}\kappa_{0}^2}{\pi}H\delta(L-L_0 (\rho))\ ,\\
{\rm or}\qquad &\partial_{\mu}(\sqrt{-G}dC_{6}^{\mu01L\psi\alpha\beta})&=- L_0' (\rho)\frac{\mu_{7}\kappa_{0}^2}{\pi}H\delta(L-L_0 (\rho))\ ,\\
&\epsilon^{ab\tilde\mu_{1}\dots\tilde\mu_{6}}\partial_{a}C_{\tilde\mu_{1}\dots\tilde\mu_{6}}&=0;~~~a,b,\tilde\mu_{1},\dots\tilde\mu_{6}\in{\cal M}_{8}\ .
\end{eqnarray}
%
One can verify that the choice:
\begin{eqnarray}
{C_{(6)}}_{01\rho\psi\alpha\beta}=\frac{1}{7}f(\rho,L,\psi),\quad {dC_{(6)}}_{L01\rho\psi\alpha\beta}=\partial_{L}f\ ,
\end{eqnarray}
is a consistent ansatz and the solution for the $C_{(6)}$ field strength can be found to be:
\begin{equation}
{dC_{(6)}}_{L01\rho\psi\alpha\beta}=\frac{\mu_{7}\kappa_{0}^2}{\pi}H\frac{\rho^3R^4}{L(\rho^2+L^2)^2}\Theta(L-L_0(\rho))\sin\psi\cos\psi\ .
\label{closed form}
\end{equation}

It is this potential which breaks the supersymmetry. It is important to note that there is no contradiction between the fact that the $B$--field that we have chosen does not break the supersymmetry of the AdS$_5\times S^5$ supergravity background, on the one hand, and the fact that the physics of the D7--brane probing that background does have supersymmetry broken by the $B$--field, on the other. This is because the physics of the probe does not back--react on the geometry.

In what follows, we will study the physics  of the D7--branes and the resulting  dual gauge theory physics. Among the solutions for the D7--brane embedding, there will be a class with non--trivial profile having zero asymptotic separation between the D3- and D7--branes. This corresponds to a non--zero quark condensate at zero bare quark mass. Therefore the non--zero background magnetic field will spontaneously break the chiral symmetry.  Geometrically this corresponds to breaking of the $SO(2)$ rotational symmetry in the $(L, \phi)$-plane \cite{Kruczenski:2003be}.

\subsection*{2.2.2 \hspace{2pt} Properties of the solution}
\addcontentsline{toc}{subsection}{2.2.2 \hspace{0.15cm} Properties of the solution}

We now proceed with the exploration of the properties of the classical D7--brane embedding.
If we consider the action (\ref{decopl}) at leading order in $\alpha'$, we get the following effective lagrangian:
\begin{equation}
{\cal L}=-\frac{\mu_{7}}{g_s}\rho^3\sin\psi\cos\psi\sqrt{1+L'^2}\sqrt{1+\frac{R^4H^2}{(\rho^2+L^2)^2}}\ .
\label{lagrangian}
\end{equation}
The equation of motion for the profile $L_0(\rho)$ of the D7--brane is given by:
\begin{equation}
\partial_{\rho}\left(\rho^3\frac{L_0'}{\sqrt{1+L_0'^2}}\sqrt{1+\frac{R^4H^2}{(\rho^2+L_0^2)^2}}\right)+ \frac{\sqrt{1+L_0'^2}}{\sqrt{1+\frac{R^4h^2}{(\rho^2+L_0^2)^2}}}\frac{2\rho^3L_0R^4H^2}{(\rho^2+L_0^2)^3}=0\ .
\label{eqnMnL}
\end{equation}
As expected for large $(L_0^2+\rho^2) \to \infty$ or $H \to 0$, we get the equation for the pure AdS$_{5}\times S^5$ background \cite{Karch:2002sh}:
\begin{eqnarray*}
\partial_{\rho}\left(\rho^3\frac{L_0'}{\sqrt{1+L_0'^2}}\right)=0\ .
\end{eqnarray*}
Therefore the solutions to equation (\ref{eqnMnL}) have the following behavior at infinity:
\begin{equation}
L_0(\rho)=m+\frac{c}{\rho^2}+\dots.
\end{equation}
where the parameters $m$ (the asymptotic separation of the D7- and D3- branes) and $c$ (the degree of bending of the D7--brane) are related to the bare quark mass $m_{q}=m/2\pi\alpha'$ and the quark condensate $\langle\bar\psi\psi\rangle\propto -c$ respectively \cite{Kruczenski:2003uq}. As we shall see below, the presence of the external magnetic field and its effect on the dual SYM  provide a non vanishing value for the quark condensate. Furthermore, the theory exhibits chiral symmetry breaking.

   Now notice that $H$ enters in (\ref{lagrangian}) only through the combination $H^2R^4$. The other natural scale is the asymptotic separation $m$. It turns out that different physical configurations can be studied in terms of  the ratio $\tilde m^2={m^2}/{(H R^2)}$: Once the $\tilde m$ dependence of our solutions  are known, the $m$ and $H$ dependence follows. Indeed, let us introduce dimensionless variables {\it via}:
  \begin{eqnarray}
  \rho=R\sqrt{H}\tilde\rho\ , \quad
  L_0=R\sqrt{H}\tilde L\ , \quad
  L_0'(\rho)=\tilde L'(\tilde\rho)\ .\label{cordchange}
  \end{eqnarray}
The equation of motion (\ref{eqnMnL})  then takes the form:
\begin{equation}
\partial_{\tilde\rho}\left(\tilde\rho^3\frac{\tilde L'}{\sqrt{1+{\tilde L}'^2}}\sqrt{1+\frac{1}{(\tilde\rho^2+\tilde L^2)^2}}\right)+ \frac{\sqrt{1+\tilde L'^2}}{\sqrt{1+\frac{1}{(\tilde\rho^2+\tilde L^2)^2}}}\frac{2\tilde\rho^3\tilde L}{(\tilde\rho^2+\tilde L^2)^3}=0
\label{eqnMnLD}
\end{equation}
The solutions for $\tilde L(\tilde\rho)$ can be expanded again to:
\begin{equation}
\tilde L(\tilde\rho)=\tilde m+\frac{\tilde c}{\tilde\rho^2}+\dots, \label{ExpansionD}
\end{equation}
and using the transformation (\ref{cordchange}) we can get:
\begin{equation}
c=\tilde c R^3H^{3/2} \label{Hdepend}\ .
\end{equation}

It is instructive to study first the properties of (\ref{eqnMnLD}) for $\tilde m\gg1$, which corresponds to weak magnetic field $H\ll{m^2}/{R^2}$,
  or equivalently large quark mass $m\gg R\sqrt{H}$.

\subsection*{2.2.3 \hspace{2pt} Weak magnetic field}
\addcontentsline{toc}{subsection}{2.2.3 \hspace{0.15cm} Weak magnetic field}

In order to analyze the case of weak magnetic field let us expand $\tilde L(\tilde\rho)=\tilde m+\eta(\tilde\rho)$ and linearize equation (\ref{eqnMnLD})
 while leaving only the leading terms in $(\tilde\rho^2+\tilde m^2)^{-1}$. The result is:
\begin{equation}
\partial_{\tilde\rho}\left(\tilde\rho^3\eta'\right)+\frac{2\tilde\rho^3\tilde m}{(\tilde\rho^2+\tilde m^2)^3}=0\ ,
\label{eqnSmA}
\end{equation}
which has the general solution:
\begin{equation}
\eta(\tilde\rho)=\frac{C_{1}}{\tilde\rho^2}-\frac{\tilde m}{4\tilde\rho^2(\tilde m^2+\tilde\rho^2)}+C_{2}\ . \label{solutionSmA}
\end{equation}
From the definition of $\eta(\tilde\rho)$ and equation~(\ref{ExpansionD}) we can see that $C_{1}=\tilde c$ and since $\eta|_{\tilde\rho\to\infty}=0$ we have
 $C_{2}=0$. Now if we consider $\tilde m$ large enough, equation (\ref{solutionSmA}) should be valid for all $\tilde\rho$. It turns out that if we require
 that our solution be finite as $\tilde\rho \to 0$ we can determine the large~$\tilde m$ behavior of $\tilde c$. Indeed, the second term in (\ref{solutionSmA})
 has the expansion:
\begin{equation}
-\frac{\tilde m}{4\tilde\rho^2(\tilde m^2+\tilde\rho^2)}=-\frac{1}{4\tilde m}\frac{1}{\tilde\rho^2}+\frac{1}{4\tilde m^3}+O(\tilde\rho^2)\ .
\end{equation}
Therefore we deduce that:
\begin{equation}
C_{1}=\tilde c=\frac{1}{4\tilde m} \label{1/m}\ ,
\end{equation}
and finally, we get for the profile of the D7--brane for $\tilde m\gg 1$:
\begin{equation}
\tilde L(\tilde\rho)=\tilde m+\frac{1}{4\tilde m}\frac{1}{\tilde\rho^2}-\frac{\tilde m}{4\tilde\rho^2(\tilde m^2+\tilde\rho^2)}\ . \label{embSmA}
\end{equation}
If we go back to dimensionful parameters we can see, using equations (\ref{Hdepend}) and (\ref{1/m}) that for weak magnetic field $H$ the theory has developed a
 quark condensate:
\begin{equation}
\langle\bar\psi\psi\rangle \propto -c =-\frac{R^4}{4m}H^2\ . \label{condSmA}
\end{equation}

However, this formula is valid only for sufficiently large $m$ and we cannot make any prediction for the value of the quark condensate at zero
quark mass. To go further, the involved form of equation (\ref{eqnMnLD}) suggests the use of numerical techniques.

\subsection*{2.2.4 \hspace{2pt} Numerical results}
\addcontentsline{toc}{subsection}{2.2.4 \hspace{0.15cm} Numerical results}

\begin{figure}[h] 
   \centering
   \includegraphics[ width=11cm]{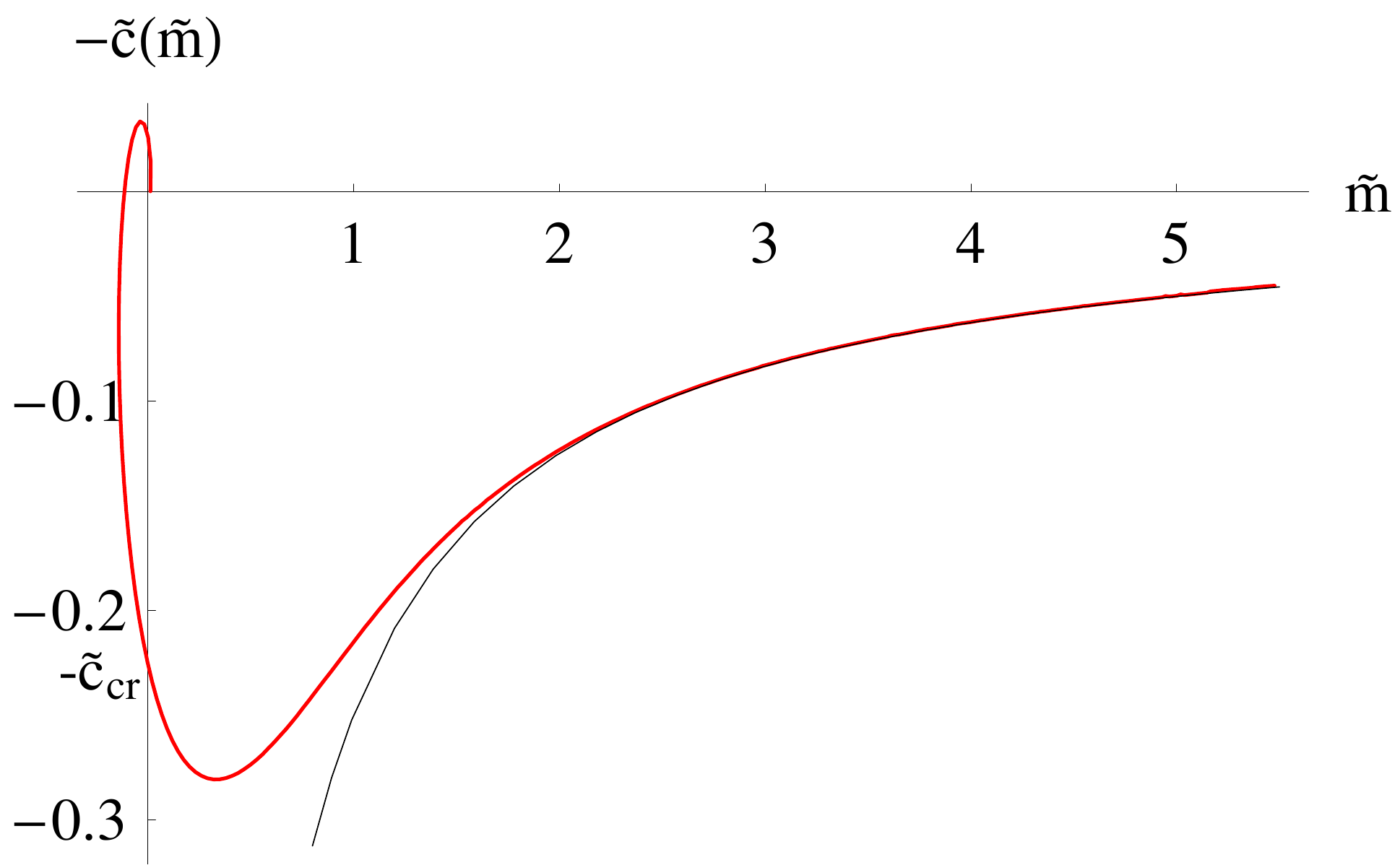}
   \caption{The black line corresponds to (\ref{1/m}), one can observe that the analytic result is valid for large $\tilde m$.
    It is also evident that for $\tilde m=0$ $\langle\bar\psi\psi\rangle\neq0$. The corresponding value of the condensate is $\tilde c_{\rm cr}=0.226$. }   \label{fig:fig1}
\end{figure}

In this subsection we solve numerically equation (\ref{eqnMnLD}) for the embedding of the D7--brane, using Mathematica. It is convenient to use initial conditions in the IR. We use the boundary condition  $\tilde L'(\tilde\rho)\vert_{\tilde\rho=0}=0$. We used  shooting
 techniques to generate the embedding of the D7--brane for a wide range of $\tilde m$. Having done so we expanded numerically
 the solutions for $\tilde L(\tilde\rho)$ as in equation (\ref{ExpansionD}) and generated the points in the  $(\tilde m,-\tilde c)$ plane corresponding to the solutions. The resulting plot is presented in Figure~\ref{fig:fig1}.

As one can see there is a non zero quark condensate for zero bare quark mass, the corresponding value of the condensate is $\tilde c_{\rm cr}=0.226$. It is also evident that the analytical expression for the condensate (\ref{1/m}) that we got in the previous section is valid for large $\tilde m$, as expected. Now using equation (\ref{Hdepend}) we can deduce the dependence of $c_{\rm cr}$ on $H$:
\begin{equation}
c_{\rm cr}=\tilde c_{\rm cr}R^3H^{3/2}=0.226R^3H^{3/2}\ . \label{Ccr}
\end{equation}

\begin{figure}[h] 
   \centering
   \includegraphics[ width=11cm]{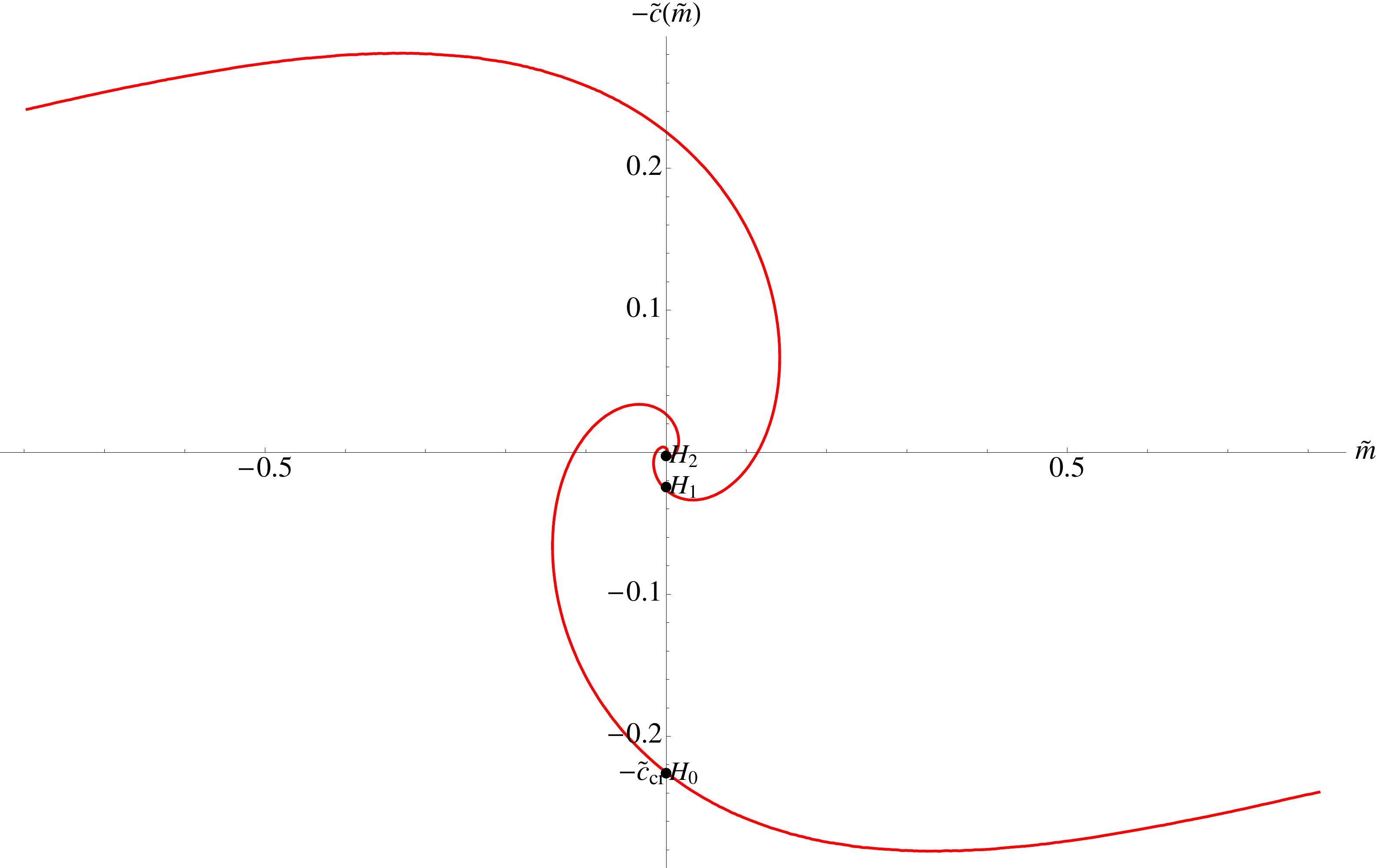}
   \caption{ A plot of the magnitude of the quark condensate at zero bare quark mass $c_{\rm cr}$ as function of $R^2H$, the black curve represents
   equation (\ref{Ccr}).}
   \label{fig:fig2}
\end{figure}

It is interesting to check the consistency of our numerical analysis by solving equation~(\ref{eqnMnL}) numerically and extracting the value of $c_{\rm cr}$
 for wide range of $R^2H$, the resulting plot fitted with equation (\ref{Ccr}) is presented in Figure~\ref{fig:fig2}.

\begin{figure}[h] 
   \centering
   \includegraphics[ width=11cm]{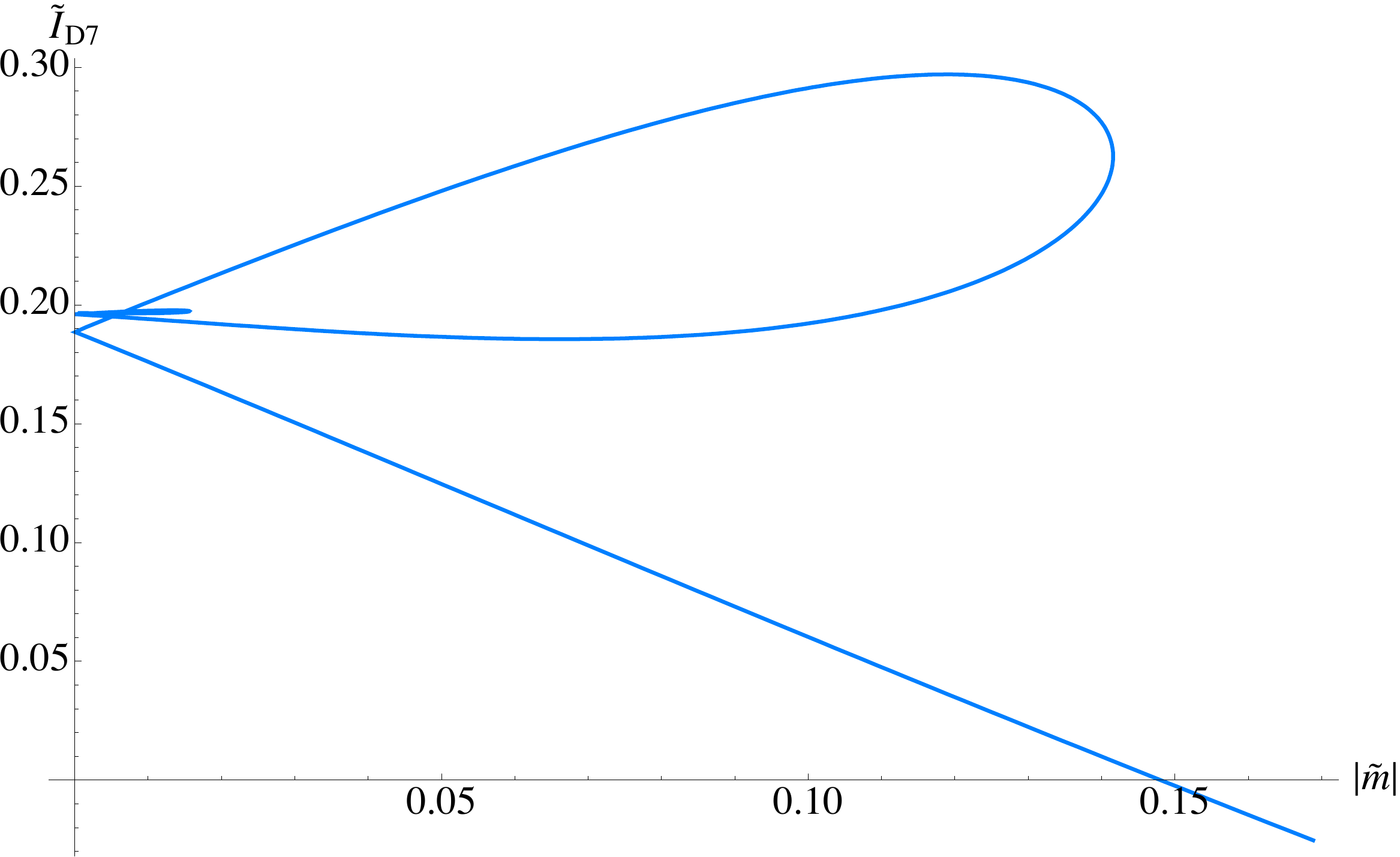}
   \caption{The lowest lying curve correspond to the positive $\tilde m$ part of the lowest branch of the spiral, suggesting that this is the stable phase of the theory. }
   \label{fig:free-energy}
\end{figure}

\begin{figure}[h] 
   \centering
   \includegraphics[ width=11cm]{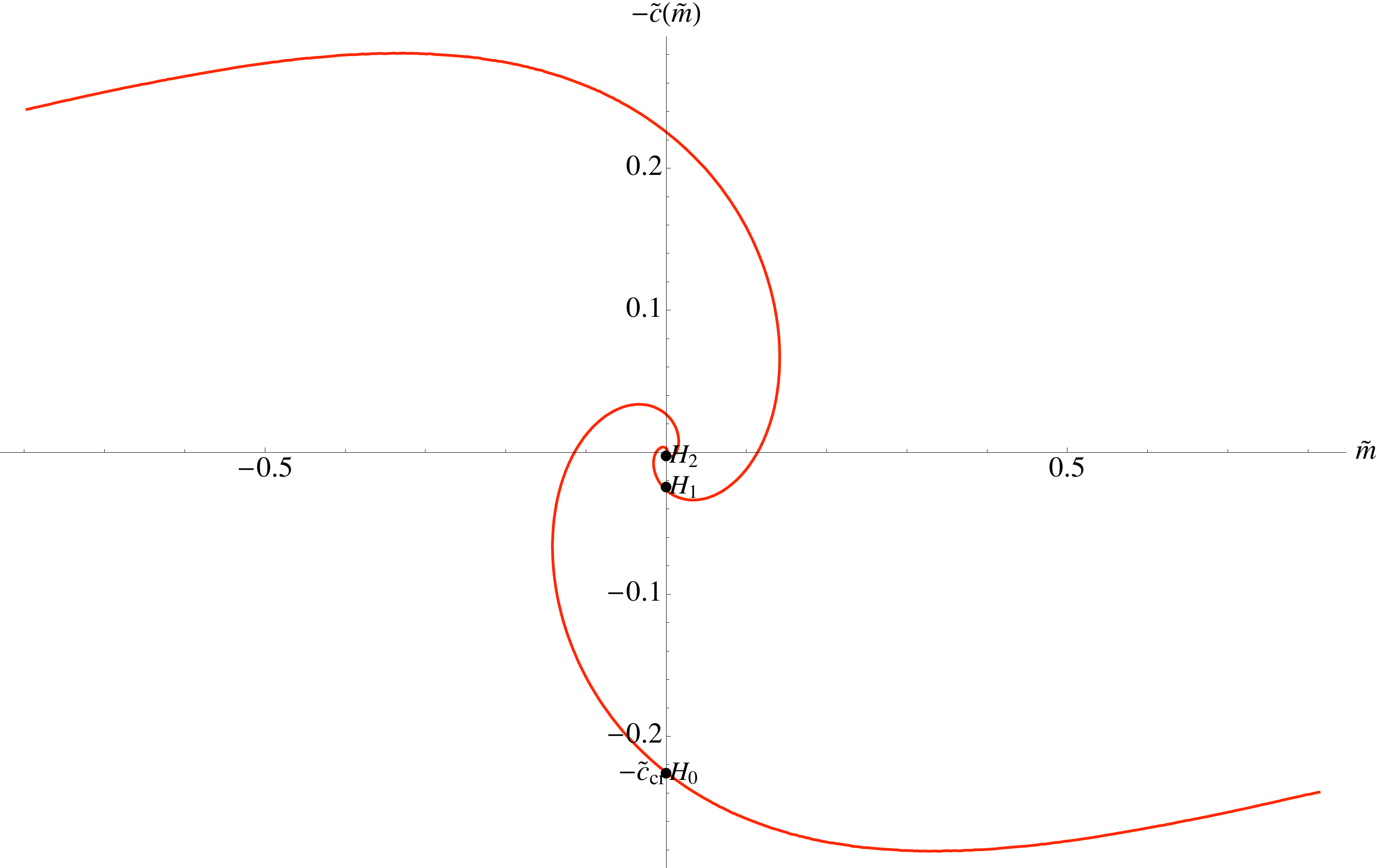}
   \caption{A magnification of Figure~\ref{fig:fig1} to show the spiral behavior near the origin of the $(-\tilde c,\tilde m)$-plane. We have added the second (left) arm of the spiral representing the $(\tilde m, -\tilde c)\to (-\tilde m,\tilde c)$ symmetry of the diagram. }
   \label{fig:spiral-revisited}
\end{figure}

Another interesting feature of our
   phase diagram is the spiral behavior near the origin of the $(\tilde m,-\tilde c)$-plane which can be seen in Figure~\ref{fig:spiral-revisited}. Note that the spiral presented in Figure~\ref{fig:spiral-revisited} has two arms, we have used the fact that any two points in the $(\tilde m,-\tilde c)$ plane related by reflection with respect to the origin describe the same physical state. A similar spiraling feature has been observed
  in ref.~\cite{Albash:2006bs}, where the authors have argued that only the lowest branch of the spiral corresponding to positive values of
   $m$ is the stable one (corresponding to the lowest energy state).  The spiral behavior near the origin signals instability of the
   embedding corresponding to $L_0\equiv 0$. If we trace the curve of the diagram in
   Figure~\ref {fig:spiral-revisited} starting from large $m$, as we go to smaller values of $m$ we will reach zero bare quark mass for some
   large negative value of the quark condensate $c_{cr}$. Now if we continue tracing along  the diagram one can verify numerically that all other points correspond to embeddings of the D7--brane which intersect the origin of the transverse plane at least once. After further study of the right arm of the spiral, one finds  that the part of the diagram corresponding to negative values of $\tilde m$ represents solutions for the D7--brane embedding which intersect the origin of the transverse plane odd number of times, while the positive part of the spiral represents solutions which intersect the origin of the transverse plane even number of times. The lowest positive branch corresponds to solutions which don't intersect the origin of the transverse plane and is the stable one, while the upper branches have correspondingly $2,4, {\it etc.,}$ intersection points and are ruled out after evaluation of the free energy.
   Indeed, let us explore the stability of the spiral by calculating the regularized free energy of the system. We identify the free energy of the dual gauge theory \cite{{Albash:2007bk},{Erdmenger:2007bn}} with the wick rotated and regularized on-shell action of the D7--brane:
\begin{eqnarray}
&&F=2\pi^2N_fT_{D7}R^4H^2\tilde I_{D7}\ ,\\
&&\tilde I_{D7}=\int\limits_{0}^{\tilde\rho_{max}}d\tilde\rho\left({\tilde\rho}^3\sqrt{1+\frac{1}{({\tilde\rho}^2+{\tilde L}^2)}}\sqrt{1+{\tilde L}'^2}-\tilde\rho\sqrt{{\tilde\rho}^4+1}\right)\label{freeenergy}
\end{eqnarray}
The second term under the sign of the integral in (\ref{freeenergy}), corresponds to the subtracted free energy of the $\tilde L(\tilde\rho)\equiv 0$ embedding and serves as a regulator.
Now we can evaluate numerically the integral in (\ref{freeenergy}) for the first several branches of the spiral. The corresponding plot is presented in Figure~\ref{fig:free-energy}. Note that we have plotted $\tilde I_{D7}$ versus $|\tilde m |$, since the bare quark mass depends only on the absolute value of the parameter $\tilde m$. The lowest curve on the plot corresponds to the lowest positive branch of the spiral, as one can see it has the lowest energy and thus corresponds to the stable phase of the theory.

In the next section we will provide more detailed analysis of the spiral structure from Figure~\ref{fig:spiral-revisited} and explore the discrete self-similarity associated to it.


\subsection*{2.2.5 \hspace{2pt} Criticality and spontaneous chiral symmetry breaking}
\addcontentsline{toc}{subsection}{2.2.5 \hspace{0.15cm} Criticality and spontaneous chiral symmetry breaking}
In the following section we analyze the spiral structure described in the previous subsection. The technique that we employ is similar to the one used in ref.~\cite{Frolov:2006tc} and \cite{Mateos:2006nu}, where the authors studied merger transitions in brane/black hole systems.

Let us explore the asymptotic form of the equation of motion of the D7--brane probe (\ref{eqnMnLD}) in the near horizon limit $\tilde \rho^2+\tilde L^2\to 0$. To this end we change coordinates to:
\begin{eqnarray}
\tilde\rho\to\lambda\hat\rho;~~~ \tilde L\to \lambda\hat L;
\label{rescaling}
\end{eqnarray}
and consider the limit $\lambda\to0$. The resulting equation of motion is:
\begin{equation}
\partial_{\hat\rho}(\frac{\hat\rho^3}{\hat\rho^2+\hat L^2}\frac{\hat L'}{\sqrt{1+\hat L'^2}})+2\sqrt{1+\hat L'^2}\frac{\hat\rho^3\hat L}{(\hat\rho^2+\hat L^2)^2}=0\ .
\label{rescaled}
\end{equation}
Equation (\ref{rescaled}) enjoys the scaling symmetry:
\begin{equation}
\hat \rho\to \mu\hat\rho;~~~\hat L\to \mu\hat L\ .
\end{equation}
In the sense that if $\hat L=f(\hat\rho)$ is a solution to the E.O.M. then $\frac{1}{\mu}f(\mu\hat\rho)$ is also a solution.
Next we focus on the region of the parameter space, close to the trivial $L\equiv 0$ embedding, by considering the expansion:
\begin{equation}
\hat L=0+(2\pi\alpha')\hat\chi
\end{equation}
and linearizing the E.O.M. . The resulting equation of motion is:
\begin{equation}
\hat\rho\partial_{\hat\rho}(\hat\rho\partial_{\hat\rho}\hat\chi)+2\hat\chi=0
\end{equation}
and has the solution :
\begin{equation}
\hat\chi=A\cos(\sqrt{2}\ln\hat\rho)+B\sin(\sqrt{2}\ln\hat\rho)\ .
\label{linafter}
\end{equation}
Now under the scaling symmetry $\hat\rho\to\mu\hat\rho$ the constants of integration $A$ and $B$ transform as:
\begin{equation}
\begin{pmatrix}A \\ B\\ \end{pmatrix}\to\frac{1}{\mu}\begin{pmatrix}\cos\sqrt{2}\ln\mu  &\sin\sqrt{2}\ln\mu \\ -\sin\sqrt{2}\ln\mu &\cos\sqrt{2}\ln\mu \end{pmatrix}\begin{pmatrix}A\\ B \end{pmatrix}\ .
\label{scaling1}
\end{equation}
The above transformation defines a class of solutions represented by a logarithmic spiral in the parameter space $(A,B)$ generated by some $(A_{in},B_{in})$, the fact that we have a discrete symmetry $\chi\to-\chi$ suggests that $(-A_{in},-B_{in})$ is also a solution and therefore the curve of solutions in the parameter space is a double spiral symmetric with respect to the origin. Actually as we are going to show there is a linear map from the parameter space $(A,B)$ to the plane $(\tilde m,-\tilde c)$, which explains the spiral structure, a subject of our study.
To show this let us consider the linearized E.O.M. before taking the $\lambda\to 0$ limit :
\begin{eqnarray}
\tilde\rho\sqrt{1+\tilde\rho^4}\partial_{\tilde\rho}(\tilde\rho\sqrt{1+\tilde\rho^4}\partial_{\tilde\rho}\tilde\chi)+2\tilde\chi=0;~~~
\tilde\chi=\lambda\hat\chi\ ,
\end{eqnarray}
with the solution:
\begin{equation}
\tilde\chi=\tilde A\cos\sqrt{2}\ln\frac{\tilde\rho}{\sqrt{1+\sqrt{1+\tilde\rho^4}}}+\tilde B\sin\sqrt{2}\ln\frac{\tilde\rho}{\sqrt{1+\sqrt{1+\tilde\rho^4}}}\ .
\label{linbefore}
\end{equation}
Expanding at infinity:
\begin{eqnarray}
\tilde\chi=\tilde m+\frac{\tilde c}{\tilde\rho^2}+\dots=\tilde A-\frac{\tilde B}{\sqrt{2}}\frac{1}{\tilde\rho^2}+\dots,
\end{eqnarray}
we get:
\begin{equation}
\begin{pmatrix}\tilde m \\ \tilde c\end{pmatrix}=\begin{pmatrix}\tilde A\\-{\tilde B}/{\sqrt{2}}\end{pmatrix}\ .
\end{equation}
Now if we match our solution (\ref{linbefore}) with the solution in the $\tilde\rho\to 0$ limit (\ref{linafter}) we should identify $(\tilde A,\tilde B)$ with the parameters $(A, B)$. Combining the rescaling property of $(A,B)$ with the linear map to $(\tilde m,-\tilde c)$ we get that the embeddings close to the trivial embedding $L\equiv 0$ are represented in the $(\tilde m,-\tilde c)$ plane by a double spiral defined {\it via} the transformation:
\begin{equation}
\begin{pmatrix}\tilde m\\ \tilde c\\ \end{pmatrix}\to\frac{1}{\mu}\begin{pmatrix}\cos\sqrt{2}\ln\mu &-\sqrt{2}\sin\sqrt{2}\ln\mu \\ \frac{1}{\sqrt{2}}\sin\sqrt{2}\ln\mu & \cos\sqrt{2}\ln\mu \end{pmatrix}\begin{pmatrix}\tilde m\\ \tilde c \end{pmatrix}\ .
\label{scaling2}
\end{equation}
Note that the spiral is double because we have the symmetry $(\tilde m,-\tilde c)\to(-\tilde m,\tilde c)$. This implies that in order to have similar configurations at scales $\mu_{1}$ and $\mu_{2}$ we should have:
\begin{equation}
\cos\sqrt{2}\ln\mu_1=\pm\cos\sqrt{2}\ln\mu_2
\end{equation}
and hence :
\begin{equation}
\sqrt{2}\ln\frac{\mu_2}{\mu_{1}}=-n\pi,
\end{equation}
which is equivalent to:
\begin{equation}
\frac{\mu_2}{\mu_1}=e^{-n\pi/\sqrt{2}}=q^n\ .
\end{equation}
Therefore we obtain that the discrete self-similarity is described by a rescaling by a factor of:
\begin{equation}
q=e^{-\pi/\sqrt{2}}\approx 0.10845\ .
\label{q}
\end{equation}
This number will appear in the next subsection where we will study the meson spectrum. As one may expect the meson spectrum also has a self-similar structure.

 It is interesting to confirm numerically the self-similar structure of the spiral and to calculate the scaling exponents of the bare quark mass and the quark condensate. It is convenient to use the separation of the D3-- and D7--branes at $\tilde\rho=0$, $\tilde L_{in}=\tilde L(0)$ as an order parameter.  There is a discrete set of initial separations $L_{in}$, corresponding to the points $H_0, H_1, H_2, \dots$ in Figure~\ref{fig:spiral-revisited} for which the corresponding D7--brane embeddings asymptote to $\tilde m=\tilde L_{\infty}=0$ as $\tilde\rho\to\infty$. The trivial $\tilde L\equiv 0$ embedding has ${\tilde L}_{in}=0$ and is the only one which has a zero quark condensate $(\tilde c=0)$, the rest of the states have a non zero $\tilde c$ and hence a chiral symmetry is spontaneously broken. Each such point determines separate branch of the spiral where $\tilde c=\tilde c(\tilde m)$ is a single valued function. On the other side, each such branch has both positive $\tilde m$ and negative $\tilde m$ parts. The symmetry of the double spiral from Figure~\ref{fig:spiral-revisited}, suggests that the states with negative $\tilde m$ are equivalent to positive $\tilde  m$ states but with an opposite sign of  $\tilde c$. This implies that the positive and negative $\tilde m$ parts of each branch correspond to two different phases of the theory, with opposite signs of the condensate. As we can see from Figure~\ref{fig:free-energy} the lowest positive branch of the spiral has the lowest free energy and thus corresponds to the stable phase of the theory. In the next subsection we will analyze the stability of the spiral further by studying the light meson spectrum of the theory near the critical $\tilde L\equiv 0$ embedding.

\begin{figure}[p] 
   \centering
   \includegraphics[ width=11cm]{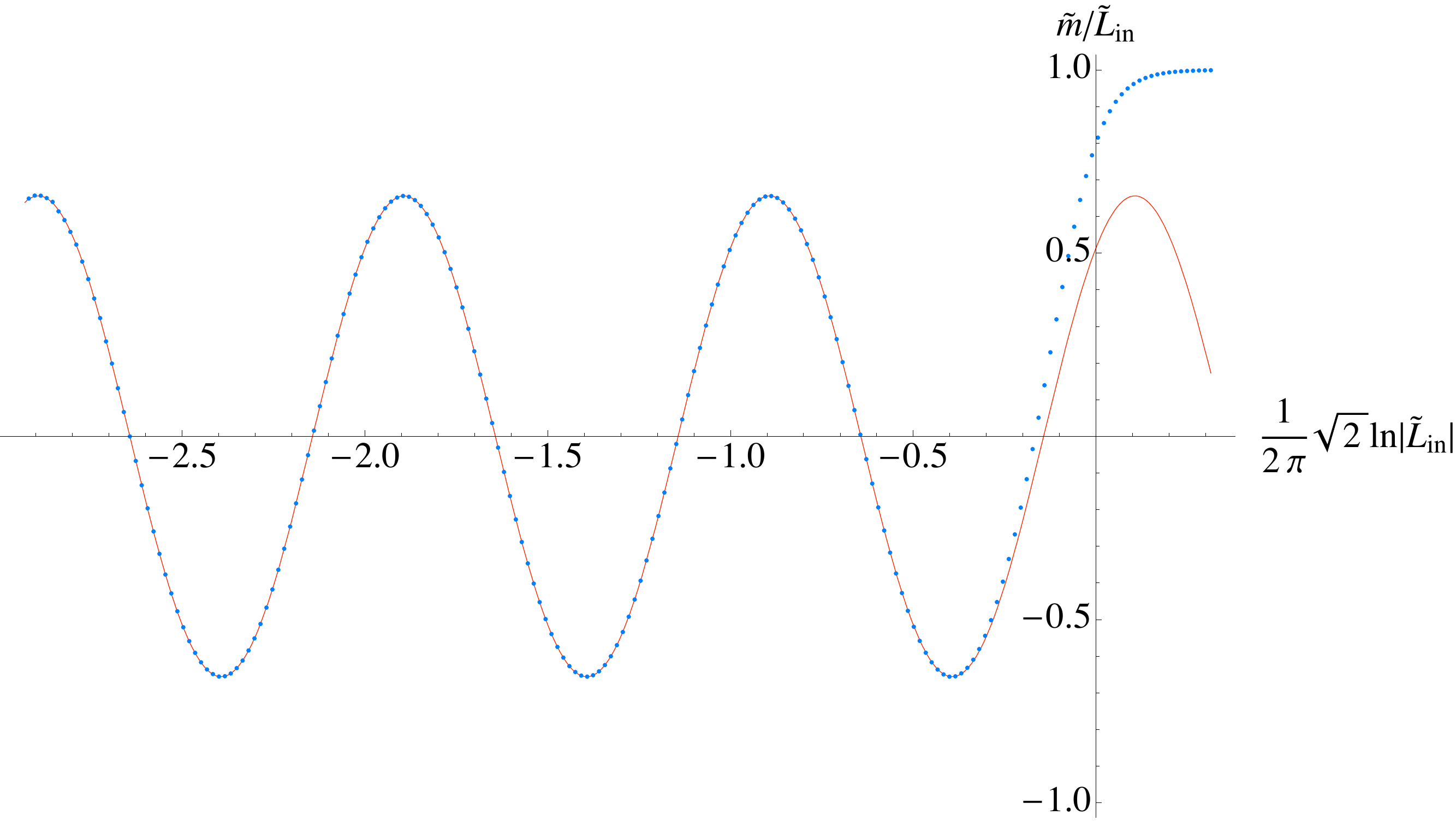}

   \includegraphics[ width=11cm]{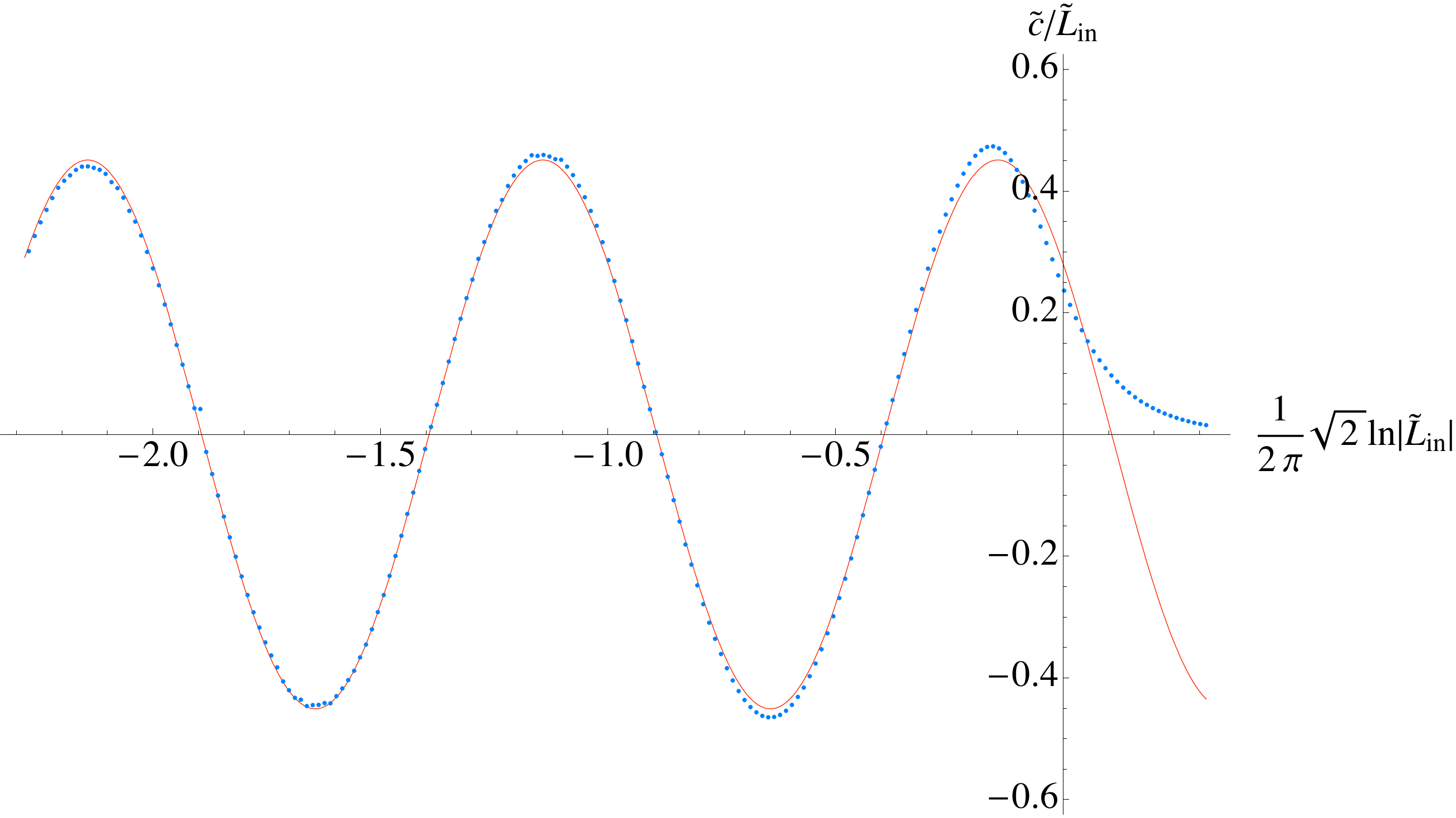}
   \caption{The red curves represent fit with trigonometric functions of unit period. For small $\tilde L_{in}$ the fit is very good, while for large $\tilde L_{in}$ we get the results for pure $AdS_{5}\times S^5$ space, namely $\tilde L=const$, $\tilde c=0$. The plots also verify that the scaling exponents of $\tilde m$ and $\tilde c$ are equal to one. }
   \label{fig:mspiral}
\end{figure}

 Here we are going to show that both the bare quark mass $\tilde m$ and the quark condensate $\tilde c$ have critical exponent one, as $\tilde L_{in} \to 0$. Indeed, let us consider the scaling property (\ref{scaling1}), (\ref{scaling2}). If we start from some $\tilde L_{in}^0$ and transform to $\tilde L_{in}=\frac{1}{\mu} \tilde L_{in}^0$, we can solve for $\mu$ and using equation (\ref{scaling2}) we can verify that the bare quark mass and the quark condensate approach zero linearly as $\tilde L_{in}\to0$. To verify numerically our analysis we generated plots of $\tilde m/\tilde L_{in}$ vs. $\sqrt{2}\log{\tilde L_{in}}/2\pi$  and $\tilde c/\tilde L_{in}$ vs. $\sqrt{2}\log{\tilde L_{in}}/2\pi$ presented in Figure~\ref{fig:mspiral}.

The red curves in Figure~\ref{fig:mspiral} represent a fit with trigonometric functions of a unit period, as one can see the fit is very good as $\tilde L_{in}\to 0$. On the other side, for large $\tilde L_{in}$ we obtain the results for a pure $AdS_{5}\times S^5$ space, namely $\tilde L=const$, $\tilde c=0$. It is also evident from the plots that the scaling exponents of $\tilde m$ and $\tilde c$ are equal to one.


\section*{2.3 \hspace{2pt} Meson spectrum}

\addcontentsline{toc}{section}{2.3 \hspace{0.15cm} Meson spectrum}

\subsection*{2.3.1 \hspace{2pt} General properties}
\addcontentsline{toc}{subsection}{2.3.1 \hspace{0.15cm} General properties}
We study the scalar meson spectrum. To do so we will consider quadratic fluctuations \cite{Kruczenski:2003be} of the embedding of the D7--brane in the
transverse $(L,\phi)$-plane. It can be shown that because of the diagonal form of the metric the fluctuation modes along the $\phi$ coordinate decouple
from the one along $L$. However, because of the non--commutativity introduced by the $B$--field we may expect the scalar fluctuations to couple to the vector
fluctuations. This has been observed in ref.~\cite{Arean:2005ar}, where the authors considered the geometric dual to non--commutative super Yang Mills. In our case the mixing will be even stronger,
because of the non--trivial profile for the D7--brane embedding, resulting from the broken supersymmetry.

Let us proceed with obtaining the action for the fluctuations. To obtain the contribution from the DBI part of the action we consider the expansion:
\begin{eqnarray}
L=L_0(\rho)+2\pi\alpha'\chi,\label{fluct}\quad\phi=0+2\pi\alpha'\ ,
\end{eqnarray}
where $L_0(\rho)$ is the classical embedding of the D7--brane solution to equation (\ref{eqnMnL}). To second order in $\alpha'$ we have the following expression:
\begin{equation}
E_{ab}=E^{0}_{ab}+2\pi\alpha'E^{1}_{ab}+(2\pi\alpha')^2E^{2}_{ab}\ ,
\label{33}
\end{equation}
where $E^0,E^1,E^2$ are given by:
\begin{eqnarray}
&&\hspace{-0.6cm}E^{0}_{ab}=G_{ab}(\rho,L_0(\rho),\psi)+B_{ab},\nonumber\\
&&\hspace{-0.6cm}E^{1}_{ab}=\frac{R^2{L_0}'}{\rho^2+L_0^2}\left(\partial_a\chi\delta_{b}^{\rho}+\partial_b\chi\delta_{a}^{\rho}\right)+\partial_{L_0}G_{ab}\chi+F_{ab}\label{ExpnMetrc}\\
&&\hspace{-0.6cm}E^{2}_{ab}=\frac{R^2}{\rho^2+L_0^2}\left(\partial_{a}\chi\partial_{b}\chi+L_0^2\partial_{a}\Phi\partial_{b}\Phi\right)-\frac{2R^2L_0L_0'}{(\rho^2+L_0^2)^2}\left(\partial_{a}\chi\delta_{b}^{\rho}+\partial_{b}\chi\delta_{a}^{\rho}\right)\chi+\frac{1}{2}\partial_{L_0}^{2}G_{ab}\chi^2\ . \nonumber
\end{eqnarray}
Here $G_{ab}$ and $B_{ab}$ are the induced metric and B field on the D7--brane's world volume. Now we can substitute equation (\ref{ExpnMetrc}) into equation (\ref{decopl}) and expand to second order in $\alpha'$. It is convenient \cite{Arean:2005ar} to introduce
the following matrices:

\begin{equation}
||{E_{ab}^0}||^{-1}=S+J,
\end{equation}
where $S$ is diagonal and $J$ is antisymmetric:
\begin{eqnarray}
&||S^{ab}||&={\rm diag}\{-G_{11}^{-1},G_{11}^{-1},\frac{G_{11}}{G_{11}^{2}+H^2},\frac{G_{11}}{G_{11}^{2}+H^2},G_{\rho\rho}^{-1},G_{\psi\psi}^{-1},G_{\alpha\alpha}^{-1},G_{\beta\beta}^{-1}\}\ ,\label{S}\\
&J^{ab}&=\frac{H}{G_{11}^{2}+H^2}(\delta_{3}^{a}\delta_{2}^{b}-\delta_{3}^{b}\delta_{2}^{a})\ ,\label{J}\\
&G_{11}&=\frac{\rho^2+{L_0}^2}{R^2}\ ;~~~G_{\rho\rho}=R^2\frac{(1+{L'_0}^2)}{\rho^2+L_0^2}\ ;~~~G_{\psi\psi}=\frac{R^2\rho^2}{\rho^2+L_0^2};\nonumber\\
&G_{\alpha\alpha}&=\cos^2\psi G_{\psi\psi}\ ;~~~G_{\beta\beta}=\sin^2\psi G_{\psi\psi}\ .
\end{eqnarray}

Now it is straightforward to get the effective action. At first order in $\alpha'$ the action for the scalar fluctuations is the first variation of
the classical action (\ref{decopl}) and is satisfied by the classical equations of motion. The equation of motion for the gauge field at first order
was considered in Section~2.2 for the computation of the $C_{(6)}$ potential induced by the $B$- field. Therefore we focus on the second order contribution from the DBI action.

After integrating by parts and taking advantage of the Bianchi identities for the gauge field, we end up with the following terms.
For $\chi$:
\begin{eqnarray}
&{\cal L_{\chi}} \propto \frac{1}{2}\sqrt{-E^0}\frac{R^2}{\rho^2+{L_0}^2}\frac{S^{ab}}{1+{L'_0}^2}\partial_{a}\chi\partial_{b}\chi+\left[\partial_{L_0}^2\sqrt{-E^0}-\partial_\rho\left(\partial_{L_0}\sqrt{-E^0}\frac{L'_0}{1+{L'_0}^2}\right)\right]\frac{1}{2}\chi^2\ ,\nonumber\\ 
&\label{Schi}
\end{eqnarray}
and for $F$:
\begin{eqnarray}
{\cal L}_{F} \propto\frac{1}{4}\sqrt{-E^0}S^{aa'}S^{bb'}F_{ab}F_{a'b'}\ ,
\label{SF}
\end{eqnarray}
and the mixed $\chi$--$F$ terms:
\begin{eqnarray}
&{\cal L}_{F\chi}&\propto \frac{\sin2\psi}{2}f\chi F_{23}\ ,
\label{SFchi}
\end{eqnarray}
and for $\Phi$:
\begin{equation}
{\cal L}_{\Phi}\propto\frac{1}{2}\sqrt{-E^0}\frac{R^2{L_0}^2}{\rho^2+L_0^2} S^{ab}\partial_{a}\Phi\partial_{b}\Phi\ ,
\label{Sphi}
\end{equation}
where the function $f$ in (\ref{SFchi}) is given by:
\begin{eqnarray}
&f(\rho)&=\partial_{\rho}\left(g(\rho)\frac{L'_0}{1+{L_0}'^{2}}J^{23}\right)+J^{32}\partial_{L_0}g(\rho)+2g(\rho)J^{23}S^{22}\partial_{L_0}G_{11}\ ,\\
{\rm with}\quad&g(\rho)&=\frac{\sqrt{-E^0}}{\sin\psi\cos\psi}=\rho^3\sqrt{1+{L_0}'^2}\sqrt{1+\frac{R^4H^2}{(\rho^2+L_0^2)^2}}\ .\nonumber
\label{Fchi}
\end{eqnarray}

As can be seen from equation (\ref{SFchi}) the $A_2,A_3$ components of the gauge field couple to the scalar field $\chi$ via the function $f$. Note that since
for $\rho \to \infty$ and $L\to\infty$, we see that $J^{23}\to 0$, the mixing of the scalar and vector field decouples asymptoticly. In order
to proceed with the analysis we need to take into account the contribution from the Wess-Zumino part of the action. The relevant terms to second order
in $\alpha'$ are \cite{Arean:2005ar}:
\begin{equation}
S_{WZ}=\frac{(2\pi\alpha')^2}{2}\mu_{7}\int{F_{(2)}\wedge F_{(2)}\wedge C_{(4)}}+(2\pi\alpha')\mu_{7}\int F_{(2)}\wedge B_{(2)}\wedge \tilde P[C_{(4)}]\ ,
\label{WZ}
\end{equation}
where $C_{(4)}$ is the background R-R potential given in equation~(\ref{AdS}) and $\tilde C_{(4)}$ is the pull back of its magnetic dual. One can show
that:
\begin{equation}
\tilde C_{4}=\frac{R^4}{g_{s}}\frac{2\rho^2+L^2}{(\rho^2+L^2)^2}L^2 \sin\psi\cos\psi d\psi\wedge d\alpha\wedge d\beta\wedge d\phi\ .
\end{equation}
Writing $\phi=2\pi\alpha'\Phi$  we write for the pull back $P[\tilde C_{(4)}]$:
\begin{equation}
P[\tilde C_{(4)}]=-\frac{2\pi\alpha'}{g_s}\frac{\sin2\psi}{2}K(\rho)\partial_{a}\Phi d\psi\wedge d\alpha\wedge d\beta\wedge dx^a,
\label{pulC}
\end{equation}
where we have defined:
\begin{equation}
K(\rho)=-R^4L_0^2\frac{2\rho^2+{L_0}^2}{(\rho^2+L_0^2)^2}
\end{equation}
Now note that the $B$--field has components only along $x^2$ and $x^3$, therefore $dx^a$ in equation (\ref{pulC}) can be only $d\rho,dx^0$ or $dx^1$.
This will determine the components of the gauge field which can mix with $\Phi$. However, after integrating by parts and using the Bianchi identities
one can get the following simple expression for the mixing term:
\begin{equation}
-(2\pi\alpha')^2\frac{\mu_7}{g_s}\int d^8\xi\frac{\sin2\psi}{2}H\partial_{\rho}K\Phi F_{01}\  ,
\label{SPhiF}
\end{equation}
resulting in the following contribution to the complete lagrangian:\\
\begin{equation}
{\cal L}_{F\Phi}\propto \frac{\sin2\psi}{2}H\partial_{\rho}K\Phi F_{01}\  .
\label{mixing}
\end{equation}
Note that this means that only the $A_0$ and $A_1$ components of the gauge field couple to the scalar field $\Phi$. Next the contribution from the first term in (\ref{WZ}) is given by:
\begin{equation}
(2\pi\alpha')^2\frac{\mu_{7}}{g_s}\int d^8\xi\frac{(\rho^2+L_0^2)^2}{8R^4}F_{ab}F_{cd}\epsilon^{a b c d}\ ,
\end{equation}
where the indices take values along the $\rho,\psi,\alpha,\beta$ directions of the world volume. This will contribute to the equation of motion for
$A_{\rho},A_{\psi},A_{\alpha}$ and $A_{\beta}$ which do not couple to the scalar fluctuations. In this section we will be interested in analyzing the
spectrum of the scalar modes, therefore we will not be interested in the components of the gauge field transverse to the D3--branes world volume.
However, although there are no sources for these components from the scalar fluctuations, they still couple to the components along the D3--branes as
a result setting them to zero will impose constraints on the $A_{0}\dots A_3$. Indeed, from the equation of motion for the gauge field along the
transverse direction one gets:
\begin{equation}
\sum\limits_{a=0}^{3}S^{aa}\partial_b\partial_a{A_a}=0,~~b=\rho,\psi,\alpha,\beta\ ,
\label{Lorenzbr}
\end{equation}
(Here, no  summation on repeated indices is intended.)
However, the non--zero $B$--field explicitly breaks the Lorentz symmetry along the D3--branes' world volume. In particular we have:
\begin{eqnarray}
S^{00}=-S^{11}\ ,\quad S^{22}=S^{33}\neq S^{11}\ ,
\end{eqnarray}
which suggests that we should impose:
\begin{eqnarray}
-\partial_0{A_0}+\partial_1{A_1}=0\label{constrA}\ ,\quad\partial_2{A_2}+\partial_{3}{A_{3}}=0\ .
\end{eqnarray}
We will see that these constraints are consistent with the equations of motion for $A_{0}\dots A_3$. Indeed, with this constraint the equations of
 motion for $\chi$, $\Phi$ and $A_{\mu},\mu=0\dots 3$ are,
 for $\chi$:
\begin{eqnarray}
&&\frac{1+{L'_0}^2}{g}\partial_\rho\left(\frac{g\partial_\rho\chi}{(1+{L'_0}^2)^2}\right)+\frac{\Delta_{\Omega_3}\chi}{\rho^2}
+\frac{R^4}{(\rho^2+L_0^2)^2}\widetilde{\Box}\chi+\label{EMCHI}\\
&&+\frac{1+{L'_0}^2}{g}\left(-\partial_{\rho}\left(\frac{\partial{g}}{\partial L_0}\frac{L'_0}{1+{L'_0}^2}\right)+\frac{\partial^2{g}}{\partial L_0^2}\right)\chi+\frac{1+{L'_0}^2}{g}f
F_{23}=0\ ,\nonumber
\end{eqnarray}
and for $\Phi$:
\begin{eqnarray}
\frac{1}{g}\partial_\rho\left(\frac{{g}L_0^2\partial_\rho\Phi}{1+{L'_0}^2}\right)+\frac{L_0^2\Delta_{\Omega_3}\Phi}{\rho^2}+
\frac{R^4L_0^2}{(\rho^2+L_0^2)^2}\widetilde{\Box}\Phi-\frac{H\partial_\rho
K}{g}F_{01}=0\ ,
\label{eqnPhi}
\end{eqnarray}
and  finally for $A_a$:
\begin{eqnarray}
\frac{1}{g}\partial_\rho\left(\frac{{g}\partial_\rho{A_0}}{1+{L'_0}^2}\right)+\frac{\Delta_{\Omega_3}{A_0}}{\rho^2}+
\frac{R^4}{(\rho^2+L_0^2)^2}\widetilde{\Box}{A_0}+\frac{H\partial_\rho
K}{g}\partial_1\Phi&=&0\ ,\label{EqGauge}\\
\frac{1}{g}\partial_\rho\left(\frac{{g}\partial_\rho{A_1}}{1+{L'_0}^2}\right)+\frac{\Delta_{\Omega_3}{A_1}}{\rho^2}+
\frac{R^4}{(\rho^2+L_0^2)^2}\widetilde{\Box}A_1+\frac{H\partial_\rho
K}{g}\partial_0\Phi&=&0\ ,\nonumber\\
\frac{1}{g}\partial_\rho\left(\frac{{g}\partial_\rho{A_2}}{(1+{L'_0}^2)(1+\frac{R^4H^2}{(\rho^2+L_0^2)^2})}\right)+
\frac{R^4}{(\rho^2+L_0^2)^2+R^4H^2}\widetilde{\Box}{A_2}&+&\frac{\Delta_{\Omega_3}{A_2}}{\rho^2(1+\frac{R^4H^2}{(\rho^2+L_0^2)^2})}\nonumber\\  -\frac{f}{g}\partial_3\chi&=&0\ ,\nonumber\\
\frac{1}{g}\partial_\rho\left(\frac{{g}\partial_\rho{A_3}}{(1+{L'_0}^2)(1+\frac{R^4H^2}{(\rho^2+L_0^2)^2})}\right)+
\frac{R^4}{(\rho^2+L_0^2)^2+R^4H^2}\widetilde{\Box}{A_3}&+&\frac{\Delta_{\Omega_3}{A_3}}{\rho^2(1+\frac{R^4H^2}{(\rho^2+L_0^2)^2})}\nonumber\\ +\frac{f}{g}\partial_2\chi&=&0\  .\nonumber
\end{eqnarray}
We have defined:
\begin{equation}
\widetilde\Box=-\partial_0^2+\partial_1^2+\frac{\partial_2^2+\partial_3^2}{1+\frac{R^4H^2}{(\rho^2+L_0^2)^2}}\ .
\end{equation}
As one can see the spectrum splits into two independent components, namely the vector modes $A_0,A_1$ couple to the scalar fluctuations along $\Phi$, while the vector modes $A_2,A_3$ couple to the scalar modes along $\chi$. However, it is possible to further simplify the equations of motion for the gauge field. Focusing on the equations of motion for $A_0$ and $A_1$ in equation~(\ref{EqGauge}), it is possible to rewrite them as:
\begin{eqnarray}
&&\frac{1}{g}\partial_\rho\left(\frac{{g}\partial_\rho{F_{01}}}{1+{L'_0}^2}\right)+\frac{\Delta_{\Omega_3}{F_{01}}}{\rho^2}+
\frac{R^4}{(\rho^2+L_0^2)^2}\widetilde{\Box}{F_{01}}-\frac{H\partial_\rho
K}{g}(-\partial_0^2+\partial_1^2)\Phi=0\label{Elec}\\
&&\frac{1}{g}\partial_\rho\left(\frac{{g}\partial_\rho{(-\partial_0{A_0}+\partial_1{A_1})}}{1+{L'_0}^2}\right)+\frac{\Delta_{\Omega_3}{(-\partial_0{A_0}+\partial_1{A_1})}}{\rho^2}\\ \nonumber&&\hspace{6.3cm}+
\frac{R^4}{(\rho^2+L_0^2)^2}\widetilde{\Box}{(-\partial_0{A_0}+\partial_1{A_1})}=0\ .\nonumber
\end{eqnarray}
Note that the first constraint in (\ref{constrA}) trivially satisfies the second equation in (\ref{Elec}). In this way we are left with the first equation in (\ref{Elec}). Similarly one can show that using the second constraint in (\ref{constrA}) the equations of motion in (\ref{EqGauge}) for $A_2$ and $A_3$ boil down to a single equation for $F_{23}$:
\begin{eqnarray}
\frac{1}{g}\partial_\rho\left(\frac{{g}\partial_\rho{F_{23}}}{(1+{L'_0}^2)(1+\frac{R^4H^2}{(\rho^2+L_0^2)^2})}\right)&+&
\frac{R^4}{(\rho^2+L_0^2)^2+R^4H^2}\widetilde{\Box}{F_{23}}\label{teschi}\nonumber\\
&+&\frac{\Delta_{\Omega_3}{F_{23}}}{\rho^2(1+\frac{R^4H^2}{(\rho^2+L_0^2)^2})}+\frac{f}{g}(\partial_2^2+\partial_3^2)\chi=0\ .
\end{eqnarray}
Now let us proceed with a study of the fluctuations along $\Phi$.
\subsection*{2.3.2 \hspace{2pt} Fluctuations along $\Phi$ for a weak magnetic field}
\addcontentsline{toc}{subsection}{2.3.2 \hspace{0.15cm} Fluctuations along $\Phi$ for a weak magnetic field}

To proceed, we have to take into account the $F_{01}$ component of the gauge field strength and solve the coupled equations of motion. Since the classical solution for the embedding of the D7--brane is known only numerically we have to rely again on numerics to study the meson spectrum. However, if we look at equation (\ref{eqnMnL}) we can see that the terms responsible for the non--trivial parts  of the equation of motion are of order $H^2$. On the other hand, the mixing of the scalar and vector modes due to the term (\ref{mixing}) appear at first order in $H$. Therefore it is possible to extract some non--trivial properties of the meson spectrum even at linear order in $H$ and as it turns out, we can observe a Zeeman--like  effect: A splitting of states that is proportional to the magnitude of the magnetic field. To describe this, let us study the approximation of weak magnetic field.

To first order in $H$ the classical solution for the D7--brane profile is given by:
\begin{equation}
L_{0}=m+O(H^2),
\end{equation}
where $m$ is the asymptotic separation of the D3-- and D7--branes and corresponds to the bare quark mass. In this approximation the expressions for $g(\rho)$ and $\partial_{\rho}K(\rho)$, become:
\begin{eqnarray}
g(\rho)=\rho^3\ ,\quad\partial_{\rho}K(\rho)=\frac{4m^2R^4\rho^3}{(\rho^2+m^2)^3}\ ,\nonumber
\end{eqnarray}
and the equations of motion for $\Phi$ and $F_{01}$, equations (\ref{eqnPhi}) and (\ref{Elec}), simplify to:
\begin{eqnarray}
&&\frac{1}{\rho^3}\left(\rho^3m^2\partial_{\rho}\Phi\right)+\frac{m^2\Delta_{\Omega_{3}}}{\rho^2}\Phi+\frac{m^2R^4}{(\rho^2+m^2)^2}\Box\Phi-4H\frac{m^2R^4}{(\rho^2+m^2)^3}F_{01}=0\ ,\\
&&\frac{1}{\rho^3}\partial_\rho\left(\rho^3\partial_\rho F_{01}\right)+\frac{\Delta_{\Omega_3}{F_{01}}}{\rho^2}+\frac{R^4}{(\rho^2+m^2)^2}\Box{F_{01}}-4H\frac{m^2R^4}{(\rho^2+m^2)^3}{\cal P}^2\Phi=0\nonumber\ ,\\
&&{\rm where}\quad\Box=-\partial_{0}^2+\partial_{1}^2+\partial_{2}^2+\partial_{3}^2,~~~{\cal P}^2=-\partial_{0}^2+\partial_{1}^2\nonumber\ .
\label{simplified}
\end{eqnarray}
This system has become similar to the system studied in ref.~\cite{Arean:2005ar} and in order to decouple it we can define the fields:
\begin{equation}
\phi_{\pm}=F_{01}\pm m{\cal P}\Phi\ ,
\end{equation}
where ${\cal P}=\sqrt{-\partial_{0}^2+\partial_{1}^2}$. The resulting equations of motion are:
 \begin{equation}
\frac{1}{\rho^3}\partial_{\rho}(\rho^3\partial_{\rho}\phi_{\pm})+\frac{\Delta_{\Omega_{3}}}{\rho^2}\phi_{\pm}+\frac{R^4}{(\rho^2+m^2)^2}\Box\phi_{\pm}\mp H\frac{4R^4m}{(\rho^2+m^2)^3}{\cal P}\phi_{\pm}=0\ .
\label{eqnMotSimpl}
\end{equation}
Note that ${\cal P}^2$ is the Casimir operator in the $(x_{0},x_{1})$ plane only, while $\Box$ is the Casimir operator along the D3--branes' world volume. If we consider a plane wave $e^{ix.k}$ then we can define:
\begin{equation}
\Box e^{ix.k}=M^2 e^{ix.k},~~~{\cal P}^2 e^{ix.k}=M_{01}^2e^{ix.k}\nonumber\ ,
\end{equation}
and we have the relation:
\begin{equation}
M^2=M_{01}^2-k_{2}^2-k_{3}^2\ .
\end{equation}
The corresponding spectrum of $M^2$ is continuous in $k_{2}, k_{3}$. However, if we restrict ourselves to motion in the $(x_{0}, x_{1})$-plane the spectrum is discrete. Indeed, let us consider the ansatz:
\begin{equation}
\phi_{\pm}=\eta_{\pm}(\rho)e^{-ix_{0}k_{0}+ik_{1}x_{1}}\ .
\end{equation}
Then we can write:
 \begin{eqnarray}
&&\frac{1}{\rho^3}\partial_{\rho}(\rho^3\partial_{\rho}\eta_{\pm})+\frac{R^4}{(\rho^2+m^2)^2}M_{\pm}^2\eta_{\pm}\mp H\frac{4R^4m}{(\rho^2+m^2)^3}{M_{\pm}}\eta_{\pm}=0\ ,\label{eqn01Sm}\\
&&M_{\pm}\equiv{M_{01}}_{\pm}\nonumber\  .
\end{eqnarray}
Let us analyze equation (\ref{eqn01Sm}). It is convenient to introduce:
\begin{eqnarray}
&& y=-\frac{\rho^2}{m^2};~~~\bar M_{\pm}=\frac{R^2}{m}M_{\pm};~~~P_{\pm}(y)=(1-y)^{\alpha_{\pm}}\eta_{\pm};\label{varchan}\\
&& 2\alpha_{\pm}=1+\sqrt{1+\bar M_{\pm}^2};~~~\epsilon=H\frac{R^2}{m^2}\ .\nonumber
\end{eqnarray}
With this change of variables equation (\ref{eqn01Sm}) is equivalent to:
\begin{equation}
y(1-y)P_{\pm}''+2(1-(1-\alpha_{\pm})y)P'-\alpha_{\pm}(\alpha_{\pm-1})P_{\pm}\pm\epsilon\frac{\bar M_{\pm}}{(1-y)^2}P_{\pm}=0\ .
\label{hyper}
\end{equation}
Next we can expand:
\begin{eqnarray}
&&P_{\pm}=P_{0}\pm\epsilon P_{1}+O(\epsilon^2)\ ;~~~\alpha_{\pm}=\alpha_{0}\pm\epsilon\alpha_1+O(\epsilon^2)\ ;\label{expansions}\\
&&\bar M_{\pm}=\bar M_{0}\pm\epsilon\alpha_{1}\frac{(4\alpha_{0}+2)}{\bar M_{0}}+O(\epsilon^2)\ ;~~~\bar M_{0}=2\sqrt{\alpha_{0}(\alpha_{0}+1)} \ .\nonumber
\end{eqnarray}
leading to the following equations for $P_{0}$ and $P_{1}$:
\begin{eqnarray}
y(1-y)P_{0}''+2(1-(1-\alpha_{0})y)P_{0}'-\alpha_{0}(\alpha_{0}-1)P_{0}&=&0\ ,\label{EqPert}\\
y(1-y)P_{1}''+2(1-(1-\alpha_{0})y)P_{1}'-\alpha_{0}(\alpha_{0}-1)P_{1}&=&(\alpha_{1}(2\alpha_{0}-1)\nonumber \\ \quad &-&\frac{\bar M_{0}}{(1-y)^2})P_{0}-2\alpha_{1}y P_{0}'\ .\nonumber
\end{eqnarray}
The first equation in (\ref{EqPert}) is the hypergeometric equation and corresponds to the fluctuations in pure $AdS_{5}\times S^5$. It has the regular solution \cite{Kruczenski:2003be}:
\begin{equation}
P_{0}(y)=F(-\alpha_{0},1-\alpha_{0},2,y)\ .
\end{equation}
Furthermore, regularity of the solution for $\eta(\rho)$ at infinity requires \cite{Kruczenski:2003be} that $\alpha_{0}$  be discrete, and hence the spectrum of $\bar M_{0}$:
\begin{eqnarray}
 &&1-\alpha_{0}=-n,~~~n=0,1,\dots\label{degen}\\
 &&\bar M_{0}=2\sqrt{(n+1)(n+2)}\nonumber\ .
\end{eqnarray}
The second equation in (\ref{EqPert}) is an inhomogeneous hypergeometric equation. However, for the ground state, namely $n=0$, $P_{0}=F(-1,0,2,y)=1$ and one can easily get the solution:
\begin{equation}
P_{1}(y)=\frac{\bar M_{0}}{6}\ln(1-y)+(6\alpha_{1}-\bar M_{0})(\ln(-y)+\frac{1}{y})-\frac{\bar M_{0}}{4(1-y)}\ .
\end{equation}
On the other hand, using the definition of $P_{\pm}(y)$ in (\ref{varchan})  to first order in $\epsilon$ we can write:
\begin{equation}
\eta_{\pm}=\frac{1}{(1-y)^{\alpha_{0}}}\left(1\mp\epsilon\frac{\alpha_{1}}{\alpha_0}\ln(1-y)\right)\left(1\pm\epsilon P_{1}(y)\right)\ ,
\end{equation}
for the ground state $\alpha_{0}=1$ and we end up with the following expression for $\eta_{\pm}$:
\begin{equation}
\eta_{\pm}=\frac{1}{1-y}\pm\epsilon\frac{\bar M_{0}}{4(1-y)^2}\pm\frac{\epsilon}{1-y}(6\alpha_1-\bar M_0)\left(\ln(-y)+\frac{1}{y}-\frac{\ln(1-y)}{6}\right)\ .
\label{groundstate}
\end{equation}
Now if we require that our solution is regular at $y=0$ and goes as $1/\rho^2\propto1/y $ at infinity, the last term in (\ref{groundstate}) must vanish. Therefore we have:
\begin{equation}
\alpha_{1}=\frac{\bar M_0}{6}\ .
\label{correction}
\end{equation}
\begin{figure}[h] 
   \centering
   \includegraphics[ width=11cm]{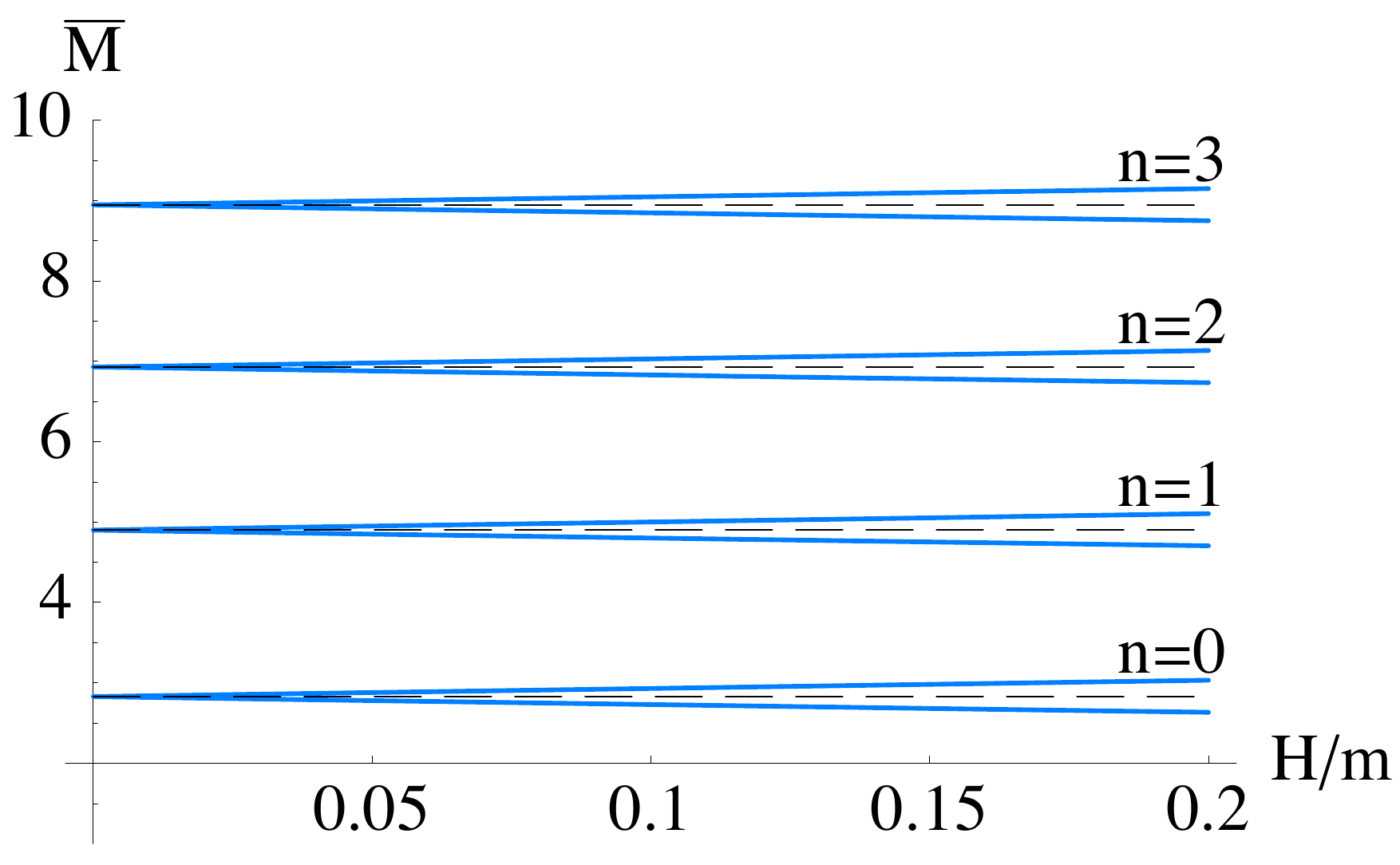}
   \caption{Plot of $\bar M=M{R^2}/{m}$ vs. $H/m$ for the first three states. The dashed black lines correspond to the spectrum given by equation~(\ref{degen}). }
   \label{fig: Zeeman}
\end{figure}
After substituting in (\ref{expansions}) and (\ref{varchan}) we end up with the following correction to the ground sate:
\begin{equation}
M_{\pm}=M_{0}\pm\frac{H}{m}\ .
\label{Zeeman}
\end{equation}
We observe how the introduction of an external magnetic field breaks the degeneracy of the spectrum given by equation (\ref{degen}) and results in Zeeman splitting of the energy states, proportional to the magnitude of $H$. Although equation (\ref{Zeeman}) was derived using the ground state it is natural to expect that the same effect takes place for higher excited states. To demonstrate this it is more convenient to employ numerical techniques for solving equation (\ref{eqn01Sm}) and use the methods described in ref.~\cite{Babington:2003vm} to extract the spectrum. The resulting plot is presented in Figure~\ref{fig: Zeeman}. As expected we observe Zeeman splitting of the higher excited states. It is interesting that equation (\ref{Zeeman}) describes well not only the ground state, but also the first several excited states.

It turns out that one can easily generalize equation (\ref{Zeeman}) to the case of non--zero momentum in the $(x_{2},x_{3})$-plane. Indeed, if we start from equation (\ref{eqnMotSimpl}) and proceed with the ansatz:
\begin{equation}
\phi_{\pm}=\tilde {\eta}_{\pm}(\rho)e^{-ix.k}\ ,
\end{equation}
we end up with:
\begin{eqnarray}
&&\frac{1}{\rho^3}\partial_{\rho}(\rho^3\partial_{\rho}\tilde\eta_{\pm})+\frac{R^4}{(\rho^2+m^2)^2}M_{\pm}^2\tilde\eta_{\pm}\mp H\frac{4R^4m}{(\rho^2+m^2)^3}{{M_{01}}_{\pm}}\tilde\eta_{\pm}=0\ ,\label{eqn01Sm+mom}\\
&&{M_{01}}_{\pm}=\sqrt{M_{\pm}^2+k_{23}^2};~~~k_{23}\equiv\sqrt{k_2^2+k_3^2}\ .\nonumber
\end{eqnarray}
After going through the steps described in equations (\ref{varchan})-(\ref{groundstate}), equation (\ref{correction}) gets modified to:
\begin{equation}
\alpha_{1}=\frac{\bar M_0}{6}\sqrt{1+\frac{k_{23}^2}{M_0^2}}\ .
\end{equation}
Note that validity of the perturbative analysis suggests that $\alpha_1$ is of the order of $\alpha_0$ and therefore we can trust the above expression as long as $k_{23}$ is of the order of $M_0$. Now it is straightforward to obtain the correction to the spectrum:
\begin{equation}
M_{\pm}=M_{0}\pm\frac{H}{m}\sqrt{1+\frac{k_{23}^2}{M_0^2}}\ .
\label{ZeemanGen}
\end{equation}
We see that the addition of momentum along the $(x_{2}-x_{3})$-plane enhances the splitting of the states. Furthermore, the spectrum depends continuously on $k_{23}$.

\subsection*{2.3.3 \hspace{2pt} Fluctuations along $\Phi$ for a strong magnetic field}
\addcontentsline{toc}{subsection}{2.3.3 \hspace{0.15cm} Fluctuations along $\Phi$ for a strong magnetic field}

\begin{figure}[h] 
   \centering
  \includegraphics[ width=11cm]{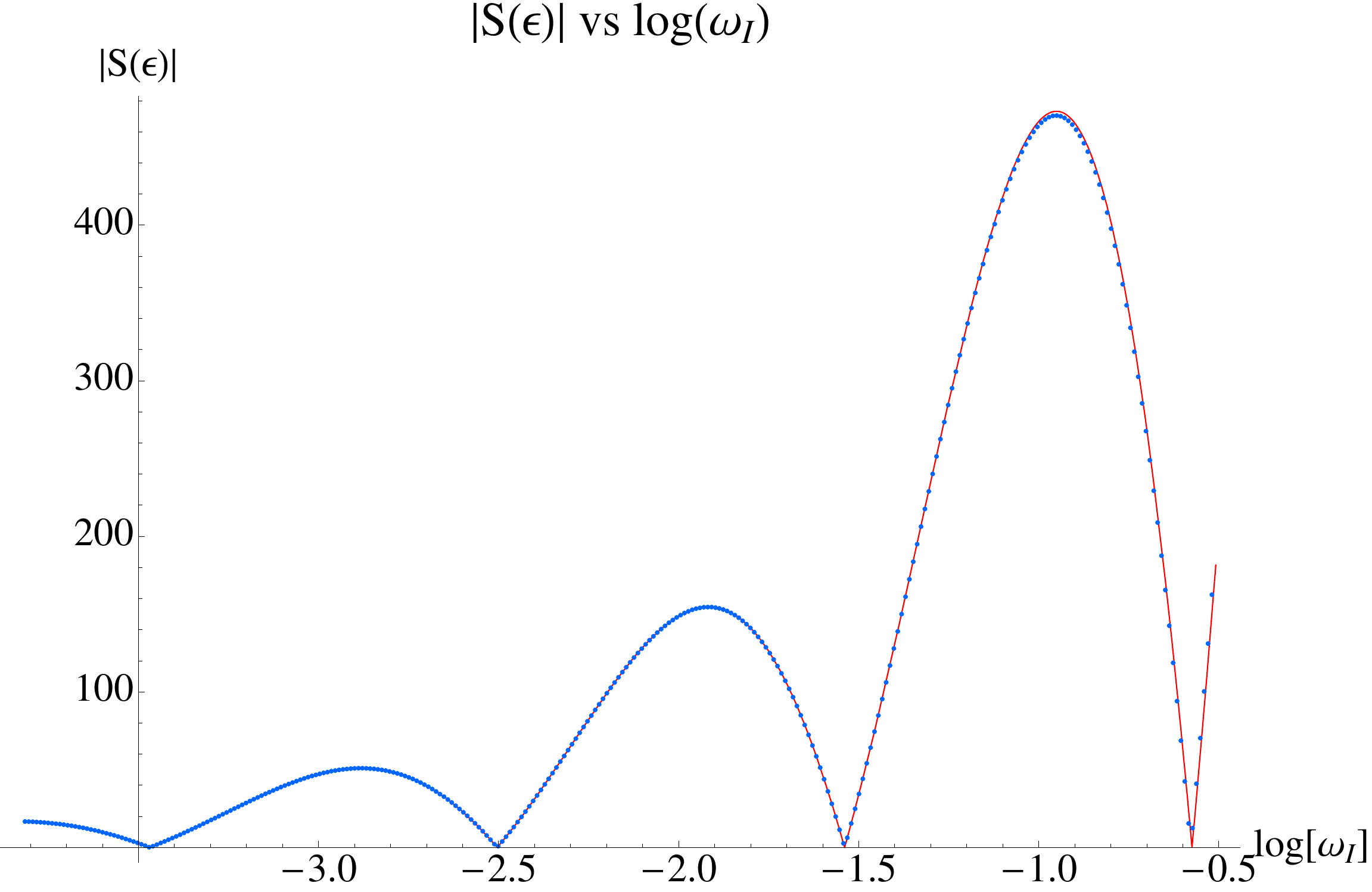}
  \caption{Spectrum of $\tilde m_{23}$ vs. $\tilde m$. The dashed line represents the lowest level of the meson spectrum
   for pure AdS$_{5}\times S^5$ space. }
   \label{fig:fig4}
\end{figure}

For strong magnetic field we have to take into account terms of order $H^2$, which means that we no longer have an expression for $L_{0}(\rho)$ in a closed form and we have to rely on numerical calculations only. Furthermore, there is no obvious way to decouple equations (\ref{eqnPhi}) and~(\ref{Elec}). However, it is still possible to extract information about the spectrum of the scalar modes if we restrict ourselves to fluctuations along the $(x^2, x^3)$ plane. In this way there is no source term in equation (\ref{EqGauge}), and we can consistently set $F_{01}$ equal to zero. We consider time independent fluctuations satisfying the ansatz $e^{-m_{23}r_{23}}$, (where $r_{23}$ is the radial coordinate in the $(x_2-x_3)$-plane). The damping factor in the exponent can be thought of as the mass of the scalar meson in 2 Euclidean dimensions. Indeed, let us consider the ansatz:
\begin{equation}
\Phi=h(\rho)e^{-ik_2x^2-ik_3x^3}Y_l(S^3)\ ,
\end{equation}
where $Y_{l}(S^3)$ are the spherical harmonics on the $S^3$ sphere satisfying: $\Delta_{\Omega_3}Y_l=-l(l+2)Y_l$.  With this set-up the equation of motion for $\Phi$, equation (\ref{eqnPhi}), reduces to equation for $h(\rho)$:
\begin{eqnarray}
\frac{1}{g}\partial_\rho\left(\frac{{g}L_0^2\partial_\rho h(\rho)}{1+{L'_0}^2}\right)-\frac{L_0^2 l(l+2)}{\rho^2}h(\rho)+
\frac{R^4L_0^2m_{23}^2}{(\rho^2+L_0^2)^2+R^4H^2}h(\rho)=0\ ,
\label{eqnPhiN}
\end{eqnarray}
where we have defined:
\begin{equation}
m_{23}^2=-k_2^2-k_3^2\ .
\end{equation}
Before we proceed with the numerical analysis of equation (\ref{eqnPhiN}) let us introduce dimensionless variables by performing the transformation (\ref{cordchange}) and defining:
\begin{equation}
\tilde m_{23} =\frac{R}{\sqrt{H}}m_{23}\ .
\label{m23}
\end{equation}
The resulting equation is:
\begin{eqnarray}
&&\frac{\tilde\rho^2+\tilde L^2}{{\tilde\rho}^3\sqrt{1+\tilde{L'}^2}(1+({\tilde\rho}^2+\tilde L^2)^2)^{1/2}}\partial_{\tilde\rho}\left(\tilde\rho^3\left(1+\frac{1}{(\tilde\rho^2+\tilde L^2)^2}\right)^{1/2}\frac{\tilde L^2}{\sqrt{1+L'^2}}\partial_{\tilde\rho}h(\tilde\rho)\right)+\nonumber\\
&-&\frac{\tilde L^2}{\tilde\rho^2}l(l+2)h(\tilde\rho)+\frac{\tilde L^2\tilde m_{23}^2}{(\tilde\rho^2+\tilde L^2)^2+1}h(\tilde\rho)=0\ .
\label{eqnPhiD}
\end{eqnarray}
In order to study the spectrum we look for normalizable solutions which have asymptotic behavior $h(\tilde\rho)\propto 1/\tilde\rho^2$ for large $\tilde\rho$ and satisfy the following boundary conditions at $\tilde\rho=0$:
\begin{equation}
h'(0)=0;~~h(0)=1\ .
\label{B.C}
\end{equation}
Let us consider first the lowest level of the spectrum. The spectrum that we get as a function of the bare quark mass is plotted in Figure~\ref{fig:fig4}.

\begin{figure}[h] 
   \centering
  \includegraphics[ width=11cm]{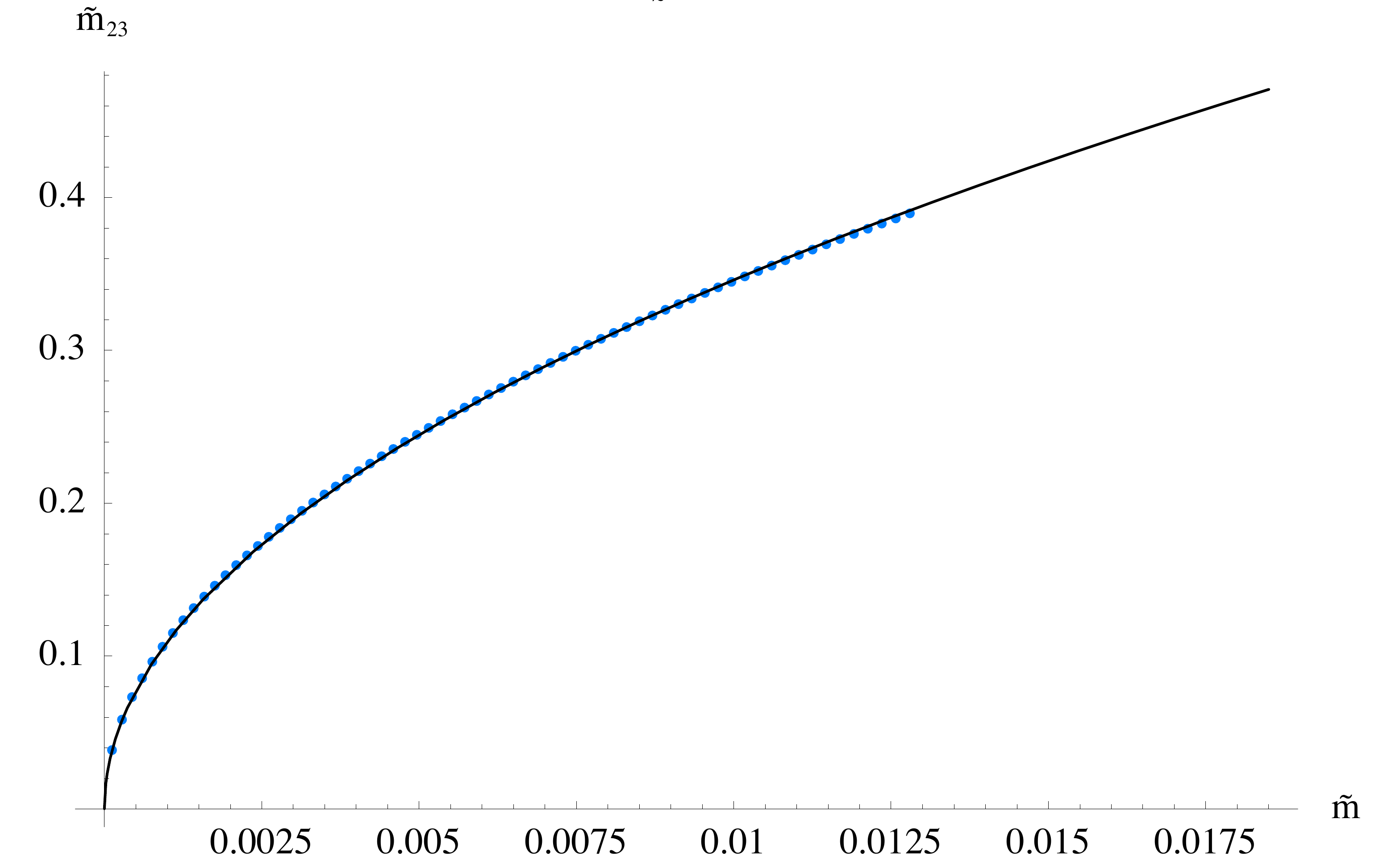}
   \caption{Enlargement of part of the spectrum of $\tilde m_{23}$ vs $\tilde m$ from Figure~\ref{fig:fig4}. The black solid curve shows the  $\propto\sqrt{\tilde m}$ fit. }
   \label{fig:Mparal-zoomed}
\end{figure}

 \begin{figure}[h] 
   \centering
  \includegraphics[ width=11cm]{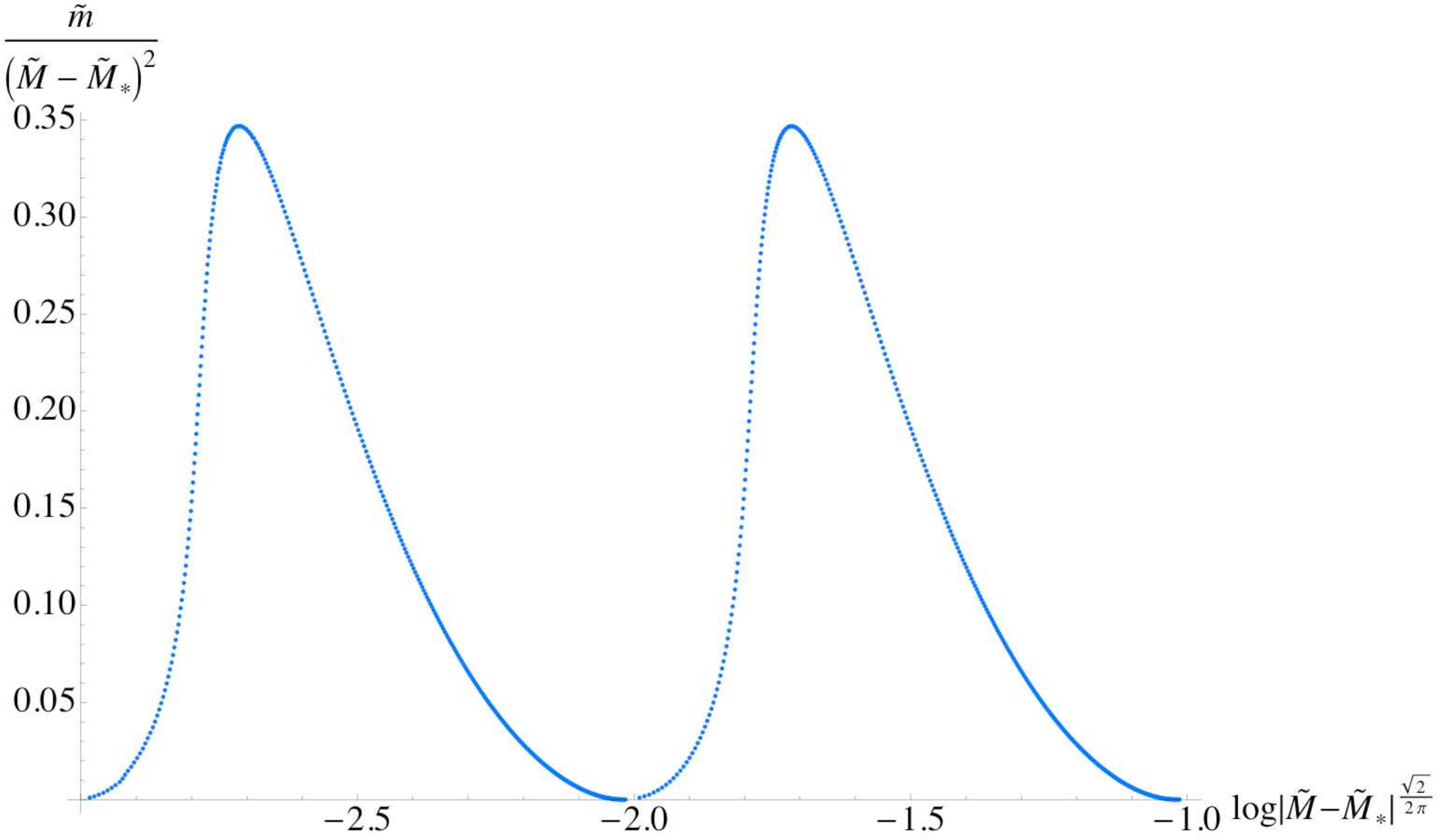}
   \caption{Spectrum of $\tilde m_{23}$ vs $\tilde m$ for $n=0\dots 4$. The dashed lines represent the spectrum for $AdS_5\times S^5$ space. }
   \label{fig:m23-n0-n4}
\end{figure}

For large $\tilde m$ the spectrum asymptotes (the dashed line in Figure~\ref{fig:fig4}) to the one  for pure $AdS_{5}\times S^5$ space obtained
by ref.~\cite{Kruczenski:2003be}
\begin{equation}
M_0=\frac{2m}{R^2}\sqrt{(n+l+1)(n+l+2)}\ ,
\label{spectrAds}
\end{equation}
with the substitution $n=0, l=0$, to obtain our case. Therefore we are describing the lowest possible state of the meson spectrum. In figure {\ref{fig:Mparal-zoomed}} we have zoomed in the area near the origin of the
 $(\tilde m,\tilde m_{23})$-plane, one can see that for small values of $\tilde m=2\pi\alpha' m_q/R\sqrt{H}$ we observe $\propto\sqrt{m_{q}}$ dependence of the ground state on the bare quark mass $m_q$, which is to be expected since the chiral symmetry associated with the spinor representation of SO(2) is spontaneously broken \cite{GellMann:1968rz}.

It is interesting to look for modes corresponding to higher excited states (non--zero $n$). In Figure~\ref{fig:m23-n0-n4} we have presented a plot of some of these. Again, the dashed line correspond to the pure $AdS_{5}\times S^5$ spectrum given by (\ref{spectrAds}) for $l=0$.  For small values of $\tilde m$ one can observe the qualitative difference of the behavior of the spectrum corresponding to the $n=0$ state from that of the higher excited states. Indeed, as $\tilde m\to 0 $ the $n=0$  states follow the $\sqrt{\tilde m}$ behavior plotted in Figure~\ref{fig:Mparal-zoomed}, while the excited states tend to some finite values at zero bare quark mass. The $n=0$ states merge into the Goldstone boson of the spontaneously broken chiral symmetry.

\subsection*{2.3.4 \hspace{2pt} The critical $\tilde L\equiv 0$ embedding}
\addcontentsline{toc}{subsection}{2.3.4 \hspace{0.15cm} The critical $\tilde L\equiv 0$ embedding}

In this section we study the $\tilde L\equiv 0$ embedding and in particular the spectrum of the fluctuations along the $\tilde L$ coordinate. Let us go back to dimensionful coordinates and consider the following change of coordinates in the transverse $R^6$ space:

\begin{eqnarray}
\rho=u\cos\theta\ ,\\
L=u\sin\theta\nonumber\ .
\end{eqnarray}
In these coordinates the trivial embedding corresponds to $\theta\equiv 0$ and in order to study the quadratic fluctuations we perform the expansion:
\begin{eqnarray}
\theta=0+(2\pi\alpha')\delta\theta(t,u)\ ,\\
\delta\theta=e^{-i\Omega t}h(u)\ .
\end{eqnarray}

Note that in order to study the mass spectrum we restrict the D7--brane to fluctuate only in time. In a sense this corresponds to going to the rest frame. Note that due to the presence of the magnetic field there is a coupling of the scalar spectrum to the vector one. However, for the fluctuations along $\theta$ the coupling depends on the momenta in the $(x_2,x_3)$ plane and this is why considering the rest frame is particularly convenient .

\begin{figure}[h] 
   \centering
   \includegraphics[ width=11cm]{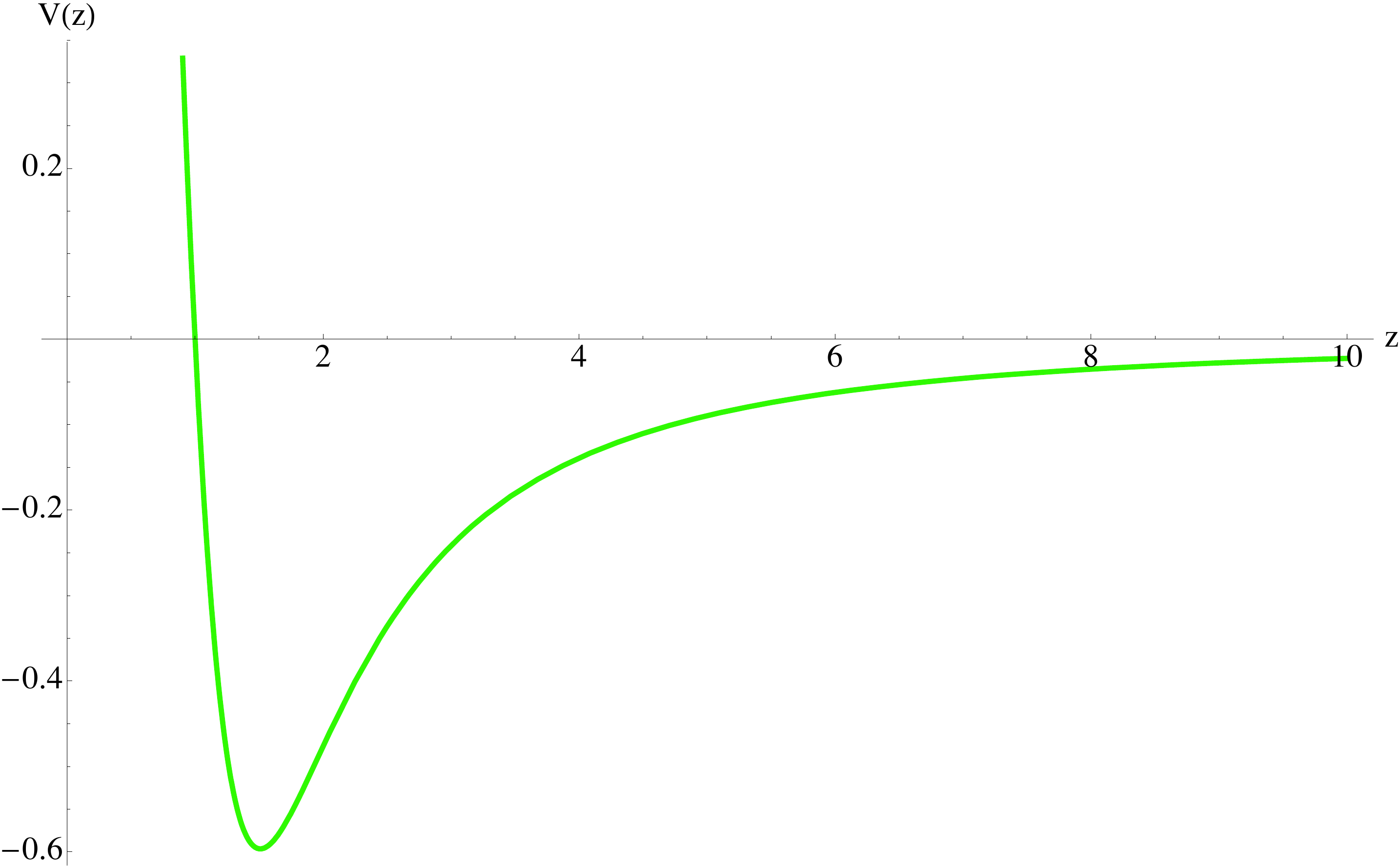}
   \caption{A plot of the effective potential $V(z)$ given in equation (\ref{Potential}). }
   \label{fig:Potential}
\end{figure}

Our analysis follows closely the one considered in ref.~\cite{Hoyos:2006gb}, where the authors have calculated the quasinormal modes of the D7--brane embedding in the AdS-black hole background by imposing an in-going boundary condition at the horizon of the black hole. Our case is the $T\to0$ limit and the horizon is extremal. However, the $\theta\equiv 0$ embedding can still have quasinormal excitations  with imaginary frequencies, corresponding to a real wave function so that there is no flux of particles falling into the zero temperature horizon.
The resulting equation of motion is:
\begin{equation}
h''+\left(\frac{3}{u}+\frac{2 u^3}{u^4+R^4H^2}\right)h'+\left(\frac{R^4}{u^4}\omega^2+\frac{3}{u^2}\right)h=0\ .
\end{equation}
It is convenient to introduce the following dimensionless quantities:
\begin{equation}
z=\frac{R}{u}\sqrt{H};~~~\omega=\frac{\Omega R}{\sqrt{H}}\ ,
\end{equation}
and make the substitution \cite{Hoyos:2006gb}
\begin{equation}
h(z)=\sigma(z)f(z);~~~\frac{\sigma'(z)}{\sigma(z)}=\frac{1}{2z}+\frac{1}{z(1+z^4)}\ ,
\end{equation}
leading to the equation for the new variable $f(z)$:
\begin{equation}
f''(z)+\left(\omega^2-V(z)\right)f(z)=0\ .
\label{Shr}
\end{equation}
Where the effective potential is equal to:
\begin{equation}
V(z)=\frac{3}{4z^2}\frac{(1+3z^4)(1-z^4)}{(1+z^4)^2}\ .
\label{Potential}
\end{equation}
The potential in (\ref{Potential}) goes as $\frac{3}{4z^2}$ for $z\to 0$ and as $-\frac{9}{4z^2}$ for $z\to\infty$ and is presented in Figure~\ref{fig:Potential}. As it was discussed in ref.~\cite{Hoyos:2006gb} if the potential gets negative the imaginary part of the frequency may become negative. Furthermore, the shape of the potential suggests that there might be bound states with a negative $\omega^2$. To obtain the spectrum we look for regular solutions of  (\ref{Shr}) imposing an in-falling boundary condition at the horizon ($z\to\infty$).

The asymptotic form of the equation of motion at $z\to\infty$ is that of the harmonic oscillator:
\begin{equation}
f''(z)+\omega^2f(z)=0\ ,
\end{equation}
with the solutions $e^{\pm i\omega z}$, the in-falling boundary condition implies that we should choose the positive sign. In our case the corresponding spectrum turns out to be tachyonic and hence the exponents are real. Therefore the in-falling boundary condition simply means that we have selected the regular solution at the horizon: $z\to \infty$. We look for a solution of the form:
\begin{equation}
f(z)=e^{+i\omega z}S(z)\ .
\end{equation}
The resulting equation of motion for $S(z)$ is:
\begin{equation}
(-3-6z^4+9z^8)S(z)+4z^2(1+z^4)^2\left(2i\omega S'(z)+S''(z)\right)=0\ .
\label{eqnS}
\end{equation}

Next we study numerically equation (\ref{eqnS}). After solving the asymptotic form of the equation at the Horizon, we impose the following boundary condition at $z=1/\epsilon$, where $\epsilon$ is a numerically small number typically $\epsilon=10^{-9}$ :
\begin{equation}
S(1/\epsilon)=1-\frac{9i\epsilon}{8\omega};~~~S'(1/\epsilon)=\frac{9i\epsilon^2}{8\omega}\ ,
\label{initialcond}
\end{equation}
after that we explore the solution for a wide range of $\omega=i\omega_{I}$. We look for regular solutions which have $|S(\epsilon)|\approx 0$, this condition follows from the requirement that $\chi \propto z^3$ as $z\to 0$. It turns out that regular solutions exist for a discrete set of positive $\omega_{I}\ll1$. The result for the first six modes that we obtained is presented in table \ref{tab:1}.

    \begin{table}[h]
\begin{center}
\begin{tabular}{|c|c|c|}
\hline
 $n$&$\omega_{I}^{(n)}$&$\omega_{I}^{(n)}/\omega_{I}^{(n-1)}$\\\hline
 0&$2.6448\times10^{-1}$&-\\\hline
 1&$2.8902\times10^{-2}$&0.10928\\\hline
 2&$3.1348\times10^{-3}$&0.10846\\\hline
 3&$3.3995\times10^{-4}$&0.10845\\\hline
 4&$3.6865\times10^{-5}$&0.10844\\\hline
 5&$3.9967\times10^{-6}$&0.10841\\\hline
\end{tabular}
\end{center}
 \caption{Numerical data for the first six quasi-normal modes of the critical embedding. The data suggests that as $n\to\infty$ the states organize in a decreasing geometrical series. }
\label{tab:1}
\end{table}%

\begin{figure}[h] 
   \centering
   \includegraphics[ width=11cm]{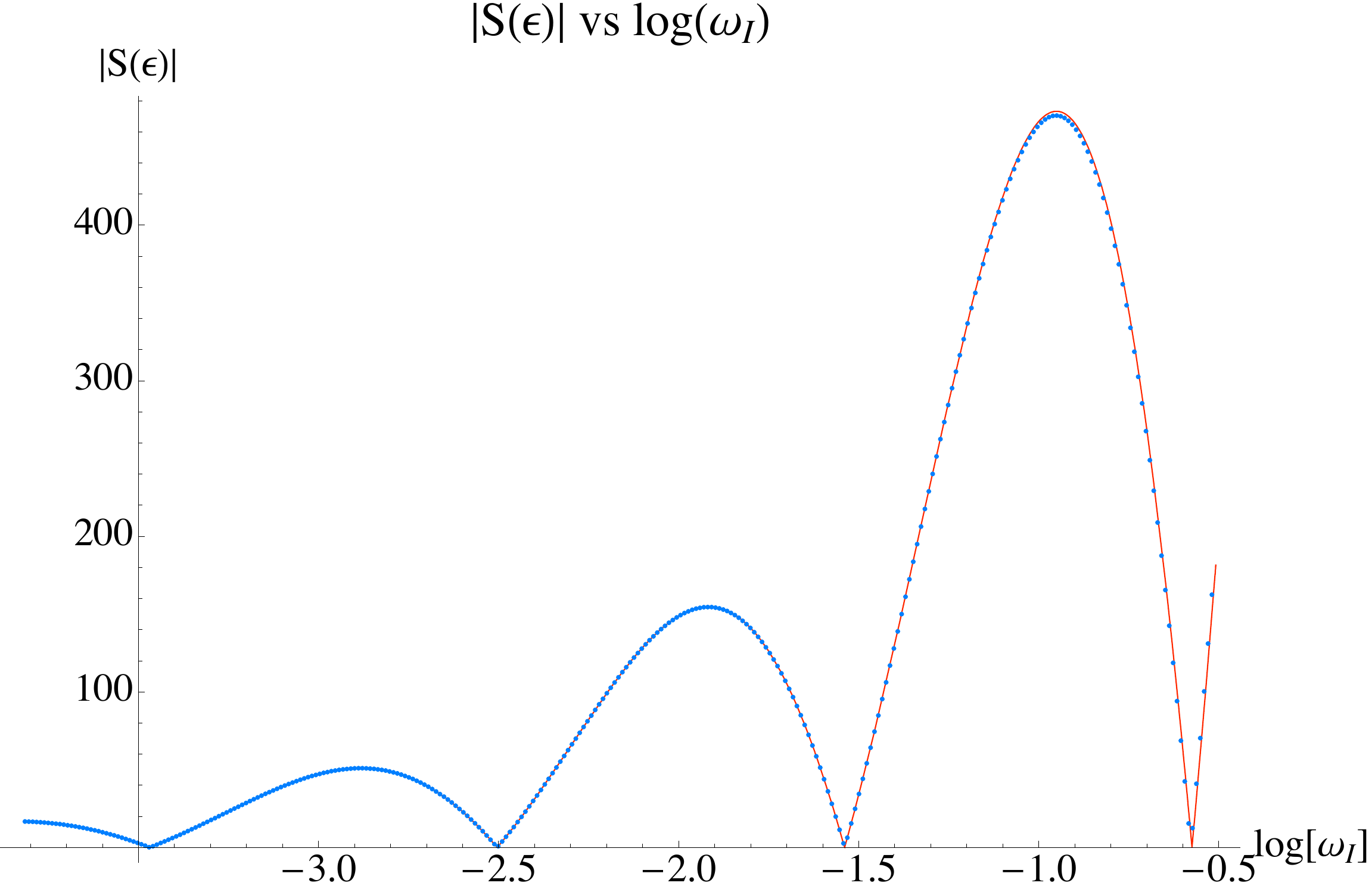}
   \caption{The dotted blue curve corresponds to the numerical solution of equation (\ref{eqnS}) while the thick red curve is the one determined by equation (\ref{Shat2}). The plots are scaled to match along the vertical axis. }
   \label{fig:spectrum}
\end{figure}

The data suggests that as $\omega_{I}\to 0$ the states organize in a decreasing geometrical series with a factor $q\approx0.1084$. Up to four significant digits, this is the number from equation (\ref{q}) which determines the period of the spiral. We can show this analytically. To this end let us consider the rescaling of the variables in equation (\ref{eqnS}) given by:
\begin{eqnarray}
z=\lambda \hat z;~~~\hat\omega=\omega/\lambda;~~~\lambda\to\infty\ .
\end{eqnarray}
This is leading to:
\begin{equation}
9\hat S(\hat z)+4\hat z^2(2i\hat\omega\hat S'(\hat z)+\hat S''(\hat z))+O({\lambda}^{-4})=0\ .
\label{small}
\end{equation}
The solution consistent with the initial conditions at infinity  (\ref{initialcond}) can be found to be:
\begin{equation}
\hat S(\hat z)=\frac{1+i}{2} e^{-i\frac{\pi}{\sqrt{2}}}e^{-i\hat z\hat\omega}\sqrt{\pi\hat z\hat\omega}H_{i\sqrt{2}}^{(1)}(\hat z\hat\omega);~~~\hat\omega=i\hat\omega_{I}\ ,
\label{Shat}
\end{equation}
where $H_{i\sqrt{2}}^{(1)}$ is the Hankel function of the first kind. Our next assumption is that in the $\omega_I\to 0 $ limit, this asymptotic form of the equation describes well enough the spectrum. To quantize the spectrum we consider some $\hat z_0=z_0/\lambda\ll1$, where we have $1\ll z_0\ll\lambda$ so that the simplified form of equation (\ref{small}) is applicable and impose:
\begin{equation}
\hat S(\hat z_0)=0\ .
\label{Shat2}
\end{equation}
Using that $\hat z\hat\omega=iz\omega_I$ this boils down to:
\begin{equation}
H_{i\sqrt{2}}^{(1)}(i\omega_I z_0)=0\ .
\end{equation}
Now using that $\omega_I z_0\ll1$ for a sufficiently small $\omega_I$, we can make the expansion:
\begin{equation}
H_{i\sqrt{2}}^{(1)}(i\omega_I z_0)\approx-A_1\left((\omega_I z_0)^{i\sqrt{2}}-(\omega_I z_0)^{-i\sqrt{2}}\right)+iA_2\left((\omega_I z_0)^{i\sqrt{2}}+(\omega_I z_0)^{-i\sqrt{2}}\right)\ ,
\end{equation}
where $A_1$ and $A_2$ are real numbers defined {\it via}:
\begin{equation}
A_1+iA_2=-\frac{1}{\pi}{i(i/2)^{-i\sqrt{2}}\Gamma(i\sqrt{2})}\ .
\end{equation}
This boils down to:
\begin{equation}
\cos(\sqrt{2}\ln(\omega_I z_0)+\phi)=0;~~~\phi\equiv\pi/2-\arg(A_1+iA_2)\ .
\label{20}
\end{equation}
The first equation in (\ref{20}) leads to:
\begin{equation}
\omega_I^{(n)}=\frac{1}{z_0}e^{-\frac{\pi/2+\phi}{\sqrt{2}}}e^{-n\frac{\pi}{\sqrt{2}}}=\omega_I^{(0)}q^n\ ,
\label{geom}
\end{equation}
suggesting that:
\begin{equation}
q=e^{-\frac{\pi}{\sqrt{2}}}\approx 0.10845\ .
\label{an_q}
\end{equation}
This is the number given in (\ref{q}). Note that the value of $z_0$ is a free parameter that we can fix by matching equation (\ref{geom}) to the data in table \ref{tab:1}. On the other side, $\hat S(\hat z)$ given in equation (\ref{Shat}) depends only on $\hat z\hat\omega=i\omega_I z$ and therefore once we have fixed $z_0$ we are left with a function of $\omega_I$ which zeroes determine the spectrum, equation (\ref{Shat2}). It is interesting to compare it to the numerically obtained plot of $|S(\epsilon)|$ vs. $\omega_I$, that we have used to determine the spectrum numerically. The result is presented in Figure~\ref{fig:spectrum}, where we have used the $n=3$ entry from table \ref{tab:1} to fix $z_0$. One can see the good agreement between the spectrum determined by equation (\ref{Shat2}), the red curve in Figure~\ref{fig:spectrum} and the numerically determined one, the dotted blue curve.

\subsection*{2.3.5 \hspace{2pt} The spectrum near criticality}
\addcontentsline{toc}{subsection}{2.3.5 \hspace{0.15cm} The spectrum near criticality}

In this section we study the light meson spectrum of the states forming the spiral structure in the $(\tilde m, -\tilde c)$ plane, Figure~\ref{fig:spiral-revisited}. In particular we focus on the study of the fluctuations along $ L$. The corresponding equation of motion is given in (\ref{teschi}). The effect of the magnetic field $H$ is to mix the vector and the meson parts of the spectrum. However, if we consider the rest frame by allowing the fluctuations to depend only on the time direction of the D3--branes' world volume, the equation of motion for the fluctuations along $ L$ decouple from the vector spectrum. To this end we expand:
\begin{eqnarray}
L=L_0(\rho)+(2\pi\alpha')\chi(\rho,t)\ ,\\
\chi=h(\rho)\cos{M t}\nonumber\ .
\end{eqnarray}
Here $L_0(\rho)$ is the profile of the D7--brane classical embedding. The resulting equation of motion for $h(\rho)$ is:
\begin{eqnarray}
&\partial_\rho(g\frac{h'}{(1+L_0'^2)^2})+\left(g\frac{R^4}{(\rho^2+L_0^2)^2}\frac{M^2}{1+L_0'^2}-\frac{\partial^2 g}{\partial L_0^2}+\partial_\rho(\frac{\partial g}{\partial L_0}\frac{L_0'}{1+L_0'^2})\right)h=0\ ,\\
{\rm where}\quad&g(\rho,L_0,L_0')=\rho^3\sqrt{1+{L_0}'^2}\sqrt{1+\frac{R^4H^2}{(\rho^2+L_0^2)^2}}\ .\nonumber
\end{eqnarray}
It is convenient to introduce the dimensionless variables:
\begin{equation}
\tilde h=\frac{h}{R\sqrt{H}};~~\tilde L_{0}=\frac{L_0}{R\sqrt{H}};~ \tilde\rho=\frac{\rho}{R\sqrt{H}};~\tilde M=\frac{M R}{\sqrt{H}}\ ,
\label{dimensionless}
\end{equation}
leading to:
\begin{eqnarray}
&\partial_{\tilde\rho}(\tilde g\frac{\tilde h'}{(1+\tilde L_0'^2)^2})+\left(\tilde g\frac{1}{(\tilde\rho^2+\tilde L_0^2)^2}\frac{\tilde M^2}{1+\tilde L_0'^2}-\frac{\partial^2 \tilde g}{\partial \tilde L_0^2}+\partial_{\tilde\rho}(\frac{\partial \tilde g}{\partial \tilde L_0}\frac{\tilde L_0'}{1+\tilde L_0'^2})\right)\tilde h=0\ ,\label{fluctPsi}\\
{\rm with}\quad&\tilde g(\tilde\rho,\tilde L_0,\tilde L_0')=\tilde\rho^3\sqrt{1+{\tilde L_0}'^2}\sqrt{1+\frac{1}{(\tilde\rho^2+\tilde L_0^2)^2}}\nonumber\ .
\end{eqnarray}

We study the normal modes of the D7--brane described by equation (\ref{fluctPsi}) by imposing Neumann boundary conditions at $\tilde\rho=0$. Since our analysis is numerical we solve the equation of motion (\ref{fluctPsi}) in terms of a power series for small $\tilde\rho$ and impose the appropriate initial conditions for the numerical solution at $\tilde\rho=\epsilon$, where $\epsilon$ is some very small number. In order to quantize the spectrum we look for numerical solutions which are normalizable and go as $1/\tilde\rho^2$ at infinity.

\begin{figure}[htbp] 
   \centering
   \includegraphics[ width=11cm]{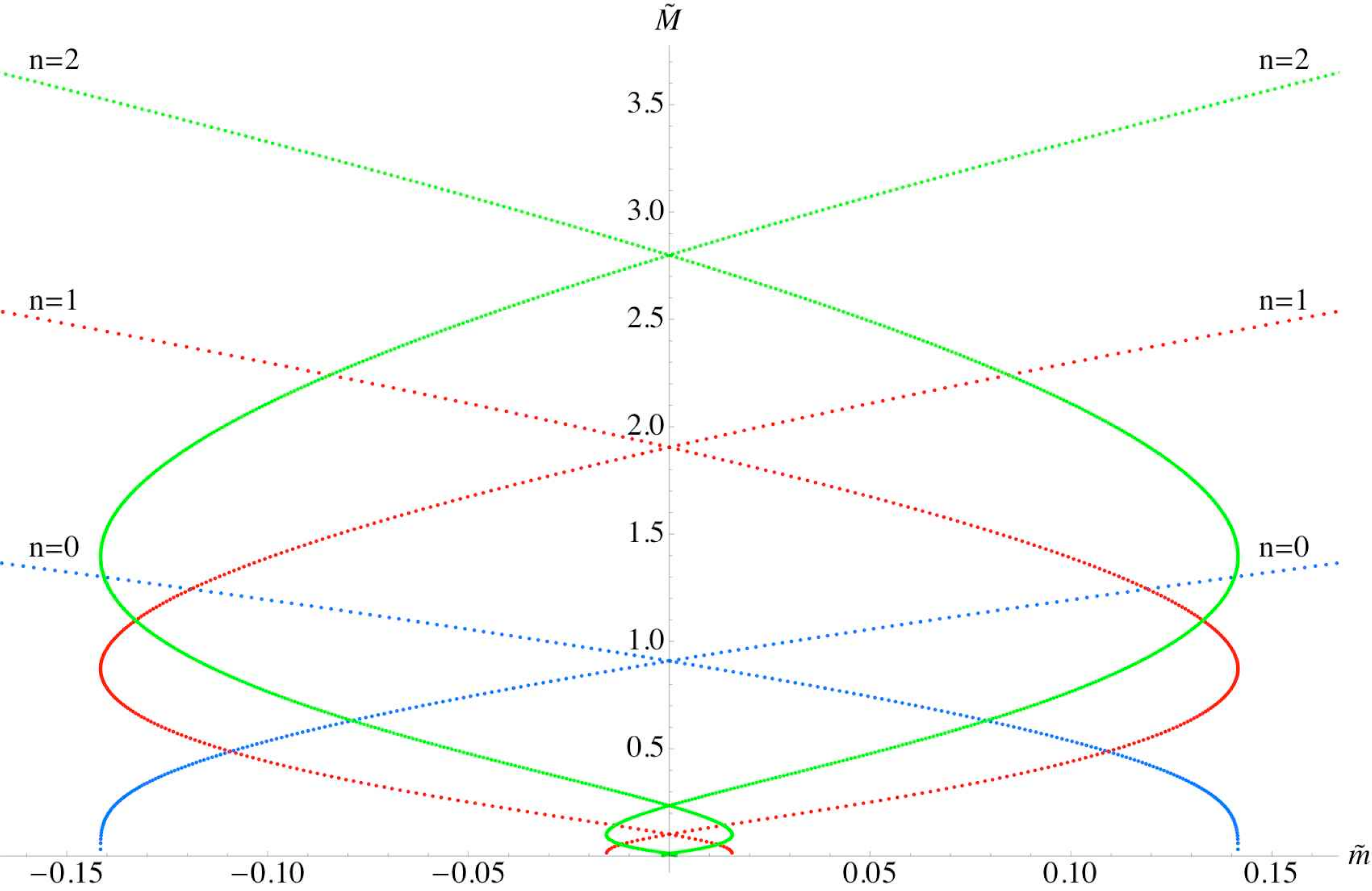}
     \includegraphics[ width=11cm]{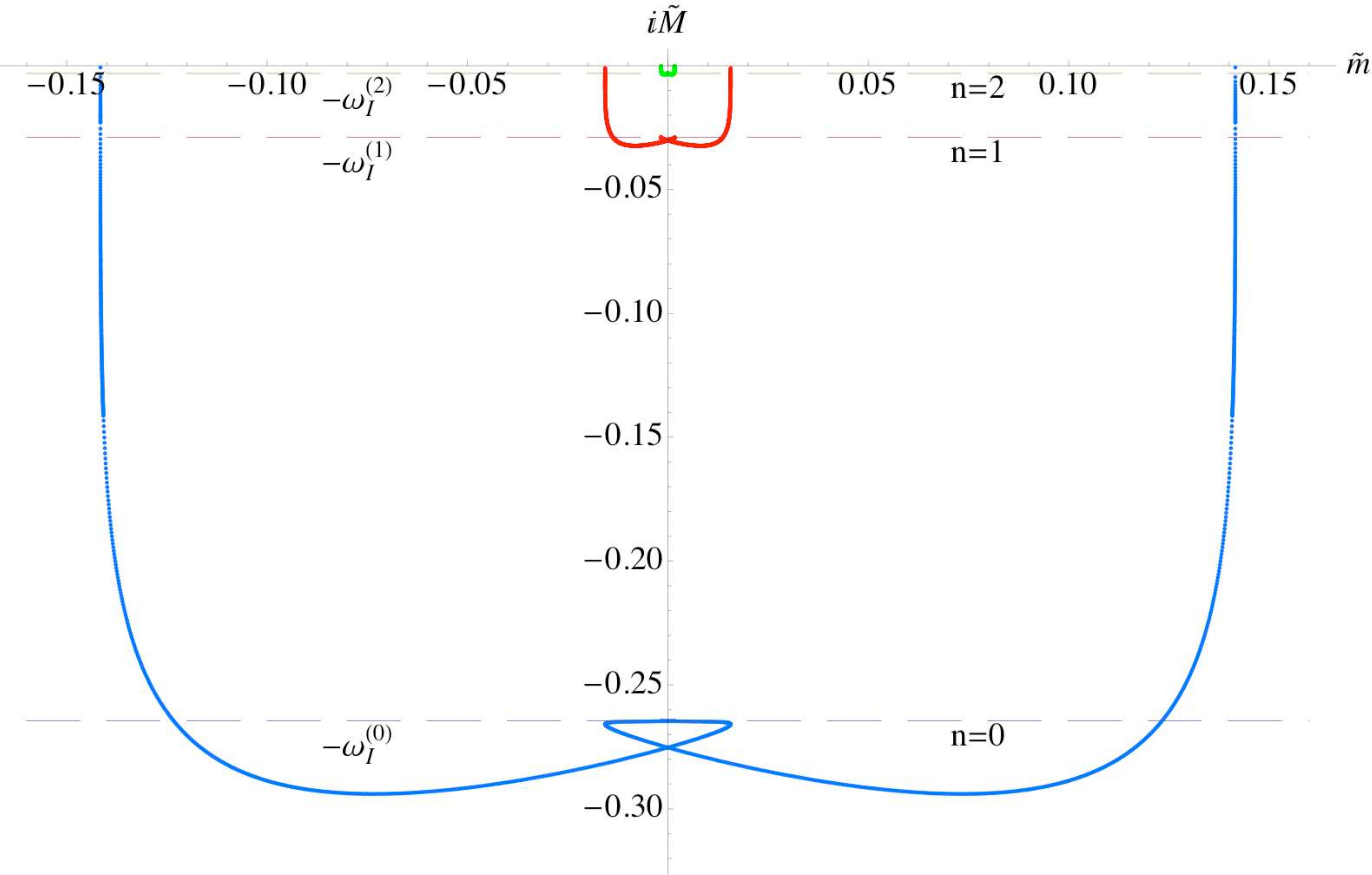}
   \caption{A plot of the meson spectrum corresponding to the two arms of the spiral structure at the origin of the $(\tilde m,-\tilde c)$ plane. The ground state ($n=0$) becomes tachyonic for the inner branches of the spiral while only the lowest branch is a tachyon free one. The tachyon sector of the diagram reveals the self-similar structure of the spectrum. }
\label{fig:spectrum-spiral}
\end{figure}

 Let us study the dependence of the spectrum of $\tilde M$ on the bare quark mass $\tilde m$, for the states corresponding to the spiral structure from Figure~\ref{fig:spiral-revisited}. A plot of the spectrum of the first three excited states is presented in Figure~\ref{fig:spectrum-spiral}. The classification of the states in terms of the quantum number $n$ is justified because at large $\tilde m$ the equation of motion for the fluctuations asymptotes to the equation of motion for the pure $AdS_5\times S^5$ space, considered in ref.~\cite{Kruczenski:2003be}, where the authors obtained the spectrum in a closed form. Note that the diagram has a left-right symmetry. This is because we plotted the spectrum for both arms of the spiral in order to emphasize its self-similar structure, physically only one side of the diagram is sufficient.

 Let us trace the blue curve corresponding to the $n=0$ state starting from the right-hand side. As $\tilde m$ decreases the mass of the meson decreases and at $\tilde m=0$ it has some non--zero value. This part of the diagram corresponds to the lowest positive branch of the spiral from Figure~\ref{fig:spiral-revisited} (the vicinity of point $H_0$).
 It is satisfying to see that the lowest positive $\tilde m$ branch of the spiral is tachyon free and therefore stable under quantum fluctuations. Note that despite that the negative $\tilde m$ part of the lowest branch has no tachyonic modes in its fluctuations along $L$, it has a higher free energy (as can be seen from Figure~\ref{fig:free-energy}) and is thus at best metastable.

 One can also see that the spectrum drops to a zero and becomes tachyonic exactly at the point where we start exploring the upper branch of the spiral. This proves that all inner branches correspond to true instability of the theory and cannot be reached by super-cooling. As we go deeper into the spiral, the $n=0$ spectrum remains tachyonic and spirals to some critical value. The dashed line denoted by $\omega_I^{(0)}$ in Figure~\ref{fig:spiral-revisited}  corresponds to the first entry in table \ref{tab:1}. As one can see this is the critical value approached by the spectrum.

 Now let us comment on the $n=1,2$ levels of the spectrum represented by the red and green curves, respectively. As one can see the $n=1$ spectrum becomes tachyonic when we reach the third branch of the spiral (the vicinity of point $H_2$ in Figure~\ref{fig:spiral-revisited}) and after that follows the same pattern as the $n=0$ level, spiraling to the second entry $\omega_I^{(1)}$ from table \ref{tab:1}. The $n=2$ level has a similar behavior, but it becomes tachyonic at the next turn of the spiral and it approaches the next entry from table \ref{tab:1}. Similar feature was reported in ref.~\cite{Mateos:2007vn} where the authors  studied topology changing transitions. The above analysis suggests that at each turn of the spiral, there is one new tachyonic state appearing. It also suggests that the structure of the $n$-th level is similar to the structure of the $n+1$-th level and in the $n\to\infty$ limit this similarity becomes an exact discrete self-similarity. The last feature is apparent from the tachyonic sector of the diagram in the second plot in Figure~\ref{fig:spectrum-spiral}, the blue, red and green curves are related by an approximate scaling symmetry, the analysis of the spectrum of the critical $L\equiv 0$ embedding suggests that this symmetry becomes exact in the $n\to\infty$ limit with a scaling factor of $q$ given  in equation (\ref{q}).

 It is interesting to analyze the way the meson mass $\tilde M$ approaches its critical value and compute the corresponding critical exponent. Let us denote the critical value of $\tilde M$ by $\tilde M_*$ and consider the bare quark mass $\tilde m$ as an order parameter, denoting its critical value by $\tilde m_*$. We are interested in calculating the critical exponent $\alpha$ defined by:
 \begin{equation}
| \tilde M-\tilde M_* |\propto|\tilde m-\tilde m_*|^\alpha\ .
\label{critexp}
\end{equation}
We will provide a somewhat heuristic argument that $\alpha=2$ and will confirm this numerically. To begin with let us consider the energy density of the gauge theory $\tilde E$ as a function of the bare quark mass $\tilde m$. Now let us consider a state close to the critical one, characterized by:
\begin{equation}
 \tilde M=\tilde M_*+\delta\tilde M;~~~\tilde m=\tilde m_*+\delta\tilde m;~~~ \tilde E=\tilde E_*+\delta\tilde E\ .
\end{equation}
 Next we assume that as we approach criticality the variation of $\tilde E$ and $\tilde M$ are proportional to the variation of the energy scale and hence $\delta \tilde E\propto\delta\tilde M$. Therefore we have:
 \begin{equation}
 \frac{\delta \tilde M}{\delta \tilde m}\propto \frac{\delta\tilde E}{\delta \tilde m}\propto \tilde c\ ,
 \label{relation}
 \end{equation}
 where $\tilde c$ is the quark condensate. The second relation in (\ref{relation}) was argued in ref.~\cite{Kruczenski:2003uq}. In the previous section we argued that the critical exponent of the condensate is one and since the critical embedding has a zero condensate it follows that $\tilde c\propto |\tilde m-\tilde m_{*}|$. Therefore we have:
 \begin{equation}
  \frac{\delta \tilde M}{\delta \tilde m}\propto \alpha|\tilde m-\tilde m_*|^{\alpha-1}\propto |\tilde m-\tilde m_*|
  \end{equation}
 and hence $\alpha=2$.

 Now let us go back to Figure~\ref{fig:spectrum-spiral}. As we discussed above, for each energy level $n$ the tachyonic spectrum spirals to the critical value $\omega_I^{(n)}$, corresponding to the center of the spiral. If we focus on the $\tilde m=0$ axis, we can see that for each level we have a tower of tachyonic states at a zero bare quark mass, corresponding to the different branches of the spiral. Let us denote by $\tilde M_k^{(n)}$ the imaginary part of the meson spectrum, corresponding to the $k$-th tachyonic state of the $n$-th energy level, at a zero bare quark mass $\tilde m$. As we go deeper into the spiral, $k\to\infty$ and $\tilde M_k^{(n)}\to \tilde M_*^{(n)}$, the data in Figure~\ref{fig:spectrum-spiral} suggests that $\tilde M_*^{(n)}=\omega_I^{(n)}$. On the other side, if the meson spectrum has a critical exponent of two, one can show that for a large $k$:
 \begin{equation}
 \frac{\tilde M_k^{(n)}-\tilde M_*^{(n)}}{\tilde M_{k-1}^{(n)}-\tilde M_*^{(n)}}=q^2\ ,
 \label{geom2}
 \end{equation}
 where $q$ is given by equation (\ref{q}). We can solve for $\tilde M_{*}^{(n)}$:
 \begin{equation}
 \tilde M_{*}^{(n)}=\tilde M_{k-1}+\frac{\tilde M_k^{(n)}-\tilde M_{k-1}^{(n)}}{1-q^2}\ .
 \label{M*}
  \end{equation}

Now assuming that for $k=1,2$ the approximate geometrical series defined {\it via} (\ref{geom2}) is already exact we calculate numerically $\tilde M_1^{(n)}, \tilde M_2^{(n)}$ for the $n=0,1,2$ levels and compare the value of $\tilde M_{*}^{(n)}$ obtained by equation (\ref{M*}) to the first three entries in table \ref{tab:1}. The results are presented in table \ref{tab:2}.

\begin{table}[h]
    \begin{center}
\begin{tabular}{|c|c|c|c|c|}
\hline
 $n$&$\tilde M_{1}^{(n)}$&$\tilde M_{2}^{(n)}$&$\tilde M_{*}^{(n)}$&$\omega_I^{(n)}$\\\hline
 0&$2.7530\times10^{-1}$&$2.6460\times10^{-1}$&$2.6447\times10^{-1}$&$2.6448\times10^{-1}$\\\hline
 1&$3.0162\times10^{-2}$&$2.8917\times10^{-2}$&$2.8902\times10^{-2}$&$2.8902\times10^{-2}$\\\hline
 2&$3.2715\times10^{-3}$&$3.1363\times10^{-3}$&$3.1347\times10^{-3}$&$3.1348\times10^{-3}$\\\hline

\end{tabular}
\end{center}
\caption{Comparison  between the critical value $\tilde M_*^{(n)}$ for the n-th energy level of the meson spectrum and the corresponding quasi-normal  mode $\omega_I^{(n)}$ of the critical embedding. The data suggests that they match. }
\label{tab:2}
\end{table}%

\begin{figure}[h] 
    \centering
    \includegraphics[ width=11cm]{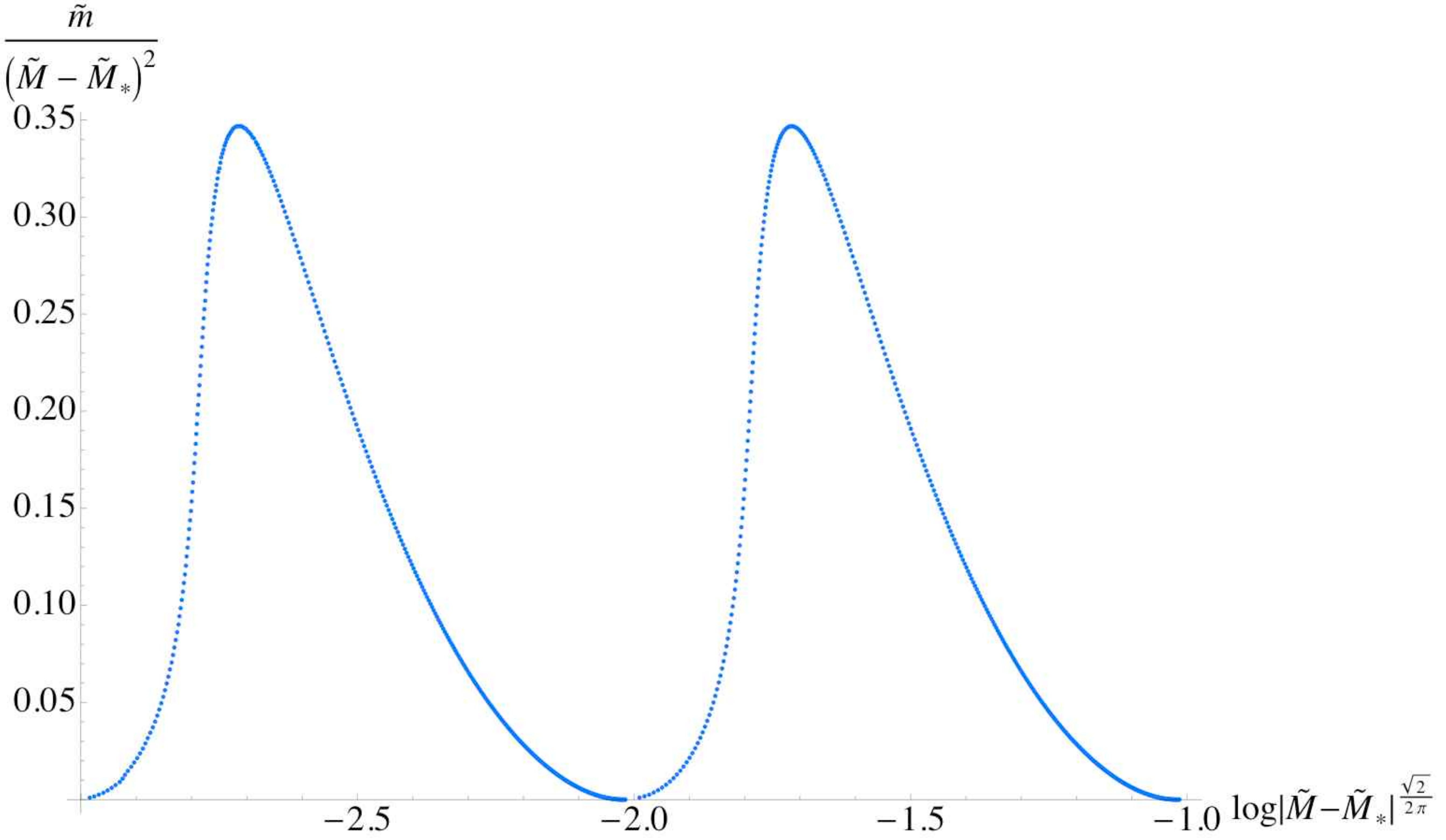}
    \caption{ A plot of the bare quark mass meson vs. the meson spectrum, in an appropriate parameterization, determined by the scaling exponents of $\tilde m$ and $\tilde M$. The discrete self-similar structure of the spectrum is manifested by the periodicity of the plotted function. }
    \label{fig:messpiral}
 \end{figure}

 One can see that up to four significant digits the critical value of the meson spectrum is given by the imaginary part of the quasi normal modes presented in table \ref{tab:1}. This supports the above argument that the meson spectrum has a critical exponent of two. Another way to justify this, is to generate a plot of the meson spectrum similar to the one presented in Figure~\ref{fig:mspiral} for the bare quark mass $\tilde m$ and the fermionic spectrum $\tilde c$. Notice that $\tilde M$ approaches criticality from above while the parameter $\tilde m$ oscillates around the critical value $\tilde m_*=0$. This suggests to use $\tilde M$ as an order parameter and to generate a plot of $\tilde m/(\tilde M-\tilde M_*)^2$ vs. $\sqrt{2}\log{|\tilde M-\tilde M_*|}/{2\pi}$. Note that according to equation (\ref{geom2}) the plot should represent periodic function of an unit period. The resulting plot for the $n=0$ level, using $\tilde M_*^{(0)}$ from table \ref{tab:2} as a critical value, is presented in Figure~\ref{fig:messpiral}.

\subsection*{2.3.6 \hspace{2pt} The stable branch of the spiral}
\addcontentsline{toc}{subsection}{2.3.6 \hspace{0.15cm} The stable branch of the spiral}
In this subsection we consider the spectrum corresponding to the states far from the origin of the $(\tilde m, -\tilde c)$ which define the outermost branch of the spiral ending at point $H_0$ from Figure~\ref{fig:spiral-revisited}. The fluctuations of the D7--brane corresponding to the massless scalar $\phi$ were studied in Subsection~2.3.3 and some features consistent with the spontaneous chiral symmetry breaking, such as a characteristic $\sqrt{m}$ behavior \cite{GellMann:1968rz} were reported.

 Here we complement the analysis by presenting the results for the fluctuations along the $\tilde L$ coordinate. Since this is the massive field in the spontaneous chiral symmetry breaking scenario, we  expect a $\sqrt{const+\tilde m}$ behavior of the meson spectrum for small values of $\tilde m$. Note that such a behavior simply means that the spectrum of the $\tilde L$ fluctuations has a mass gap at zero bare quark mass and that the slope of the spectrum vs. the bare quark mass function is finite. It is satisfying that our results are in accord with this expectations.

\begin{figure}[h] 
   \centering
   \includegraphics[ width=11cm]{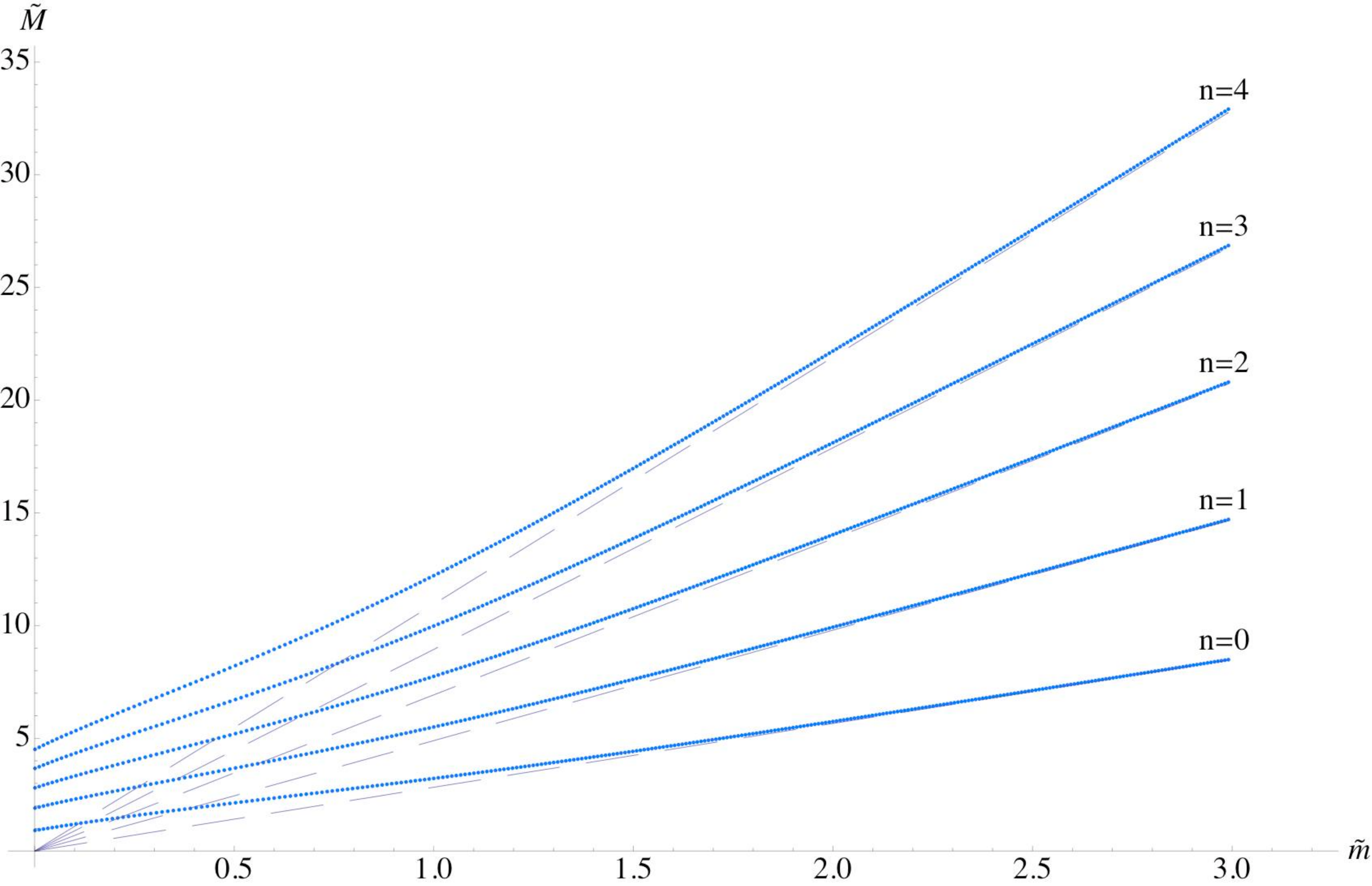}
   \caption{A plot of the meson spectrum corresponding to the stable branch of the spiral. The black dashed lines correspond to equation (\ref{dmlAdS}), one can see that for large $\tilde m$ the meson spectrum asymptotes to the result for pure $AdS_5\times S^5$ space. One can also see that at zero bare quark mass $\tilde m$ there is a mass gap in the spectrum. }
   \label{fig:final}
\end{figure}

 To obtain the spectrum, we solve numerically equation (\ref{fluctPsi}) imposing Neumann boundary conditions at $\tilde\rho=0$. A plot of the first five energy levels is presented in Figure~\ref{fig:final}. As one can see at large $\tilde m$ the spectrum approximates that of the pure ${\cal N}=2$ Flavored Yang Mills theory given in equation (\ref{spectrAds}) which we replicate here:
 \begin{equation}
M_0=\frac{2m}{R^2}\sqrt{(n+l+1)(n+l+2)}\ .
\label{spectrAds1}
\end{equation}
Here $l$ is the quantum number corresponding to the angular modes along the internal $S^3$ sphere wrapped by the D7--brane and is zero in our case. After introducing the dimensionless variables defined in (\ref{dimensionless}), equation (\ref{spectrAds1}) boils down to:
\begin{equation}
\tilde M_0= 2\sqrt{(n+1)(n+2)}\tilde m\ .
\label{dmlAdS}
\end{equation}
The black dashed lines in Figure~\ref{fig:final} represent equation (\ref{dmlAdS}). The fact that the meson spectrum asymptotes to the one described by (\ref{dmlAdS}) justifies the use of the quantum number $n$ to classify the meson spectrum. One can also see that as expected the spectrum at zero bare quark mass has a mass gap.

\section*{2.4 \hspace{2pt} Concluding remarks}
\addcontentsline{toc}{section}{2.4 \hspace{0.15cm} Concluding remarks}

In this chapter we performed a detailed analysis of the spiral structure at the origin of the condensate versus bare quark mass diagram. We revealed the discrete self-similar behavior of the theory near criticality and calculated the corresponding critical exponents for the bare quark mass, the quark condensate, and the meson spectrum.

 Our study of the meson spectrum confirmed the expectations based on thermodynamic considerations that the lowest positive $\tilde m$ branch of the spiral corresponds to a stable phase of the theory and that  the inner branches are real instabilities characterized by a tachyonic ground state and cannot be reached by a supercooling. The lowest negative $\tilde m$ branch of the spiral is tachyon free and thus could be metastable.

The supercooling mentioned above could be attempted by considering the finite temperature background, namely the AdS-Black hole geometry, in the presence of an external magnetic field. We could prepare the system in the phase corresponding to the trivial $\tilde L\equiv 0$ embedding and then take the $T\to0$ limit. If some of the inner branches of the spiral were metastable the theory could end up in the corresponding phase. The study of the finite temperature case is of a particular interest. Due to the additional scale introduced by the temperature, the theory has two dimensionless parameters and is described by a two dimensional phase diagram. The effect of the temperature is to restore the chiral symmetry and is competing with that of the external magnetic field. On the other side, the magnetic field affects the melting of the mesons \cite{Albash:2007bk}. We will come back to that in Chapter~4 of this work.

\section*{2.5 \hspace{2pt} Appendix: Calculating the condensate of the theory}
\addcontentsline{toc}{section}{2.5 \hspace{0.15cm} Appendix: Calculating the condensate of the theory}

Let us consider the on-shell action:
\begin{equation}
S= -2\pi^2\tau_7N_f\int\limits_0^{\rho_{max}}d\rho\rho^3\sqrt{1+\frac{R^4H^2}{\rho^2+L(\rho)^2}}\sqrt{1+L'(\rho)^2}
\label{Saction}
\end{equation}
it diverges as $\rho_{max}\to\infty$. To rectify this we regularize the action by subtracting the action for the $L\equiv 0$ embedding:
\begin{eqnarray}
&&\frac{S_{sub}}{2\pi^2\tau_7N_f}=-\int\limits_0^{\rho_{max}}d\rho\rho^3\sqrt{1+\frac{R^4H^2}{\rho^4}}\frac{1}{4}\rho_{max}^2\sqrt{\rho_{max}^4+R^4H^2}+\\
&&\frac{1}{4}R^4H^2\ln{\left(\frac{\rho_{max}^2+\sqrt{R^4H^2+\rho_{max}^4}}{R^2H}\right)}\nonumber\ .
\end{eqnarray}
 This results to the following regularized action:
\begin{equation}
S_{reg}[L,\rho_{max}]=S-S_{sub}\ .
\label{Sreg}
\end{equation}
On the other-side the hamiltonian density of the theory can be written as \cite{Kruczenski:2003uq}:
\begin{equation}
{\cal H}=\int d^2\theta m_q{\tilde Q}Q+{\cal H}_{0}\ ,
 \end{equation}
 where ${\cal H}_0$ is the mass independent part of the hamiltonian and $Q,\tilde Q$ are the two chiral fields of the hypermultiplet in ${\cal N}=1$ notations. Now by making the identification:
 \begin{equation}
 \langle{\cal H}\rangle=-\lim_{\rho_{max}\to\infty}S_{reg}\ ,
 \end{equation}
and using equations (\ref{Saction})-(\ref{Sreg}) and the asymptotic of $L(\rho)$ as $\rho_{max}\to\infty$:
\begin{equation}
L(\rho)=m+\frac{c}{\rho_{max}^2}+\dots\ ,
\end{equation}
we can obtain:
 \begin{eqnarray}
\langle\frac{\delta{\cal H}}{\delta m_q}\rangle&=&\langle\bar\psi\psi\rangle=-2\pi\alpha'\lim_{\rho_{max}\to\infty}\frac{\delta S_{reg}}{\delta L}=\\
&=&\lim_{\rho_{max}\to\infty}4\pi^3\alpha'\tau_7 N_f\rho_{max}^3L'(\rho_{max})=-8\pi^3\alpha'\tau_7N_f c\ .\nonumber
 \end{eqnarray}
 And hence we finally get:
\begin{equation}
 \langle\bar\psi\psi\rangle=-\frac{N_f}{(2\pi\alpha')^3g_{YM}^2} c\ .
 \label{corresp1}
\end{equation}

\section*{2.6 \hspace{2pt} Appendix: Numerical techniques}
\addcontentsline{toc}{section}{2.6 \hspace{0.15cm} Appendix: Numerical techniques}

We present the Mathematica 6 code that can be used to generate some of the main numerical results reported in Chapter~2. In fact variations of the packages bellow can generate most of the numerical results reported in this thesis and are quite representative for our research. 

Let us begin by presenting the code used to solve numerically equation~(\ref{eqnMnL}) and generate the $-\tilde c$ versus $\tilde m$ plot in Figure~\ref{fig:code1}. The first two lines clear the varibles that we use and define equation~(\ref{eqnMnL}). The assignment for $\rm{b},\rm{R},\rm{h}=1$, is equivalent to the introduction of dimensionless variables. The variables $\rm{mmin}$ and $\rm{mmax}$ specify the range of initial conditions for $\tilde L(0)$ that we explore. The rest of the code solves numerically equation~(\ref{eqnMnLD}) and reads off the asymptotic behavior of $\tilde L(\tilde\rho)$ at $\tilde\rho=\rm{Rhomax}=300$ which is our numerical approximation for $\infty$. The main Mathematica's command in this code is $\rm{NDSolve}$. The last few lines of the code generate the plot of $-\tilde c$ versus $\tilde m$ presented in Figure~\ref{fig:fig1}. Modification of the code presented in Figure~\ref{fig:code1} can be used to generate virtually all of the equation of states $\tilde c(\tilde m)$ reported in this thesis. One need only change the equation of motion and the way the initial conditions are imposed. 
\begin{figure}[h] 
   \centering
   \includegraphics[ width=14cm]{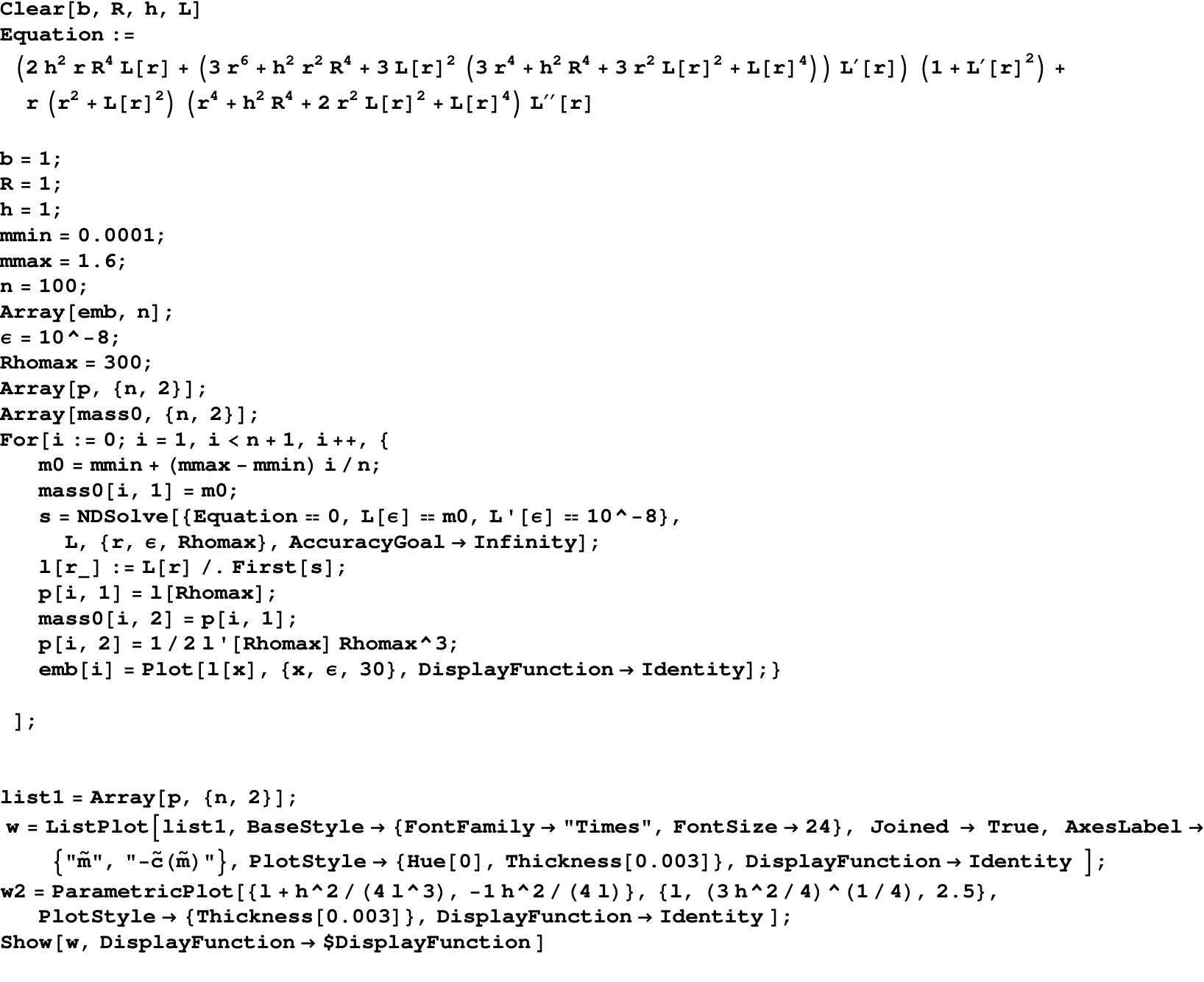}
   \caption{Mathematica 6 code used to generate the $-\tilde c$ versus $\tilde m$ plot reported in Figure~\ref{fig:fig1}. }
   \label{fig:code1}
\end{figure}
The last is sometimes non--trivial from physics point of view because of the variety of vanishing loci that we may have. However, numerically it results to the same procedure.

Let us focus now on the analysis of the free energy presented in Figure~\ref{fig:free-energy}.
\begin{figure}[h] 
   \centering
   \includegraphics[ width=14cm]{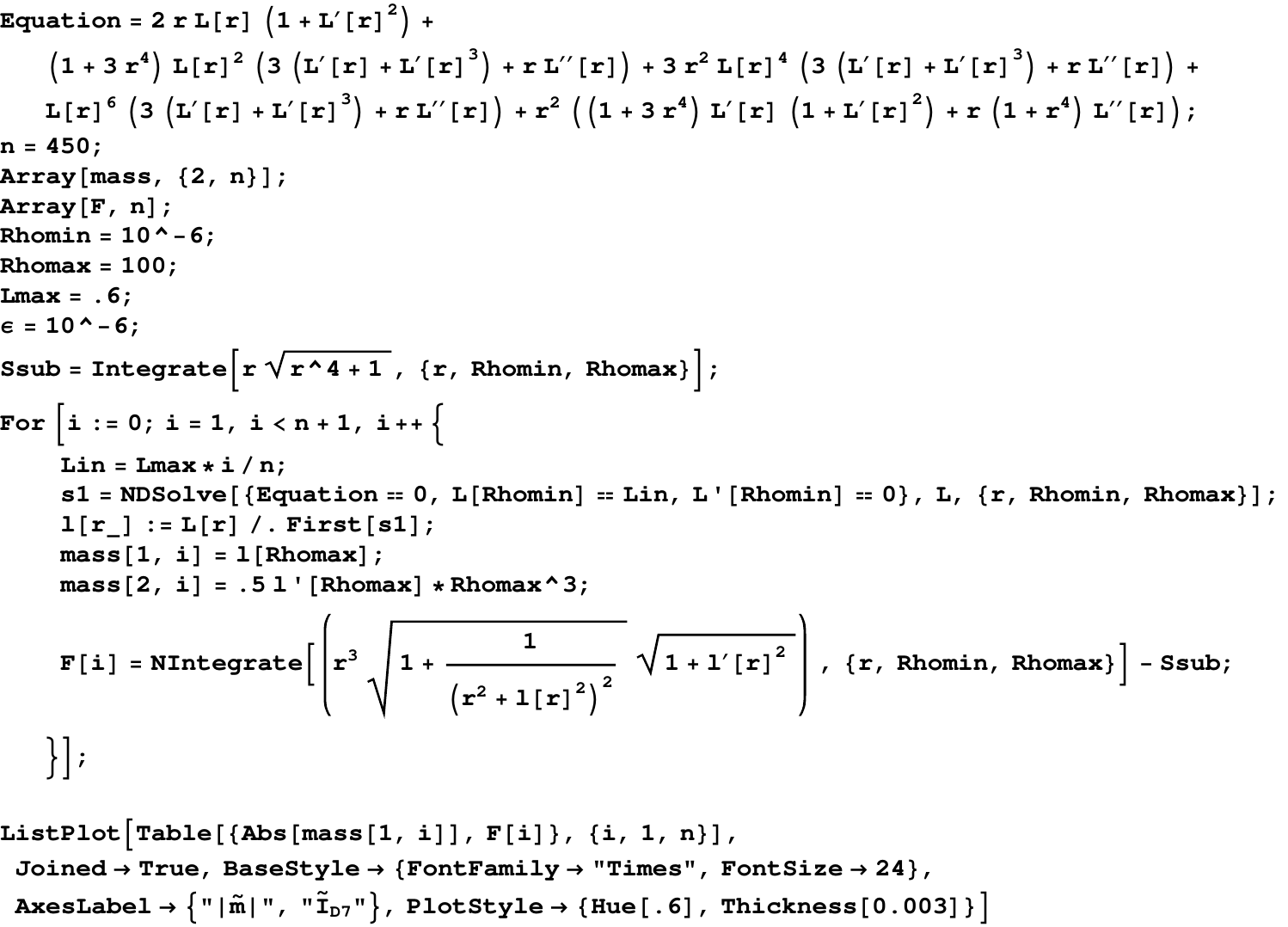}
   \caption{Mathematica 6 code used to generate the free energy $\tilde I_{\rm{D7}}$ versus $|\tilde m|$ plot reported in Figure~\ref{fig:free-energy}. }
   \label{fig:code1}
\end{figure}
The code has the same structure as the one presented in Figure~\ref{fig:code1}. The main difference is the addition of an extra array $\rm{F}[\rm{i}]$ which corresponds to the regularized on-shell action of the D7--brane for a particular value of the bare quark mass parameter $\tilde m$. The two most important Mathematica commands are $\rm{NDSolve}$ and $\rm{NIntegrate}$. Again variation of this code can be used to study the free energy for all different physical systems considered in this document. 

\begin{figure}[h] 
   \centering
   \includegraphics[ width=14cm]{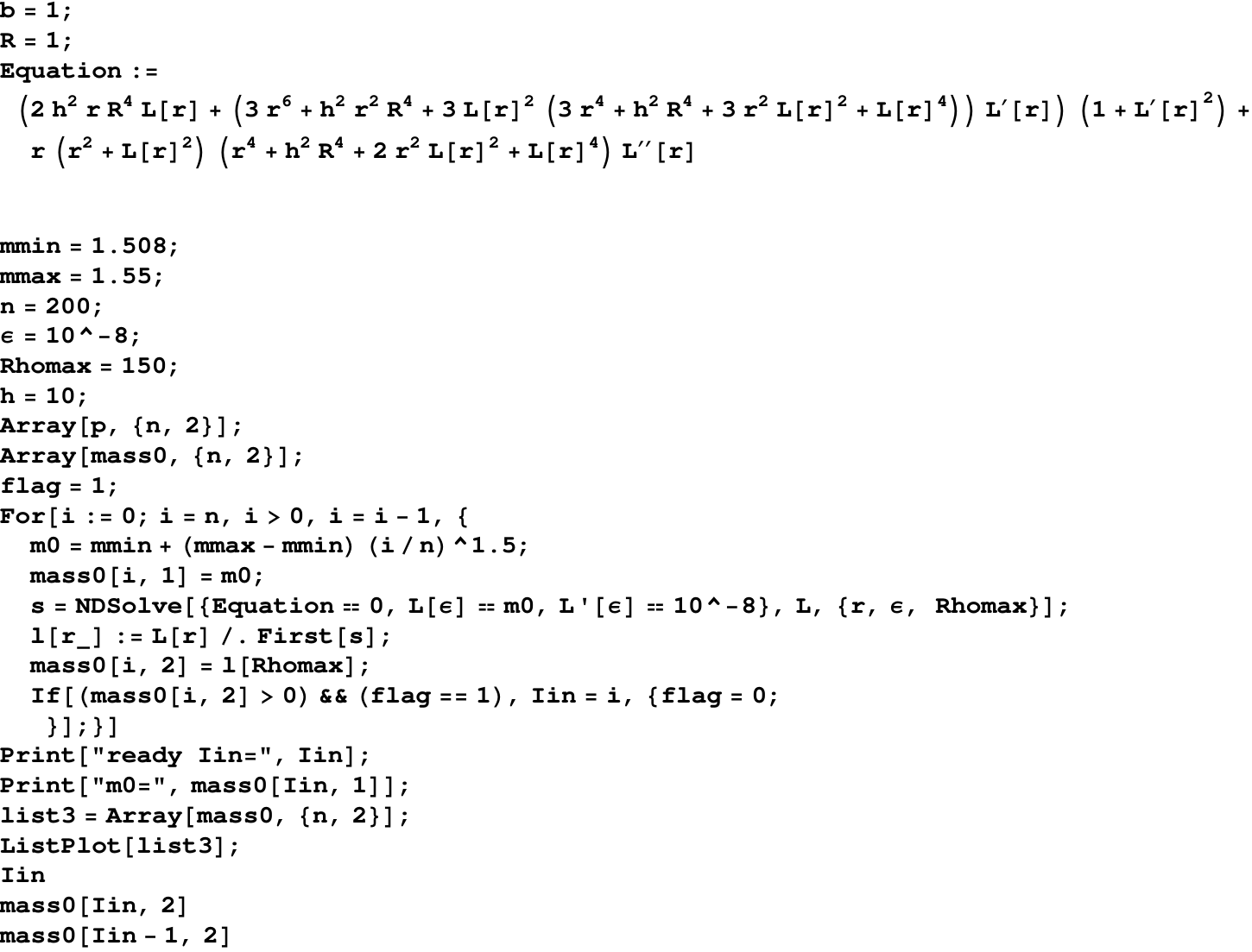}
   \caption{Mathematica 6 code used to extract the initial condition $\tilde L_{cr}(0)$ for the embedding corresponding to the critical quark condensate $\tilde c_{\rm cr}\approx 0.226$. }
   \label{fig:code3}
\end{figure}

And finally let us provide the code used to demonstrate the Gell-Mann-Oaks-Renner relation $\tilde m_{23}\propto \sqrt{\tilde m}$ demonstrated in Figure~\ref{fig:fig4}. The code runs in two parts. The first part extracts the value of the initial condition $\tilde L(0)$, corresponding to the D7--brane embedding which has zero separation at infinity ($\tilde m=0$) and does not intersect the $\tilde\rho$ axis of the $(\tilde\rho,\tilde L)$-plane. This embedding corresponds to the critical quark condensate $\tilde c_{\rm cr}\approx 0.226$, given in equation (\ref{Ccr}), and breaks the rotational $U(1)$ symmetry in the transverse to the D7--brane plane. Thus, providing a holographic description of spontaneous symmetry breaking. This part of the code is reported in Figure~\ref{fig:code3}.

Here $\rm{mmin}$ and $\rm{mmax}$ specify the interval where we expect to find the initial condition $\tilde L_{cr}(0)$. We need a very high precision because we are trying to extract the behavior of the meson spectrum for very small $\tilde m$. The main output of this code is the position $\rm{Iin}$ in the array of initial conditions $\rm{mass0[i,1]}$ for which ${\rm{mass0[Iin,1]}}=\tilde L_{cr}(0)$.

\begin{figure}[h!t] 
   \centering
   \includegraphics[ width=14cm]{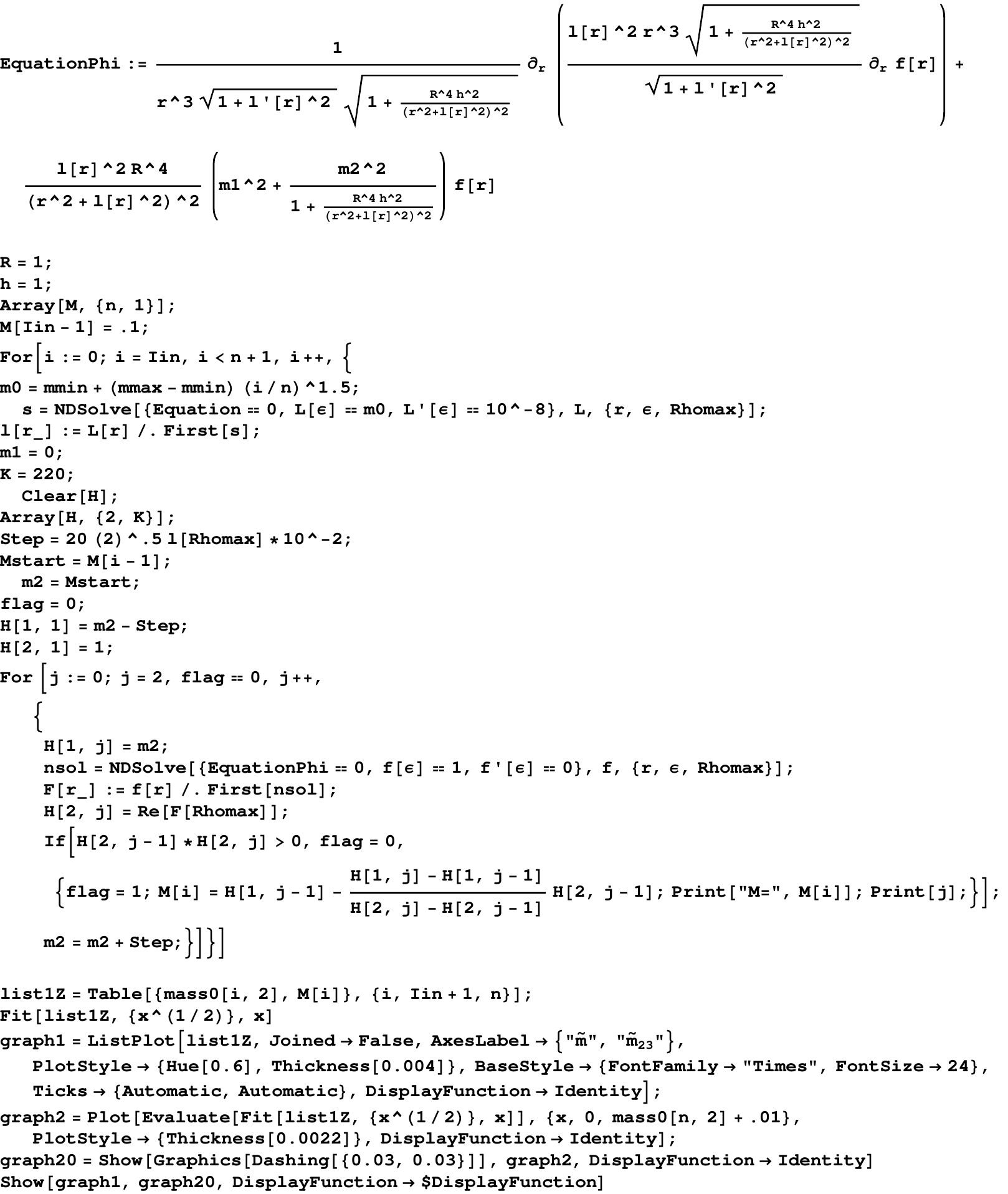}
   \caption{Mathematica 6 code used to obtain the $\tilde m_{23}\propto\sqrt{\tilde m}$ relation reported in Figure~\ref{fig:fig4}. }
   \label{fig:code4}
\end{figure}
\vspace{2cm}

In Figure~\ref{fig:code4} we have presented the second part of the code generating the $\tilde m_{23}\propto\sqrt{\tilde m}$ relation. The code uses the value of $\rm{Iin}$ calculated in the part presented in Figure~\ref{fig:code3} and the same array of initial conditions $\rm{mass0[i,1]}$ to scan only through embeddings corresponding to positive $\tilde m$. For each initial condition it solves the equation of motion for the classical embedding (the variable, $\rm{equation}$, defined in the code in Figure~\ref{fig:code3}) of the D7--brane and uses to solve the equation for the fluctuations along $\phi$ (given by the variable, $\rm{EquationPhi}$, defined in the top line in Figure~\ref{fig:code4}). The code then extracts the meson spectrum by requiring regularity of the numerical solution for the fluctuations at $\tilde\rho=\rm{Rhomax}$. This is the purpose of the inner cycle which is varying the integer $\rm{j}$. It is to be noted that the code presented in Figure~\ref{fig:code4}, after minor modifications, can be applied to other studies of meson spectrum. The main part of the code is the $\rm{j}$-cycle which verifies numerically the normalizability of the quadratic fluctuations and quantizes the spectrum.


\chapter*{Chapter 3: \hspace{1pt} Thermal properties of large $N_c$ flavored Yang--Mills theory}
\addcontentsline{toc}{chapter}{Chapter 3:\hspace{0.15cm}
Thermal properties of large $N_c$ flavored Yang--Mills theory}

\section*{3.1 \hspace{2pt} Introductory remarks}
\addcontentsline{toc}{section}{3.1 \hspace{0.15cm} Introductory remarks}

In this chapter we perform a holographic study of the thermal properties of large $N_c$ flavored Yang--Mills theory. The presented material is based on a paper written in a very close collaboration with Tameem Albash, Arnab Kundu and Clifford Johnson~\cite{Albash:2006ew}, similar result was also reported in ref.~\cite{Mateos:2006nu}. Furthermore, the main result was presented in the Ph.D. thesis of Ingo Kirsch \cite{Kirsch:2006he}. This is why this chapter should be regarded as an introduction to the ideology of Chapters~4,5,6 and 7, where the reported results are more original.

We study the geometry of AdS$_5$--Schwarzschild $\times S^5$ which is
the decoupled/near--horizon geometry of $N_c$ D3--branes, where $N_c$
is large and set by the (small, for reliability) curvature of the
geometry. The physics of closed type~IIB string theory in this
background is dual to the physics of ${\cal N}=4$ supersymmetric
$SU(N_c)$ gauge theory in four dimensions, with the supersymmetry broken by being at finite
temperature\cite{Witten:1998zw}. The temperature is set by the horizon
radius of the Schwarzschild black hole, as we will recall below.

We introduce a D7--brane probe into the background. Four of the
brane's eight world--volume directions are parallel with those of the
D3--branes, and three of them wrap an $S^3\subset S^5$. The remaining
direction lies in the radial direction of the asymptotically AdS$_5$
geometry.

Such a D3--D7 configuration controls the physics of
the $SU(N_c)$ gauge theory with a dynamical quark in the fundamental
representation\cite{Karch:2002sh}.  The configuration (at zero temperature)
preserves ${\cal N}=2$ supersymmetry in $D=4$, and the quark is part
of a hypermultiplet.  Generically, we will be studying the physics at finite
temperature, so supersymmetry will play no explicit role here.

We are studying the D7--brane as a probe only, corresponding to taking the $N_c \gg N_f$ limit, and therefore there is no backreaction on the background geometry. This is roughly analogous
to the quenched approximation in lattice QCD.  The quark mass and other flavor physics--such as the
vacuum expectation value (vev) of a condensate and the  spectrum of mesons
that can be constructed from the quarks--are all physics which are
therefore invisible in the background geometry. We will learn nothing
new from the background; our study is of the response of the probe D7--branes to the background, and this is where the new physics emerges from.

We carefully study the physics of the probe itself as
it moves in the background geometry. The coordinates of the probe in
the background are fields in an effective D7--brane world--volume
theory, and the geometry of the background enters as couplings controlling the dynamics of
those fields. One such coupling in the effective model represents the
local separation, $L(u)$, of the D7--brane probe from the D3--branes,
where $u$ is the radial AdS$_5$--Schwarzschild coordinate.

In fact, the asymptotic value of the separation between the D3--branes and
D7--brane for large $u$ yields the bare quark mass $m_q$ and the
condensate vacuum expectation value (vev) $ \langle \bar{\psi} \psi
\rangle$ as follows \cite{Kruczenski:2003uq,Polchinski:2000uf}:
\begin{equation} \label{eqt: L}
\lim_{u \rightarrow \infty} L(u) = m + \frac{c}{u^2} + \dots\ ,
\end{equation}
where $m = 2 \pi \alpha' m_q$ and $-c = \langle \bar{\psi} \psi \rangle / (8 \pi^3 \alpha' N_f \tau_{\mathrm{7}})$, in this chapter we will set $N_f = 1$. The fundamental string tension is defined as $T=1/(2\pi\alpha^\prime)$ here, and the D7--brane tension is $\tau_{\mathrm{7}}=(2\pi)^{-7}(\alpha^\prime)^{-4}$. The
zero temperature behavior of the D7--branes in the geometry is
simple. The D7--brane world--volume actually {\it vanishes} at finite~$u$, corresponding to the part of the brane wrapped on the $S^3$
shrinking to zero size.  The location in $u$ where this vanishing
happens encodes the mass of the quark, or equivalently, the separation
of the probe from the D3--branes.  In addition, in the zero temperature background, the only value of $c$ allowed is zero, meaning no condensate is allowed to form, as is expected from supersymmetry.

The finite temperature physics introduces an important new feature.
As is standard \cite{Gibbons:1979xm}, finite temperature is studied by
Euclideanizing the geometry and identifying the temperature with the
period of the time coordinate.  The horizon of the background geometry
is the place where that $S^1$ shrinks to zero size. The D7--brane
 is also wrapped on this $S^1$, so it can vanish at
the horizon, if it has not vanished due to the shrinking of the $S^3$.
For large quark mass compared to the temperature (horizon size), the
$S^3$ shrinking will occur at some finite $u>u_{\rm{H}} $, and the physics
will be similar to the zero temperature situation. However, for small quark
mass, the world--volume will vanish due to the shrinking of
the $S^1$ corresponding to the D7--branes going into the horizon. This
is new physics of the flavor sector.

The authors of ref.~\cite{Babington:2003vm} explored some of the
physics of this situation (the dependence of the condensate and of the meson mass on the bare quark mass), and predicted that a phase transition
should occur when the topology of the probe D7--brane changes.  However, they were not able to explicitly see this transition because of poor data resolution in the transition region, coming from using UV boundary conditions on the scalar fields on the D7--brane world--volume.  The
origin of this phase transition, as we shall see, is as follows: The
generic behavior of an allowed solution for $L(u)$ as in
equation~(\ref{eqt: L}), is not enough to determine whether the
behavior corresponds to an $S^3$--vanishing D7--brane or an
$S^1$--vanishing D7--brane. The choices of branch of solutions have
different values of $c$, generically. In other words, for a given
value of $m$ there can be more than one value of $c$. There are
therefore two or more candidate solutions potentially controlling the
physics. The actual physical solution is the one which has the lowest
value for the D7--brane's free energy. The key point is that, at a
certain value of the mass, the lowest energy solution may suddenly
come from a different branch, and, as the corresponding value of the
condensate changes discontinuously in moving between branches, we find
that the system therefore undergoes a first order phase transition.  On the gauge theory side, we can imagine a similar situation occurring; two different branches of solution are competing, and the lowest energy branch is always picked.  
We are able to uncover this physics by doing a careful numerical analysis of the equations of motion for the probe dynamics on the gravity side of the AdS/CFT correspondence, by using IR boundary conditions instead of UV boundary conditions.

\section*{3.2 \hspace{2pt} Holographic meson melting}
\addcontentsline{toc}{section}{3.2 \hspace{0.15cm} Holographic meson melting}

\subsection*{3.2.1 \hspace{2pt} General set up}
\addcontentsline{toc}{subsection}{3.2.1 \hspace{0.15cm} General set up}
We begin by reviewing the physics of the D7--brane probe in the
AdS$_5$--Schwarzschild background solution\cite{Babington:2003vm}.  The metric is
given by:
\begin{equation} \label{eqt: our metric}
ds^2 = - \frac{f(u)}{R^2} dt^2 + \frac{R^2}{f(u)}  du^2 + \frac{u^2}{R^2} d\vec{x}\cdot d\vec{x} + R^2 d\theta^2 + R^2 \cos^2 \theta d \Omega_3^2  + R^2 \sin^2\theta d\phi^2\ ,
\end{equation}
where $\vec{x}$ is a three vector,
\begin{eqnarray}
f(u) &=& u^2 - \frac{b^4}{u^2} \nonumber \ ,
\end{eqnarray}
and the quantity $R^2$ is given by:
\begin{eqnarray}
R^2 &=& \sqrt{4 \pi g_s N_c} \alpha' \nonumber \ ,
\end{eqnarray}
where $g_s$ is the string coupling (which, with the inverse string
tension $\alpha^\prime$ sets for example, Newton's constant).  The quantity $b$ is related to the mass of the black-hole, $b^2=8 G_5 m_\mathrm{b.h.} / (3 \pi)$.  The
temperature of the black hole can be extracted using the standard
Euclidean continuation and requiring regularity at the horizon.
Doing this in the metric given by equation~(\ref{eqt: our metric}), we
find that $\beta^{-1}=b/\pi R^2$.  Therefore, by picking the value of $b$, we are choosing at what temperature we are holding the theory.  We choose to embed the D7--brane probe
transverse to $\theta$ and $\phi$.  In order to study the embeddings with the lowest value of the on--shell action (and hence the lowest free energy), we choose an ansatz of the form $\phi = 0$ and $\theta = \theta(u)$.  The asymptotic separation of the D3-- and D7--branes is given by $L(u) = u \sin \theta$.  Given this particular choice of embedding, the world--volume of the D7--brane is given by:
\begin{eqnarray}
 \sqrt{-g} &=& u^2 \cos^3 \theta(u) \sqrt{\det S^3} \sqrt{u^2 +\left(u^4-b^4 \right)\theta'(u)^2} \ ,
\end{eqnarray}
where $g$ is the determinant of the induced metric on the D7--brane given by the pull--back of the space--time metric $G_{\mu \nu}$.  We are interested in two particular cases.  First, there is the case where $u$ goes to~$b$,
which corresponds to the D7--brane probe falling into the event horizon.  In the Euclidean section, this case corresponds to the shrinking of the $S^1$ of periodic time.
we will refer to these solutions as the `black hole' embeddings. Second, there is the case where $\theta$ goes to~$\pi/2$, which corresponds to the shrinking of the $S^3$.  We will name these Minkowski embeddings.  It is this change in topology ($S^1$ versus $S^3$ shrinking) between the different solutions that will correspond to a phase transition.
The classical equation of motion for $\theta(u)$ is:
\begin{eqnarray} \label{eqt: eqt of motion}
\frac{d}{d u} \left( \frac{u^2\left(u^4-b^4 \right) \theta' \cos^3 \theta} {\sqrt{u^2 +\left(u^4-b^4 \right){\theta'}^2}} \right) + 3 u^2 \cos^2 \theta\sin \theta\sqrt{u^2 +\left(u^4-b^4 \right){\theta'}^2} &=& 0\ .
\end{eqnarray}
When $u$ goes to infinity and the background metric becomes asymptotically AdS$_5 \times S^5$, the equation of motion reduces to:
\begin{eqnarray}
\frac{d}{d u} \left(u^5 \theta'(u) \right) + 3 u^2 \theta(u) &=& 0,
\end{eqnarray}
which has solution:
\begin{eqnarray}
\theta(u) &=& \frac{1}{u} \left(m + \frac{c}{u^2} \right) \ .
\end{eqnarray}
These two terms are exactly the non--normalizable and normalizable terms corresponding to a dimension 3 operator ($\bar{\psi} \psi$) in the dual field theory with source $m$ and vacuum expectation value (vev) $c$.

Next we solve equation~(\ref{eqt: eqt of motion}) numerically using a shooting technique.  Shooting from infinity towards the horizon, physical solutions are those that have a finite value at the horizon.  This will only be accomplished for a particular $m$ and $c$ value from equation
(\ref{eqt: L}) which would have to be delicately chosen by hand.
Therefore, if we instead start from the horizon with a finite
solution, it will shoot towards the physical asymptotic solutions we
desire.  In order to be able to analyze the phase transition between
the black hole and Minkowski solutions, we shoot from the
horizon for the condensate solutions and from $\theta= \pi/2$ for the
Minkowski solutions.  This technique avoids having to correctly
choose the boundary conditions at infinity, which is a sensitive
procedure, allowing us to have many more data points to analyze the
phase transition.  We impose the boundary condition that, at
our starting point--the horizon, we have:
\begin{eqnarray} \label{eqt: bc}
\theta'(u) \big|_{S^1\to 0} &=&  \frac{3 b^2}{8}  \tan \theta(b) \nonumber \\
\theta'(u) \big|_{S^3 \to 0} &=& \infty
\end{eqnarray}
We argue that this is the physical boundary condition to take; the first is simply a result of taking the limit of $u \to b$ in the equation of motion in equation~(\ref{eqt: eqt of motion}), whereas the second is a result of requiring no conifold singularity as the $S^3$ shrinks to zero size \cite{Karch:2006bv}.

\begin{figure}[!ht]
\begin{center}
\includegraphics[angle=0,
width=11cm]{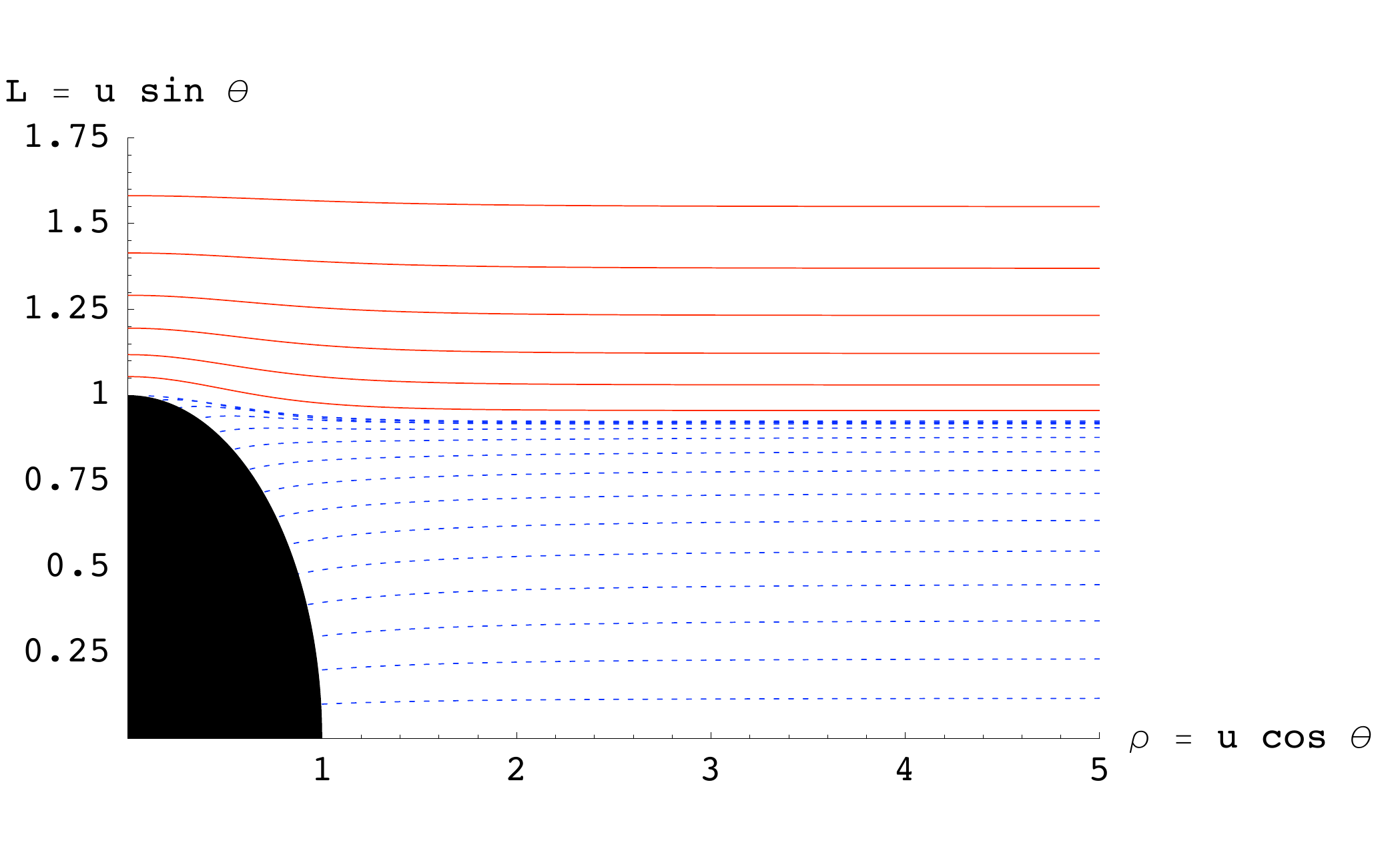}
\caption{Solutions for the D7--brane probe in the black hole background. }
\label{fig:solutionsinLandrho}
\end{center}
\end{figure}

We solve the equation of motion, equation~(\ref{eqt: eqt of motion}),
numerically using $b=1$ and $R=1$.  These numerical choices correspond to fixing the temperature and to measuring lengths in units of the radius of the AdS space; the latter condition also means we are choosing a particular relationship between the t'Hooft coupling and the dual quantities:
\begin{equation}
\lambda = g_{\text{YM}}^2 N_c = \frac{1}{2 \alpha^{\prime 2}} \ , \quad \beta^{-1} = \frac{1}{\pi} \ .
\end{equation}
Several D7--brane embedding solutions are shown in Figure~\ref{fig:solutionsinLandrho}.  The red (solid) lines correspond to Minkowski solutions, and the blue (dashed) lines correspond to black hole solutions.  From each of these solutions, we can extrapolate the bare quark mass and quark condensate vev.

\subsection*{3.2.2 \hspace{2pt} The first order phase transition}
\addcontentsline{toc}{subsection}{3.2.2 \hspace{0.15cm} The first order phase transition}

\begin{figure}[!ht]
\begin{center}
\subfigure[] {\includegraphics[angle=0,
width=0.45\textwidth]{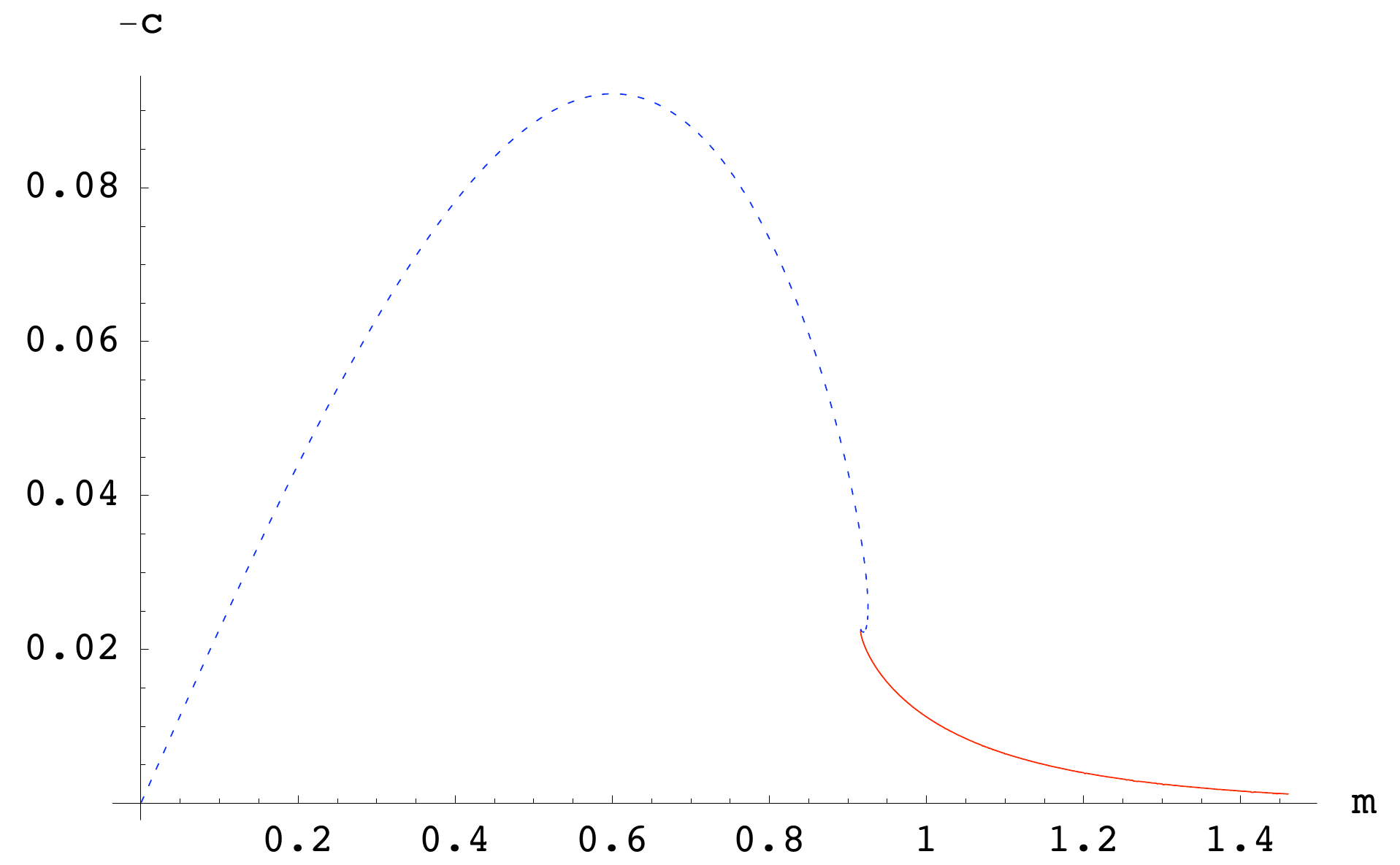} \label{fig: c vs m}}
\subfigure[] {\includegraphics[angle=0,
width=0.45\textwidth]{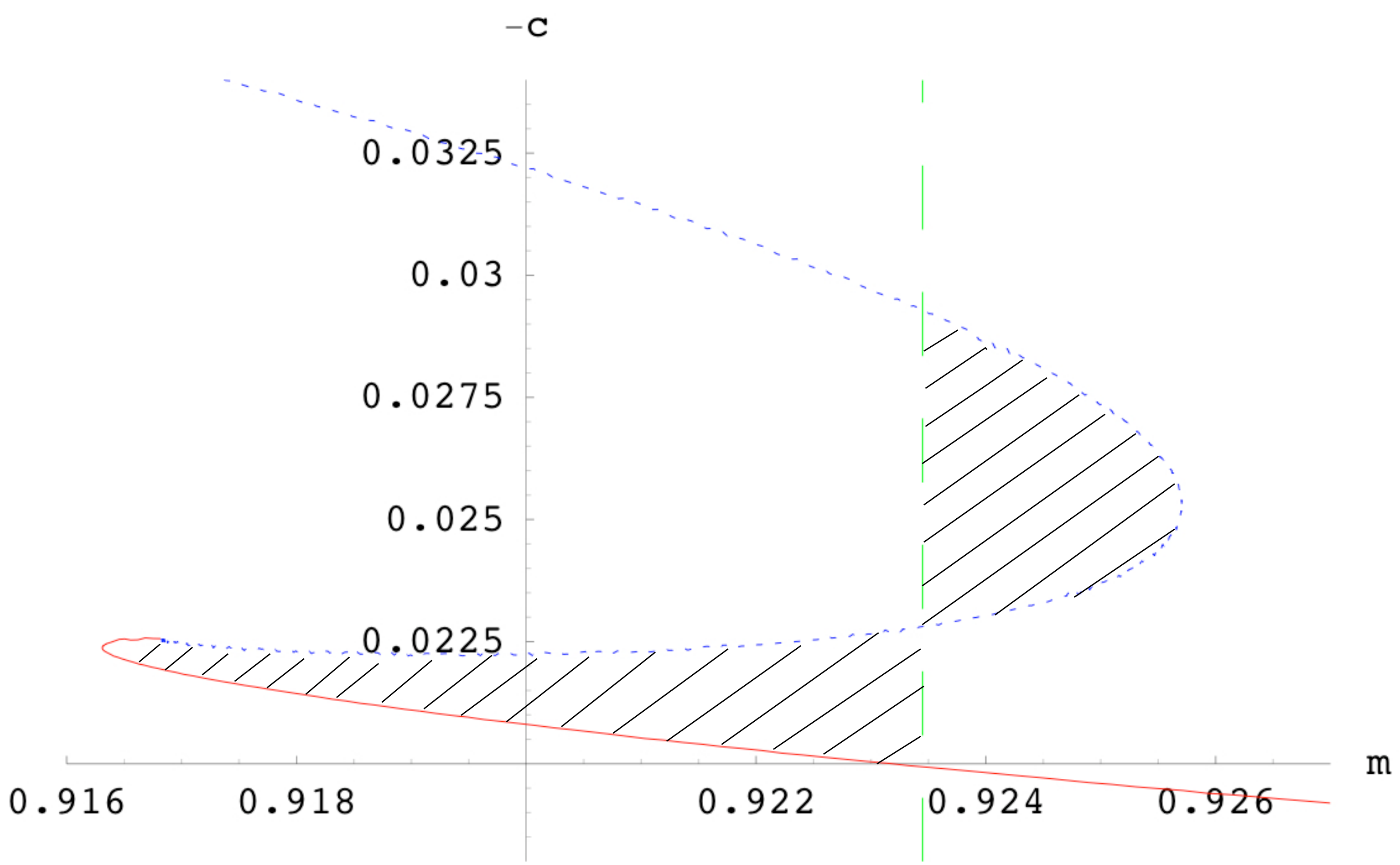} \label{fig: c vs m zoom}}
\subfigure[] {\includegraphics[angle=0,
width=0.45\textwidth]{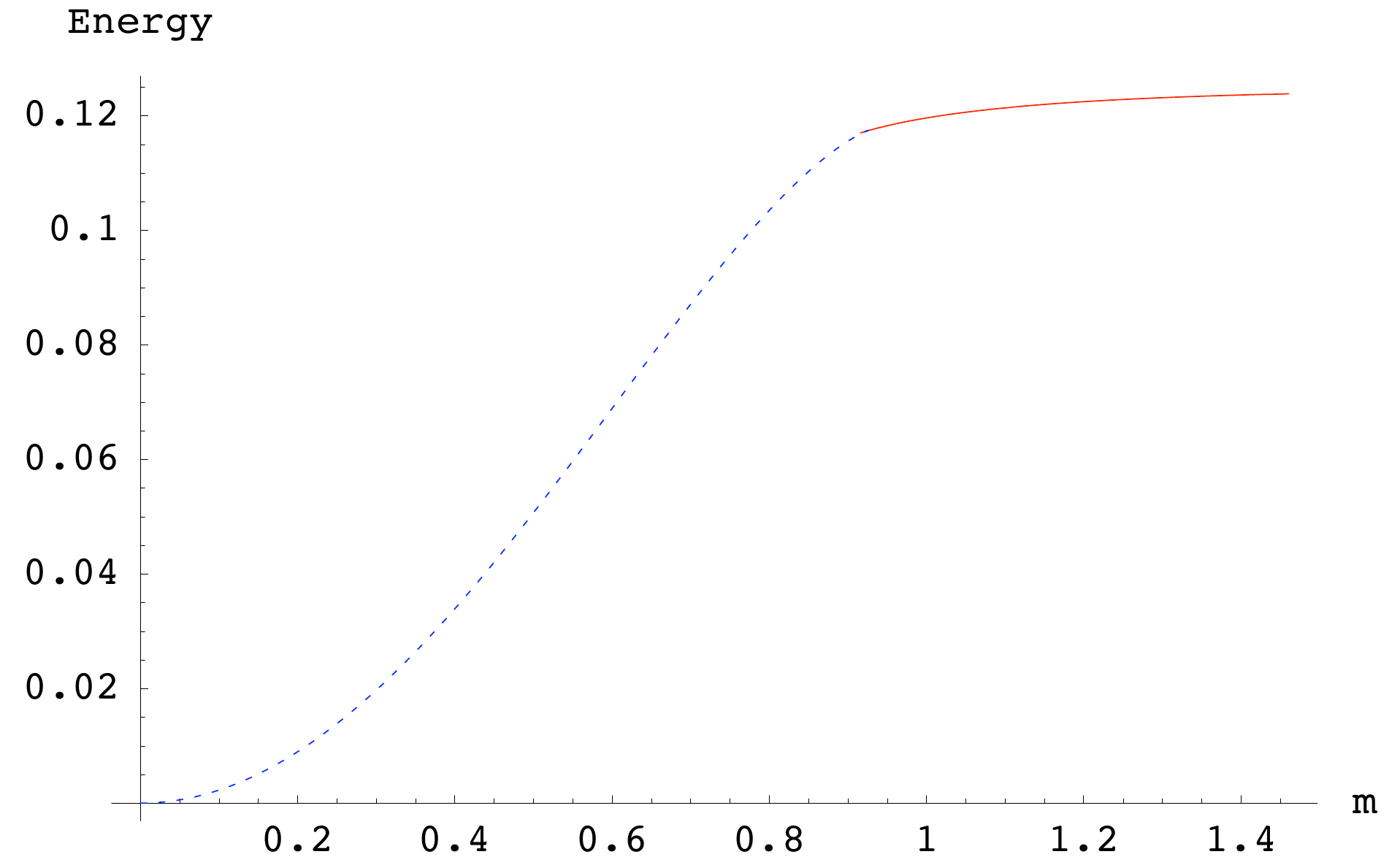} \label{fig: energy}}
\subfigure[] {\includegraphics[angle=0,
width=0.45\textwidth]{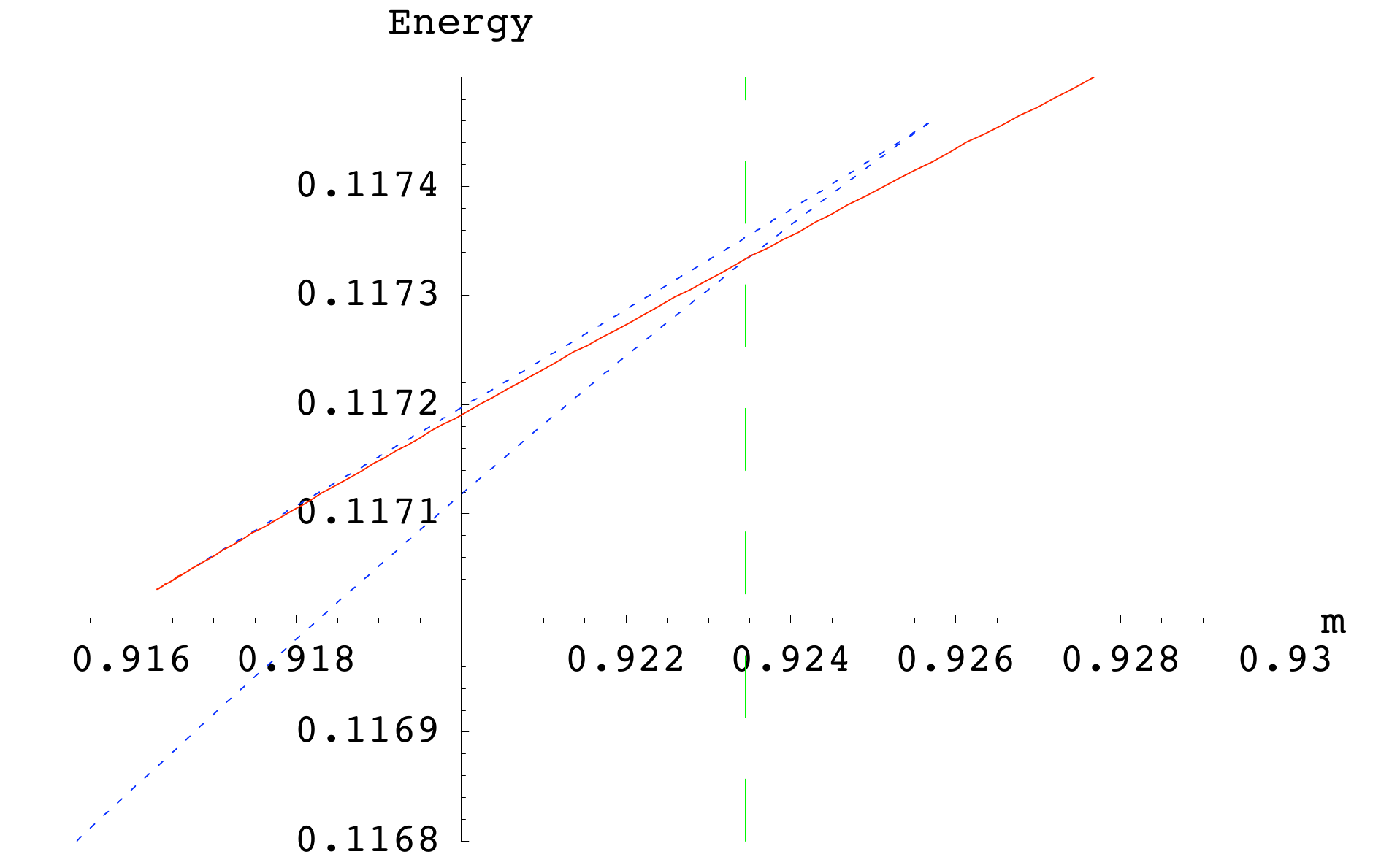} \label{fig: energy zoom}}
\caption{Probe free energy (in units of $2 \pi  \tau_7 N_f b/ R^2)$ and condensate vev at the phase transition. See text for more details. }
\end{center}
\end{figure}

We plot the $c$ values as a function of $m$ in Figure~\ref{fig: c vs m}.  When enlarged, as shown in Figure~\ref{fig: c vs m zoom}, we find, as anticipated in the introductory remarks, the multi--valuedness in $c$ for a given $m$. Physics will choose just one answer for $c$. There is therefore the possibility of a transition from one branch to another as one changes $m$.

In order to determine exactly where the transition takes place, we have to calculate the free energy of the D7--brane.  In the semi-classical limit that we are considering, the free energy is given by the on--shell action times $\beta^{-1}$.  For our case, this is simply given by \cite{Kruczenski:2003uq}:
\begin{eqnarray}
\mathcal{F} &=& \beta^{-1} \tau_7 N_f \int d^4 x \ d\Omega_3 \ du \, \sqrt{-\det g}  \ ,
\end{eqnarray}
where here $N_f = 1$.  We calculate this integral numerically using our solutions for $\theta(u)$, and we plot the results for the energy in Figure~\ref{fig: energy} after we regulate the result by subtracting off the energy from the $\theta=0$ solution.  There is again multivaluedness at the same value of $m$ as before, and we zoom in on this neighborhood in Figure~\ref{fig: energy zoom}.  If one follows the solutions of lowest energy for a given $m$, one can clearly see that there is a crossover from one branch to another as $m$ changes.  This was also observed in ref.~\cite{Ghoroku:2005tf}.
Therefore, we find a first order phase transition ---at $m\approx 0.92345$--- where the condensate's vev jumps discontinuously.  This had been deduced on other grounds in ref.~\cite{Apreda:2005yz}.  We show where the jump between curves occurs with the dashed green line in figures~\ref{fig: c vs m zoom} and~\ref{fig: energy zoom}.  We note that the shaded areas in Figure~\ref{fig: c vs m zoom} are equal; this is to be expected since $c$ and $m$ are thermodynamically conjugate to each other, and the area of the graph has the interpretation as a free energy difference $d\mathcal{E} \sim - c \, dm$. This first order phase transition was studied in details in ref.~\cite{Hoyos:2006gb}, where the authors developed an appropriate formalism to study the meson spectrum of the theory before and after the phase transition. They were able to show that after the phase transition there are no bound meson states, but only quasi-normal excitations corresponding to a melting mesons. This is why the present understanding of the observed first order phase transition is that it corresponds to the confinement/deconfinement phase transition of the fundamental matter which in the case of consideration is the meson melting phase transition.

\section*{3.3 \hspace{2pt} Concluding remarks}
\addcontentsline{toc}{section}{3.3 \hspace{0.15cm} Concluding remarks}
We expect that, while the details of this construction will not persist
in a ``realistic'' QCD string dual, the phase transition itself
represents strongly coupled dynamics that may well persist as part of
the full story of the QCD phase diagram. The transition is exciting in
itself, of course (particularly since its dual involves a change of
topology in the D7--brane world--volume), but much further work is
needed on several questions. For example, the robustness of the phase
transition against $1/{N_c}$ corrections would be interesting to
study. In fact it is believed that the observed first order phase transition is an artifact of the large $N_c$ limit. We will come back to this issue in Chapter~7, when we study the critical behavior of the theory. Once again a peculiar feature of the theory will turn out to be an artifact of the large $N_c$ limit.

\chapter*{Chapter 4: \hspace{1pt} Phase structure of finite temperature large $N_c$ flavored Yang--Mills theory in an external magnetic field}
\addcontentsline{toc}{chapter}{Chapter 4:\hspace{0.15cm} Phase structure of finite temperature large $N_c$ flavored Yang--Mills theory in an external magnetic field}

\section*{4.1 \hspace{2pt} The string background}
\addcontentsline{toc}{section}{4.1 \hspace{0.15cm} The string background}
%
Consider the AdS$_5$--Schwarzschild$\times S^5$ solution that we studied in the previous section given by:
\begin{eqnarray}
ds^2/\alpha'&=&-\frac{u^4-b^4}{R^2u^2}dt^2+\frac{u^2}{R^2}d\vec{ x}^2+\frac{R^2u^2}{u^4-b^4}du^2+R^2d\Omega_{5}^2 \ , \\ \mathrm{where}\quad
d\Omega_5^2&=&d\theta^2+\cos^2\theta d\Omega_3^2+\sin^2\theta d\phi^2\nonumber \ ,\\ \mathrm{and}\quad
d\Omega_{3}^2&=&d\psi^2+\cos^2\psi d\beta+\sin^2\theta d\gamma^2 \ .\nonumber
\end{eqnarray}
The dual gauge theory will inherit the time and space coordinates
$t\equiv x^0$ and $\vec{x}\equiv(x^1,x^2,x^3)$ respectively. Also, in
the solution above, $u\in[0,\infty)$ is a radial coordinate on the
asymptotically AdS$_5$ geometry and we are using standard polar
coordinates on the $S^5$. The scale $R$ determines the gauge theory
't Hooft coupling according to $R^2=\alpha^\prime\sqrt{g_{\rm YM}^2 N_c}$. For
the purpose of our study it will be convenient \cite{Babington:2003vm}
to perform the following change of variables:
\begin{eqnarray} \label{eqt:changeofcoordinates}
&r^2&=\frac{1}{2}(u^2+\sqrt{u^4-b^4})=\rho^2+L^2 \ ,\\ \mathrm{with}\quad
&\rho&=r\cos\theta\ , \,\, L=r\sin\theta\nonumber \ .
\end{eqnarray}
The expression for the metric now takes the form:
\begin{eqnarray}
ds^2/\alpha'=-\left(\frac{(4r^4-b^4)^2}{4r^2R^2(4r^4+b^4)}\right)dt^2+\frac{4r^4+b^4}{4R^2r^2}d\vec{x}^2+\frac{R^2}{r^2}(d\rho^2+\rho^2d\Omega_{3}^2+dL^2+L^2d\phi^2)  \ .  \nonumber
\end{eqnarray}
Following ref.~\cite{Karch:2002sh}, we introduce fundamental matter
into the gauge theory by placing D7--brane probes into the dual
supergravity background.  The probe brane is parameterized by the
coordinates $\{x_{0},x_{1},x_{2},x_{3},\rho,\psi,\beta,\gamma\}$ with the
ansatz from equation (\ref{ansatzEmb}) that we replicate here:
\begin{eqnarray}
\phi\equiv \mathrm{const},\quad L\equiv L(\rho)\nonumber \label{ansatzEmb1} \ .
\end{eqnarray}
In order to introduce an external magnetic field, we excite a pure
gauge $B$--field along the $(x^2,x^3)$ directions as in Chapter~2:
\begin{equation}
B=Hdx^2\wedge dx^3,
\end{equation}
where $H$ is a real constant. As explained in
Chapter 2, while this does not change the
supergravity background, it has a non--trivial effect on the physics of
the probe, which is our focus. To study the effects on the probe, let
us consider the general (Abelian) DBI action:
\begin{eqnarray}
S_{DBI}=- N_f  T_{D7} \int\limits_{{\cal M}_{8}}d^{8}\xi \ \mathrm{det} ^{1/2}(P[G_{ab}+B_{ab}]+2\pi\alpha' F_{ab})\ , \label{DBI1}
\end{eqnarray}
where $T_{D7}=\mu_7 / g_s = [(2\pi)^7\alpha'^4 g_s]^{-1}$ is the
D7--brane tension, $P[G_{ab}]$ and $P[B_{ab}]$ are the induced metric
and induced $B$--field on the D7--branes' world--volume, $F_{ab}$ is
the world--volume gauge field, and $N_f=1$ here. It was shown in
Chapter~2 that, for the AdS$_5\times S^5$ geometry, we
can consistently set the gauge field $F_{ab}$ to zero to leading order
in $\alpha'$, and the same argument applies to the finite temperature
case considered here. The resulting Lagrangian is:
\begin{equation}
{\cal L}=-\rho^3\left(1-\frac{b^8}{16 \left(\rho^2+L(\rho)^2\right)^4}\right) \left\{1+\frac{16 H^2 \left(\rho^2+L(\rho)^2\right)^2 R^4}{\left(b^4+4 \left(\rho^2+L(\rho)^2\right)^2\right)^2}\right\}^{\frac12}
   \sqrt{1+L'(\rho)^2} \ .
 \end{equation}
For large $\rho \gg b$, the Lagrangian asymptotes to:
\begin{equation}
{\cal L}\approx-\rho^3\sqrt{1+L'(\rho)^2} \ ,
\end{equation}
which suggests the following asymptotic behavior for the embedding
function $L(\rho)$:
\begin{equation}
L(\rho)=m+\frac{c}{\rho^2}+\dots \ ,
\label{asymptote}
\end{equation}
where the parameters $m$ (the asymptotic separation of the D3-- and
D7--branes) and $c$ (the degree of transverse bending of the D7--brane
in the $(\rho,\phi)$ plane) are related to the bare quark mass
$m_{q}=m/2\pi\alpha'$ and the quark condensate
$\langle\bar\psi\psi\rangle\propto -c$ respectively
\cite{Kruczenski:2003uq} (this calculation was repeated in the Appendix of Chapter~2). It was shown in
Chapter~2 that the presence of the external magnetic
field spontaneously breaks the chiral symmetry of the dual gauge
theory (it generates a non--zero $\langle\bar\psi\psi\rangle$ at zero
$m$). However \cite{Babington:2003vm}, the effect of the finite
temperature is to melt the mesons and restore the chiral symmetry at
zero bare quark mass. Therefore, we have two competing processes
depending on the magnitudes of the magnetic field $H$ and the
temperature $T=b/\pi R^2$.  This suggests an interesting two
dimensional phase diagram for the system which we shall study in
detail later.

To proceed, it is convenient to define the following dimensionless
parameters:
 \begin{eqnarray}
 \tilde\rho&=&\frac{\rho}{b} \ , ~~~\eta=\frac{R^2}{b^2}H \ , \quad {\tilde m}=\frac{m}{b}\ ,\\
 \tilde L(\tilde\rho)&=&\frac{L(b\tilde\rho)}{b}=\tilde m+\frac{\tilde c}{\tilde \rho^2}+\dots \ .\nonumber
 \label{formula1}
 \end{eqnarray}
This leads to the Lagrangian:
\begin{equation}
\tilde{\cal L}=-\tilde\rho^3\left(1-\frac{1}{16 \left(\tilde\rho^2+\tilde L(\tilde\rho)^2\right)^4}\right) \left\{1+\frac{16\left(\tilde\rho^2+\tilde L(\tilde\rho)^2\right)^2 \eta^2}{\left(1+4 \left(\tilde\rho^2+\tilde L(\tilde\rho)^2\right)^2\right)^2}\right\}^{\frac12}\sqrt{1+\tilde L'(\tilde\rho)^2} \ .
\label{LagrangianM}
\end{equation}
For small values of $\eta$, the analysis of the second order,
non--linear differential equation for $\tilde L(\tilde\rho)$ derived
from equation (\ref{LagrangianM}) follows closely that performed in
refs.~\cite{Albash:2006ew,Babington:2003vm,Mateos:2006nu}. The
solutions split into two classes: the first class are solutions
corresponding to embeddings that wrap a shrinking $S^3$ in the $S^5$
part of the geometry and (when the $S^3$ vanishes) closes at some
finite radial distance $r$ above the black hole's horizon which is
located at $r=b/\sqrt{2}$.  These embeddings are referred to as
`Minkowski' embeddings.  The second class of solutions correspond to
embeddings falling into the black hole, since the $S^1$ of the
Euclidean section, on which the D7--branes are wrapped, shrinks away
there.  These embeddings are referred to as `black hole' embeddings.
There is also a critical embedding separating the two classes of
solutions which has a conical singularity at the horizon, where the
$S^3$ wrapped by the D7--brane shrinks to zero size, along with the
$S^1$. If one calculates the free energy of the embeddings, one can
show \cite{Albash:2006ew,Babington:2003vm,Mateos:2006nu} that it is a
multi--valued function of the asymptotic separation $m$, which amounts
to a first order phase transition of the system (giving a jump in the
condensate) for some critical bare quark mass $m_{\rm cr}$. (For fixed
mass, we may instead consider this to be a critical temperature.) We
show in this chapter that the effect of the magnetic field is to
decrease this critical mass, and, at some critical magnitude of the
parameter $\eta_{\rm cr}$, the critical mass drops to zero.  For $\eta >
\eta_{\rm cr}$ the phase transition disappears, and only the Minkowski
embeddings are stable states in the dual gauge theory, possessing a
discrete spectrum of states corresponding to quarks and anti--quarks
bound into mesons. Furthermore, at zero bare quark mass, we have a
non--zero condensate and the chiral symmetry is spontaneously broken.
 \section*{4.2 \hspace{2pt} Properties of the solution}
\addcontentsline{toc}{section}{4.2 \hspace{0.15cm} Properties of the solution}

 \subsection*{4.2.1 \hspace{2pt} Exact results at large mass}
\addcontentsline{toc}{subsection}{4.2.1 \hspace{0.15cm} Exact results at large mass}
 It is instructive to first study the properties of the solution for
 $\tilde m \gg 1$. This approximation holds for finite temperature,
 weak magnetic field, and large bare quark mass $m$, or, equivalently,
 finite bare quark mass $m$, low temperature, and weak magnetic field.

 In order to analyze the case $\tilde m\gg 1$, let us write $\tilde
 L(\tilde{\rho})=\tilde a +\zeta(\tilde{\rho})$ for $\tilde{a} \gg 1$
 and linearize the equation of motion derived from equation
(\ref{LagrangianM}), while leaving only the first two leading terms in
 $(\rho^2+\tilde m^2)^{-1}$. The result is:
\begin{equation}
\partial_{\tilde\rho}(\tilde\rho^3\zeta')-\frac{2\eta^2}{(\tilde m^2+\tilde\rho^2)^3}\tilde m+\frac{2(\eta^2+1)^2-1}{2(\tilde m^2+\tilde\rho^2)^5}\tilde m+O(\zeta)=0 \ .
\label{bigm}
\end{equation}
 Ignoring the $O(\zeta)$ terms in equation (\ref{bigm}), the general solution takes the form:
 \begin{equation}
 \zeta(\tilde \rho)=-\frac{\eta^2}{4\rho^2(\tilde m^2+\tilde\rho^2)}\tilde m+\frac{2(\eta^2+1)^2-1}{96\tilde\rho^2(\tilde m^2+\tilde\rho^2)^3}\tilde m \ ,
 \label{weakL}
 \end{equation}
where we have taken $\zeta'(0) = \zeta(0) = 0$.  By studying the asymptotic behavior of this solution, we can extract the following:
\begin{eqnarray}
\tilde{m} &=& \tilde{a} - \frac{\eta^2}{4 \tilde{a}^3} + \frac{1+ 4 \eta^2 +2 \eta^4}{32 \tilde{a}^7} + O\left(\frac{1}{\tilde{a}^{7}} \right) \nonumber \ , \\
\tilde{c} &=& \frac{\eta^2}{4 \tilde{a}} - \frac{1+ 4\eta^2 +2 \eta^4}{96 \tilde{a}^5} + O\left(\frac{1}{\tilde{a}^7} \right)\ .
\end{eqnarray}
By inverting the expression for $\tilde{m}$, we can express $\tilde{c}$ in terms of $\tilde{m}$:
\begin{eqnarray}
\tilde{c} &=& \frac{\eta^2}{4 \tilde{m}} - \frac{1+ 4 \eta^2 + 8 \eta^4}{96 \tilde{m}^5} + O \left(\frac{1}{\tilde{m}^7} \right) \label{weakcond} \ .
\end{eqnarray}
Finally, after going back to dimensionful parameters, we can see that the theory has developed a fermionic  condensate:
\begin{equation}
\langle\bar\psi\psi\rangle\propto-c=-\frac{R^4}{4m}H^2+\frac{b^8 + 4 b^4 R^4 H^2 + 8 R^8 H^4}{96m^5} \ .
\label{weakc}
\end{equation}
The results of the above analysis can be trusted only for finite bare
quark mass and sufficiently low temperature and weak magnetic field.
As can be expected, the physically interesting properties of the
system should be described by the full non--linear equation of motion
of the D7--brane. To explore these we need to use numerical
techniques.
 %
\subsection*{4.2.2 \hspace{2pt} Numerical analysis}
\begin{figure}[h]
   \centering
   \includegraphics[ width=11cm]{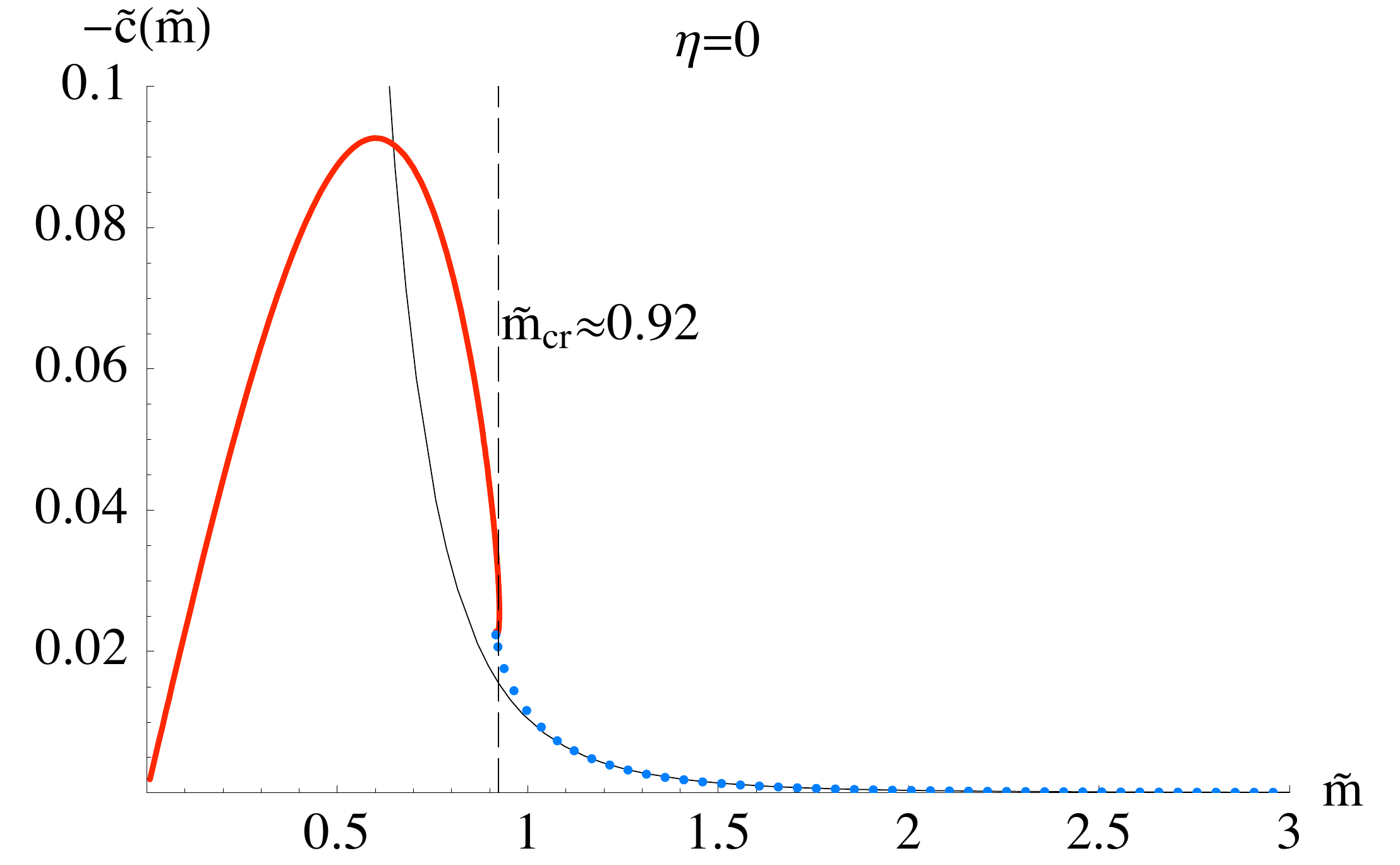}
   \caption{The solid curve starting far left (red) represents solutions falling into the black hole, the dotted  (blue) curve represents solutions with shrinking $S^3$. The vertical dashed line corresponds to the critical value of $\tilde m$ at which the first order phase transition takes place. The solid black curve dropping sharply from  above is the function derived in  equation (\ref{weakcond}), corresponding to the large mass limit. }
   \label{fig:1}
\end{figure}

\begin{figure}[h]
   \centering
   \includegraphics[ width=11cm]{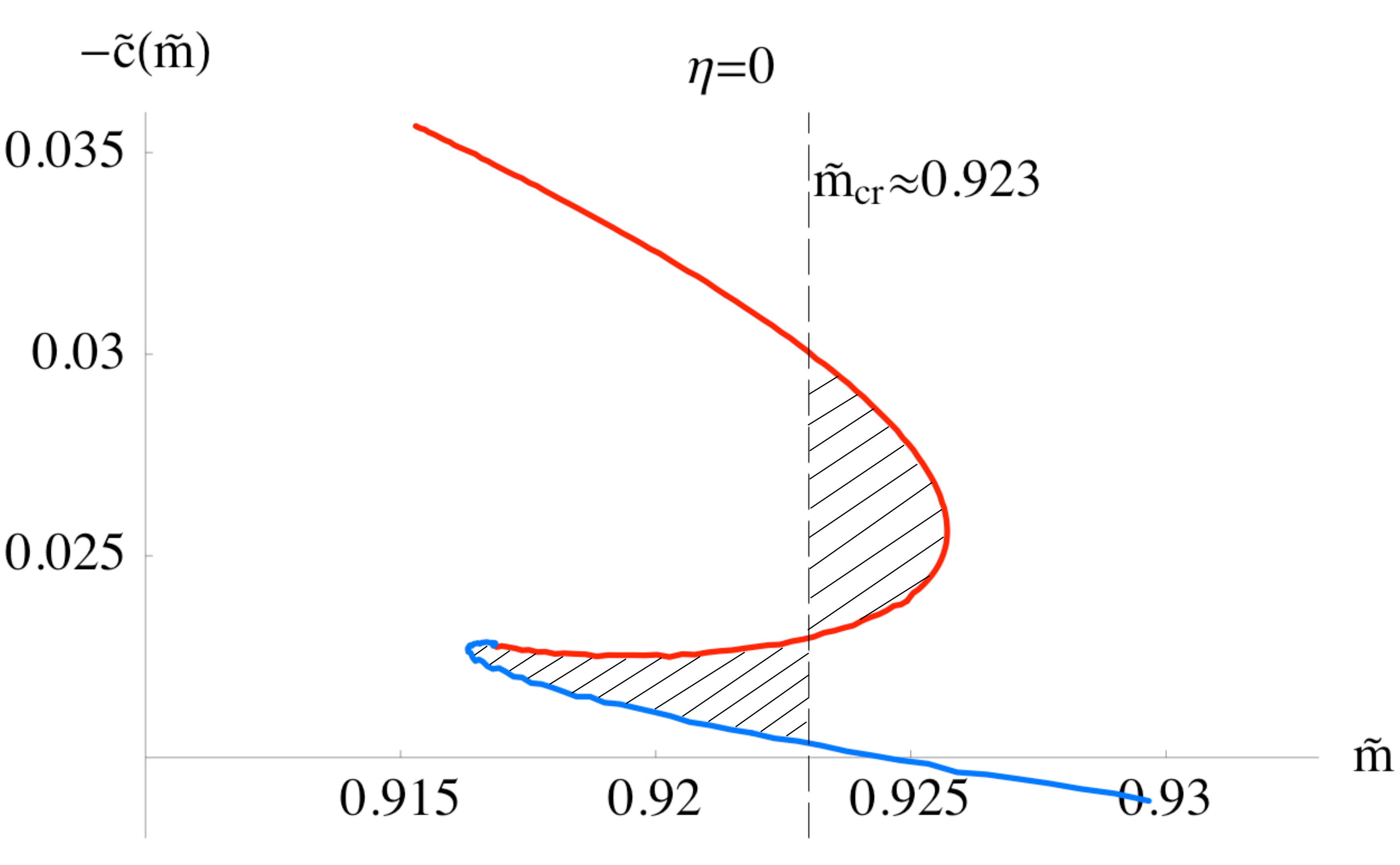}
   \caption{The area below the $(-{\tilde c}, {\tilde m})$ curve has the interpretation of the free energy of the D7--brane; thus the phase transition pattern obeys the ``equal area law''--- the area of the shaded regions is equal. }
   \label{fig:2}
\end{figure}

\addcontentsline{toc}{subsection}{4.2.2 \hspace{0.15cm} Numerical analysis}  
 We solve the differential equation derived from equation
 (\ref{LagrangianM}) numerically using Mathematica. It is convenient to
 use infrared initial conditions \cite{Albash:2006bs,Albash:2006ew}.
 For the Minkowski embeddings, based on symmetry arguments, the
 appropriate initial conditions are:
\begin{equation}
\tilde L(\tilde\rho)|_{\tilde\rho=0}=L_{\rm in}, \quad \tilde L'(\tilde\rho)|_{\tilde\rho=0}=0 \ .
\end{equation}
For the black hole embeddings, the following initial conditions:
\begin{equation}
\tilde L(\tilde\rho)|_{\rm e.h.}=\tilde L_{\rm in}, \quad \tilde L'(\tilde\rho)|_{\rm e.h.}=\left.\frac{\tilde L}{\tilde\rho}\right|_{\rm e.h.} \ ,
\end{equation}
ensure regularity of the solution at the event horizon. After solving
numerically for $\tilde L(\tilde\rho)$ for fixed value of the
parameter $\eta$, we expand the solution at some numerically large
$\tilde\rho_{\rm max}$, and, using equation (\ref{asymptote}), we generate
the plot of $-\tilde c$ vs $\tilde m$. It is instructive to begin our
analysis by revisiting the case with no magnetic field ($\eta=0)$,
familiar from refs. \cite{Albash:2006ew,Babington:2003vm,Mateos:2006nu}. The
corresponding plot for this case is presented in Figure~\ref{fig:1}.
Also in the figure is a plot of the large mass analytic result of
equation~(\ref{weakcond}), shown as the thin black curve in the figure,
descending sharply downwards from above; it can be seen that it is
indeed a good approximation for $\tilde m> \tilde m_{\rm cr}$.  Before
we proceed with the more general case of non--zero magnetic field, we
review the techniques employed in ref.~\cite{Albash:2006ew} to
determine the critical value of $\tilde m$. In Figure~\ref{fig:2}, we
have presented the region of the phase transition considerably
magnified.
Near the critical value $\tilde m_{\rm cr}$, the condensate $\tilde c$ is
a multi-valued function of $\tilde m$, and we have three competing
phases.  The parameter $\tilde c$ is known \cite{Kruczenski:2003uq} to
be proportional to the first derivative of the free energy of the
D7--brane, and therefore the area below the curve of the $-\tilde c$
vs $\tilde m$ plot is proportional to the free energy of the brane.
Thus, the phase transition happens where the two shaded regions in
Figure~\ref{fig:2} have equal areas; furthermore, for $\tilde m <
\tilde m_{\rm cr}$, the upper--most branch of the curve corresponds to the
stable phase, and the lower--most branch of the curve corresponds to a
meta--stable phase.  For $\tilde m > \tilde m_{\rm cr}$, the lower--most
branch of the curve corresponds to the stable phase, and the
upper--most branch of the curve corresponds to a metastable phase.
At $m=m_{\rm cr}$ we have a first order phase transition.  It
should be noted that the intermediate branch of the curve corresponds
to an unstable phase.

\begin{figure}[h] %
   \centering
   \includegraphics[ width=11cm]{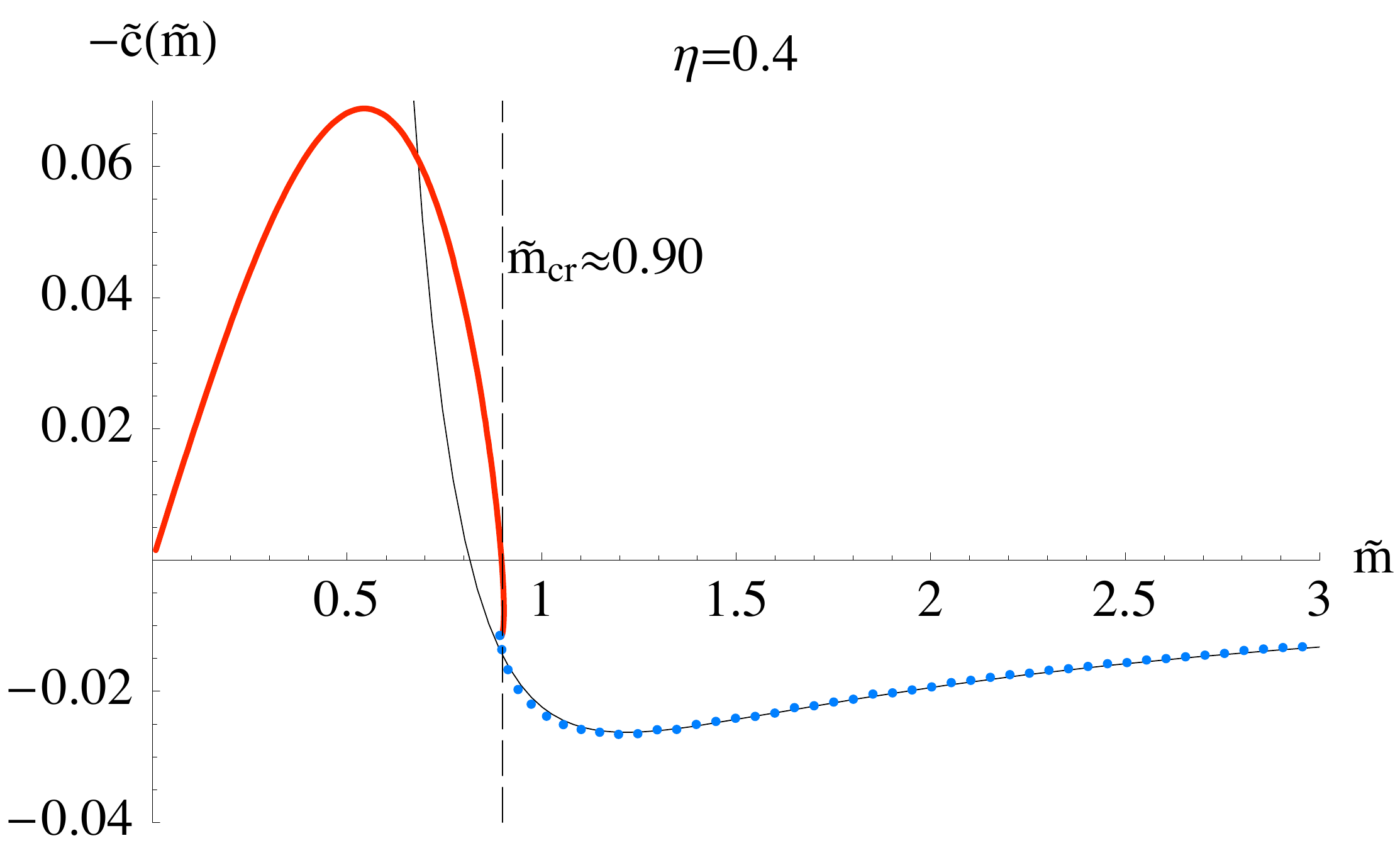}
   \caption{The effect of the weak magnetic field is to decrease the values of $\tilde m_{\rm cr}$ and the condensate. Equation (\ref{weakcond}) is still a good approximation for $\tilde m>\tilde m_{\rm cr}$. }
   \label{fig:3}
\end{figure}

\begin{figure}[h] %
   \centering
   \includegraphics[ width=11cm]{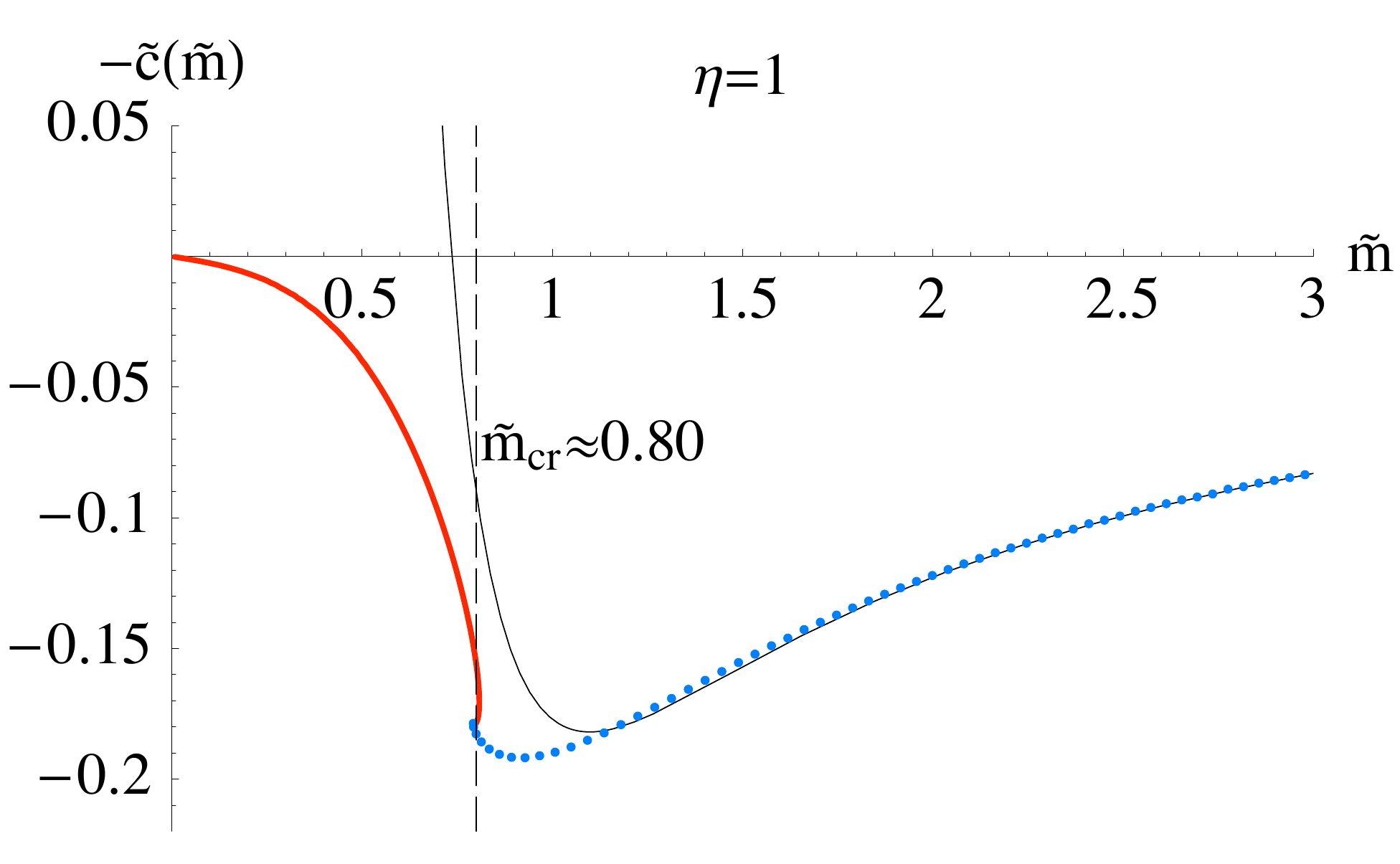}
      \includegraphics[ width=11cm]{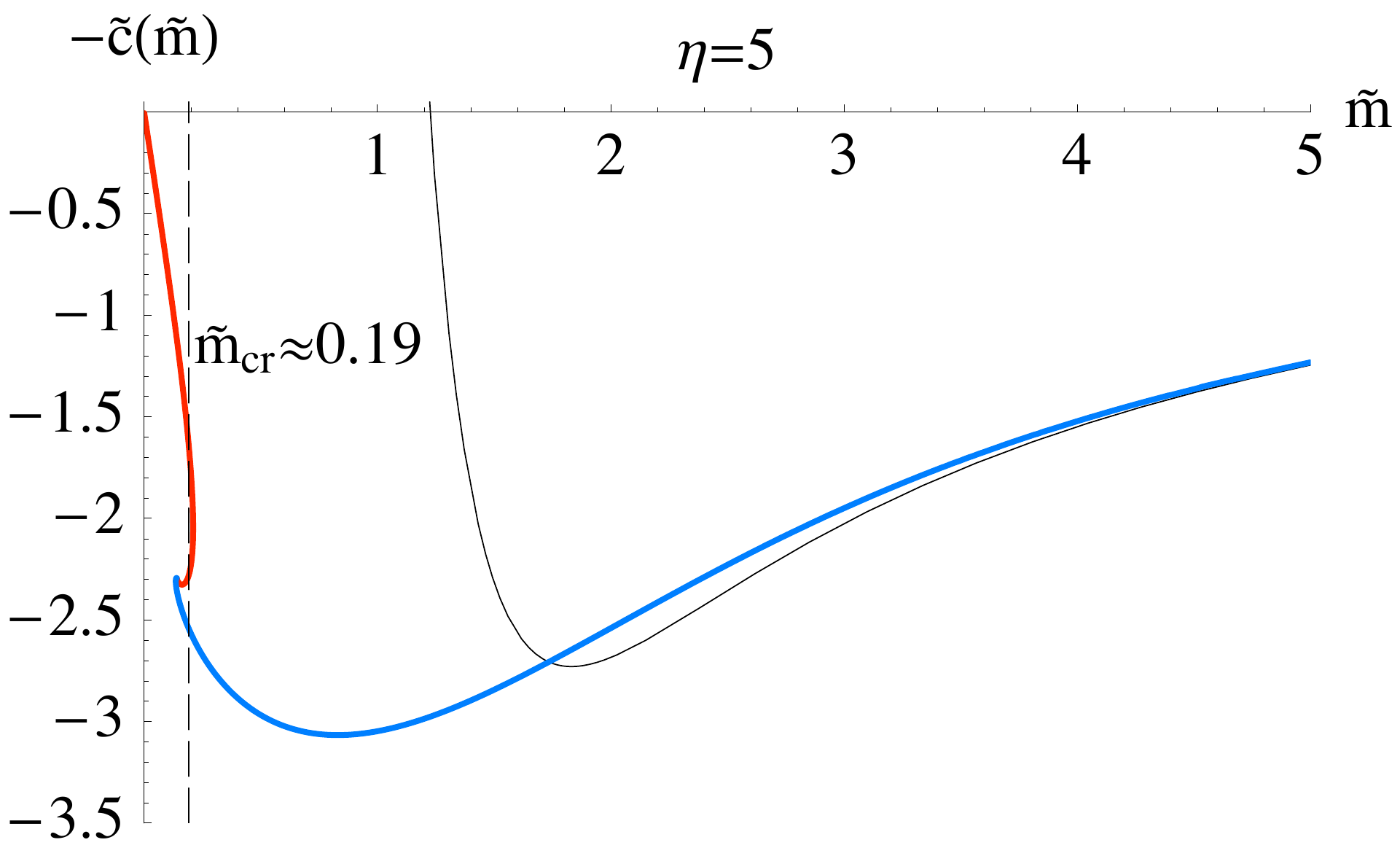}
   \caption{For strong magnetic  field the condensate is negative. The value of $\tilde m_{\rm cr}$ continues to drop as we increase $\eta$. }
   \label{fig:4}
   \end{figure}

\begin{figure}[h]
   \centering
   \includegraphics[ width=11cm]{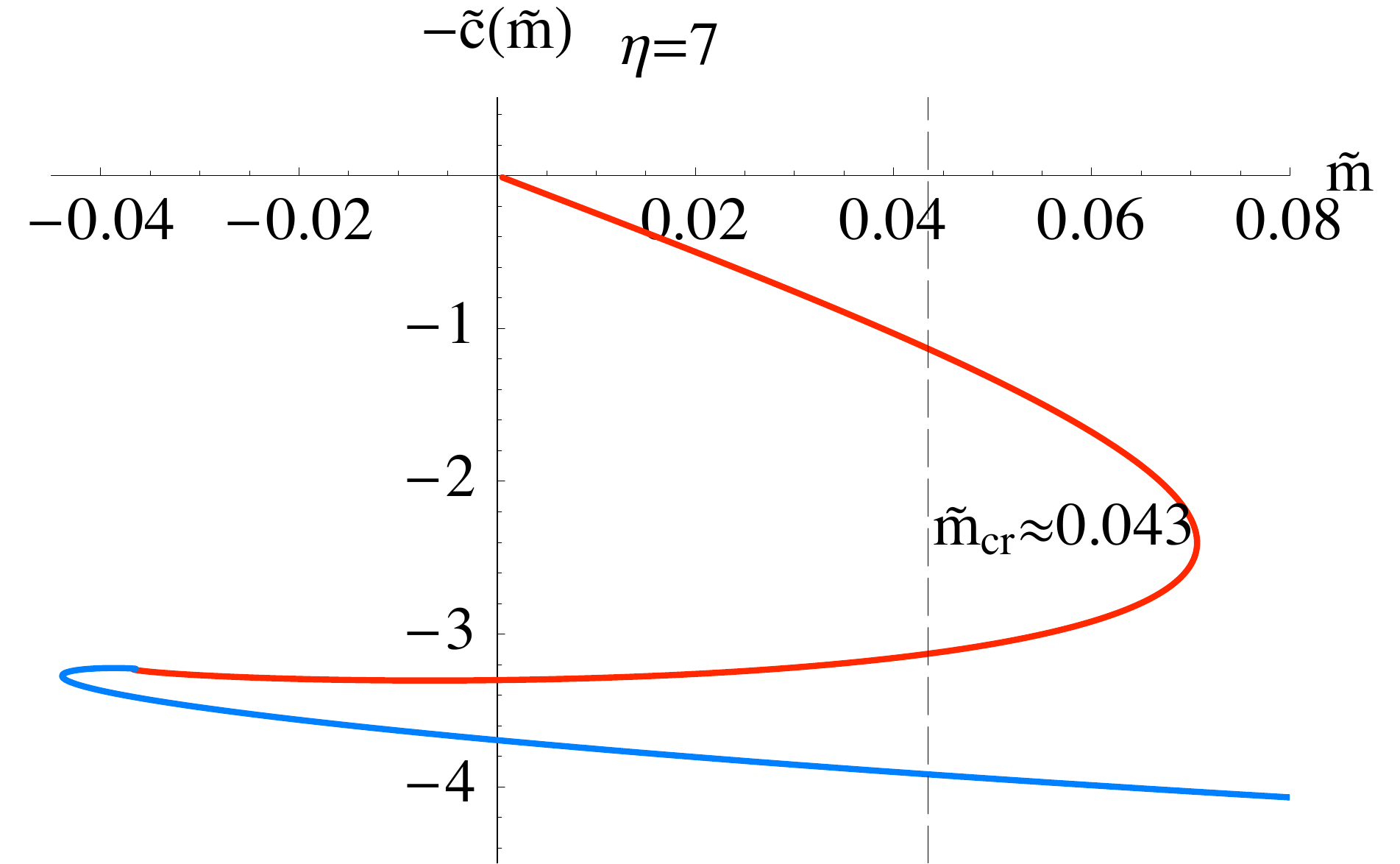}
   \caption{For sufficiently high values of $\eta$ there are states with negative $\tilde m$ which are considered non--physical. However, the equal area law is still valid as long as $m_{\rm cr}>0$. }
   \label{fig:5}
\end{figure}

\begin{figure}[h]
   \centering
   \includegraphics[ width=11cm]{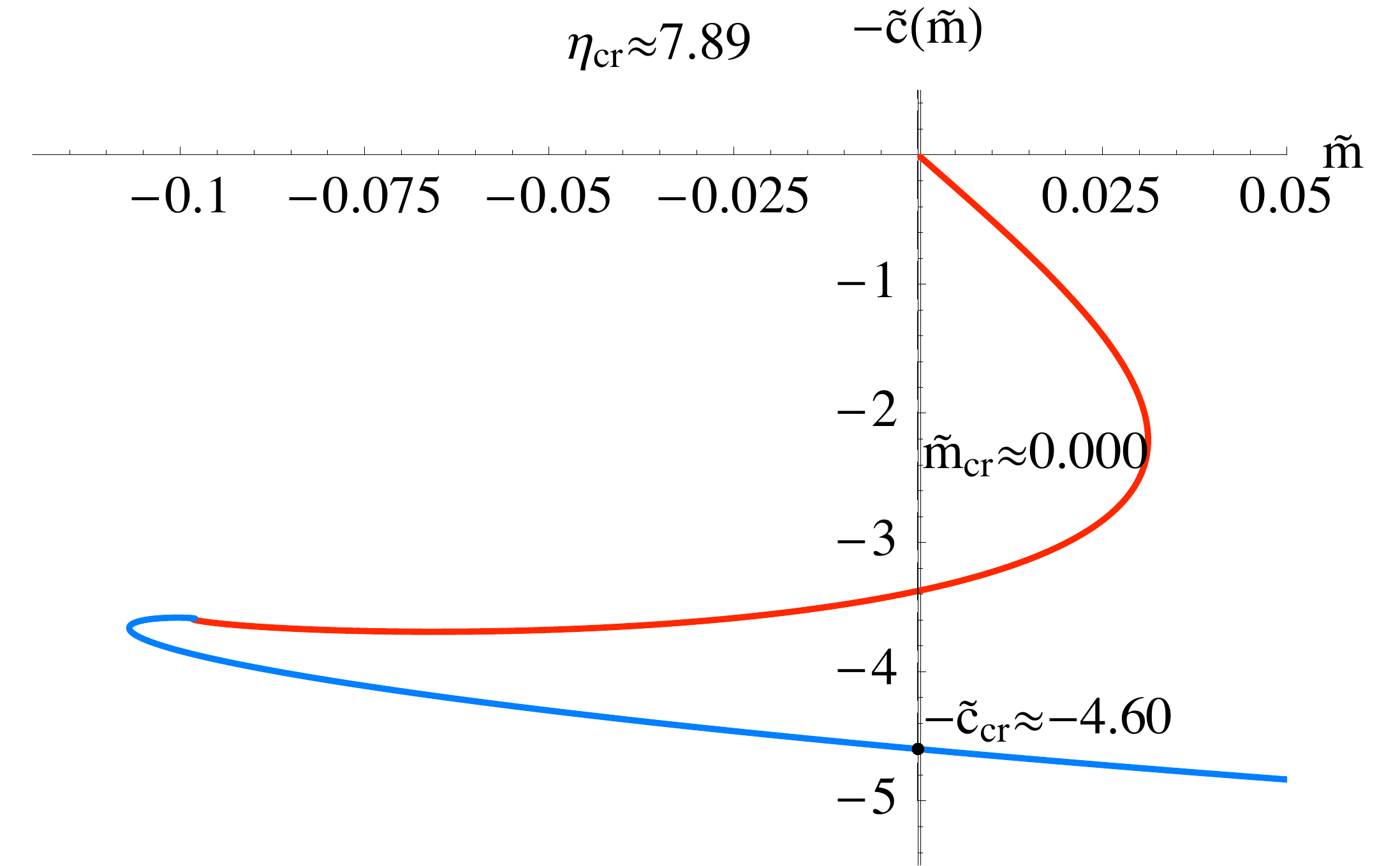}
   \caption{For $\eta=\eta_{\rm cr}$ the critical parameter $m_{\rm cr}$ vanishes. There are two $\tilde m=0$ states with equal energies, one of them has non--vanishing condensate $-\tilde c_{\rm cr}\approx -4.60$ and therefore spontaneously breaks the chiral symmetry. }
   \label{fig:6}
\end{figure}

\begin{figure}[h]
   \centering
   \includegraphics[ width=11cm]{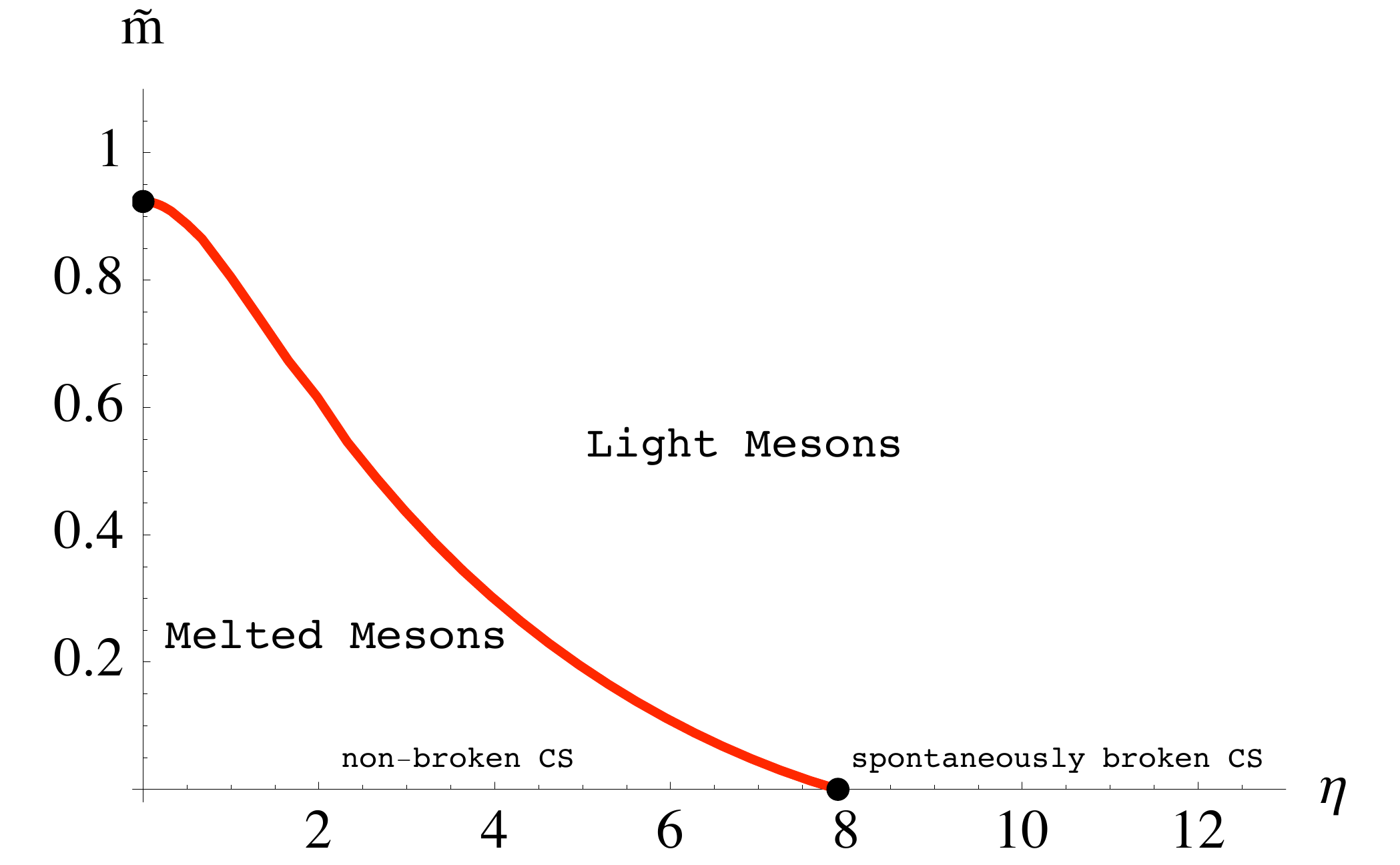}
   \caption{The curve separates the two phases corresponding to discrete meson spectrum (light mesons) and continuous meson spectrum (melted mesons). }
   \label{fig:phase diagram}
\end{figure}

Now, let us turn on a weak magnetic field.  As one can see from Figure~\ref{fig:3}, the effect of the magnetic field is to decrease the
magnitude of $\tilde m_{\rm cr}$.  In addition, the condensate now becomes
negative for sufficiently large $\tilde m$ and approaches zero from
below as $\tilde m\to\infty$. It is also interesting that equation
(\ref{weakcond}) is still a good approximation for $\tilde m>\tilde
m_{\rm cr}$.
For sufficiently strong magnetic field, the condensate has only
negative values and the critical value of $\tilde m$ continues to
decrease, as is presented in Figure~\ref{fig:4}.
   If we further increase the magnitude of the magnetic field, some
   states start having negative values of $\tilde m$, as shown in
   Figure~\ref{fig:5}. The negative values of $\tilde m$ do not mean
   that we have  negative bare quark masses; rather, it implies that
   the D7--brane embeddings have crossed $L = 0$ at least once. It was
   argued in ref.~\cite{Babington:2003vm} that such embeddings are not
   consistent with a holographic gauge theory interpretation and are
   therefore to be considered unphysical.  We will adopt this
   interpretation here, therefore taking as physical only the $\tilde
   m>0$ branch of the $-\tilde c$ vs $\tilde m$ plots.  However, the
   prescription for determining the value of $m_{\rm cr}$ continues to be
   valid, as long as the obtained value of $\tilde m_{\rm cr}$ is
   positive.  Therefore, we will continue to use it in order to
   determine the value of $\eta\equiv\eta_{\rm cr}$ for which $\tilde
   m_{\rm cr}=0$.
As one can see in Figure~\ref{fig:6}, the value of $\eta_{\rm cr}$ that we
obtain is $\eta_{\rm cr}\approx 7.89$. Note also that, for this value of
$\eta$, the Minkowski $\tilde m=0$ embedding has a non--zero fermionic
condensate $\tilde c_{\rm cr}$, and hence the chiral symmetry is
spontaneously broken. For $\eta>\eta_{\rm cr}$, the stable solutions are
purely Minkowski embeddings, and the first order phase transition
disappears; therefore, we have only one class of solutions (the blue
curve) that exhibit spontaneous chiral symmetry breaking at zero bare
quark mass.  Some black hole embeddings remain meta--stable, but
eventually all black hole embeddings become unstable for large enough
$\eta$.  This is confirmed by our study of the meson spectrum which
we present in later sections of the chapter.
The above results can be summarized in a single two dimensional phase
diagram which we present in Figure~\ref{fig:phase diagram}.
The curve separates the two phases corresponding to a discrete meson
spectrum (light mesons) and a continuous meson spectrum (melted
mesons) respectively. The crossing of the  curve is associated with
the first order phase transition corresponding to the melting of the
mesons. If we cross the  curve along the vertical axis, we have the
phase transition described in refs.~\cite{Albash:2006ew,Babington:2003vm,
  Mateos:2006nu}. Crossing the curve along the
horizontal axis corresponds to a transition from unbroken to
spontaneously broken chiral symmetry, meaning the
parameter $\tilde c$ jumps from zero to $\tilde c_{\rm cr}\approx 4.60$,
resulting in non--zero quark condensate of the ground state.
It is interesting to explore the dependence of the fermionic
condensate at zero bare quark mass on the magnetic field. From
dimensional analysis it follows that:
\begin{equation}
c_{\rm cr}=b^3\tilde c_{\rm cr}(\eta)=\frac{\tilde c_{\rm cr}(\eta)}{\eta^{3/2}}R^3H^{3/2} \ .
\label{ccrit}
\end{equation}
\begin{figure}[h]
   \centering
   \includegraphics[ width=11cm]{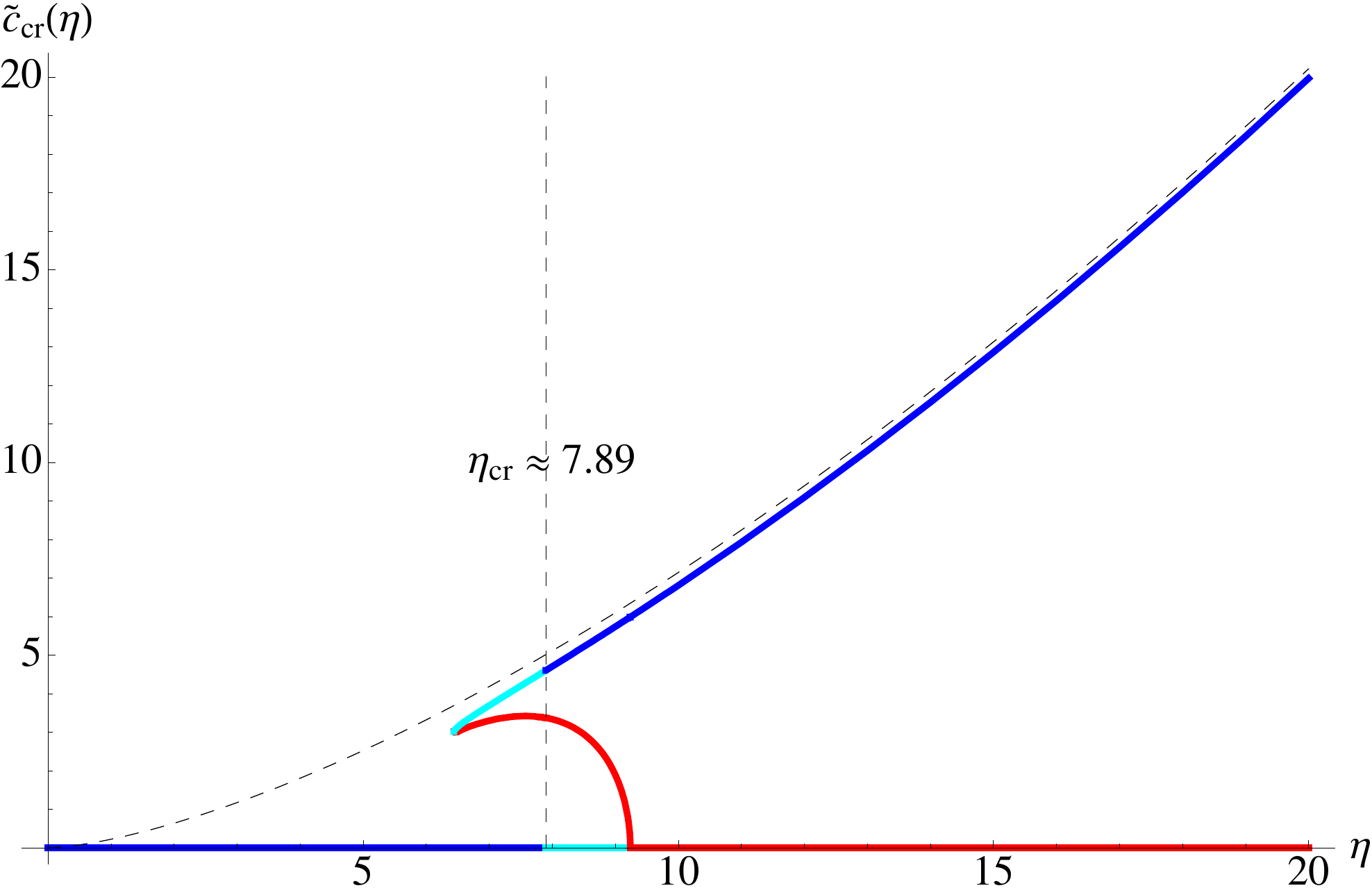}
   \caption{The solid curves are the numerically extracted dependence $\tilde c_{\rm cr}(\eta)$, while the dashed curve represents the expected large $\eta$ behavior $\tilde c_{\rm cr}(\eta)\approx 0.226\eta^{3/2}$.  The solid  curve segments at the bottom left and to the upper right (blue) are the stable states. The straight segment and the arc that joins it (lower right, red) and red are the unstable states. The rest (cyan) are meta--stable states. }
   \label{fig:ccrit}
\end{figure}
In the $T\to 0$ limit, we should recover the result from
Chapter~2: $c_{\rm cr}\approx0.226R^3H^{3/2}$, which
implies that $\tilde c_{\rm cr}(\eta)\approx 0.226\eta^{3/2}$ for $\eta\gg
1$.  The plot of the numerically extracted dependence $\tilde
c_{\rm cr}(\eta)$ is presented in Figure~\ref{fig:ccrit}; for
$\eta>\eta_{\rm cr}$, $\tilde c_{\rm cr}(\eta)$ very fast approaches the curve
$0.226\eta^{3/2}$. This suggests that the value of the chiral symmetry
breaking parameter $c_{\rm cr}$ depends mainly on the magnitude of the
magnetic field $H$, and only weakly on the temperature $T$.

\section*{4.3 \hspace{2pt} Thermodynamics}
\addcontentsline{toc}{section}{4.3 \hspace{0.15cm} Thermodynamics}

Having understood the phase structure of the system, we now turn to
the extraction of various of its important thermodynamic quantities.
\subsection*{4.3.1 \hspace{2pt} The free energy}
\addcontentsline{toc}{subsection}{4.3.1 \hspace{0.15cm} The free energy}
%
Looking at our system from a thermodynamic point of view, we must
specify the potential characterizing our ensemble. We are fixing the
temperature and the magnetic field, and hence the appropriate
thermodynamic potential density is:
\begin{equation}
dF=-SdT-\mu dH\ ,
\end{equation}
where $\mu$ is the magnetization density and $S$ is the entropy
density of the system.  Following ref.~\cite{Mateos:2007vn}, we relate
the on--shell D7--brane action to the potential density~$F$ via:
\begin{equation}
F= 2 \pi^2N_f T_{D7}I_{D7}\ ,
\label{TDpot}
\end{equation}
where (here, $N_f=1$):
\begin{eqnarray}
&&I_{D7}=b^4\int\limits_{\tilde\rho_{\rm min}}^{\tilde\rho_{\rm max}}d\tilde\rho\tilde\rho^3\left(1-\frac{1}{16\tilde r^8}\right)\left(1+\frac{16\eta^2\tilde r^4}{(4\tilde r^4+1)^2}\right)^{\frac12}\sqrt{1+\tilde L'^2}+I_{\rm bound};
\label{action}\\
&&\eta=\frac{R^2}{b^2}H;~~~\tilde r=r/b;~~~\tilde\rho=\rho/b; ~~~\tilde L=L/b;~~~r^2=\rho^2+L^2 .\nonumber
\end{eqnarray}
In principle, on the right hand side of equation (\ref{TDpot}), there
should be terms proportional to $-H^2/2$, which subtract the energy of
the magnetic field alone; however, as we comment below, the
regularization of $I_{D7}$ is determined up to a boundary term of the
form $\mathrm{const}\times H^2$.  Therefore, we can omit this term in
  the definition of $F$.
  The boundary action $I_{\rm bound}$ contains counterterms
  designed \cite{Balasubramanian:1999re} to cancel the divergent terms
  coming from the integral in equation (\ref{action}) in the limit of
  $\rho_{\rm max}\to\infty$.  A crucial observation is that the finite
  temperature does not introduce new divergences, and we have the
  usual quartic divergence from the spatial volume of the
  asymptotically AdS$_5$ spacetime~\cite{Karch:2005ms}. The presence
  of the non--zero external magnetic field introduces a new
  logarithmic divergence which can be cancelled by introducing the
  following counterterm:
\begin{equation}
- \frac{R^4}{2}\log\left(\frac{\rho_{\rm max}}{R}\right) \int d^4x\sqrt{-\gamma}\frac{1}{2!}B_{\mu\nu}B^{\mu\nu} \ ,
\end{equation}
where $\gamma$ is the metric of the 4--dimensional surface at
$\rho=\rho_{\rm max}$. Note that in our case:
\begin{equation}
\frac{1}{2!}\sqrt{-\gamma}\, B_{\mu\nu}B^{\mu\nu}=H^2\ ,
\label{counterterm}
\end{equation}
which gives us the freedom to add finite terms of the form
$\mathrm{const}\times H^2$ at no cost to the regularized action.  This
makes the computation of some physical quantities scheme dependent.
We will discuss this further in subsequent sections.
The final form of $I_{\rm bound}$ in equation (\ref{action}) is:
\begin{equation}
I_{\rm bound}=-\frac{1}{4}\rho_{\rm max}^4-\frac{1}{2}R^4H^2\log{\frac{\rho_{\rm max}}{R}} \ .
\end{equation}
It is instructive to evaluate the integral in equation (\ref{action})
for the $L\equiv0$ embedding at zero temperature.  Going back to
dimensionful coordinates we obtain:
\begin{equation}
\int\limits_{0}^{\rho_{\rm max}}d\rho\rho^3\sqrt{1+\frac{R^4H^2}{\rho^4}}=\frac{1}{4}\rho_{\rm max}^4+\frac{1}{2}R^4H^2\log{\frac{\rho_{\rm max}}{R}}+\frac{R^4H^2}{8}(1+\log{4}-\log{H^2})+O(\rho_{\rm max}^{-3}) \ .
\end{equation}
The first two terms are removed by the counter terms from $I_{\rm bound}$,
and we are left with:
\begin{equation}
F(b=0,m=0,H)= 2 \pi^2 N_f T_{D7}\frac{R^4H^2}{8}(1+\log{4}-\log{H^2}) \ .
\label{freeenergy0}
\end{equation}
This result can be used to evaluate the magnetization density of the
Yang--Mills plasma at zero temperature and zero bare quark mass. Let
us proceed by writing down a more general expression for the free
energy of the system.  After adding the regulating terms from
$I_{\rm bound}$, we obtain that our free energy is a function of $m, b,
H$:
\begin{equation}
F(b,m,H)= 2 \pi^2 N_f T_{D7}b^4\tilde I_{D7}(\tilde m,\eta^2)+F(0,0,H) \ ,
\label{freeenergy_gen}
\end{equation}
where $\tilde I_{D7}(\tilde m,\eta)$ is defined {\it via}:
\begin {eqnarray}
  \tilde I_{D7} &=&\int\limits_{\tilde\rho_{\rm min}}^{\tilde\rho_{\rm max}}d\tilde\rho\left(\tilde\rho^3\left(1-\frac{1}{16\tilde r^8}\right)\left(1+\frac{16\eta^2\tilde r^4}{(4\tilde r^4+1)^2}\right)^{\frac12}\sqrt{1+\tilde L'^2}-\tilde\rho^3\right)-\tilde\rho_{\rm min}^4/4\nonumber\\
  &&-\frac{1}{2}\eta^2\log\tilde\rho_{\rm max}-\frac{1}{8}\eta^2(1+\log4-\log\eta^2);~~~\tilde
  r^2=\tilde\rho^2+\tilde L(\tilde\rho)^2\label{tildeaction}\ .
\end{eqnarray}
In order to verify the consistency of our analysis with our numerical
results, we derive an analytic expression for the free energy that is
valid for $\tilde m\gg \sqrt{\eta}$. To do
this we use that for large $\tilde m$, the condensate $\tilde c$ is
given by equation (\ref{weakcond}) which we repeat here:
\begin{equation}
\tilde c(\tilde m,\eta^2)=\frac{\eta^2}{4\tilde m}-\frac{1+4\eta^2+8\eta^4}{96\tilde m^5}+O(1/\tilde m^7) \ ,
\end{equation}
as well as the relation ${\partial\tilde I_{D7}}/{\partial\tilde
  m}=-2\tilde c$.  We then have:
%
\begin{equation}
\tilde I_{D7}=-2\int\limits^{\tilde m}\tilde c(\tilde m,\eta)d\tilde m+\xi(\eta)=\xi(\eta)-\frac{1}{2}\eta^2\log\tilde m-\frac{1+4\eta^2+8\eta^4}{192\tilde m^4}+O(1/\tilde m^6) \ ,
\label{tildeaction1}
\end{equation}
where the function $\xi(\eta)$ can be obtained by evaluating the
expression for $\tilde I_{D7}$ from equation (\ref{tildeaction}) in the
approximation $\tilde L\approx \tilde m$. Note that this suggests
ignoring the term $\tilde L'^2$ which is of order $\tilde c^2$.
Since the leading behavior of $\tilde c^2$ at large $\tilde m$ is
$1/\tilde m^2$, this means that the results obtained by setting
$\tilde L'^2=0$ can be trusted to the order of $1/\tilde m$, and
therefore we can deduce the function $\xi(\eta)$, corresponding to the
zeroth order term. Another observation from earlier in this chapter is
that the leading behavior of the condensate is dominated by the
magnetic field and therefore we can further simplify equation~(\ref{tildeaction}):
\begin{eqnarray}
\tilde I_{D7}&\backsimeq& \lim_{\tilde\rho_{\rm max}\to\infty}\int\limits_0^{\tilde\rho_{\rm max}}d\tilde\rho\rho^3(\sqrt{1+\frac{\eta^2}{(\tilde\rho^2+\tilde m^2)^2}}-1)-\frac{1}{2}\eta^2\log\tilde\rho_{\rm max}-\frac{1}{8}\eta^2(1-\log\frac{\eta^2}{4})\nonumber\\
&=&-\frac{\eta^2}{2}\log\tilde m-\frac{\eta^2}{8}(3-\log\frac{\eta^2}{4})+O(1/\tilde m^3)\ .
\end{eqnarray}
Comparing to equation (\ref{tildeaction1}), we obtain:
\begin{equation}
\xi(\eta)=-\frac{\eta^2}{8}(3-\log\frac{\eta^2}{4}) \ ,
\end{equation}
%
and our final expression for $\tilde I_{D7}$, valid for $\tilde m\gg \sqrt{\eta}$:
\begin{equation}
\tilde I_{D7}=-\frac{\eta^2}{8}\left(3-\log\frac{\eta^2}{4}\right)-\frac{1}{2}\eta^2\log\tilde m-\frac{1+4\eta^2+8\eta^4}{192\tilde m^4}+O(1/\tilde m^6) \ .
\label{tildeaction2}
\end{equation}
%
\subsection*{4.3.2 \hspace{2pt} The entropy}
\begin{figure}[h]
 \centering
  \includegraphics[ width=11cm]{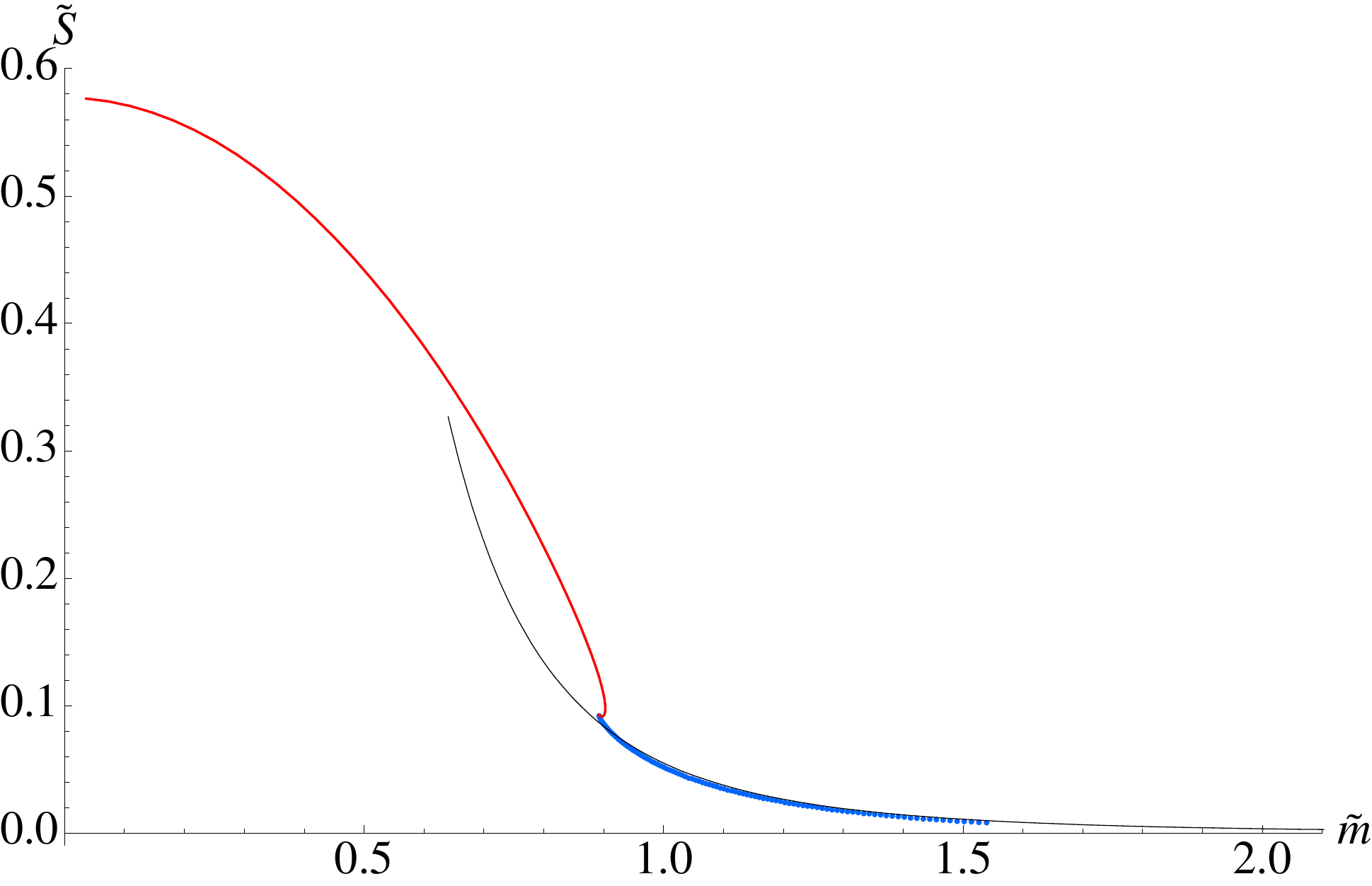}
   \caption{A plot of $\tilde S$ versus $\tilde m$ for $\eta=0.4$. The  thin (sharply descending and extending to the right) black curve corresponds to that large mass result of equation~(\ref{large m}). }
   \label{fig:entropyh=.4}
 \end{figure}

\addcontentsline{toc}{subsection}{4.3.2 \hspace{0.15cm} The entropy}
Our next goal is to calculate the entropy density of the system. Using
our expressions for the free energy we can write:
\begin{eqnarray}
S&=&-\left(\frac{\partial F}{\partial T}\right)_{H}=-\pi R^2 \frac{\partial F}{\partial b}=-2 \pi^3 R^2 N_f T_{D7}b^3\left(4\tilde I_{D7}+b\frac{\partial\tilde I_{D7}}{\partial\tilde m}\frac{\partial\tilde m}{\partial b}+b\frac{\partial\tilde I_{D7}}{\partial\eta^2}\frac{\partial\eta^2}{\partial b}\right) \nonumber \\
&=&-2 \pi^3 R^2 N_f T_{D7}b^3\left(4\tilde I_{D7}+2\tilde c\tilde m-4\frac{\partial\tilde I_{D7}}{\partial\eta^2}\eta^2\right)=2 \pi^3 R^2 N_f T_{D7}b^3\tilde S(\tilde m,\eta^2)\ . \label{entropy}
\end{eqnarray}
It is useful to calculate the entropy density at zero bare quark mass and zero quark condensate. To do this, we need to calculate the free energy density by evaluating the integral in equation (\ref{tildeaction}) for $\tilde L\equiv0$. The expression that we get for $\tilde I_{D7}(0,\eta^2)$ is:
\begin{equation}
\tilde I_{D7}(0,\eta^2)=\frac {1} {8}\left (1-2 \sqrt {1 + \eta^2}-\eta^2\log\frac{(1+ \sqrt {1 + \eta^2})^2}{\eta^2} \right)\ .
\end{equation}
The corresponding expression for the entropy density is:
\begin{equation}
S|_{m=0}=2 \pi^6 R^8 N_f T_{D7}T^3\left(-\frac{1}{2}+\sqrt{1+\frac{\pi^4 H^2}{R^4 T^4}}\right)\ .
\end{equation}
One can see that the entropy density is positive and goes to zero as
$T\to 0$. Our next goal is to solve for the entropy density at finite
$\tilde m$ for fixed $\eta$. To do so, we have to integrate
numerically equation (\ref{entropy}) and generate a plot of $\tilde S$
versus $\tilde m$. However, for $\tilde m\gg \sqrt{\eta}$ we can
derive an analytic expression for the entropy.  After substituting the
expression from equation (\ref{tildeaction2}) for $\tilde I_{D7}$ into
equation (\ref{entropy}) we obtain:
 \begin{equation}
 \tilde S(\tilde m,\eta^2)=\frac{1+2\eta^2}{24\tilde m^4}+\dots \ ,
 \label{large m}
 \end{equation}
 or if we go back to dimensionful parameters:
 \begin{equation}
 S(b,m,H)=2 \pi^3R^2 N_f T_{D7}b^3\left(\frac{b^4+2R^4H^2}{24m^4}\right)\ .
 \end{equation}
 One can see that if we send $T\to 0$, while keeping $\eta$ fixed we
 get the $T^7$ behavior described in ref.~\cite{Mateos:2007vn}, and
 therefore the (approximate; $N_f/N\ll1$) conformal behavior is
 restored in this limit.  In Figure~\ref{fig:entropyh=.4}, we present
 a plot of $\tilde S$ versus $\tilde m$ for $\eta=0.4$ . The solid
 smooth black curve corresponds to equation (\ref{large m}).  For this
 $\tilde S$ is positive and always a decreasing function of $\tilde
 m$.  Hence, the entropy density at fixed bare quark mass $m=\tilde m
 b$, given by $S=2 \pi^3 R^2 N_f T_{D7}m^3\tilde S/\tilde m^3$, is
 also a decreasing function of $\tilde m$ and therefore an increasing
 function of the temperature, except near the phase transition (the
 previously described crossover from black hole to Minkowski
 embeddings) where an unstable phase appears that is characterized by
 a negative heat capacity.
 %
\subsection*{4.3.3 \hspace{2pt} The magnetization}
\begin{figure}[h]
   \centering
   \includegraphics[ width=11cm]{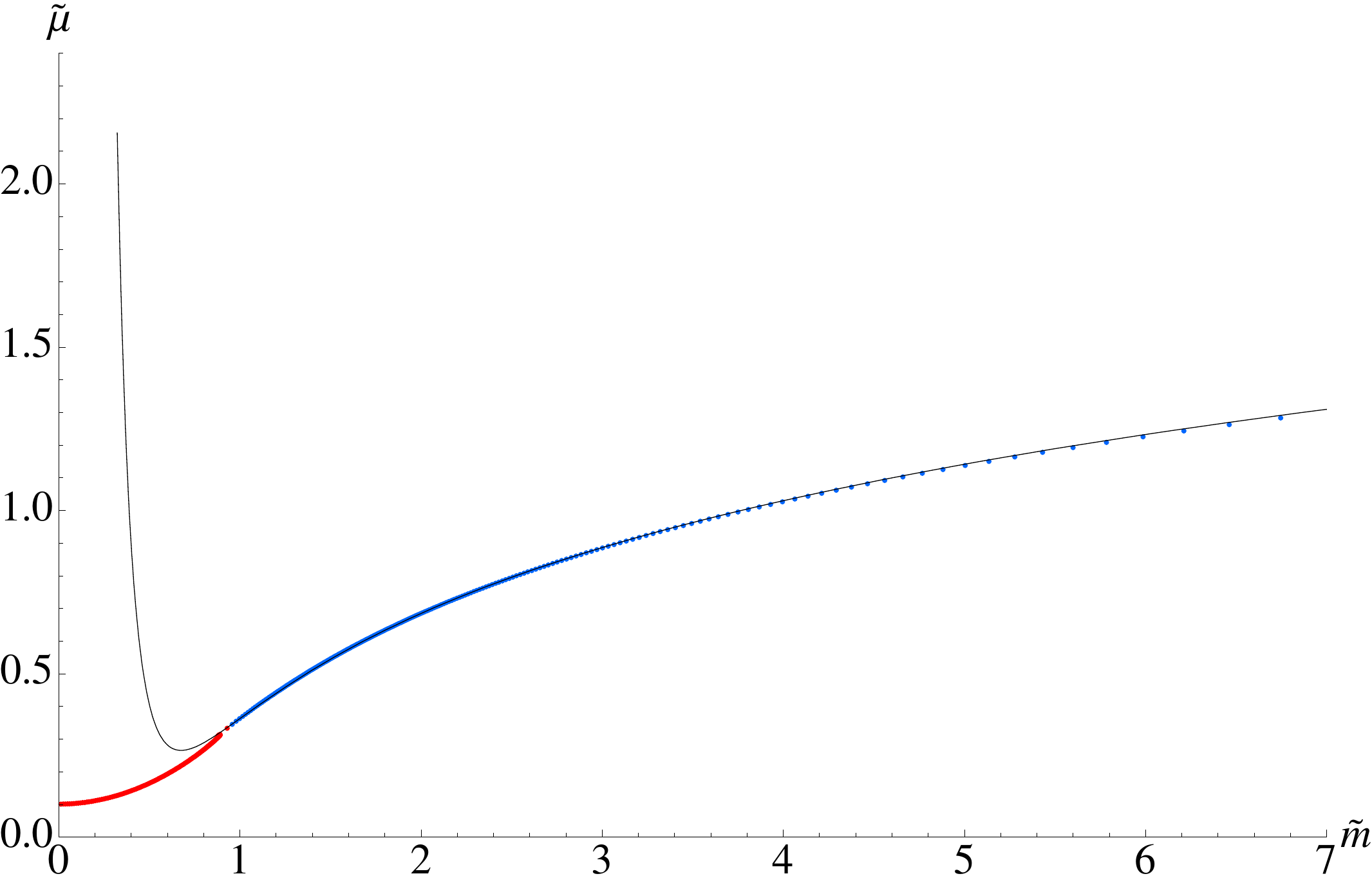}
   \caption{A plot of the dimensionless relative magnetization $\tilde\mu$ versus $\tilde m$ for $\eta=0.5$. The thin black curve (starting with a steep descent) corresponds to the large mass  result of equation (\ref{magn-anal}). }
   \label{fig:magnh=0.5}
\end{figure}

\addcontentsline{toc}{subsection}{4.3.3 \hspace{0.15cm} The magnetization}
 Let us consider equation (\ref{freeenergy0}) for the free energy
 density at zero temperature and zero bare quark mass. The
 corresponding magnetization density is given by:
\begin{equation}
\mu_0=-\left(\frac{\partial F}{\partial H}\right)_{T,m=0}=2 \pi^2 R^4 N_f T_{D7}\frac{H}{2}\log\frac{H}{2}.
\end{equation}
Note that this result is scheme dependent because of the freedom to
add terms of the form $\mathrm{const}\times H^2$ to the boundary action that we
discussed earlier. However, the value of the relative magnetization is
given by:
\begin{equation}
\mu-\mu_0=-\left(\frac{\partial F}{\partial H}\right)_T-\mu_0=-2 \pi^2 R^2 N_f T_{D7}b^2 \left(\frac{\partial\tilde I_{D7}}{\partial\eta}\right)_{\tilde{m}}= 2 \pi^2 R^2 N_f T_{D7}b^2\tilde\mu \ ,
\label{magn}
\end{equation}
is scheme independent and is the quantity of interest in the section.
In equation (\ref{magn}), we have defined
$\tilde\mu=-{\partial\tilde
    I_{D7}}/{\partial\eta}|_{\tilde{m}}$ as a dimensionless
parameter characterizing the relative magnetization. The expression for $\tilde\mu$ follows directly from equation
(\ref{tildeaction}):
\begin{equation}
\tilde\mu=\lim_{\tilde\rho_{\rm max}\to\infty}-\int\limits_{\tilde\rho_{\rm min}}^{\tilde\rho_{\rm max}}d\tilde\rho\frac{\tilde\rho^3(4\tilde r^4-1)}{\tilde r^4\sqrt{(4\tilde r^4+1)^2+16\eta\tilde r^4}}+\eta\log\tilde\rho_{\rm max}-\frac{\eta}{2}\log\frac{\eta}{2}.
\label{tildemagnetization}
\end{equation}
For the large $\tilde m$ region we use the asymptotic expression for
$\tilde I_{D7}$ from equation (\ref{tildeaction2}) and obtain the
following analytic result for $\tilde\mu$:
\begin{equation}
 \tilde\mu=\frac{\eta}{2}-\frac{\eta}{2}\log\frac{\eta}{2}+\eta\log\tilde m+\frac{\eta(1+4\eta^2)}{24\tilde m^4}+O(1/\tilde m^6)\ .
\label{magn-anal}
\end{equation}
We evaluate the above integral numerically and generate a plot of
$\tilde\mu$ versus $\tilde m$.  A plot of the dimensionless relative
magnetization $\tilde\mu$ versus $\tilde m$ for $\eta=0.5$ is
presented in Figure~\ref{fig:magnh=0.5}. The black curve corresponding
to equation (\ref{magn-anal}) shows good agreement with the asymptotic
behavior at large $\tilde m$.  It is interesting to verify the
equilibrium condition ${\partial\tilde\mu}/{\partial T}>0$. Note
that since $\mu_0$ does not depend on the temperature, the value of
this derivative is a scheme independent quantity. From equations
(\ref{magn}) and equation (\ref{magn-anal}), one can obtain:
\begin{equation}
\frac{\partial\mu}{\partial T}=2 \pi^3 R^4 N_f T_{D7}b\left(2\tilde\mu-\frac{\partial\tilde\mu}{\partial\tilde m}\tilde m-2\frac{\partial\tilde\mu}{\partial\eta}\eta\right)=2 \pi^3 R^6 N_f T_{D7}\frac{Hb^3}{6m^4}>0\ ,
\end{equation}
which is valid for large $m$ and weak magnetic field $H$. Note that
the magnetization seems to increase with the temperature. Presumably
this means that the temperature increases the ``ionization'' of the
Yang--Mills plasma of mesons even before the phase transition occurs.
\section*{4.4 \hspace{2pt} Meson spectrum}
\addcontentsline{toc}{section}{4.4 \hspace{0.15cm} Meson spectrum}
In this section, we calculate the meson spectrum of the gauge theory.
The mesons we are considering are formed from quark--antiquark pairs,
so the relevant objects to consider are 7--7 strings.  In our
supergravity description, these strings are described by fluctuations
(to second order in $\alpha'$) of the probe branes' action about the
classical embeddings we found in the previous
sections \cite{Kruczenski:2003be}.  Studying the meson spectrum serves
two purposes.  First, tachyons in the meson spectrum from fluctuations
of the classical embeddings indicate the instability of the embedding.
Second, a massless meson satisfying a Gell-Mann-Oakes-Renner (GMOR)
relation \cite{GellMann:1968rz} will confirm that spontaneous chiral symmetry breaking has
occurred.  As a reminder, in ref.~\cite{Kruczenski:2003be}, the exact
meson spectrum for the AdS$_5 \times S^5$ background was found to be
given by:
\begin{eqnarray}
M(n,\ell) &=& \frac{2 m}{R^2} \sqrt{(n+\ell + 1)(n + \ell + 2)} \ ,
\end{eqnarray}
where $\ell$ labels the order of the spherical harmonic expansion, and $n$ is a positive integer that represents the order of the mode.
The relevant pieces of the action to second order in $\alpha'$ are:
\begin{eqnarray} \label{eqt:fluctuation}
S/N_f&=& -  T_{D7} \int d^8 \xi \sqrt{g_{ab} + B_{a b}+ 2 \pi \alpha' F_{a b}} + \left(2\pi \alpha'\right) \mu_7 \int_{\mathcal{M}_8} F_{(2)} \wedge B_{(2)} \wedge P\left[\tilde{C}_{(4)} \right]  \nonumber \\
&&  +\left(2\pi \alpha' \right)^2 \mu_7 \frac{1}{2} \int_{\mathcal{M}_8} F_{(2)} \wedge F_{(2)} \wedge P\left[C_{(4)}\right] \ , \\
C_{(4)} &=& \frac{1}{g_s} \frac{u^4}{R^4} dt \wedge dx^1 \wedge dx^2 \wedge dx^3  \ ,\end{eqnarray}
\begin{equation}
\tilde{C}_{(4)} = -\frac{R^4}{g_s}  \left(1- \cos^4\theta \right) \sin\psi \cos\psi\ d \psi \wedge d \phi_2 \wedge d \phi_3 \wedge d\phi _1\ ,
\end{equation}
where $P\left[C_{(4)}\right]$ is the pull--back of the 4--form
potential sourced by the stack of $N_c$ D3--branes,
$P\left[\tilde{C}_{(4)}\right]$ is the pull--back of the 4--form
magnetic dual to $C_{(4)}$, and $F_{(2)}$ is the Maxwell 2--form on
the D7--brane world--volume.  At this point, we resort to a different
set of coordinates than we have been using.  Instead of using the
coordinates $(\rho, L)$ introduced in
equation~(\ref{eqt:changeofcoordinates}), we return to the coordinates
$(z=1/u^2, \theta)$ because the analysis is simpler.
We consider fluctuations of the form:
\begin{eqnarray}
\theta &=& \theta_0(z) + 2 \pi \alpha' \chi(\xi^a) \ , \label{eqt:ansatz1} \\
\phi_1 &=& 2 \pi \alpha' \Phi(\xi^a) \ , \label{eqt:ansatz2}
\end{eqnarray}
where the indices $a, b = 0 \dots 7$ run along the world--volume of the
D7--brane.  $\theta_0(z)$ corresponds to the classical embedding from
the classical equations of motion.  Plugging the ansatz in
equations~(\ref{eqt:ansatz1}) and~(\ref{eqt:ansatz2}) into the action
and expanding to second order in $\left(2 \pi \alpha' \right)$, we get
as second order terms in the lagrangian:
\begin{eqnarray}
-\mathcal{L}_{\chi^2} &=& \frac{1}{2} \sqrt{-E} S^{a b} R^2 \partial_a \chi \partial_b \chi -\frac{1}{2} \sqrt{-E} R^4 \left(\theta_0' \right)^2 E^{z z} S^{a b} \partial_a \chi \partial_b \chi \\
&& + \frac{1}{2} \chi^2 \left[ \partial_\theta^2 \sqrt{-E} - \partial_z \left(E^{zz} R^2 \theta_0' \partial_\theta \sqrt{-E} \right) \right] \nonumber \ , \\
-\mathcal{L}_{\Phi^2}&=&  \frac{1}{2} \sqrt{-E} S^{a b} R^2 \sin^2 \theta_0 \partial_a \Phi \partial_b \Phi \ , \nonumber \\
-\mathcal{L}_{F^2} &=& \frac{1}{4} \sqrt{-E} S^{a b} S^{c d} F_{b c} F_{a d}  \nonumber \ , \\
-\mathcal{L}_{F-\chi} &=& \chi F_{23} \left[ \partial_z \left(\sqrt{-E} R^2 \theta_0' E^{zz} J^{23} \right) - J^{23} \partial_\theta \sqrt{-E} \right] = \chi F_{2 3} f \nonumber \ , \\
\mathcal{L}_{F^2}^{\mathrm{WZ}} &=&  \frac{1}{8} \frac{1}{z^2 R^4} F_{m n} F_{o p} \epsilon^{m n o p} \nonumber \ , \\
\mathcal{L}_{F-\Phi}^{\mathrm{WZ}} &=& -\Phi F_{01} B_{23}  R^4 \sin\psi \cos\psi \partial_z \left( 1- \cos^4 \theta_0 \right) = - \Phi F_{0 1} B_{2 3} R^4 \sin \psi \cos \psi \partial_z K \ .\nonumber
\end{eqnarray}
We have taken $E_{a b} = g^{(0)}_{a b} + B_{a b}$ to be the zeroth
order contribution from the DBI action.  In addition, we use that
$E^{a b} = S^{a b} + J^{a b}$, where $ S^{a b} = S^{ b a}$ and $J^{a
  b} = - J^{b a}$.  We use this notation for brevity.  The indices $m,
n, o, p = 4 \dots 9$ run in the transverse directions to the
D3--branes.  From these lagrangian terms, we derive the equation of
motion for $\chi$ to be:
\begin{eqnarray}
\hspace{-0.6cm}&&0= \partial_a \left(\sqrt{-E} S^{a b} R^2 \left(\frac{1+4b^4 z^4 \left(\theta_0'\right)^2}{1+4 z^2 \left(\theta'_0\right)^2} \right) \partial_b \chi \right) -\chi  \left[ \partial_\theta^2 \sqrt{-E} - \partial_z \left(E^{zz} R^2 \theta_0' \partial_\theta \sqrt{-E} \right) \right]  \nonumber \\
\hspace{-0.6cm}&&~~~~~~ - F_{23} \left[ \partial_z \left(\sqrt{-E} R^2 \theta_0' E^{zz} J^{23} \right) - J^{23} \partial_\theta \sqrt{-E}  \right]  \ .
\end{eqnarray}
The equation of motion for $\Phi$ is given by:
\begin{eqnarray}
\partial_a \left( \sqrt{-E} S^{a b} R^2 \sin^2 \theta_0 \partial_b \Phi \right) - F_{01} B_{23}  R^4 \sin\psi \cos\psi \partial_z K &=& 0 \ .
\end{eqnarray}
The equation of motion for $A_b$ is given by:
\begin{eqnarray}
 \partial_a \left( - \sqrt{-E} S^{a a'} S^{b b'} F_{a' b'} - \chi f  \left(\delta^a_2 \delta^b_3 - \delta^a_3 \delta^b_2 \right) + B_{23} \Phi \partial_z K \left(\delta^a_0 \delta^b_1 - \delta^a_1 \delta^b_0 \right) \right. &&  \\
 \left. + \frac{1}{2} \frac{1}{z^2 R^4} \epsilon^{ m n o p} \delta^a_m \delta^b_n F_{o p} \right) &=& 0 \ . \nonumber
\end{eqnarray}
We are allowed to set $A_m = 0$ with the constraint (using that $S^{22} = S^{33}$):
\begin{eqnarray}
S^{00} \partial_m \partial_0 A_0 + S^{11} \partial_m \partial_1 A_1 + S^{22} \partial_m \left(\partial_2 A_2 +  \partial_3 A_3\right) &=& 0
\end{eqnarray}
Therefore, we can consistently take $A_0 = \partial_1 A_1 = 0$, $\partial_2 A_2 = -\partial_3 A_3$.  With this particular choice, we have as equations of motion for the gauge field:
\begin{eqnarray}
- \partial_0 \left(\sqrt{-E} S^{00} S^{11} \partial_0 A_1 \right) + \partial_z K B_{2 3} \partial_0 \Phi - \partial_z \left(\sqrt{-E} S^{zz} S^{11} \partial_z A_1 \right) &&\\
- \partial_{\tilde{m}} \left(\sqrt{-E} S^{\tilde{m} \tilde{n}} S^{11} \partial_{\tilde{n}} A_1 \right) &=& 0 \ ,\nonumber\\
- \partial_0 \left(\sqrt{-E} S^{00} S^{22} \partial_0 A_2 \right) + f \partial_3 \chi - \partial_z \left(\sqrt{-E} S^{zz} S^{22} \partial_z A_2 \right) &&\nonumber\\
- \partial_{\tilde{m}} \left(\sqrt{-E} S^{\tilde{m} \tilde{n}} S^{22} \partial_{\tilde{n}} A_2 \right) &=& 0 \ , \nonumber\\
- \partial_0 \left(\sqrt{-E} S^{00} S^{33} \partial_0 A_3 \right) - f \partial_2 \chi - \partial_z \left(\sqrt{-E} S^{zz} S^{33} \partial_z A_3 \right) &&\nonumber\\
- \partial_{\tilde{m}} \left(\sqrt{-E} S^{\tilde{m} \tilde{n}} S^{33} \partial_{\tilde{n}} A_3 \right) &=& 0 \ ,\nonumber
\end{eqnarray}
where the indices $\tilde{m}, \tilde{n}$ run over the $S^3$ that the D7--brane wraps.  If we assume that $\partial_i \chi = 0$, we find that the equations for $A_2$ and $A_3$ decouple from $\chi$.  Therefore, we can consistently take $F_{2 3} = 0$, or, in other words, $A_2 = A_3 = 0$.   This simplifies the equations of motion that we need to consider to:
\begin{eqnarray}
0 &=& \partial_a \left[\sqrt{-E} S^{a b} R^2 \left(\frac{1+4b^4 z^4 \left(\theta_0'\right)^2}{1+4 z^2 \left(\theta'_0\right)^2} \right) \partial_b \chi \right] -\chi  \left[ \partial_\theta^2 \sqrt{-E} \right. \nonumber \\
&& \left. - \partial_z \left(E^{zz} R^2 \theta_0' \partial_\theta \sqrt{-E} \right) \right]  \ , \label{eqt:eom_theta} \\
0 &=&- \partial_0 \left(\sqrt{-E} S^{00} S^{11} \partial_0 A_1 \right) + \partial_z K B_{2 3} \partial_0 \Phi - \partial_z \left(\sqrt{-E} S^{zz} S^{11} \partial_z A_1 \right)  \nonumber \\
&&- \partial_{\tilde{m}} \left(\sqrt{-E} S^{\tilde{m} \tilde{n}} S^{11} \partial_{\tilde{n}} A_1 \right)  \ , \label{eqt:eom_phi_A2} \\
0 &=& \partial_a \left( \sqrt{-E} S^{a b} R^2 \sin^2 \theta_0 \partial_b \Phi \right) - F_{01} B_{23}  R^4 \sin\psi \cos\psi \partial_z K  \ . \label{eqt:eom_phi_A1}
\end{eqnarray}
In the proceeding sections, we will work out the solutions to these
equations numerically using a shooting method.  With an appropriate
choice of initial conditions at the event horizon, which we explain
below, we numerically solve these equations as an initial condition
problem in Mathematica.  Therefore, the D.E. solver routine ``shoots''
towards the boundary of the problem, and we extract the necessary data
at the boundary.
\subsection*{4.4.1 \hspace{2pt} The $\chi$ meson spectrum}
\addcontentsline{toc}{subsection}{4.4.1 \hspace{0.15cm} The $\chi$ meson spectrum}
In order to solve for the meson spectrum given by equation (\ref{eqt:eom_theta}), we consider an ansatz for the field $\chi$ of the form:
\begin{eqnarray}
\chi = h(\tilde{z}) \exp \left(- i \tilde{\omega} t \right) \ ,
\end{eqnarray}
where we are using the same dimensionless coordinates as before, with the addition that:
\begin{equation*}\begin{array}{rclcrcl}
z &=&  b^{-2} \tilde{z} & \ , \quad & \omega &=& R^{-2} b \ \tilde{\omega} \ . \\
\end{array}
\end{equation*}
In these coordinates, the event horizon is located at $\tilde{z} = 1$.  Since there are two different types of embeddings, we analyze each case separately.  We begin by considering black hole embeddings.  In order to find the appropriate infrared initial conditions for the shooting method we use, we would like to understand the behavior of $h(\tilde{z})$ near the horizon.  The equation of motion in the limit of $\tilde{z} \to 1$ reduces to:
\begin{eqnarray} \label{eqt:quasinormal}
h''(\tilde{z}) + \frac{1}{\tilde{z}-1} h'(\tilde{z}) + \frac{\tilde{\omega}^2}{16 \left(\tilde{z}-1 \right)^2} h(\tilde{z}) &=& 0 \ .
\end{eqnarray}
The equation has solutions of the form $(1-\tilde{z})^{\pm i
  \tilde{\omega} /4}$, exactly of the form of quasinormal modes
\cite{Starinets:2002br}.  Since the appropriate fluctuation modes are
in--falling modes \cite{Hoyos:2006gb}, we require only the solution of
the form $(1-\tilde{z})^{ - i \tilde{\omega} /4}$.  This is our
initial condition at the event horizon for our shooting method.  In
order to achieve this, we redefine our fields as follows:
\begin{eqnarray*}
h(\tilde{z}) &=& y(\tilde{z}) (1-\tilde{z})^{ - i \tilde{\omega} /4} \ ,
\end{eqnarray*}
which then provides us with the following initial condition:
\begin{eqnarray}
y(\tilde{z} \to 1)  &=& \epsilon \ ,
\end{eqnarray}
where $\epsilon$ is chosen to be vanishingly small in our numerical
analysis.  The boundary condition on $y'(\tilde{z} \to 1)$ is
determined from requiring the equation of motion to be regular at the
event horizon.  The solution for the fluctuation field $y(\tilde{z})$
must be comprised of only a normalizable mode, which in turn
determines the correct value for $\tilde{\omega}$.  Since we are
dealing with quasinormal modes for the black hole embeddings,
$\tilde{\omega}$ will be complex; the real part of $\tilde{\omega}$
corresponds to the mass of the meson before it melts, and the
imaginary part of $\tilde{\omega}$ is the inverse lifetime (to
a factor of 2) \cite{Hoyos:2006gb}.
We begin by considering the trivial embedding $\theta_0(\tilde{z}) =
0$ (a black hole embedding).  This embedding corresponds to having a
zero bare quark mass.  The equation of motion (\ref{eqt:eom_theta})
simplifies tremendously in this case:
\begin{eqnarray}
h''(\tilde{z}) + \left( \frac{2 \tilde{z}}{\tilde{z}^2 - 1} - \frac{1}{\tilde{z} \left(1+ \tilde{z}^2 \eta^2 \right)} \right) h'(\tilde{z}) + \frac{3 + \tilde{z} \left(-3 \tilde{z} + \tilde{\omega}^2 \right)}{ 4 \tilde{z}^2 \left(\tilde{z}^2 - 1\right)^2} h(\tilde{z}) &=& 0\ .
\end{eqnarray}
We show solutions for $\tilde{\omega}$ in Figure~\ref{fig:trivial
  fluctuations in theta} as a function of the magnetic field $\eta$.
In particular, we find the same additional mode discussed in
ref.~\cite{Filev:2007qu}.  This mode becomes massless and eventually
tachyonic at approximately $\eta \approx 9.24$.  This point was
originally presented in Figure~\ref{fig:ccrit}, where the intermediate
unstable phase joins the trivial embedding.  This is exactly when the
$-\tilde{c}$ vs $\tilde{m}$ plot has negative slope for all black hole
embeddings.
\begin{figure}[h]
\begin{center}
\includegraphics[ width=11cm]{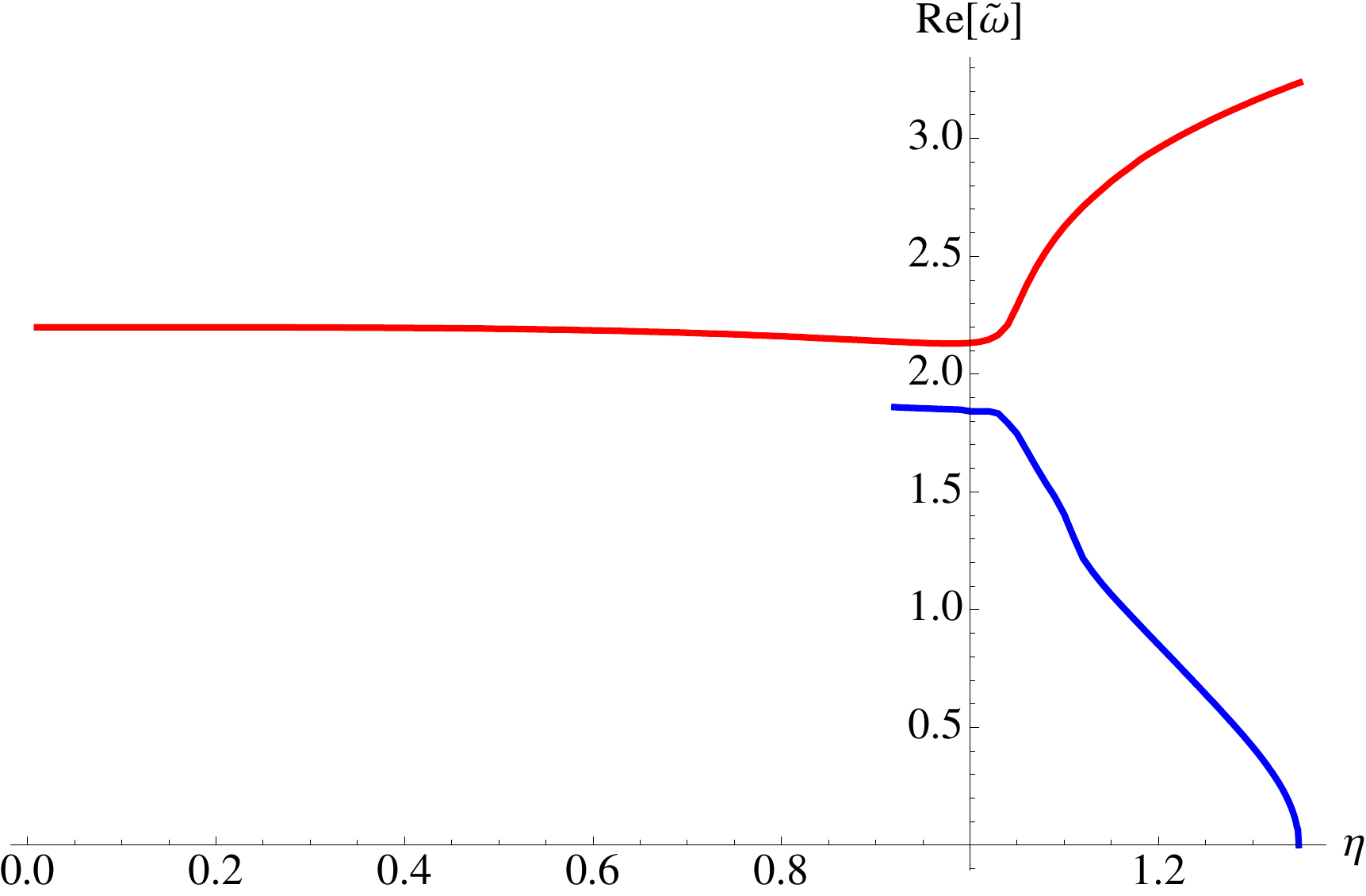}
\end{center}
\caption{The $\chi$ meson mass as a function of magnetic field for the trivial embedding.  The upper (red) curve is the generalization of the mode described in ref.~\cite{Hoyos:2006gb}.  The lower (blue) curve is the generalization the mode discussed in ref.~\cite{Filev:2007qu}.  We do not extend this second curve to small $\eta$ because the numerics become unreliable. }  \label{fig:trivial fluctuations in theta}
\end{figure}
\\
We now consider embeddings with non--zero bare quark mass.  This means
solving the full equation (\ref{eqt:eom_theta}).  We have both
embeddings to consider; for the black hole embeddings, we will follow
the same procedure presented above to solve for the complex
$\tilde{\omega}$.  We can still use the same procedure because in the
limit of $\tilde{z} \to 1$, the equation of motion still reduces to
equation (\ref{eqt:quasinormal}).  For the Minkowski embeddings, we do
not have quasinormal modes, and $\tilde{\omega}$ is purely real.
Therefore, we use as initial conditions:
\begin{eqnarray}
\chi(\tilde{z} \to \tilde{z}_{\mathrm{max}})  &=& \epsilon \ , \\
\chi'(\tilde{z} \to \tilde{z}_{\mathrm{max}}) &=& \infty \ .
\end{eqnarray}
In Figure~\ref{fig:fluctuations in theta eta1} and
Figure~\ref{fig:fluctuations in theta eta10}, we show solutions for very
different magnetic field values.  In the former case, $\eta$ is small,
and we do not have chiral symmetry breaking; in the latter, $\eta$ is
large, and we have chiral symmetry breaking.  It is important to note
that in neither of the graphs do we find a massless mode at zero bare
quark mass.  In Figure~\ref{fig:fluctuations in theta eta1}, we find
that fluctuations about both the black hole and Minkowski embeddings
become massless and tachyonic (we do not show this in the graph).  The
tachyonic phase corresponds exactly to the regions in the $-\tilde{c}$
vs $\tilde{m}$ plot with negative slope.
\begin{figure}[h]
\begin{center}
\includegraphics[ width=11cm]{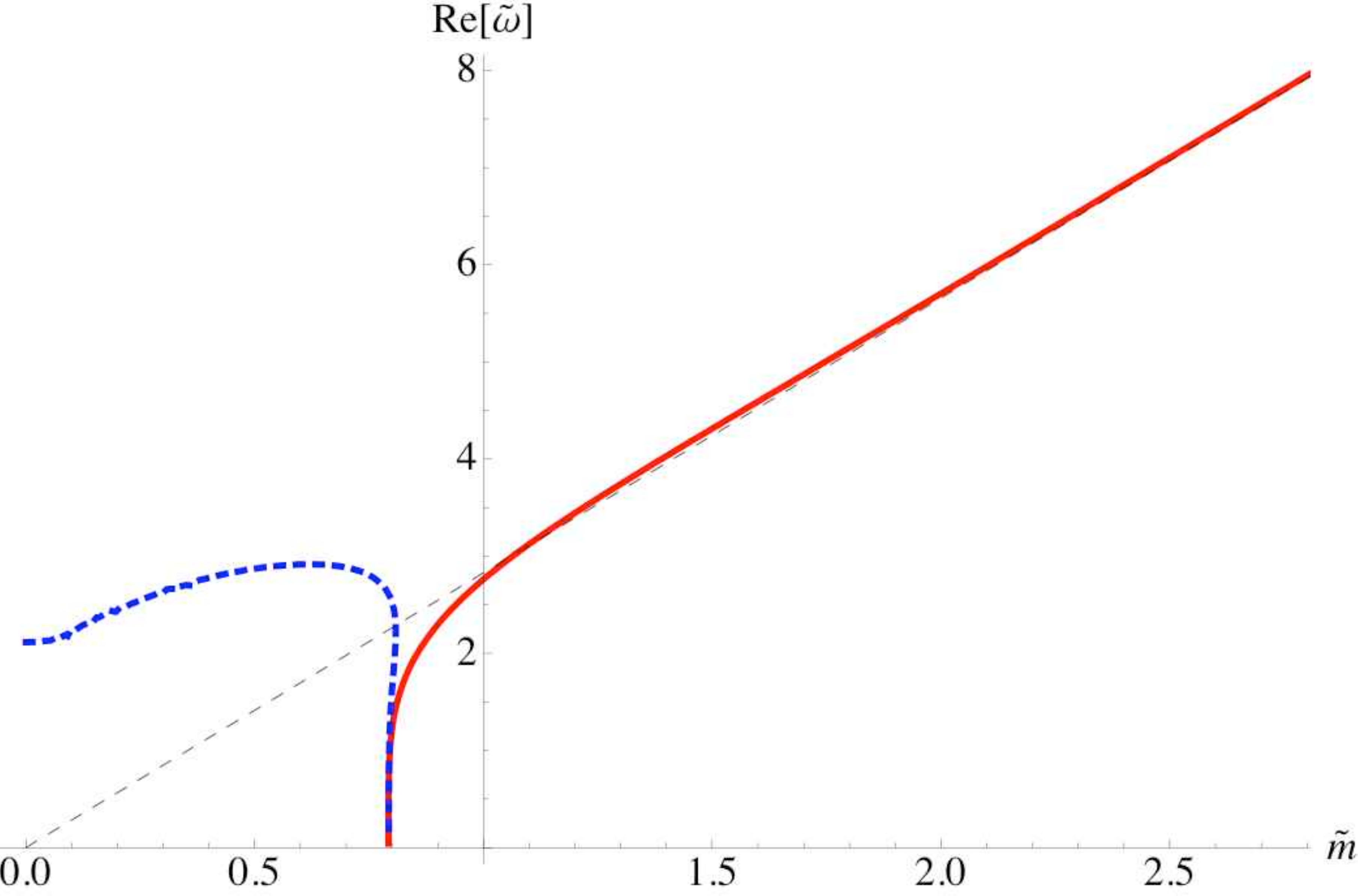}
\end{center}
\caption{The $\chi$ meson mass as a function of bare quark mass for $\eta = 1$.  The dashed (blue) curve corresponds to fluctuations about black hole embeddings.  The solid (red) line corresponds to fluctuations about Minkowski embeddings.  These modes have a purely real $\omega$.  The straight dashed (black) line corresponds to the pure AdS$_5 \times S^5$ solution. }
\label{fig:fluctuations in theta eta1}
\end{figure}
\begin{figure}[h]
\begin{center}
  \includegraphics[ width=11cm]{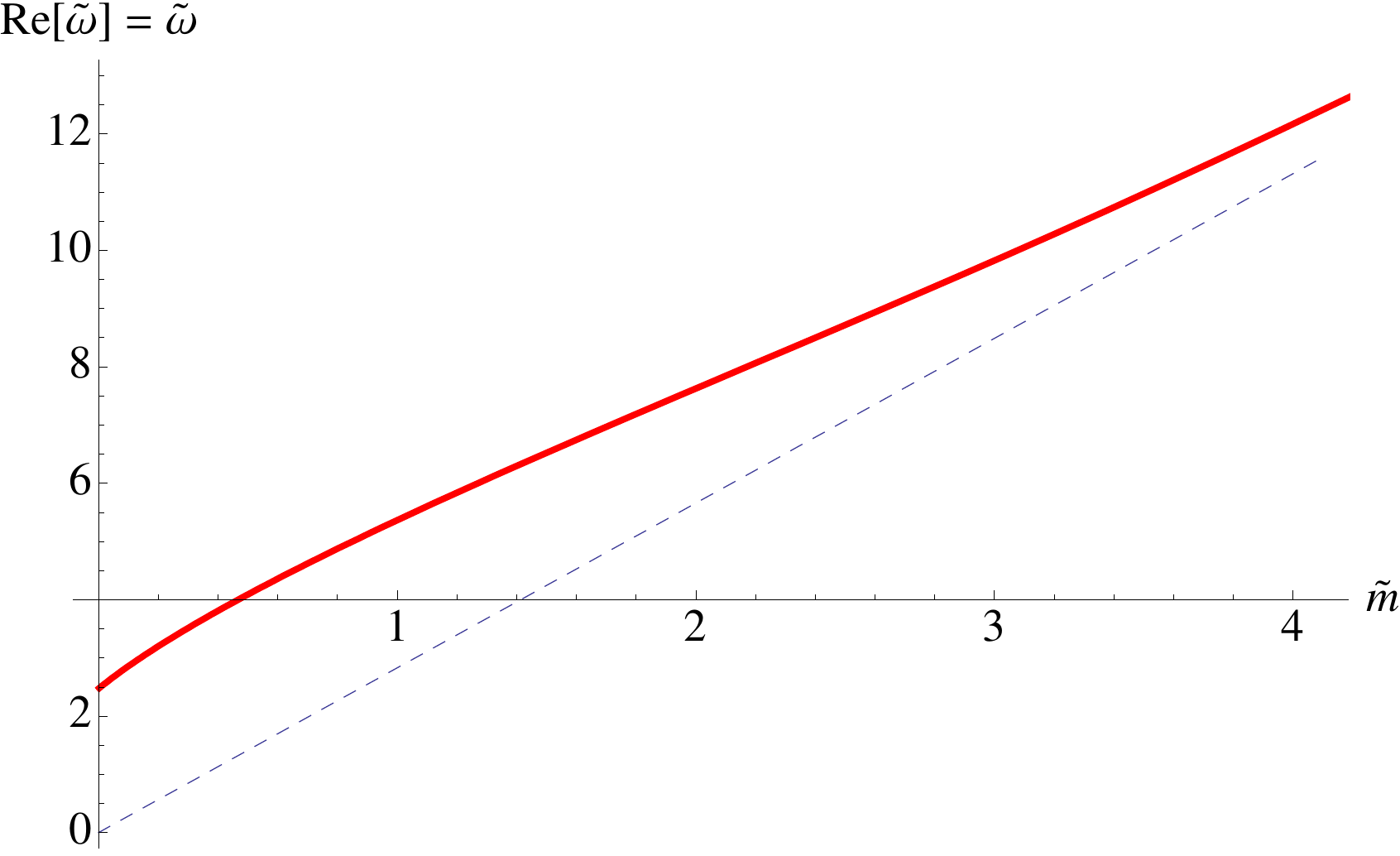}
\end{center}
\caption{The $\chi$ meson mass as a function of bare quark mass for $\eta = 10$.  The dashed (black) line corresponds to the pure AdS$_5 \times S^5$ solution. }
\label{fig:fluctuations in theta eta10}
\end{figure}
\\
\subsection*{4.4.2 \hspace{2pt} The $\Phi$ and $A$ meson spectra}
\begin{figure}[h]
\begin{center}
\includegraphics[ width=11cm]{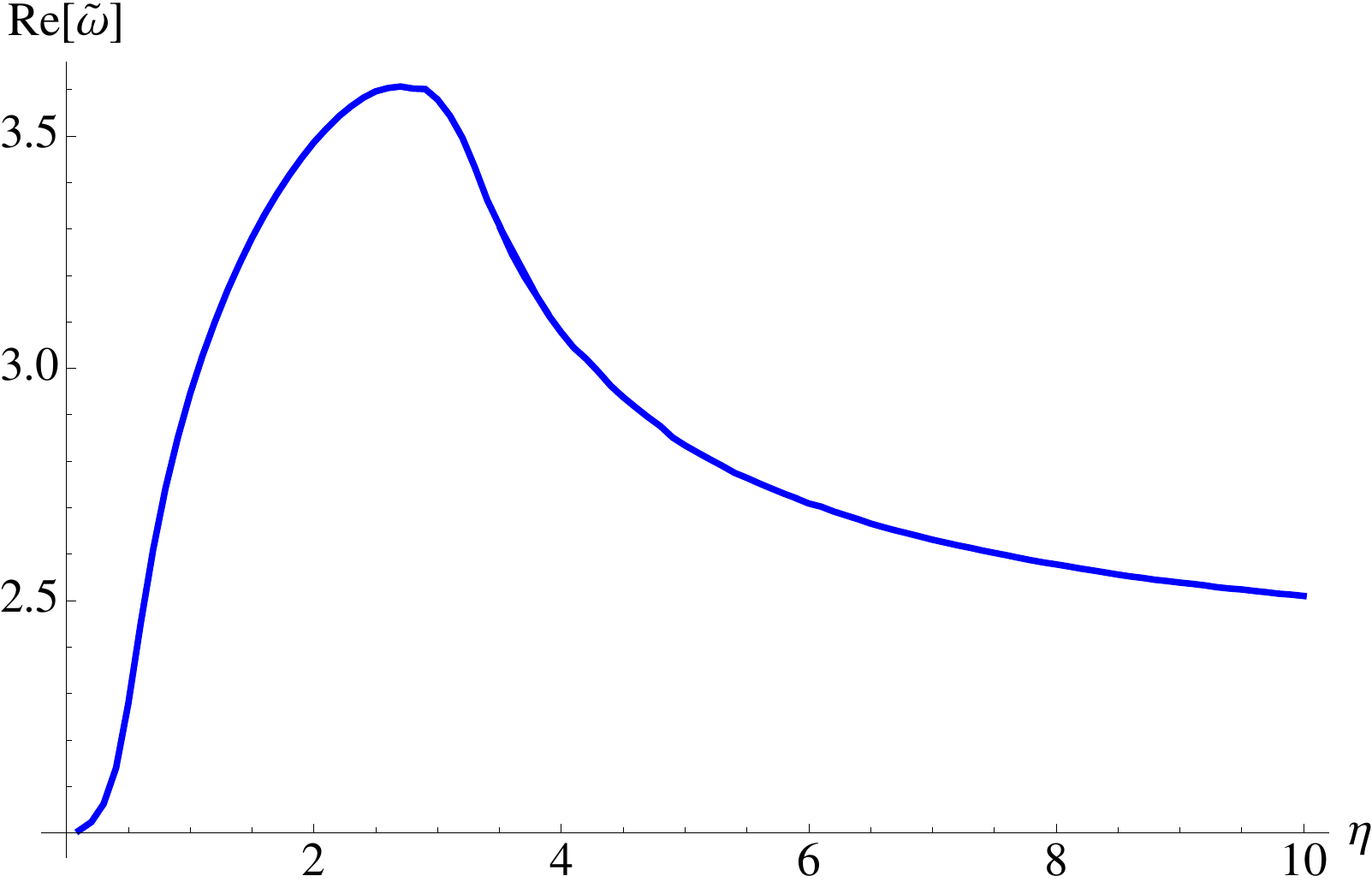}
\end{center}
\caption{The $A$ meson mass as a function of magnetic field for the trivial embedding. }  \label{fig:trivial fluctuations in A}
\end{figure}

\begin{figure}[h!]
\begin{center}
\includegraphics[ width=10cm]{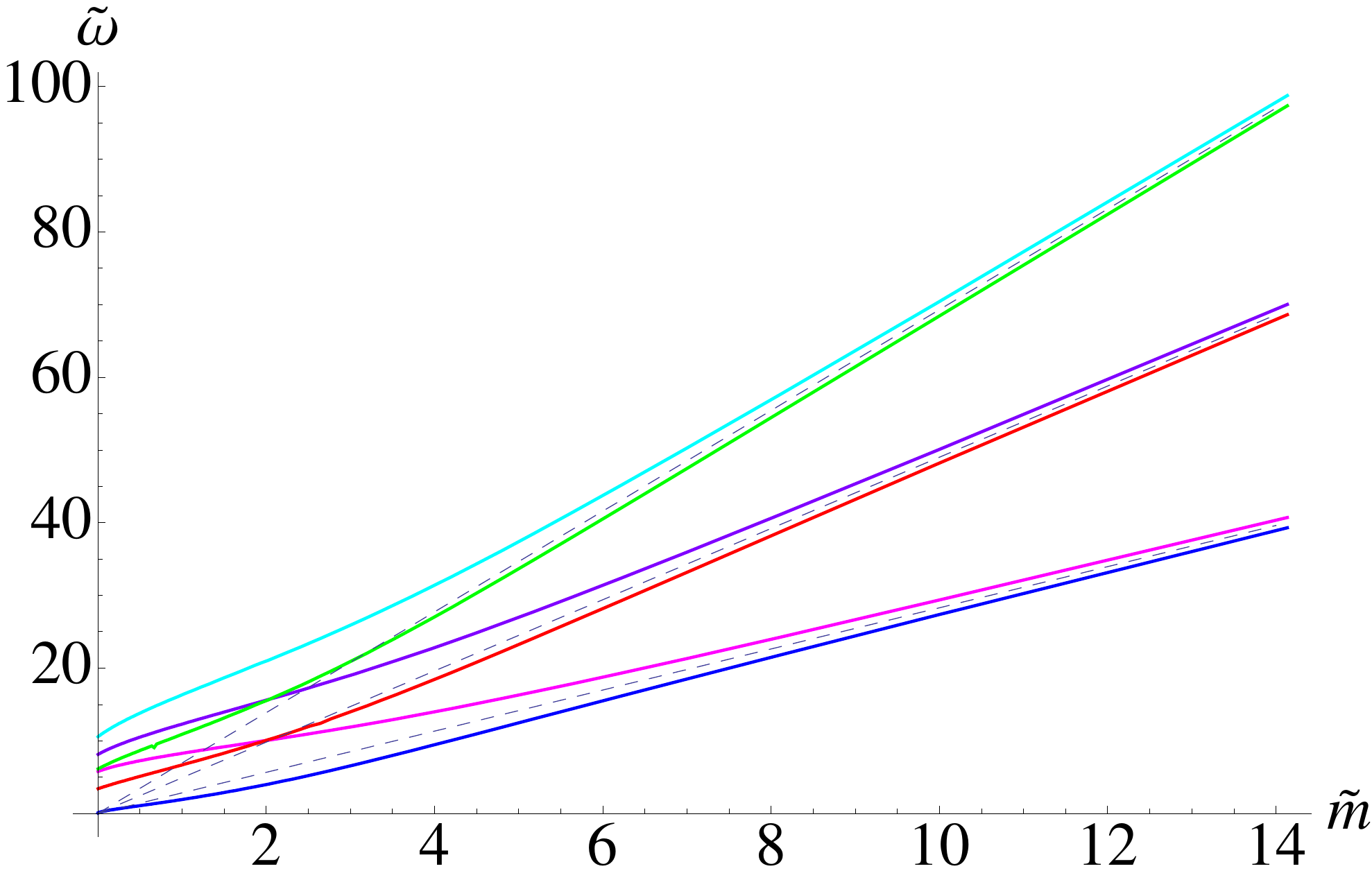}
\end{center}
\caption{Coupled $A-\phi$ fluctuations for $\eta = 10$.   The dashed (black) line corresponds to the pure AdS$_5 \times S^5$ solution. }  \label{fig:fluctuations in A-phi}
\end{figure}

\addcontentsline{toc}{subsection}{4.4.2 \hspace{0.15cm} The $\Phi$ and $A$ meson spectra}
Let us now consider the coupled fluctuations of $\Phi$ and $A$ in
equations (\ref{eqt:eom_phi_A2}) and (\ref{eqt:eom_phi_A1}).  We
consider an ansatz (as before) of the form:
\begin{eqnarray*}
\Phi &=& \phi(\tilde{z}) \exp \left(- i \tilde{\omega} t \right) \ , \\
A_1 &=& A(\tilde{z}) \exp \left(- i \tilde{\omega} t \right) \ .
\end{eqnarray*}
It is interesting to note that, for the trivial embedding
$\theta_0(\tilde{z}) = 0$, one of the coupled equations is equal to
zero, and we simply have:
\begin{eqnarray*}
A''(\tilde{z}) + \frac{\tilde{z} \left(2+\eta^2 \left( 3 \tilde{z}^2 -1 \right) \right)}{\left(\tilde{z}^2-1 \right) \left(1+ \tilde{z}^2 \eta^2 \right)} A'(\tilde{z}) + \frac{\tilde{\omega}^2}{4 \tilde{z} \left(\tilde{z}^2 - 1\right)^2 } A(\tilde{z})&=& 0 \ .
\end{eqnarray*}
Again, we note that in the limit of $\tilde{z} \to 1$, we have:
\begin{eqnarray*}
A''(\tilde{z}) + \frac{1}{\tilde{z}-1} A'(\tilde{z}) + \frac{\tilde{\omega}^2}{16 \left(\tilde{z}-1 \right)^2} A(\tilde{z}) &=& 0 \ .
\end{eqnarray*}
This is exactly the form of equation (\ref{eqt:quasinormal}), so
$A(\tilde{z})$ has the same solutions of in--falling and outgoing
solutions, which provides us with the necessary initial conditions for
our shooting method.  We show solutions for $\tilde{\omega}$ in Figure~\ref{fig:trivial fluctuations in A}.
Unfortunately, we do not know how to solve for the quasinormal modes
for other black hole embeddings.  Since the equations of motion are
coupled, we are unable to find an analytic solution for the
fluctuations near the event horizon.  This prevents us from using
infrared initial conditions for our shooting method.  However, we are
actually more interested in searching for the ``pion" of our system,
which will occur when we have chiral symmetry breaking.  In those
cases, we are only dealing with Minkowski embeddings with a pure real
$\omega$, so we may ignore the black hole embeddings.
\\
Since the equations of motion are coupled, it turns out that only for
specific initial conditions will both fluctuations only be comprised
of normalizable modes.  We represent this by a parameter $\alpha$
(which we must tune) as follows:
\begin{eqnarray}
A(\tilde{z} \to \tilde{z}_{\mathrm{max}}) &=& i \cos \alpha \ , \\
\phi(\tilde{z} \to \tilde{z}_{\mathrm{max}}) &=& \sin \alpha \ .
\end{eqnarray}
The initial conditions on $A'$ and $\phi'$ are determined from the
equations of motion.  We show several solutions in Figure~\ref{fig:fluctuations in A-phi}.  There are several important points
to notice.  First, we find that the lowest mode satisfies an GMOR
relationship given by:
\begin{eqnarray}
\tilde{\omega} &\approx& 1.1 \tilde{m}^{1/2} \ .
\end{eqnarray}
Therefore, we have found the Goldstone boson of our system related to
the breaking of chiral symmetry.  Second, the modes exhibit the
Zeeman splitting behavior that was discussed in
ref.~\cite{Filev:2007gb}.  It is interesting that the various modes
always cross each other at approximately $\tilde{m} \approx 2$.  We
have no intuitive explanation for this behavior.

\begin{figure}[h!]
\begin{center}
\includegraphics[ width=10cm]{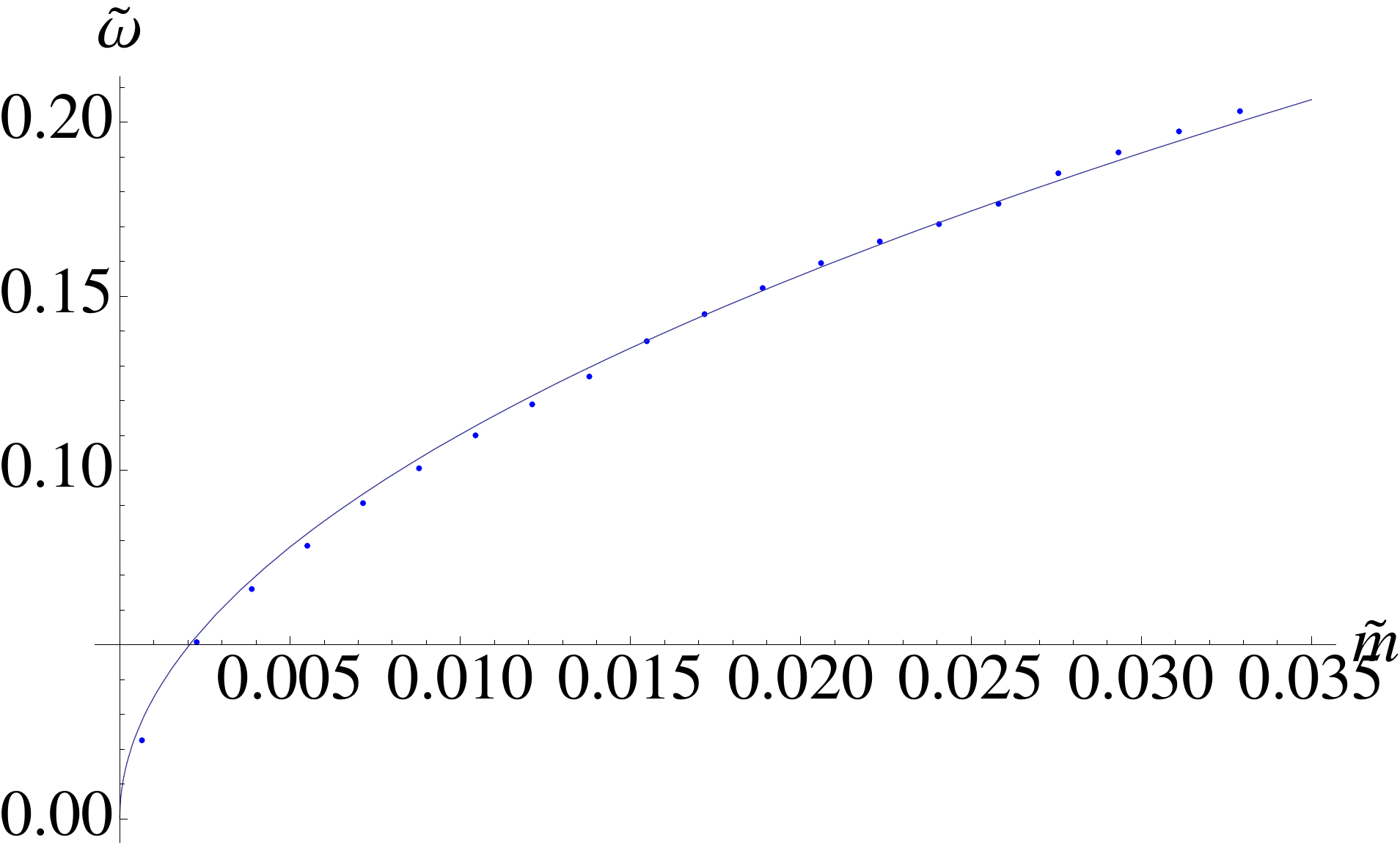}
\end{center}
\caption{Zoom near zero bare quark mass for the lowest mode. }  \label{fig:fluctuations in A-phi-pion}
\end{figure}

\section*{4.5 \hspace{2pt} Concluding remarks}
\addcontentsline{toc}{section}{4.5 \hspace{0.15cm} Concluding remarks}

We have extended the holographic study of large $N_c$ gauge theory in an
external magnetic field described in Chapter~2, to the
case of finite temperature, allowing us to study the properties of the
quark dynamics when the theory is in the deconfined plasma phase.

The meson melting phase transition exists only below a critical value
of the applied field. This is the critical value above which
spontaneous chiral symmetry breaking is triggered (in the case of zero
mass). Above this value, regardless of the quark mass (or for fixed
quark mass, regardless of the temperature) the system remains in a
phase with a discrete spectrum of stable masses. Evidently, for these
values of the field, it is magnetically favorable for the quarks and
anti--quarks to bind together, reducing the degrees of freedom of the
system , as can be seen from our computation of the entropy.
Meanwhile, the magnetization is greater in this
un--melted phase.

There have been non--perturbative studies of fermionic models in
background magnetic field before, and there is a large literature (see
{\it e.g.}, the reviews of refs.~\cite{Miransky:2002eb,Wang:2007bg},
and the discussion of ref.~\cite{Semenoff:1999xv} and references
therein).  Generally, those works use quite different methods to
examine aspects of the physics --- some primary non--perturbative
tools are the Dyson--Schwinger equations in various truncations). Our
results (and the zero temperature result obtained with these methods
in the zero temperature case \cite{Filev:2007gb}) are consistent with
the general expectations from those works, which is that strong
magnetic fields are generically expected to be a catalyst for
spontaneous chiral symmetry breaking in a wide class of models (see
{\it e.g.}, refs. \cite{Gusynin:1995nb,Semenoff:1999xv,Miransky:2002eb} for a discussion
of the conjectured universality of this result).

\chapter*{Chapter 5: \hspace{1pt} Phase structure of finite temperature large $N_c$ flavored Yang--Mills theory in an external electric field}

\addcontentsline{toc}{chapter}{Chapter 5:\hspace{0.15cm}
Phase structure of finite temperature large $N_c$ flavored Yang--Mills theory in an external electric field}

\section*{5.1 \hspace{2pt} Introductory remarks}
\addcontentsline{toc}{section}{5.1 \hspace{0.15cm} Introductory remarks}

In this chapter, we study large $N_c$ $SU(N)$ gauge theory with
non--backreacting hypermultiplet quark flavors in the
presence of a background electric field, at both zero and finite
temperature. The holographic description of this theory at finite baryon chemical potential was first studied in ref.~\cite{Karch:2007pd}, where the existence of a global electric current induced was demonstrated. Further study of this set up was considered in ref.~\cite{O'Bannon:2007in}. The phase structure of the theory at zero baryon chemical potential was studied in ref.~\cite{Albash:2007bq}.

   We may expect the behavior of the theory to be quite different from the magnetic case considered in the previous chapter. In a sense the electric field is pulling the quarks apart, decreasing the bound energy and thus we may expect that the onset of the deconfinement phase transition will happen at lower temperature, than in the absence of an electric field. Furthermore, at zero temperature we may expect that for sufficiently strong electric field the binding energy of the quarks will be completely overcome and the mesons will dissociate into their constituent quarks.

It is worth noting that the phase transition will be driven by the quantum fluctuations, which will result in a quantum phase transition. On the other side, the mesons are electrically neutral, while the dissociated quarks aren't. Hence the dissociation can be seen as a {\it conductor/insulator} phase transition \cite{Albash:2007bq}.


\section*{5.2 \hspace{2pt} General set up}
\addcontentsline{toc}{section}{5.2 \hspace{0.15cm} General set up}

In this section we describe the gravitational background and the technique employed to introduce an external electric field. %
We consider the following form for the metric of the AdS$_5$--Schwarzschild$\times S^5$ background:
\begin{eqnarray}
ds^2 /\alpha' &=&- \frac{u^4-b^4}{R^2u^2} dt^2 + \frac{u^2}{R^2} d \vec{x}^2 + \frac{R^2u^2}{u^4-b^4}  du^2 + R^2 \cos^2 \theta d \Omega_3^2 \label{backgr-el} \\
&& \hskip2cm + R^2 d \theta^2 + R^2 \sin^2 \theta d \phi^2 \ ,\nonumber \\
e^{\Phi}  &=& g_s;~~~C_{(4)}=\frac{u^4}{R^4}dt\wedge dx_1\wedge dx_2\wedge dx_3 \nonumber \ .
\end{eqnarray}
In the above, there are time and space coordinates $t\equiv x^0$ and
$\vec{x}\equiv(x^1,x^2,x^3)$ respectively (which the dual gauge theory
will also have) and also $u\in[0,\infty)$ which is a radial
coordinate. Those are the coordinates on the asymptotically AdS$_5$ geometry. We are using
standard polar coordinates on the $S^5$, with:
\begin{eqnarray}
d\Omega_5^2&=&d\theta^2+\cos^2\theta d\Omega_3^2+\sin^2\theta d\phi^2\nonumber \ ,\\ \mathrm{and}\quad
d\Omega_{3}^2&=&d\psi^2+\cos^2\psi d\beta+\sin^2\theta d\gamma^2 \ .
\end{eqnarray}
The scale $R$ determines the gauge theory 't Hooft coupling according
to $R^2=\alpha^\prime\sqrt{g_{\rm YM}^2 N}$. The parameter $b$ sets
the radius of the horizon of the black hole, which in turn sets the
temperature, according to $T=b/\pi R^2$.
Now following the usual technique to introduce flavors to the theory we add $N_f$ D7--branes to the theory. The D7--branes are extended along the $x_0,x_1,x_2,x_3,u,\psi,\beta,\gamma$ directions of the geometry and have a non--trivial profile along $\theta$, namely $\theta=\theta(u)$. In this chapter we do not study the meson spectrum of the theory, thus the relevant part of the lagrangian is:
\begin{equation}
\frac{S_{\mathrm{D7}}}{N_f}=- T_{\mathrm{D7}} \int d^8 \xi \ \mathrm{det}^{1/2} \left(P\left[ G_{a b} \right] + P \left[B_{a b} \right] + 2 \pi \alpha' F_{a b} \right)
\label{action-el}
\end{equation}
In order to introduce electric field we consider a constant $B$--field along the ($x_0,x_1$) plane:
\begin{equation}
B=Edx_0\wedge dx_1
\label{B-el}
\end{equation}
It is convenient to work with the coordinate $u$. The corresponding lagrangian is given by:
\begin{equation}
{\cal L}\propto \sqrt{\frac{u^4-u_*^4}{u^4-b^4}}u^3\cos^3\theta\sqrt{1+\frac{u^4-b^4}{u^2}\theta'^2}\cos\psi\sin\psi
\label{el-action-nh}
\end{equation}

where,

\begin{equation}
u_*^4=b^4+R^4E^2 \ .
\end{equation}
\bibliographystyle{plain}
As one can see at $u=u_*$ the action (\ref{el-action-nh}) vanish and for $u<u_*$ it becomes imaginary. Often we will refer to $u_*$ (or equivalently $r_*$ ) as the ``pseudo-horizon'' or
the ``vanishing locus''. This implies instability of the
D7--brane embedding and is a signal that we should generalize the ansatz to include non--trivial $U(1)$ gauge field on the D7--brane.
As it was shown in ref.~\cite{Karch:2007pd}, one can heal the on-shell action (\ref{el-action-nh}) by generalizing the ansatz to:
\begin{equation}
A_1=A_1(u) \ .
\end{equation}
using
Furthermore, it can be shown that the asymptotic behavior of the gauge field at infinity $u\to\infty$ can be related to the global
 electric current of the theory along $x_1$, namely $J^1=\langle\bar\psi\gamma^1\psi\rangle$.

After substituting in the lagrangian (\ref{action-el}) we obtain the following lagrangian:
\begin{equation}
{\cal L}\propto u^3\cos^3\theta\sqrt{\frac{u^4-u_*^4}{u^4-b^4}(1+\frac{u^4-b^4}{u^2}\theta'^2)+\frac{u^4-b^4}{u^4}f'^2}\sin\psi\cos\psi\ ,
\label{lagr-el}
\end{equation}
where:
\begin{equation}
f=2\pi\alpha'A_1(u)\ .
\end{equation}
Now using the fact that $f(u)$ is a cyclic field, we arrive at following the equation of motion for $f(u)$:
\begin{equation}
\partial_u\left(\frac{\partial{\cal L}}{\partial f'}\right)=0\ ,
\end{equation}
which results to:
\begin{equation}
\frac{u^3{\cos^3\theta}\left(\frac{u^4-b^4}{u^4}f'\right)}{\sqrt{\frac{u^4-u_*^4}{u^4-b^4}(1+\frac{u^4-b^4}{u^2}\theta'^2)+\frac{u^4-b^4}{u^4}f'^2}}=K=\mathrm{const}\
.\label{constK}
\end{equation}
We will show now that the constant $K$ is related to the global electric current $J^1=\langle\bar\psi\gamma^1\psi\rangle$. Indeed, the four
dimensional action of the dual gauge theory is obtained after integrating along $u,\Omega_3$ and regularizing the corresponding integrals. Therefore for the variation of
the four dimensional action we obtain:
\begin{equation}
\delta S_{\rm4D}=\int d^4x J^{\mu}\delta A_{\mu}(\infty)\ ,
\label{var1}
\end{equation}
where we have identified $A_\mu(\infty)$ with the $U(1)$ gauge field in the $\rm4D$ dual gauge theory.
On the other side, we have that:
\begin{eqnarray}
\delta S_{\rm{4D}}&=&\int d^4x d\Omega_3du\delta{\cal L}=\int d^4x d\Omega_3 du\left(\frac{\partial{\cal L}}{\partial A^{\mu}}\delta A^{\mu}+
\frac{\partial{\cal L}}{\partial(\partial_{\nu}A_{mu})}\partial_{\nu}\delta A_{\mu}\right)=\nonumber\\
&=&\int d^4x d\Omega_3 du\partial_u\left(\frac{\partial{\cal L}}{\partial(\partial_{u}A_1)}\delta A_1\right)=
\int d^4x d\Omega_3\frac{\partial{\cal L}}{\partial(\partial_{u}A_1)}\delta A_1(\infty)\ ,
\label{var2}
\end{eqnarray}
where we have considered variation such that $\delta A_1|_{\mathrm{horizon}}=0$. Now after comparing equations (\ref{var1}) and (\ref{var2})
we obtain that:
\begin{equation}
J^1=\lim_{u\to\infty}\int d\Omega_3\frac{\partial{\cal L}}{\partial(\partial_{u}A_1)}=2\pi\alpha'\int d\Omega_3\frac{\partial{\cal L}}{\partial
f'}=4\pi^3\alpha'N_fT_{D7}K\ , 
\end{equation}
therefore $K$ is indeed proportional to $J^1$. Now let us proceed by solving for $f'$ from equation (\ref{constK}) and substituting in equation (\ref{lagr-el}). The lagrangian of the resulting on-shell action is:
\begin{equation}
{\cal L}\propto u^4\cos^6\theta\sqrt{\frac{u^4-u_*^4}{u^2(u^4-b^4)\cos^6\theta-K^2}}\sqrt{1+\frac{u^4-b^4}{u^2}\theta'^2}\sin\psi\cos\psi \ .
\label{action-el-on-shell}
\end{equation}
As one can see, the on-shell lagrangian from equation (\ref{action-el-on-shell}) is regular at $u=u_*$ provided that we have:
\begin{equation}
K^2=u_*^2R^4E^2\cos^6\theta_0\ .
\label{currentK}
\end{equation}
Here $\theta_0$ is the angle at which the D7--brane enter the vanishing locus (at $u=u_*$). One can also see that if the D7--brane close at some $u_{\rm min}>u_* $ above the vanishing locus (Minkowski type of embeddings), we have to set $K=0$ to avoid imaginary action. This implies that the current $J^1$  is equal to zero. And it better vanish because Minkowski type of embeddings describe the confined (meson gas) phase of the fundamental matter and hence, there are no free charges in the system. On the other hand, as we are going to demonstrate in the next sections, embeddings which reach the vanishing locus will reach the horizon and are thus classified as a black hole embeddings, the latter describe deconfined quarks and therefore the theory is in a conductive phase.

 From equation (\ref{lagr-el}), we can derive the equation of motion for $\theta(\tilde{u})$ and substitute the expression for $f'$ which is obtained from solving equation (\ref{constK}). Its exact form is not
particularly illuminating, this is why we will not display it fully here. In the limit of large $u$, the equation of motion asymptotes to:
\begin{equation}
\frac{d}{d u} \left( u^5 \theta' (u) \right) + 3 u^3 \theta(u) = 0 \ ,
\end{equation}
which has solution:
\begin{equation}
\theta(u) = \frac{{m}}{u} + \frac{c}{u^3} \ .
\end{equation}
The constants $\tilde{m}$ and $\tilde{c}$ are related to the bare quark mass and the condensate respectively, in a manner that is by now very standard \cite{Albash:2006ew,Babington:2003vm,Karch:2002sh,Kruczenski:2003be}.  In our notation, the exact relationship is given by:
\begin{eqnarray}
m_q &=& \frac{m}{2 \pi \alpha'} \ , \qquad
\langle \bar{\psi} \psi \rangle = - 8 \pi^3  \alpha' N_f T_{D7}  c \ .
\label{corresp}
\end{eqnarray}
\section*{5.3 \hspace{2pt} Properties of the solutions}
\addcontentsline{toc}{section}{5.3 \hspace{0.15cm} Properties of the Solutions}
\subsection*{5.3.1 \hspace{2pt} Exact results at large mass}
\addcontentsline{toc}{subsection}{5.3.1 \hspace{0.15cm} Exact results at large mass}

It is instructive to study the properties of the quark condensate as a
function of bare quark mass for large mass. This corresponds to
$m\gg R^2E$. Note that this corresponds to the
so--called Minkowski embeddings for the probe D7--brane for which the constant $K$ proportional to the electric current $J^1$
vanishes by virtue of equation (\ref{currentK}) since Minkowski embeddings correspond to
$\theta=\pi/2$. To extract the behavior of quark mass and condensate
we linearize the equation of motion obtained from equation
(\ref{action-el}) in the same way as described in the previous chapter on the
magnetic case. We find the following analytic behavior:
\begin{eqnarray}
\langle \bar{\psi}\psi \rangle\propto -c=\frac{R^4E^2}{4m}+\frac{b^8-4b^4R^4E^2+8R^8E^4}{96m^5}+O\left(\frac{1}{m^7}\right).
\end{eqnarray}
This suggests that for high enough quark mass the condensate vanishes.
However, unlike the magnetic field case from the previous chapter, the vanishing occurs from the positive side of the condensate axis. We will explore these features and more in the
next section using numerical techniques.
\subsection*{5.3.2 \hspace{2pt} The case of vanishing temperature}
\addcontentsline{toc}{subsection}{5.3.2 \hspace{0.15cm} The case of vanishing temperature}

It is instructive to first study the case of vanishing temperature,
since a number of important features will already appear in this case.
For this simple case, the energy scale is set by $R\sqrt{E}$, and
therefore it is convenient to introduce the dimensionless quantities
${\hat u}$ and ${\hat m}$ {\it via}:
\begin{equation}
u=R\sqrt{E}\, {\hat u}\ ;\quad m=R\sqrt{E}\,{\hat m}\ .
\label{dimless}
\end{equation}

We solve the equation of motion for $\theta(\hat u)$ using shooting
technique and infrared boundary condition described in
refs.~\cite{Albash:2006bs,Albash:2006ew,Karch:2006bv}. For 
Minkowski embeddings we impose the following boundary condition
(motivated by the avoidance of conical
singularities~\cite{Karch:2006bv}):
\begin{equation}\label{eqt:minkow}
 \theta'(\hat u)|_{\theta = \pi/2}= \infty \ .
\end{equation}
For embeddings reaching the pseudo-horizon we find it numerically
convenient to shoot forward (towards infinity) and backward (towards
origin) from the vanishing locus ${\hat u}_*=1$.
The boundary condition ensuring the smoothness of solutions across the
vanishing locus is found from the equation of motion itself to be:
\begin{eqnarray}
&& \theta_0 \equiv \theta(1);\\
&&\frac{\partial\theta}{\partial\hat u}=-\tan\frac{{\theta}_0}{2}\nonumber
\end{eqnarray}
From the above boundary condition we can see that the embeddings
reaching the vanishing locus at $\theta=\pi/2$ has a conical
singularity.  Next we proceed to discuss our numerical findings.

\begin{figure}[ht]
   \centering
   \includegraphics[width=11cm]{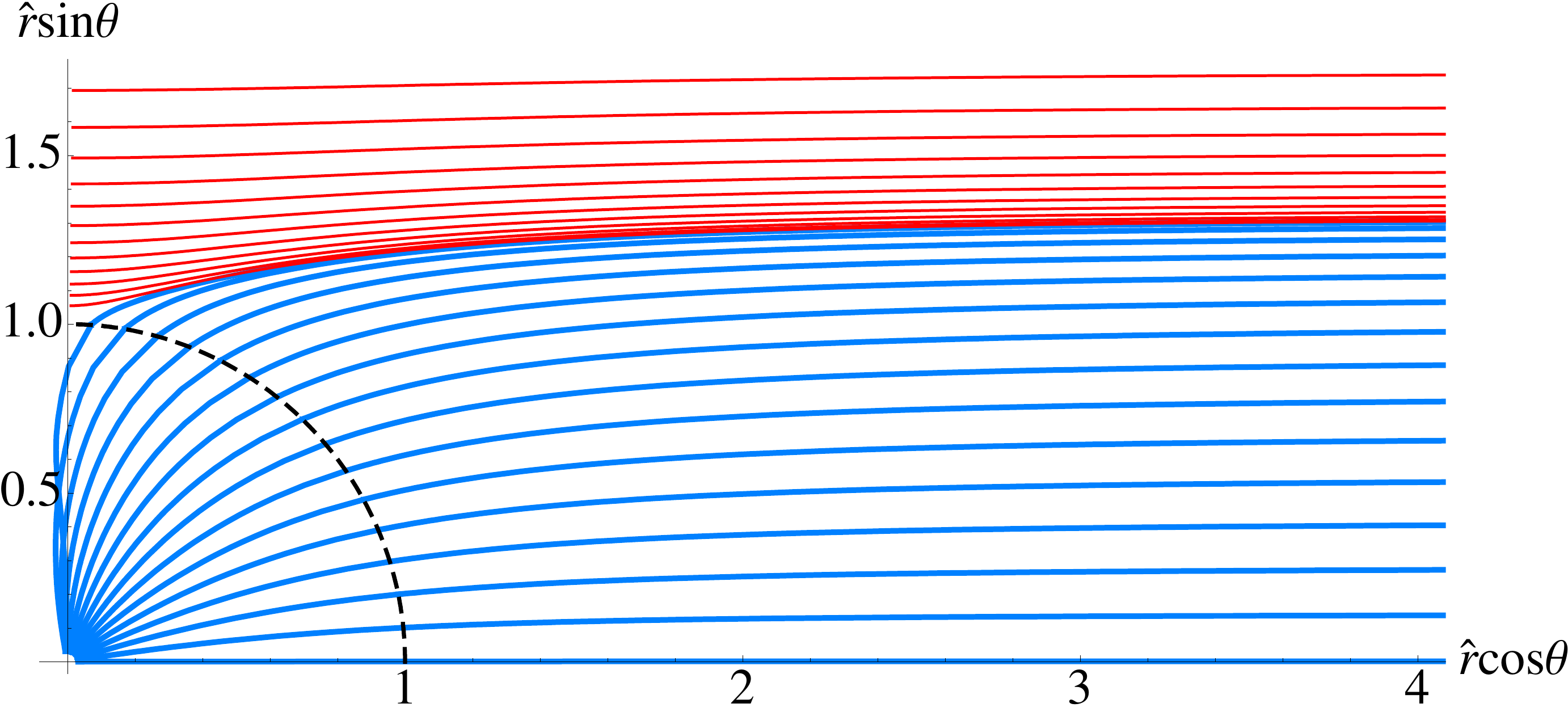}
   \caption{The solid curves  starting far left (red) represent solutions smoothly closing before reaching the vanishing locus.  The solid remaining curves (blue) correspond to embeddings passing the vanishing locus (denoted by a semi--circular dashed black curve). }
   \label{fig:embed0}
\end{figure}

\begin{figure}[h] 
  \centering \includegraphics[width=2.8in]{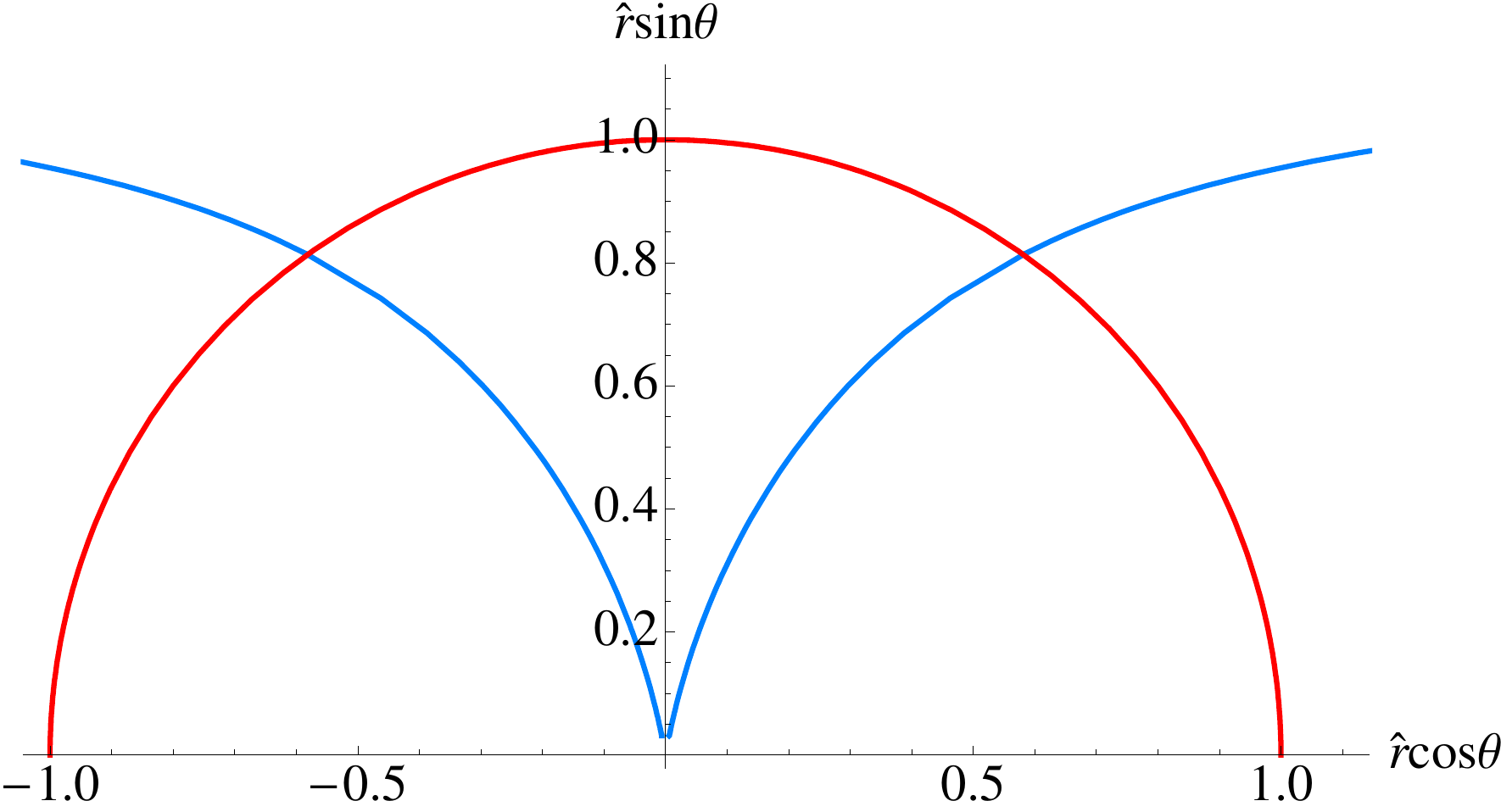}
  \includegraphics[width=2.8in]{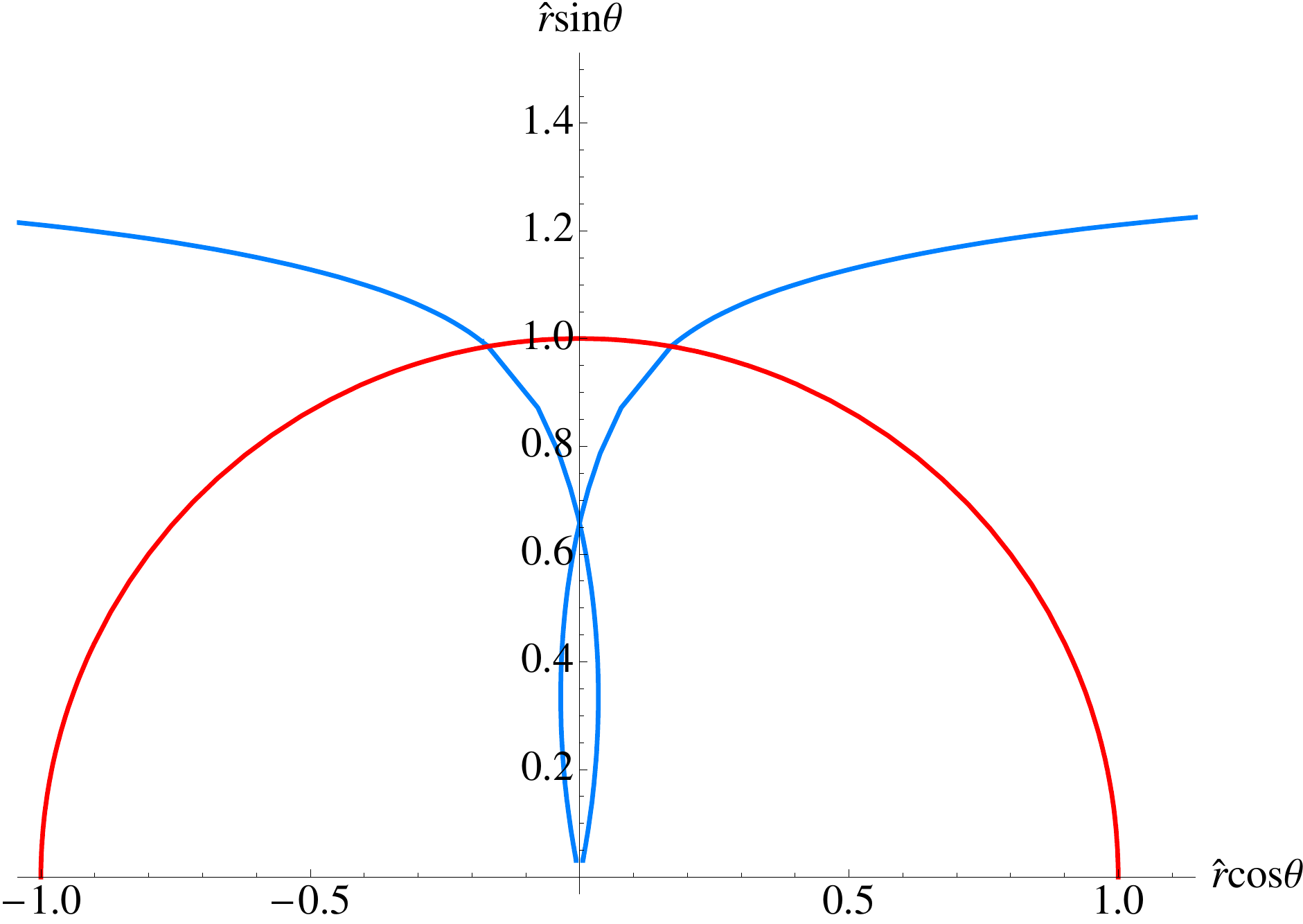}
     \caption{Two kinds of embeddings that pass the vanishing locus corresponding to D7--brane closing at the origin (left), and forming a conical singularity (right) respectively. The semi--circular (red) curve corresponds to the vanishing locus. }
     \label{fig:em02}
  \end{figure}

Numerical analysis shows that there are three different classes of
embeddings for finite electric field and zero background temperature,
which can be classified by the topology of the probe D7--brane. There
are the smooth Minkowski embeddings that close before reaching the
vanishing locus, due to shrinking of $S^3$ wrapped by the probe brane.
There are also the embeddings that reach the vanishing locus before
the $S^3$ shrinks.  These can in turn be classified in two different
categories. The first are singular solutions that reach the vanishing locus but close before reaching the origin. These therefore
have a conical singularity at the closing point. We will discuss these
further below. The second are smooth solutions that pass through the
vanishing locus and reach all the way to the origin with no singular
behavior. These different embeddings are summarized in
figures \ref{fig:embed0} and \ref{fig:em02}.

  We can explicitly see the conical singularity appearing for the
  brane (second curve in Figure~\ref{fig:em02}) since it intersects
  the vertical axis at some non--zero value of $\theta$. Depending on
  the magnitude of the electric field this class of singular solutions
  starts appearing after a certain value of
  $\left(\theta_{0}\right)_{\rm min}$, close to $\pi/2$ (which
  corresponds to the critical embedding), is reached and persists
  until $\theta_0=\pi/2$.

  The existence of these embeddings suggests the possibility of a
  transition in topology of the probe brane as a function of the
  parameters, as happened for finite temperature. Here, it can happen
  as a result of the external electric field that we've applied. The
  Minkowski embeddings simply have a shrinking $S^3$, while the
  embeddings reaching the origin are distinguished by having an $S^3$
  shrinking as well as touching the AdS horizon. However, the presence
  of singular solutions makes the nature of the transition subtle, as
  we discuss later. We will later see that finite temperature case
  also reveals similar classification of embeddings.

\begin{figure}[ht]
  \centering \includegraphics[width=6.5cm]{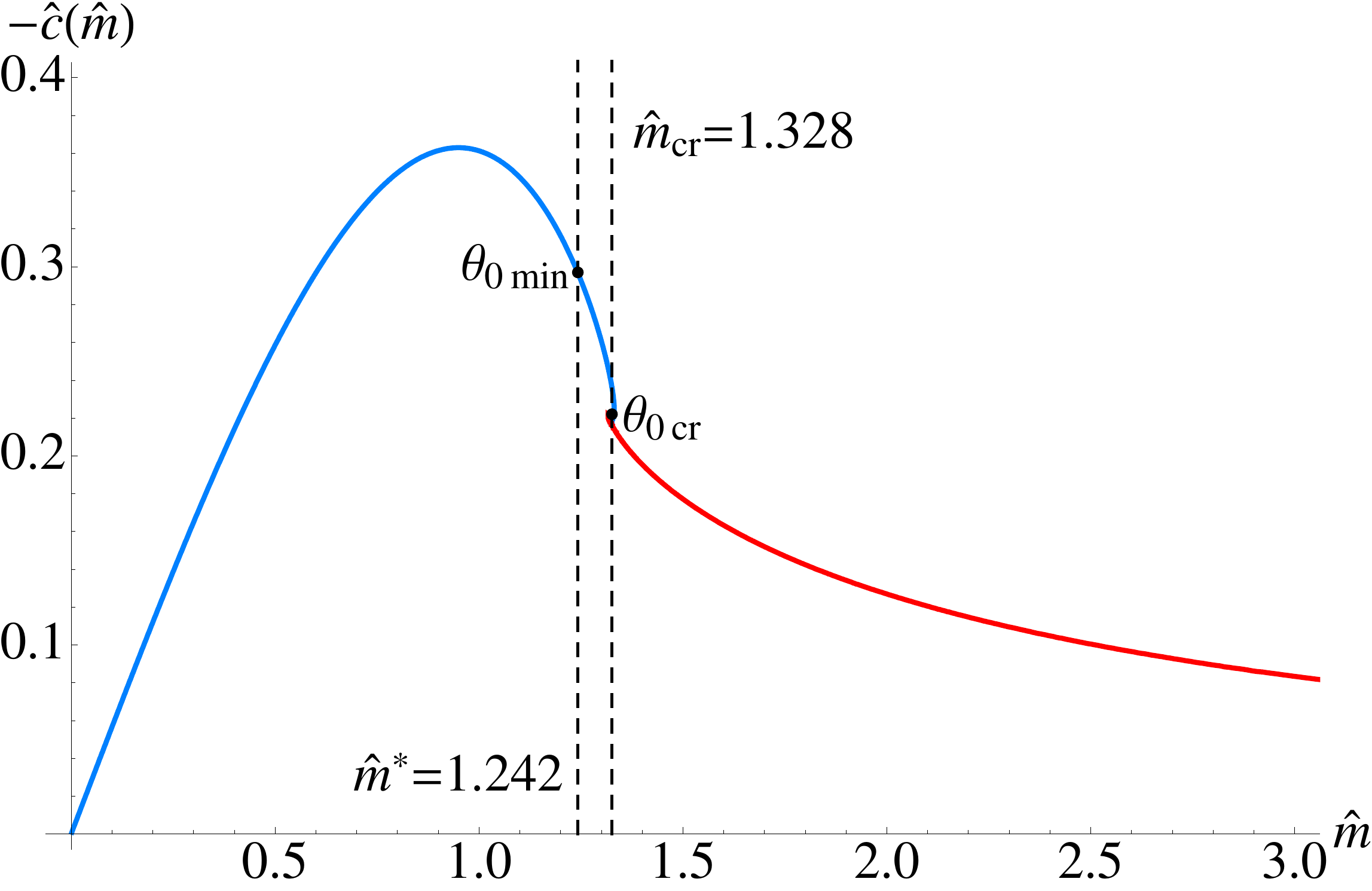}
  \includegraphics[width=6.5cm]{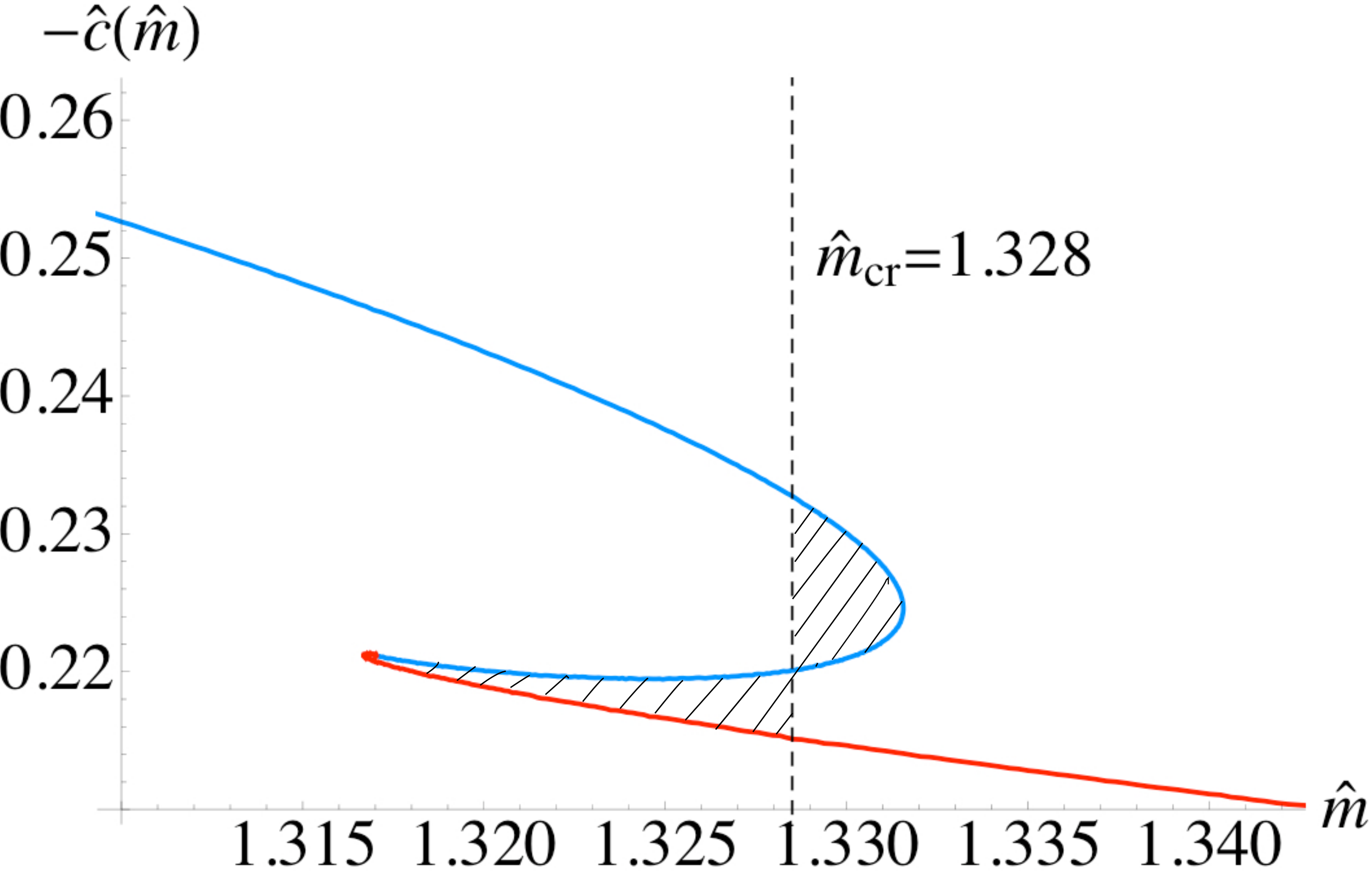}
   \caption{The condensate as a function of bare quark mass ${\hat m}$. The curve segment coming in from the left (blue)  corresponds to embeddings reaching the vanishing locus and the curve segments going out to the right (red) correspond to Minkowski embeddings. There is a family of  solutions that contain a conical singularity between the vanishing locus and the origin. They exist for $(\theta_0)_{\rm min}<{\theta}_0<(\theta_0)_{\rm cr}$. On the right is a magnification of the turn--around region where these segments join, showing multi--valuedness. An analysis of the free energy reveals a first order phase transition at ${\hat m}_{\rm cr}$ where there is a jump from one type of embedding to another. }
   \label{fig:cm0}
\end{figure}

  From the asymptotic behavior of these embeddings we can extract
  condensate as a function of the bare quark mass. In Figure~\ref{fig:cm0} we show this dependence. There are two important mass
  scales. One, ${\hat m}_{\rm cr}$, is where the phase transition
  between the two types of embedding occurs, and the other ${\hat
    m}^\ast$, is where the singular solutions first appear. These
  values are in turn determined by the parameters $(\theta_0)_{\rm
    cr}$ and $(\theta_0)_{\rm min}$, respectively. If ${\theta_0}_{\rm
    min}>(\theta_0)_{\rm cr}$, then the singular solutions are never
  thermodynamically favored.
We see the presence of a first order phase transition speculated based
on the general arguments. Furthermore, we observe that there is no
chiral symmetry breaking, since in all cases the condensate vanishes for vanishing
mass. This is to be contrasted with the case of external magnetic
field studied with these methods in the previous chapters.

This latter observation fits our intuition that the mesons in this
theory have a binding energy that grows with (it is proportional to)
the constituent quark mass~\cite{Kruczenski:2003be}. For a given quark
mass there should exist sufficiently high electric field that can
reduce the binding energy the quarks. This allows two things to happen:
First, this inhibits the formation of chiral condensate, and second,
this ultimately will dissociate the mesons into its constituent
quarks.  This dissociation is in fact a transition from an insulating
to a conducting phase, mediated by the external electric field. On the
dual gravity side this corresponds to a transition from Minkowski
embeddings that do not reach the origin  to embeddings that do.

\begin{figure}[ht]
   \centering
   \includegraphics[width=11cm]{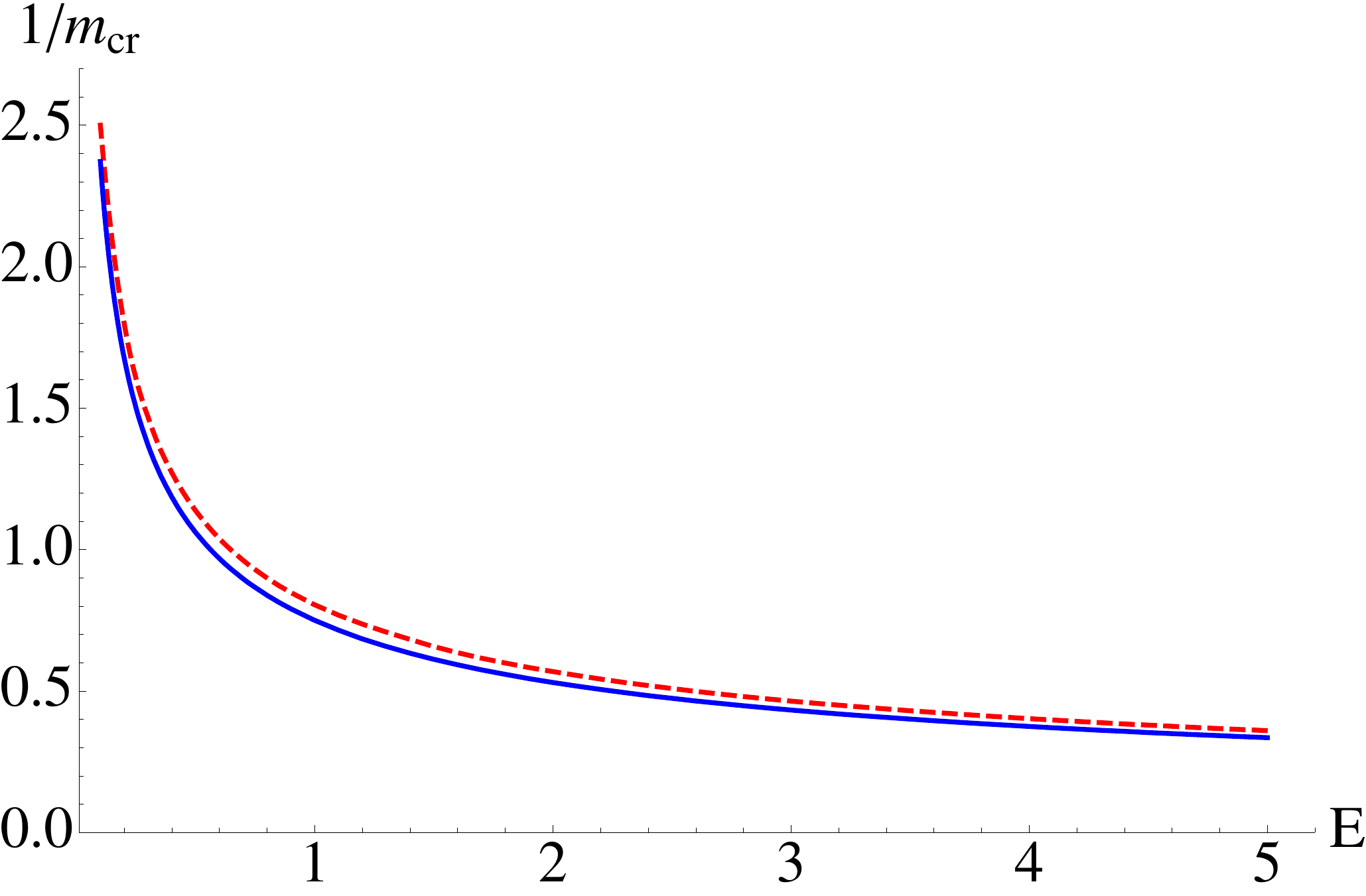}
   \caption{The phase diagram showing the electric field induced first order phase transition separating the dissociated phase from the phase with  stable mesons. The blue (dashed) curve separates the  phases. The region within the blue (solid) curve and red (dashed) curve represents the embeddings with conical singularity. }
   \label{fig:phase0}
\end{figure}

We must note that the appearance of the singular solutions (those that
have a conical singularity) are not well understood at the moment.
There is therefore the possibility that there is an as yet to be
identified intermediate phase right after the meson dissociates.
We can make our dissociation transition explicit. From
equation~(\ref{dimless}) we can deduce the exact dependence of
$m^\ast$ and $m_{\rm cr}$ on the electric field $E$ as
$m^\ast=R\sqrt{E}\, {\hat m}^\ast$ and $m_{\rm cr}=R\sqrt{E}\, {\hat
  m}_{\rm cr}$. This latter is the value of the critical mass for a
given electric field, Using the free energy considerations that we
have used in previous work (it amounts an equal--area law, see
ref.~\cite{Albash:2006ew}), we can also independently (as a test of our
methods for later) determine this dependence numerically for our
solutions, and we display this in Figure~\ref{fig:phase0}.  We also
include the critical mass for which the singular solutions appear as a
function of the field. We see that the analytic behavior deduced
above is nicely confirmed by the numerics.
Indeed, the singular solutions seem to be unavoidable in the phase
diagram (at least at the level of our analysis). However, they seem to
lie very close to the dissociation transition and in the limits
${\hat E}\to 0$ and ${\hat E}\to\infty$ the two curves tend to
merge.

Presently we do not completely understand the role played by the
singular solutions in the physics. One simple possibility (but not the
only one) is that stringy corrections smooth out the singularity in
the interior, while preserving the asymptotic behavior. This would
mean that we have a simpler phase diagram that that in
Figure~\ref{fig:phase0} ({\it i.e.}, with no solid line).

We now turn to the finite temperature case. It is worth noting that a
simple check on our results there will be that the limit of large
electric field (compared to temperature) should recover the physics
that we have seen here.


\subsection*{5.3.3 \hspace{2pt} The case of finite temperature}
\addcontentsline{toc}{subsection}{5.3.3 \hspace{0.15cm} The case of finite temperature}

\begin{figure}[h]
   \centering
   \includegraphics[width=11cm]{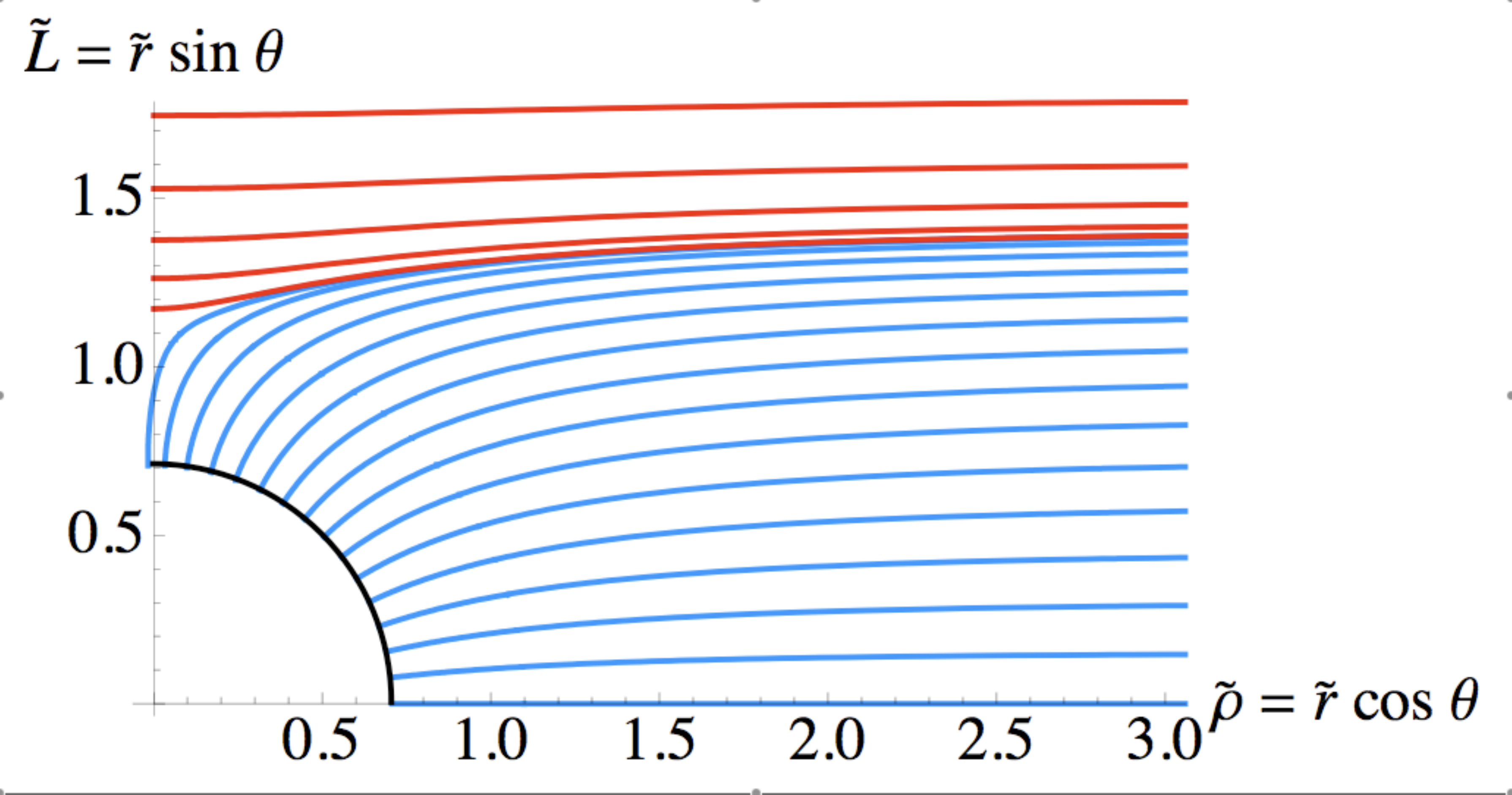}
   \caption{Several solutions for classical D7--brane embeddings with $\tilde{E}=1$. }
   \label{fig:embeddings}
\end{figure}

\begin{figure}[h]
  \centering
    \includegraphics[width=11cm]{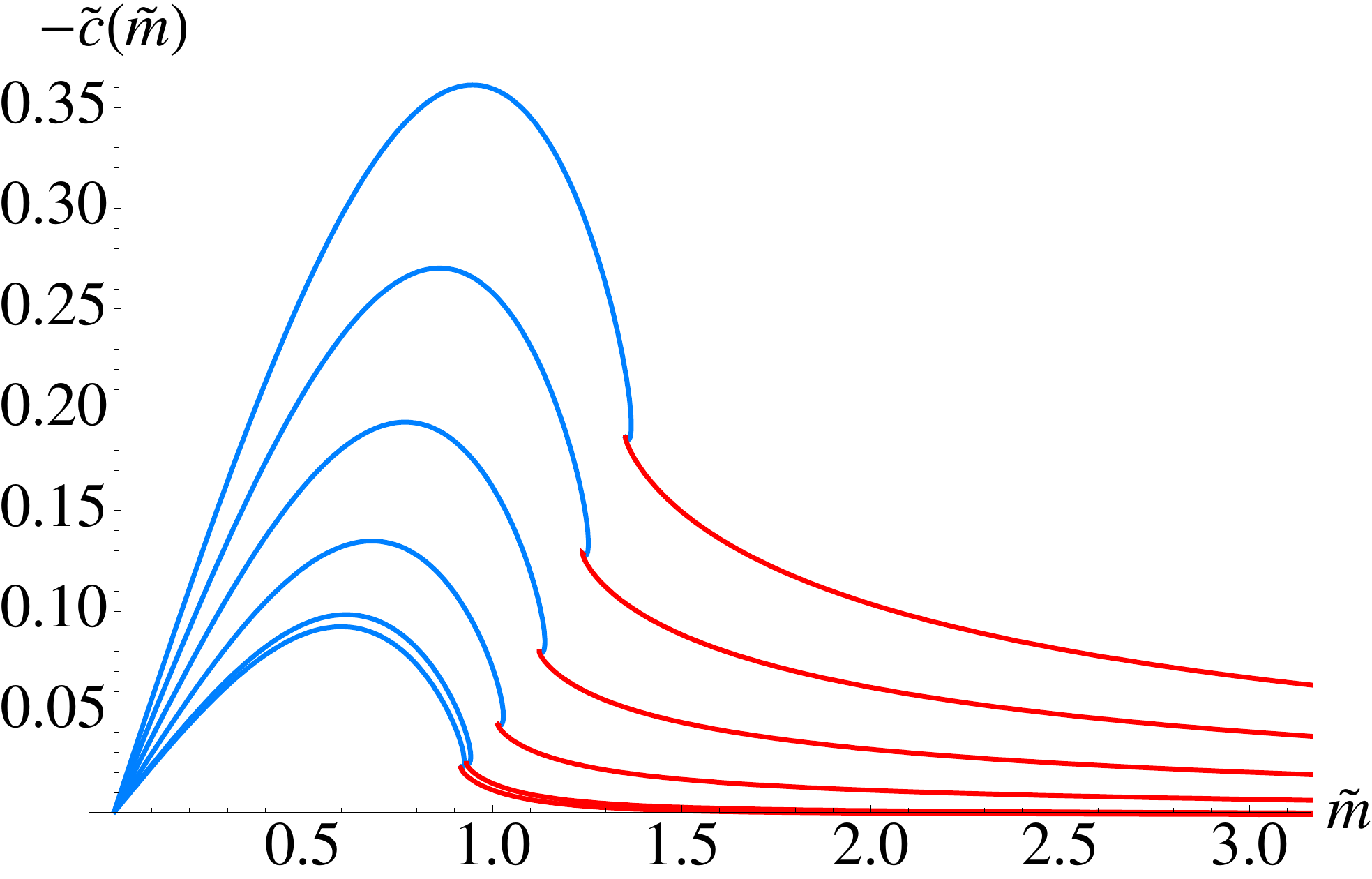}
  \caption{The condensate as a function of bare quark mass for increasing electric field.  From bottom to top, the electric field values are $\tilde{E} = 0.01,0.1,0.3,0.5,0.7,0.9$. }
   \label{fig:c vs m}
\end{figure}

\begin{figure}[h!]
  \centering
  \includegraphics[width=11cm]{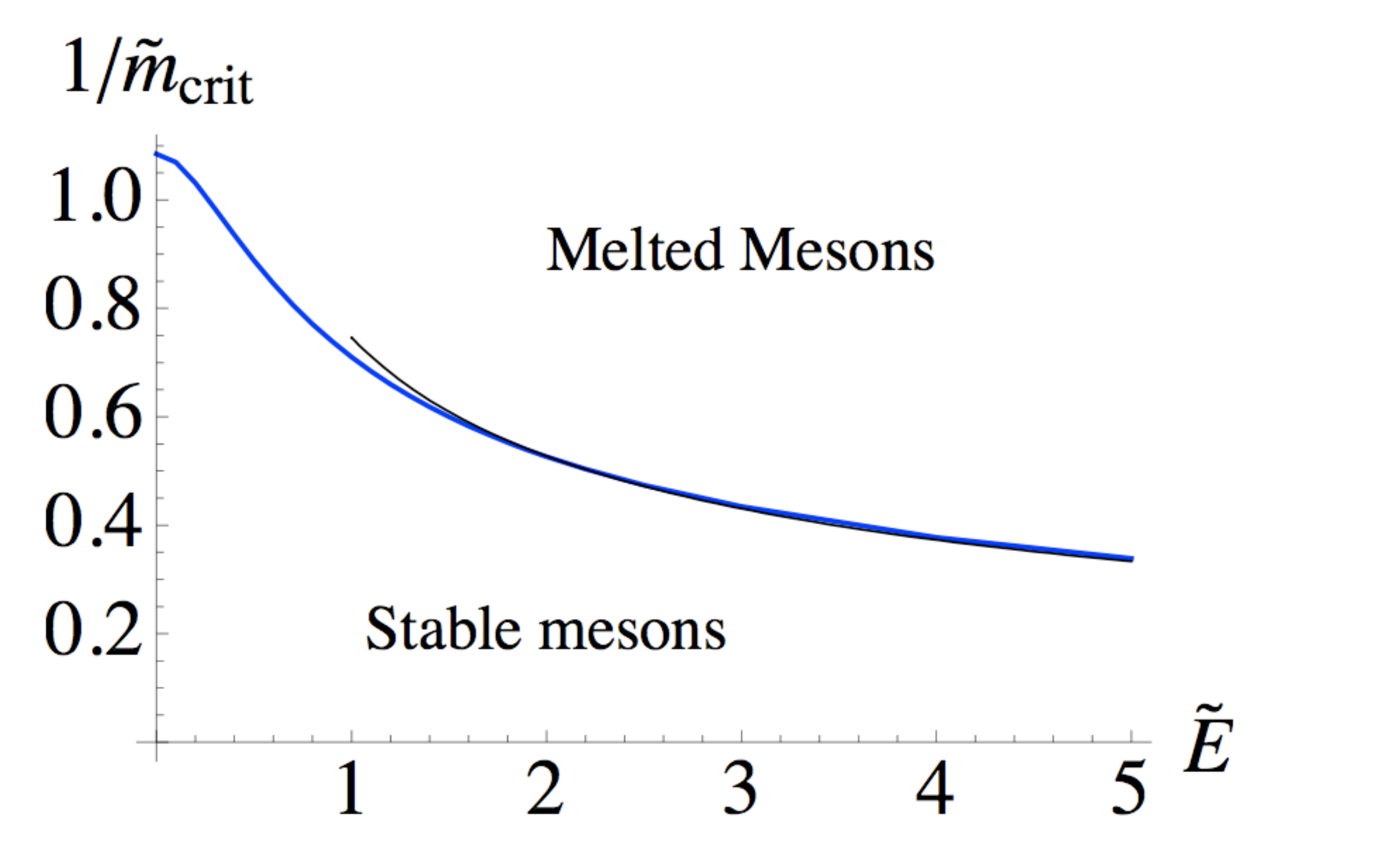}
  \caption{The phase diagram depicting the first order phase transition, denoted by a heavy line (blue) that separates the melted and stable meson phase.  The thin black curve is the best fit curve given by $\mathrm{const.} \times \tilde{E}^{-1/2}$ for large electric field. }
 \label{fig:phase}
\end{figure}

In order to proceed, it is convenient to introduce dimensionless parameters:
\begin{equation}
\tilde r=r/b;~~~\tilde u=u/b;~~~\tilde E=R^2E/b^2;~~~\tilde m=m/b;~~~\tilde c= c/b^3;
\label{dim-el}
\end{equation}

 we again solve the equation of motion for
$\theta(\tilde{u})$ numerically using Mathematica. We use similar
shooting technique and boundary conditions outlined in the previous
section. For Minkowski embeddings we impose equation
(\ref{eqt:minkow}); and for for black hole embeddings the appropriate
initial condition to ensure smoothness of the embedding when we stitch
the solutions across $\tilde{u}_\ast$ is:
\begin{eqnarray}
&& \theta_0\equiv\theta(\tilde{u}_\ast);\\
&&\frac{\partial\theta}{\partial\tilde u_*}=\frac{(3\tilde u_*^4-1)-\sqrt{(3\tilde u_*^4-1)^2+9\tilde u_*^4(\tilde u_*^4-1)\tan^2\theta_0}}{3\tilde u_*(\tilde u_*^4-1)\tan\theta_0}\nonumber
\end{eqnarray}
We show several solutions for $\tilde{E}=1$ in Figure~\ref{fig:embeddings}.
We can extract from these embeddings the condensate $\tilde{c}$ as a
function of the bare quark mass $\tilde{m}$ by analyzing the
asymptotic behavior of the solutions.  We show several of these
solutions for various dimensionless electric field values in Figure~\ref{fig:c vs m}.
The first order phase transition described in
refs.~\cite{Albash:2006ew,Babington:2003vm} persists in the presence
of an electric field.  We determine the critical mass at which this
transition occurs using the free energy analysis (equivalent to an
equal--area law) described in ref.~\cite{Albash:2006ew}, and we
present the phase diagram in Figure~\ref{fig:phase}.  The numerical
results confirm our intuition: at zero temperature, the mass of the
meson is proportional to the bare quark mass, and hence the binding
energy is proportional to the bare quark
mass~\cite{Kruczenski:2003be}.  Turning on an electric field will
decrease the binding energy (the electric field will pull the
quark--antiquark pairs apart), and only mesons built out of
sufficiently heavy quarks--the origin of the critical mass--will
survive.  At finite temperature, we can think of the increasing of the
critical mass as equivalent to the decreasing of the critical
temperature.  Therefore, lowering the binding energy causes mesons to
melt at lower critical temperatures.  In addition, since $\tilde{c} =
c / b^3$, the magnitude of the dimensionless condensate increases with
decreasing critical temperature (as we see in Figure~\ref{fig:c vs
  m}).  Furthermore, the stronger the electric field, the more initial
binding energy is required and therefore a higher critical mass is
needed.

We can deduce the asymptotic behavior of the phase diagram by
considering large electric fields; the dominant energy scale is set by
the electric field, and therefore, we are effectively at zero
temperature.  To see this, one should note that the energy scale in
the geometry is set by $u_\ast$, and for strong electric field,
$u_\ast$ is much larger than the position of the event horizon
horizon.  Therefore, we should expect to reproduce the results for
pure AdS$_5\times S^5$ in an external electric field.  In the case of
pure AdS$_5\times S^5$ in an external electric field, the only energy
scale is given by $R \sqrt{E}$, and therefore we can easily show that
the bare quark mass and condensate scale as:
\begin{eqnarray}
m & \propto& R \sqrt{E} \ , \nonumber \\
c &\propto& R^3 E^{3/2} \ .
\end{eqnarray}
This result suggests that:
\begin{eqnarray}
1/\tilde{m}_\mathrm{crit} & \propto& 1/ \sqrt{\tilde{E}} \ .
\end{eqnarray}
This dependence is checked numerically (and shown in Figure~\ref{fig:phase}) and appears to be valid for $\tilde{E} > 2$.f

  \begin{figure}[h] 
    \centering \includegraphics[width=1.2in]{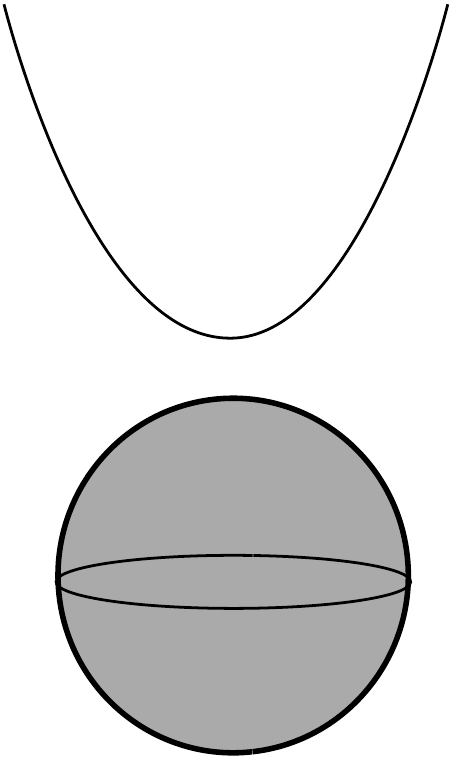}
    \includegraphics[width=1.2in]{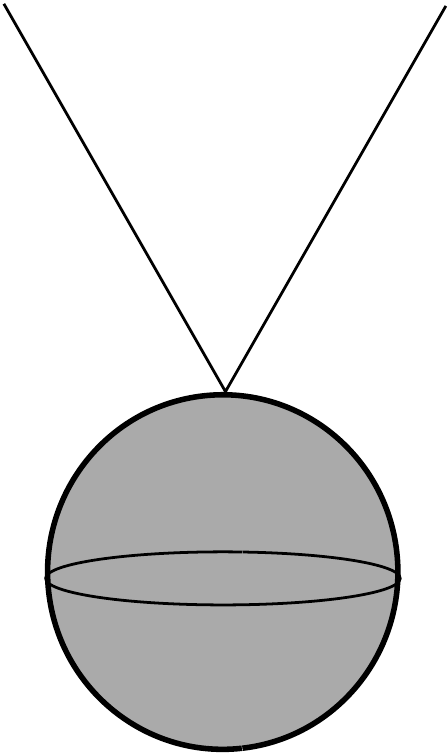}
    \includegraphics[width=1.45in]{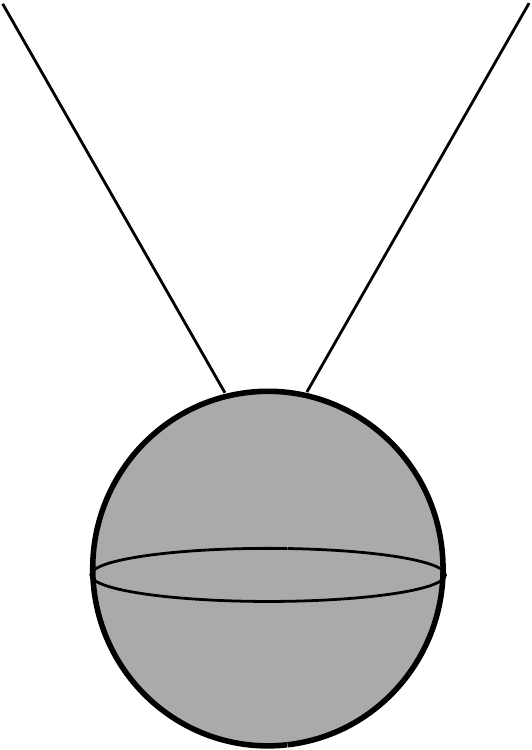}
     \caption{At zero electric field and finite temperature the Minkowski and black hole solutions are separated by a "critical" embedding which has conical singularity at the event horizon. }
     \label{fig:A1}
  \end{figure}
  It is instructive to compare the phase transition to the one for
  zero electric field and finite temperature. In the latter case the
  Minkowski and black hole embeddings are separated by a ``critical''
  embedding with a conical singularity at the event horizon as shown
  in Figure~\ref{fig:A1}. The straightforward generalization of this
  picture to the case of finite electric field suggests the existence
  of a ``critical'' embedding with a conical singularity at the
  vanishing locus ($u=u_\ast$). However, similarly to the zero
  temperature case a more detailed study of the transition from
  Minkowski to black hole embeddings reveals an interesting third
  class of embeddings. These are embeddings which enter the vanishing
  locus at $u_\ast$, and have a conical singularity above the event
  horizon. See Figure~\ref{fig:A2}.
 \begin{figure}[h] 
     \centering
     \includegraphics[width=1.051in]{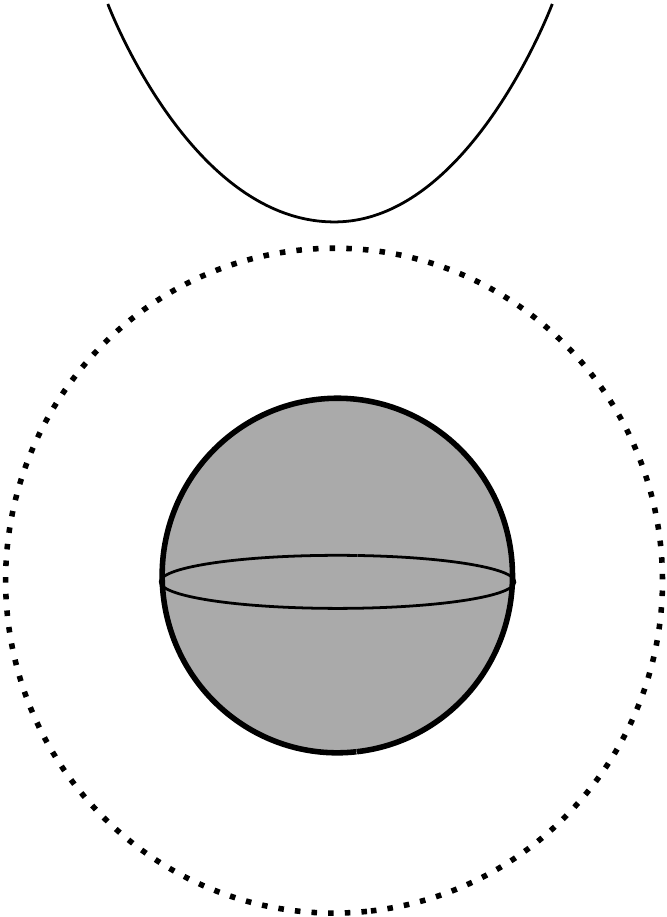}
      \includegraphics[width=1.051in]{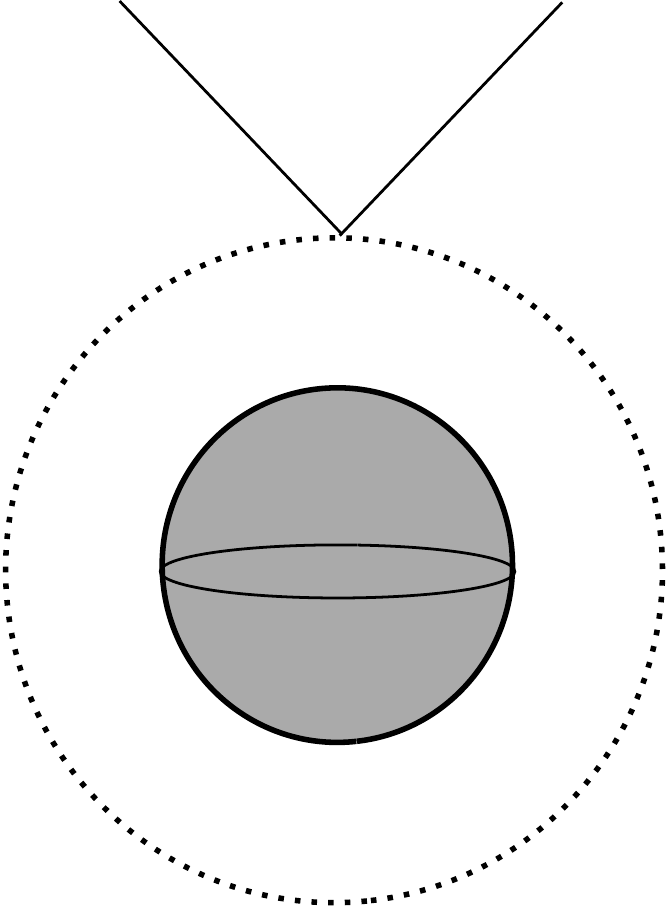}
       \includegraphics[width=1.051in]{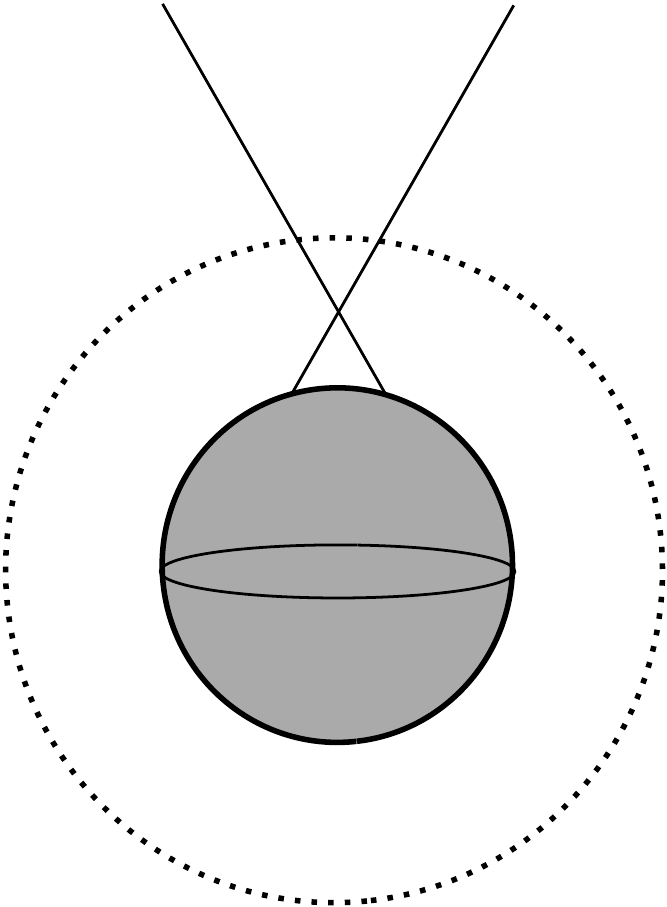}
       \includegraphics[width=1.051in]{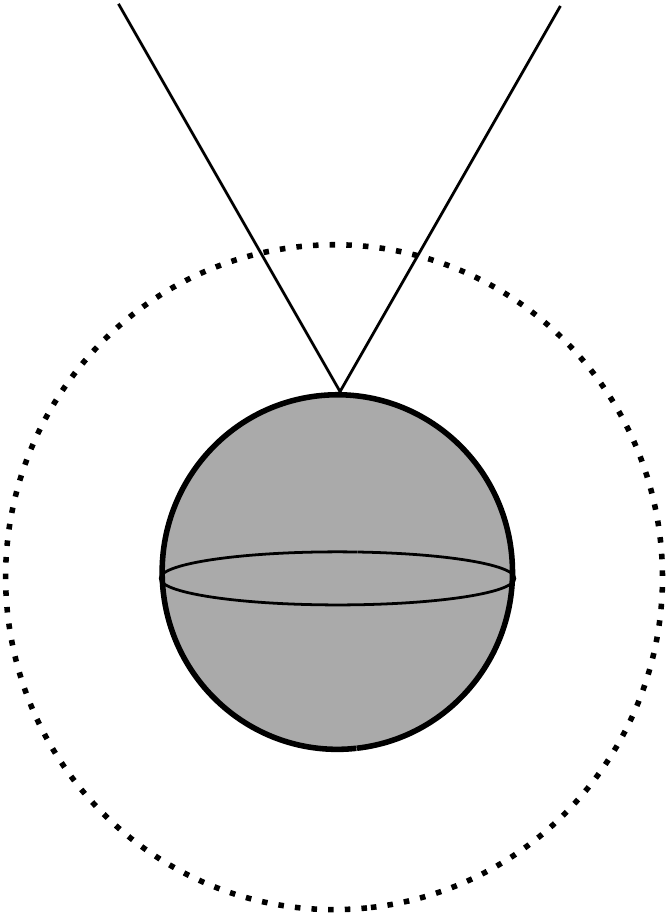}
       \includegraphics[width=1.051in]{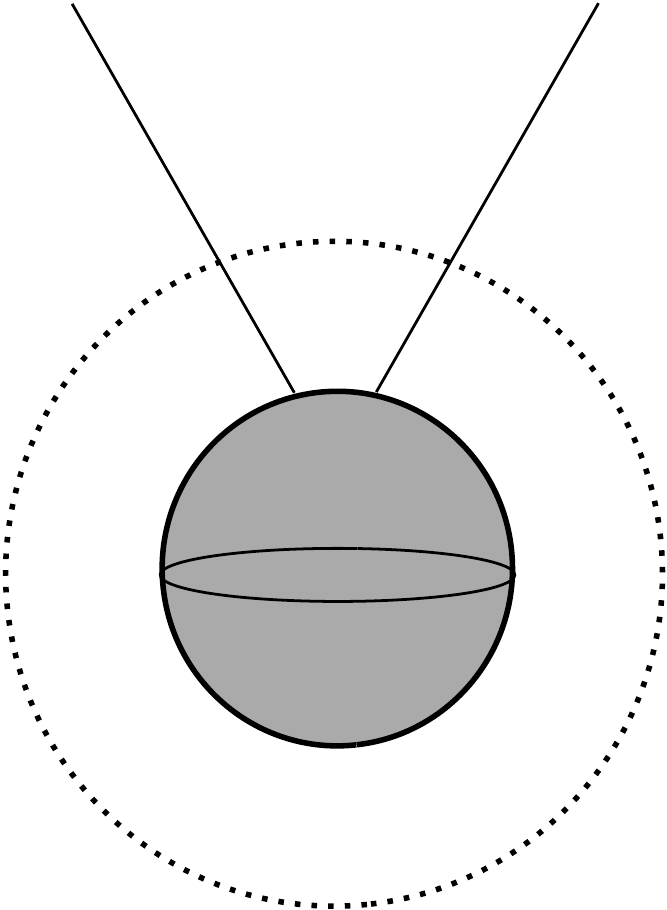}
     \caption{At finite electric field there are two "critical" solutions corresponding to embeddings with a conical singularity at the vanishing locus ($u=u_{*}$) and at the event horizon respectively. The intermediate solutions correspond to black hole embeddings which have conical singularity between the event horizon and the vanishing locus. }
     \label{fig:A2}
  \end{figure}

\begin{figure}[h] 
    \centering
    \includegraphics[width=9.5cm]{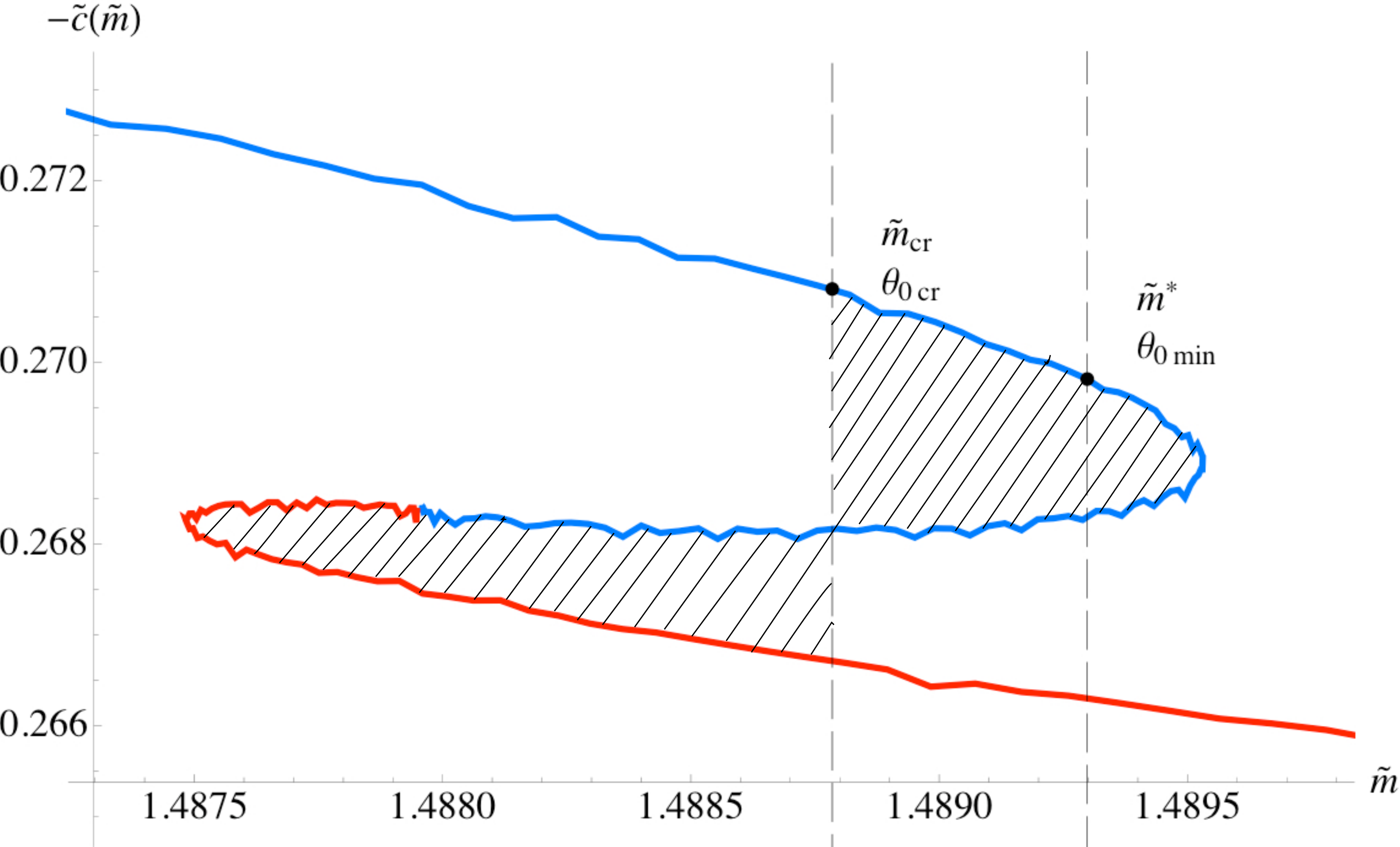}
\medskip
    \includegraphics[width=9.5cm]{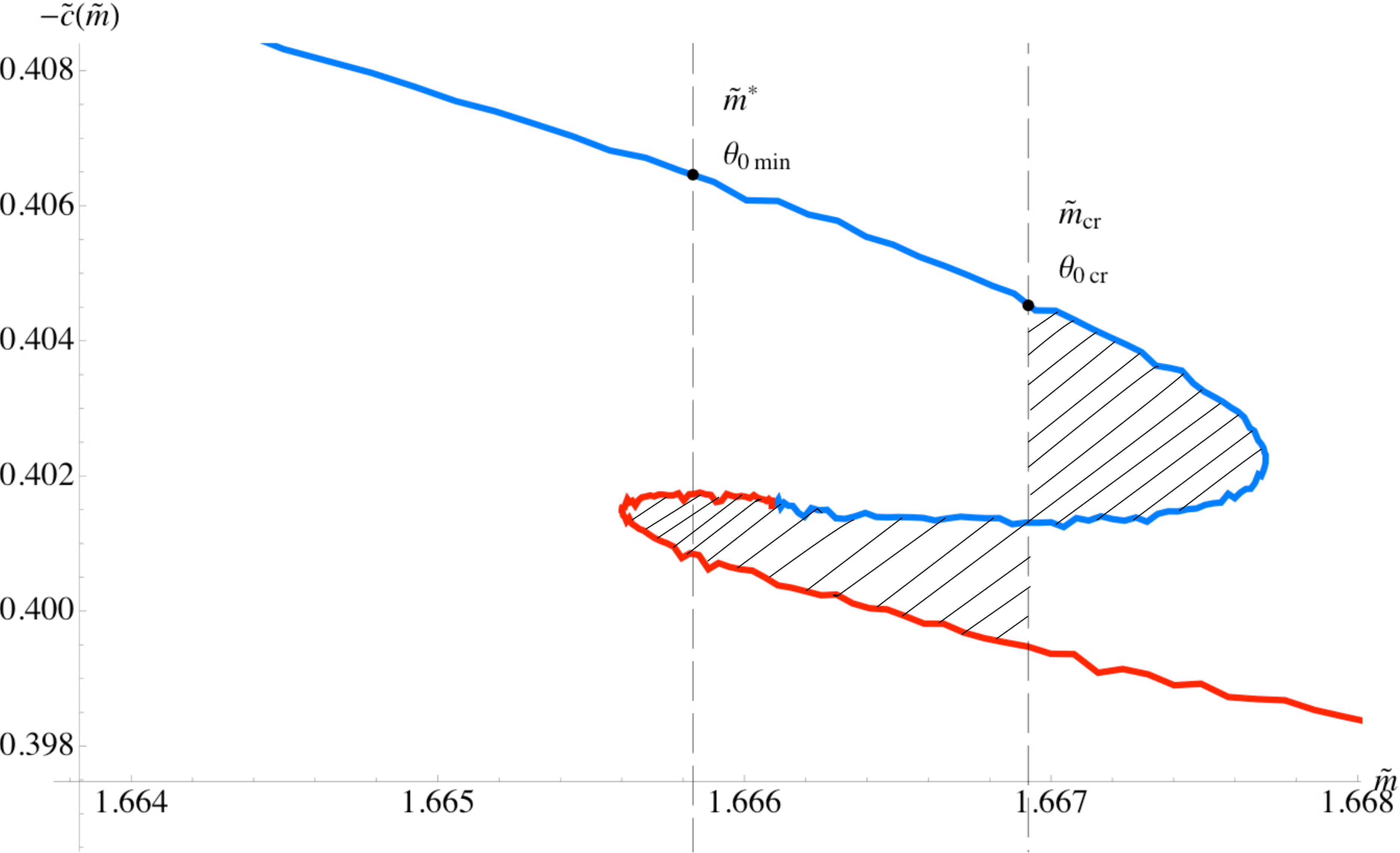}
    \caption{The physical scenarios described in the text for (top) ${\tilde E}<{\tilde E}_{\rm cr}$ and (right) ${\tilde E}>{\tilde E}_{\rm cr}$. (These are magnifications, and the  small scale wiggles are due to the onset of numerical fluctuations at the scale we are observing, and can be ignored.) At the top, we have $\theta_0<(\theta_{0})_{\rm min}$ and so the phase transition bypasses the conically singular branch of solutions. On the bottom, we have $\theta_0>(\theta_{0})_{\rm min}$, and so the conically singular solutions appear on the resulting phase diagram. See text for further discussion. }
    \label{fig:scenarios}
 \end{figure}

\begin{figure}[h] 
    \centering
    \includegraphics[width=1.8in]{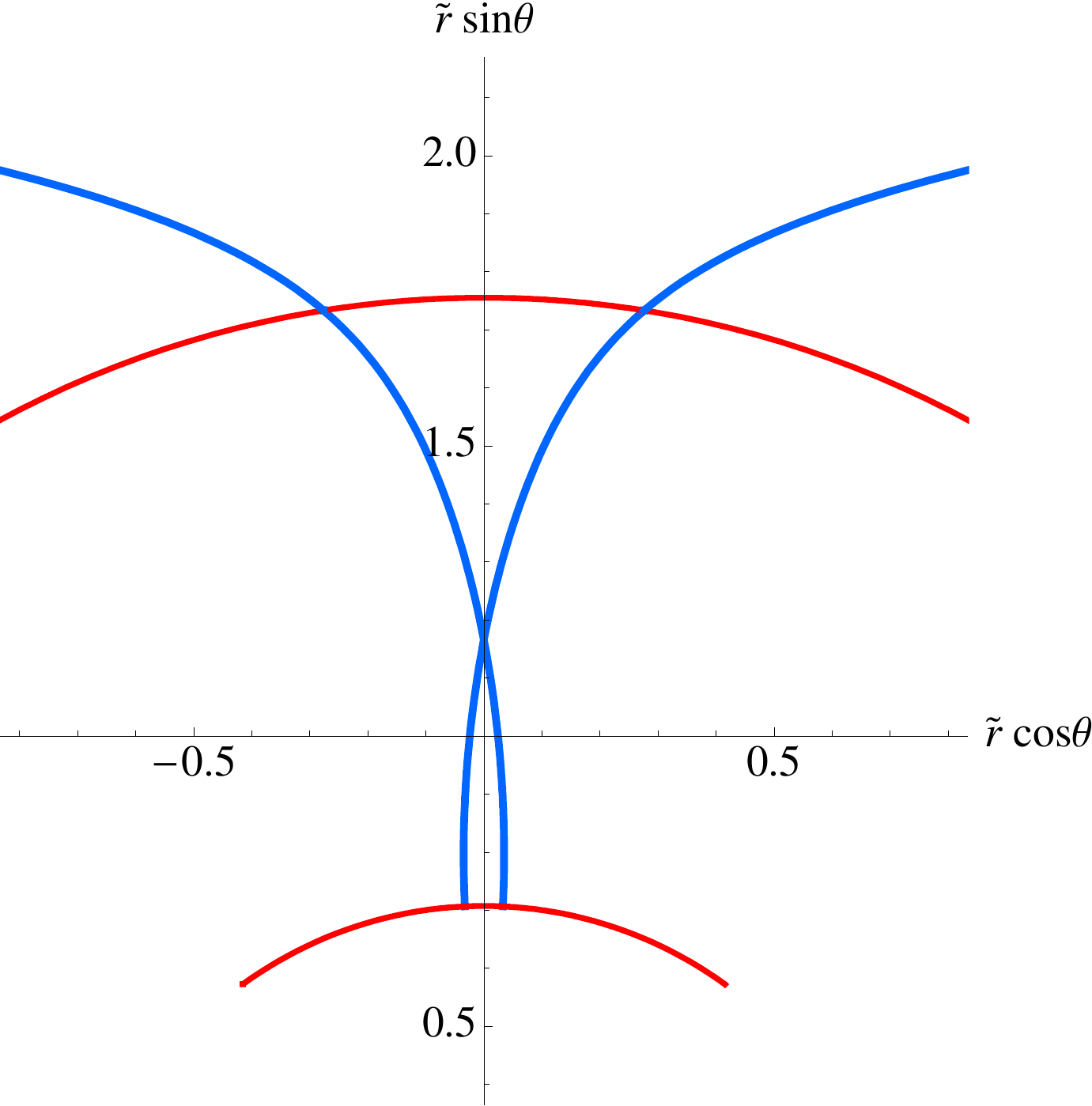}
        \includegraphics[width=1.8in]{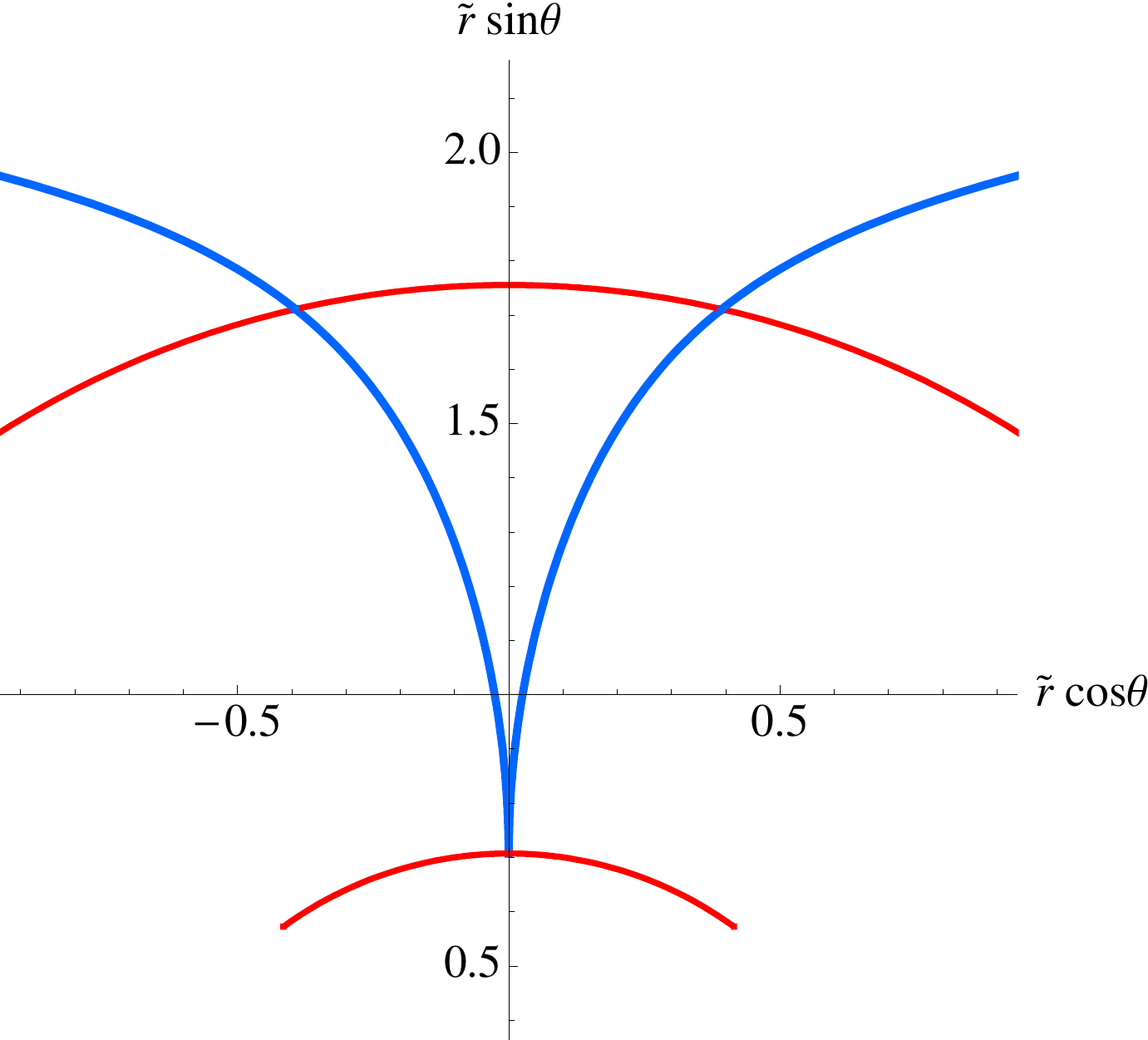}
    \includegraphics[width=1.8in]{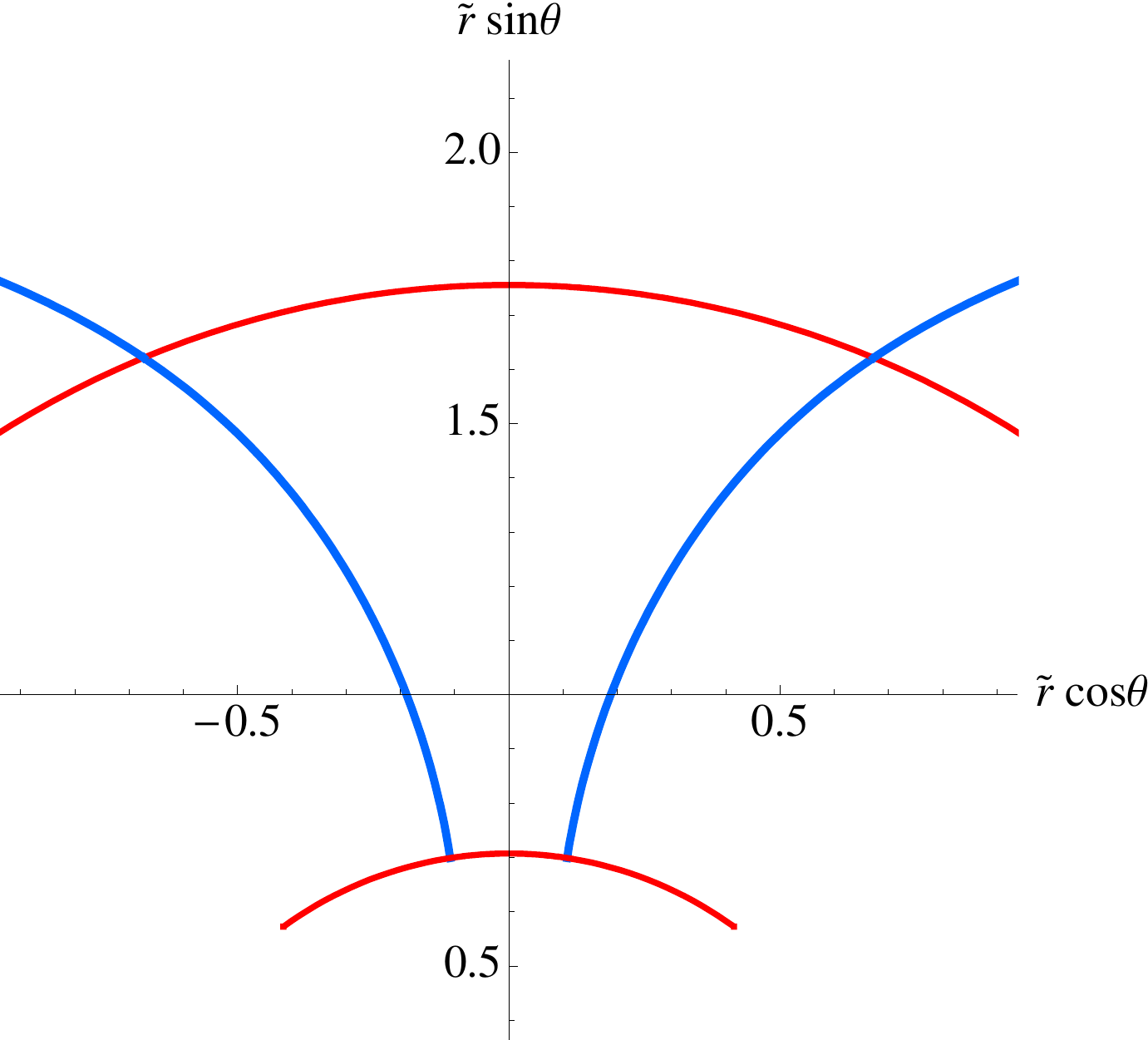}
\caption{Examples of D7--brane embeddings for $\theta_0>(\theta_{0})_{\rm min}$, $\theta_0=(\theta_{0})_{\rm min}$ and $\theta_0<(\theta_{0})_{\rm min}$ respectively. The inner semi--circular segments represent the event horizon, while the outer semi--circular arcs correspond to the vanishing locus at $u=u_*$. The numerical value of the minimal angle is $\theta_{0min}\approx 1.345$ and the parameter $\tilde E=3$. }
    \label{fig:B}
 \end{figure}

  These solutions correspond to embeddings which enter the vanishing
  locus at $\theta_0$ close to $\pi/2$. Such solutions persist until
  the angle $\theta_0$ reaches some minimal value $(\theta_{0})_{\rm
    min}$ below which the embeddings simply fall into the black hole.
  In Figure~\ref{fig:B} one can see examples of embedding for different
  values of $\theta_0$.  Finite temperature adds richness to the
  picture by adding the possibility that these singular solutions may
  or may not be bypassed by the otherwise melting transition induced by
  temperature. This depends entirely on the value of the electric
  field for a given temperature, and for sufficiently high electric
  field the singular solutions appears unavoidable (at least at this
  level of analysis --- as mentioned earlier, there may well be
  stringy corrections to the conical geometry in the interior). The two
   possible scenarios mentioned above, are presented in 
   Figure~\ref{fig:scenarios}. Let us summarize them again.

For sufficiently weak electric field or large temperature ($\tilde E< \tilde E_{\rm cr}$) the $(\theta_{0})_{\rm min}$ embedding is
 in the vicinity of the ``critical'' ($\theta_0=\pi/2$) embedding and
 is bypassed by the phase transition. The thermodynamically stable
 phases correspond to smooth Minkowski and black hole embeddings on
 gravity side or meson gas and quark gluon plasma states in the dual
 gauge theory.

\begin{figure}[h] 
     \centering
     \includegraphics[width=11cm]{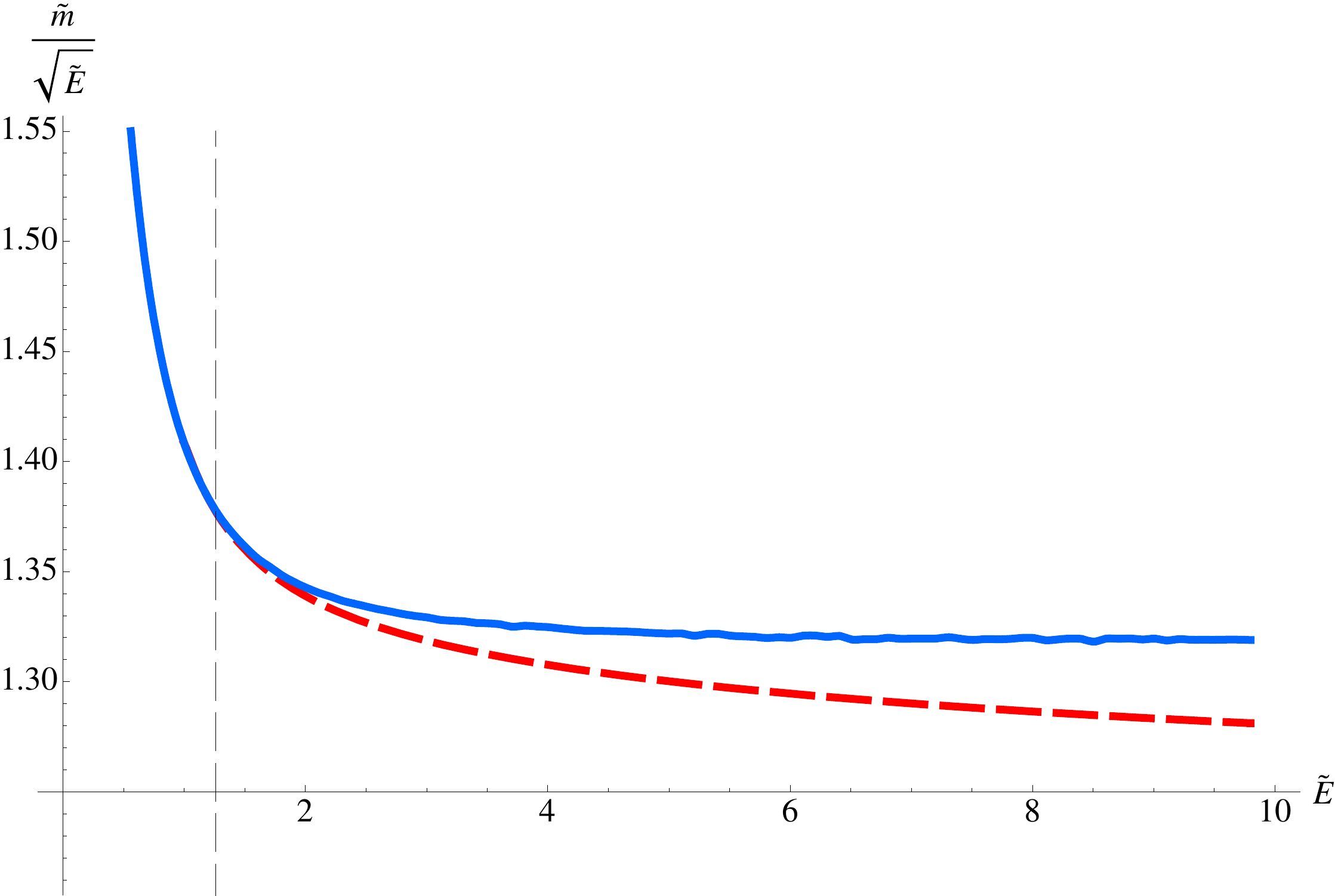}
     \caption{The solid curve correspond to the phase curve from Figure~\ref{fig:phase} and the dashed curve separates embeddings with conical singularity from smooth black-hole embeddings. The area above the solid curve is in meson gas phase and the area below the dashed curve is in quark gluon plasma phase. The area between the solid and dashed curves  corresponds to embeddings which reach the black hole horizon  but have a conical singularity  along the way. Their physical status remains to be determined. (See text.) The vertical dashed line corresponds to the critical value $\tilde E_{\rm cr}\approx 1.26$. }
     \label{fig:phase2}
  \end{figure}

 For $\tilde E>\tilde E_{\rm cr}$ the $(\theta_{0})_{\rm min}$
 embedding is thermodynamically stable and corresponds to some bare
 quark mass $\tilde m^{*}<\tilde m_{\rm cr}$. This means that some of the
 classically stable black hole solutions, namely the one corresponding
 to bare quark mass in the range $\tilde m^{*}<{\tilde m}<\tilde
 m_{\rm cr}$, will have a conical singularity.
This suggests a richer structure for the phase diagram than the one
 presented in Figure~\ref{fig:phase} and in particular the existence
 of a special point at $\tilde E=\tilde E_{\rm cr}$ and $\tilde
 m_{\rm cr}(\tilde E_{\rm cr})$. The corresponding phase diagram is presented
 in Figure~\ref{fig:phase2}. The new choice of dimensionless variables
 on the axes is made in order to emphasize the existence of the
 non--smooth ({\it i.e.,} conically singular) black hole embeddings.
 The solid curve corresponds to the phase curve from
 fig~\ref{fig:phase} (but note that the axes are different here, and
 hence the shape) and the dashed curve separates embeddings with
 conical singularity from smooth black hole embeddings (that lie below
 it). The area above the solid curve is in the stable meson phase and
 the area below the dashed curve is in the melted/dissociated phase.
 The area between the curves corresponds to the embeddings that have a
 conical singularity before ending on the horizon.  The vertical
 dashed line corresponds to the critical value $\tilde E_{\rm cr}\approx
 1.26$ below which the conical solutions are not part of the
 energetically favorable solutions and so do not complicate the
 story. Note that in the $T\to0$ limit the only energy scale is
 $R\sqrt{E}$ and the appropriate dimensionless variable is:
  \begin{equation}
  \hat m=\frac{m}{R\sqrt{E}}=\lim_{T\to0}\frac{\tilde m}{\sqrt{\tilde E}}\ ,
  \end{equation}
  while in the same limit $\tilde E\to\infty$. Therefore the phase
  curves from Figure~\ref{fig:phase2} qualitatively approach the zero
  temperature (large electric field) case of the previous section, with
  the critical parameters $\hat m^{*}\approx 1.22$ and $\hat
  m_{\rm cr}\approx 1.32$.

%
\vspace{6cm}
\section*{5.4 \hspace{2pt} Geometric aspects of the instability}
\addcontentsline{toc}{section}{5.4 \hspace{0.15cm} Geometric aspects of the instability}
As we saw in the previous sections, the addition of external electric field results in the formation of a new vanishing locus defined by $u_*^4=b^4+R^4E^2$. Minkowski embeddings which reach that vanishing locus become unstable. To rectify this, the $A_1$ component of the $U(1)$ gauge field on the D--brane should be turned on with a non--trivial profile along the holographic direction $u$. The interpretation is that for sufficiently strong electric field (equivalently sufficiently small bare quark mass) the binding energy of the quarks is completely overcome by the external electric field and mesons get dissociated to their constituent quarks, which provides free electric charges. The additional gauge field is reflecting this by supporting global electric current along the direction of the external electric field. In this way the vanishing locus at $u=u_*$ naturally reflects the existence of an {\it insulator/conductor} phase transition. 

However, it would be nice if we could shed more light on the geometrical aspects of that transition and in particular on the physical reason for the instability of the Minkowski embeddings at the vanishing locus. In this section we will show that in an appropriate T--dual description this corresponds to an over spinning of the D7--branes, causing them to become superluminal. To illustrate this let us T--dualise the gravitational background  along the $x_1$ direction. The relevant part of the Buscher transformation rules \cite{Johnson:2003gi} is:

\begin{eqnarray}
&&\tilde G_{11}=\frac{1}{G_{11}};~~~\tilde G_{\mu\nu}=G_{\mu\nu}+\frac{B_{1\mu} B_{1\nu}-G_{1\mu}G_{1\nu}}{G_{11}};   \label{Buscher}\\
&&\tilde G_{\mu 1}=\frac{B_{\mu 1}}{G_{11}};~~~\tilde B_{\mu 1}=\frac{G_{\mu 1}}{G_{11}};~~~e^{2\tilde \Phi}=\frac{e^{2\Phi}}{G_{11}}\ .\nonumber
\end{eqnarray}
Applying the transformations (\ref{Buscher}) to the background (\ref{backgr-el}) with the constant Kalb-Ramond B--field from equation (\ref{B-el}) results to the following geometry:
\begin{eqnarray}
d\tilde s^2/\alpha'&=&-\frac{u^4-u_*^4}{R^2u^2}dt^2+\frac{u^2}{R^2}(d x_2^2+dx_3^2)+\frac{R^2u^2}{u^4-b^4}du^2+R^2d\theta^2\label{el-TD}\\
&&+R^2\cos^2\theta d\Omega_3^2+R^2\sin^2\theta d\phi^2+\frac{R^2}{u^2}d\tilde x_1^2+2\frac{E R^2}{u^2}dtd\tilde x_1;\nonumber\\
e^{\tilde\Phi}&=&g_s\frac{R}{u};~~~\tilde B_{\mu\nu}=0\ .\nonumber
\end{eqnarray}
In this way we trade the non-zero $B$--field for an off diagonal component of the metric. Furthermore, the metric of the T--dual background develop an ergosphere at $u=u_*$, as can be seen from equation (\ref{el-TD}). This suggests that the D6--brane probes (the probes loose one dimension because of the T--dualization), which reach the ergosphere, will develop speed (along the $\tilde x_1$ direction) equal to the local speed of light and hence will become superluminal beyond the ergosphere, which on the other side, will result in an imaginary action. Naively it seems that the D6--branes embeddings should become non--static beyond the ergosphere and develop angular velocity along the $\tilde x_1$ direction. However, there is another possibility stemming from the fact that the D6--branes are extended objects. Indeed, for extended objects the only condition for stability that we need to satisfy is that they should not be spacelike, hence if they extend along the $\tilde x_1$ direction they can still describe static configurations. 

Let us make the discussion more explicit by writing down the DBI action for a D6--brane which is extended along the $t,x_2,x_3,u$ directions of the background and wraps an $S^3$ described by $\psi,\beta,\gamma$. In addition we will consider an ansatz of the probe having non--trivial profile along the $\theta$ and $\tilde x_1$ directions:
\begin{equation}
\theta\equiv\theta(u);~~~\tilde x_1\equiv \tilde x_1(u);
\end{equation}
After substituting into the action (\ref{action-el}) and using equation (\ref{el-TD}) we get the lagrangian:
\begin{equation}
{\cal L}\propto u^3\cos^3\theta\sqrt{\frac{u^4-u_*^4}{u^4-b^4}(1+\frac{u^4-b^4}{u^2}\theta'^2)+\frac{u^4-b^4}{u^4}\tilde x_1'^2}\ ,
\label{lagr-el-TD}
\end{equation}
if one compares equation (\ref{lagr-el-TD}) to equation (\ref{lagr-el}) one can see that they are indeed identical (as they should) provided that we define $f(u)=\tilde x_1(u)$. The rest of the analysis is completely analogous to the analysis of the lagrangian (\ref{lagr-el}). This geometric point of view can be particularly useful if one is trying to address the universal properties of the insulator/conductor phase transition subject of our study. We shall come back to that in a later chapter when we will, show that this electrically driven quantum phase transition falls into the same universality class as the quantum phase transition driven by an R--charge chemical potential.

Another point that is worth noting is that since the $B-$field is a pure gauge, it is equivalent to a fixing of the $U(1)$ gauge field of the flavor brane, actually this is the way an external electric field was introduced in ref.~\cite{Albash:2007bq,Karch:2006bv}. In the T--dual description this equivalence corresponds to the freedom to change coordinates to a rotating frame. Indeed, one can easily check that the background (\ref{el-TD}) is actually just the background T--dual to the AdS$_5$-BH$\times S^5$ background (\ref{backgr-el}), but in a rotating frame. Note that being able to work consistently in a rotating frame is due to the fact that our backgrounds are asymptotically AdS$_5\times S^5$ and the rotation is along the compact directions of the background. In this way a D6--brane probe, corotating with the frame, becomes superluminal as it goes deeper into the bulk of the geometry and is regular at infinity, which is in accord with the holographic requirements. 

\section*{5.5 \hspace{2pt} Concluding  remarks}
\addcontentsline{toc}{section}{5.5 \hspace{0.15cm} Concluding remarks}
%

It is very encouraging that non--trivial non--perturbative phenomena
resulting from external fields such as those we have seen here (a
dissociation phase transition, metal--insulator transition and the
associated response current) can be so readily extracted in this kind
of holographic study. We found that since the electric field works
together with the presence of finite temperature, the resulting phase
diagram which accounts for the effects of both is rather simple
(ignoring the complication of the special conical solutions ---see
below). 

Clearly, the story is not quite complete in this electric case, since
above a certain value of the electric field, some of our solutions
develop conical singularities in the interior. We are not entirely
sure about the nature of these solutions. As already stated, one
possibility is that the solutions are locally (near the singularity)
corrected by stringy physics, perhaps smoothing the conical points
into throats that connect to the horizon. We would expect in that case
that our phase diagram would largely remain intact, since the values
of the masses and condensates for each solution are read off at
infinity, far from where the conical singularity develops. The results
for the relative free energies of those solution (compared to the
other solutions at the same values of the mass) would be the same, and
so the complete story would be unaffected. A more drastic possibility
is simply that the physics is considerably modified by instabilities
associated with those conical solutions that our current study has not
revealed --- the appropriate part of the meson spectrum is
particularly difficult to study numerically here, and so the
tell--tale signs of tachyonic modes that represent an instability are
hard to check for in this case. Such a modification could remove quite
a significant part of the phase diagram for large enough values of the
electric field, allowing the possibility of a richer structure than we
have seen so far.

\chapter*{Chapter 6:  \hspace{1pt} Phase structure of finite temperature large $N_c$ flavored Yang- 
Mills theory with an R--charge chemical potential}
\addcontentsline{toc}{chapter}{Chapter 6:\hspace{0.15cm}
Phase structure of finite temperature large $N_c$ flavored Yang- 
Mills theory with an R--charge chemical potential}

\section*{6.1 \hspace{2pt} Introductory remarks}
\addcontentsline{toc}{section}{6.1 \hspace{0.15cm} Introductory remarks}

Not long after the birth of AdS/CFT  \cite{Gubser:1998bc,Maldacena:1997re,Witten:1998qj}, where the finite temperature phase structure of ${\cal N}=4$ $SU(N_c)$ (at large $N_c$) was understood in terms of the thermodynamics of Schwarzschild black holes in AdS \cite{Witten:1998qj,Witten:1998zw}, it was recognized \cite{Chamblin:1999tk} that progress could be made in understanding the properties of gauge theories in the presence of a global current, akin to a chemical potential for baryon number or isospin, by studying the physics of charged black holes, such as Riessner--Nordstrom \cite{Chamblin:1999tk,Chamblin:1999hg}, or more general charged black holes with non--trivial scalars \cite{Cvetic:1999rb}, in AdS.

The resulting $(\mu_{\rm{R}}, T)$ or $(q,T)$ phase diagrams (where $q$ is conjugate to $\mu_{\rm{R}}$--the analogue of Baryon number) of the ${\cal N}=4$ gauge theory were found to be rather rich, with {\it e.g.,} a first order phase transition line, and even a second order critical point in the $(q,T)$ plane \cite{Chamblin:1999tk,Chamblin:1999hg}. This led to the hope that such studies might lead to insights into the physics of the QCD phase diagram. The idea is that studies of the effect of the global $U(1)$ symmetry--even though it is not exactly baryon number--might lead to physics in the same universality class as the more realistic gauge theories, giving insight into QCD at finite temperature and density.

A more firm footing for this idea should be obtained by the study of those dynamics in the presence of fundamental flavors of quark, a key feature of QCD that distinguishes it from the ${\cal N}=4$ gauge theory (which of course only has adjoint matter of very specific types and quantities allowed by supersymmetry). To this end in this chapter we consider D7--brane probing of the spinning D3--brane geometry.

\section*{6.2 \hspace{2pt} Spinning D3--branes}
\addcontentsline{toc}{section}{6.2 \hspace{0.15cm} Spinning D3--branes}

\subsection*{6.2.1 \hspace{2pt} The string background}
\addcontentsline{toc}{subsection}{6.2.1 \hspace{0.15cm} The string background}

To source subgroups of the $U(1)^3\subset SO(6)$ global symmetries corresponding to rotations in the transverse $\IR^6$, we consider D3--branes with angular momentum.  The supergravity description/metric for spinning D3--branes is \cite{Cvetic:1999xp,Kraus:1998hv}:
\begin{eqnarray}
ds^2 &=& H_3^{-1/2}\left\{-\left(1-\frac{r_H^4}{r^4 \Delta}\right)dt^2 + d\vec{x}\cdot d\vec{x}\right\} +H_3^{1/2} \left\{\frac{\Delta dr^2}{\mathcal{H}_1 \mathcal{H}_2 \mathcal{H}_3 -\frac{r_H^4}{r^4}}\right.\nonumber\\ 
& & \left.+r^2\sum_{i=1}^3\mathcal{H}_i \left(d\mu_i^2 +\mu_i^2 d\phi_i^2\right) -\frac{2r_H^4
\cosh\left(\beta_4\right)}{r^4 H_3 \Delta}dt \left(\sum_{i=1}^3
R_i\mu_i^2 d\phi_i\right) \right.\nonumber\\
& &\left.+\frac{r_H^4}{r^4}\frac{1}{H_3
\Delta}\left(\sum_{i=1}^3 R_i \mu_i^2 d\phi_i\right)^2 \right\} \ ,
\label{spinninng D-branes}
\end{eqnarray}
where:
\begin{eqnarray}
\Delta &=& \mathcal{H}_1 \mathcal{H}_2 \mathcal{H}_3 \sum_{i=1}^3
\frac{\mu_i^2}{\mathcal{H}_i} \nonumber\\
H_3 &=& 1+\frac{r_H^4}{r^4}\frac{\sinh\left(\beta_3\right)}{\Delta}
= 1+\frac{\alpha_3 r_3^4}{\Delta r^4} \nonumber\\
\mathcal{H}_i &=& 1+\frac{\ell_i^2}{r^2} \nonumber\\
r_3^4&=&4 \pi g_s N \alpha'^2 = \alpha'^2 \hat{r}^4 \nonumber\\
\alpha_3 &=&\sqrt{1+\left(\frac{r_H}{2r_3}\right)^8}
-\left(\frac{r_H}{2r_3}\right)^4 \nonumber \ .
\end{eqnarray}
In order to explore the decoupling limit ($\alpha' \to 0$), which we are interested in, we define:
\begin{eqnarray}
r &=& \alpha ' u\ ,\quad 
r_H = \alpha' u_{\rm{H}} ,\quad 
\ell_i =\alpha' q_i \nonumber \ ,
\end{eqnarray}
such that:
\begin{eqnarray}
\frac{\alpha_3 r_3^4}{r_H^4} &\to& \frac{\hat{r}_3^4}{\alpha'^2
u_{\rm{H}} ^2}\ ,\quad 
\sinh\left(\beta_3\right) \to \frac{\hat{r}_3^2}{\alpha' u_{\rm{H}} ^2}\ ,
\nonumber\\
\mathcal{H}_i &\to& 1+\frac{q_i^2}{u^2}\ , \quad 
H_3 \to \frac{\hat{r}_3^4}{\alpha'^2 u^4 \Delta}, \quad \hat{r}_3 \to R \nonumber \ .
\end{eqnarray}
With these definitions, in the decoupling limit, the metric is:
\begin{eqnarray} 
\label{spinning}
\hspace{-2cm}ds^2/\alpha' &=& \Delta^{\frac{1}{2}}\left\{-\left(\mathcal{H}_1
\mathcal{H}_2 \mathcal{H}_3\right)^{-1} f dt^2 + f^{-1} du^2 +
\frac{u^2}{R^2}d\vec{x}\cdot d\vec{x}\right\} \nonumber\\
\hspace{-2cm}&&+\Delta^{-\frac{1}{2}}\left\{\sum_{i=1}^3\mathcal{H}_i \mu_i^2
\left(R d\phi_i - A_t^i dt\right)^2 + R^2 \mathcal{H}_i d\mu_i^2\right\} \ ,
\end{eqnarray}
where we have defined:
\begin{eqnarray}
&&f=\frac{u^2}{R^2}{\cal H}_1{\cal H}_2{\cal H}_3-\frac{u_{\rm{H}} ^4}{u^2R^2},~~~{\cal H}_i=1+\frac{q_i^2}{u^2},~~~A_t^i=\frac{u_{\rm{H}} ^2}{R}\frac{q_i}{u^2+q_i^2},\label{fields}\\~~~&&\Delta={\cal H}_1{\cal H}_2{\cal H}_3\sum_{i=1}^{3}\frac{\mu_i^2}{{\cal H}_i},~~~\quad\mu_1=\sin\theta,~~~\mu_2=\cos\theta\sin\psi,~~~\mu_3=\cos\theta\cos\psi .\nonumber
\end{eqnarray}
Of immediate interest for us is to compute the temperature of the background and to examine its dependence on the different choices of angular momenta $q_i$ in $S^5$ that we may have. As we are going to show, the case of three equal charges $q_i=q$ is the simplest scenario containing extremal horizon and hence zero temperature at finite chemical potential. This is important because this will enable us to study the first order phase transition of the fundamental matter at zero temperature, when the fluctuations of the system are purely quantum. Another interesting property of the at three equal charges case is the extension of the phase diagram of the dual flavored gauge theory to the case of non--zero R--charge chemical potential. As we are going to demonstrate later, it has the same structure as the one for the adjoint sector of the theory, corresponding to the Hawking-Page transition of the gravitational background~\cite{Chamblin:1999tk}. Let us proceed with the calculation of the temperature.

\subsection*{6.2.2 \hspace{2pt} Calculation of the temperature}
\addcontentsline{toc}{subsection}{6.2.2 \hspace{0.15cm} Calculation of the temperature}

In order to calculate the temperature of the background we could evaluate the surface gravity at the event horizon. This calculation was performed in ref.~\cite{Russo:1998by}. Here we are going to undertake a more explicit approach. First we will regularize the background (\ref{spinning}) by going to a rotating frame in which the geometry does not have an ergosphere. After that we will re-scale the coordinates to focus on geometry near the event horizon of the background, thus going to an appropriate Rindler frame.

 The radius $u_{\rm{E}} $ of the actual event horizon of the background (\ref{spinning}) is determined by the largest
root of $f(u)=0$. In addition it has an ergosphere determined by the
expression:
\begin{equation}
\Delta({\cal H}_1{\cal H}_2{\cal H}_3)^{-1}f-\sum_{i=1}^3{\cal H}_i\mu_i^2(A_t^i)^2=0\ .
\label{ergosphere}
\end{equation}
Since the background (\ref{spinning}) is asymptotically AdS$_5\times
S^5$, we can ``remove'' the ergosphere (\ref{ergosphere}), by going to a
rotating frame. This is equivalent to gauge shifting $A_t^i$
from~(\ref{fields}) such that:
\begin{equation}
{A'_t}^i=-R\mu_{\rm{R}}^i+A_t^i
\label{shifted}
\end{equation}
The parameters $\mu_{\rm{R}}^i$ are set by the condition ${A'_t}^i|_{u_{\rm{E}} }=0$ and
hence:
\begin{equation} 
\mu_{\rm{R}}^i=\frac{u_{\rm{H}} ^2}{R^2}\frac{q_i}{u_{\rm{E}} ^2+q_i^2}\ .
\label{chempot6}
\end{equation}
From the asymptotic behavior at infinity ($u\to\infty$), it is clear that
$\mu_{\rm{R}}^i$ correspond to the angular velocities of the frame along
$\phi_i$. In the dual gauge theory these correspond to having time
dependent phases of the adjoint complex scalars or equivalently to
an R--charge chemical potentials for the corresponding
scalars \cite{Chamblin:1999tk}.

Next we consider the following change of coordinates:
\begin{equation}
u=u_{\rm{E}} +a z^2;~~~\theta=\theta_{\rm{E}}-y_1;~~~\psi=\psi_{\rm{E}}-y_2\ .\
\label{zoom-R}
\end{equation}
Note that the choice of $\theta_{\rm{E}},\psi_{\rm{E}}$ is not important for the computation of the temperature and the constant $a$ is to be determined. In order to compute the temperature we should focus on the following two terms in the metric (\ref{spinning}):
\begin{equation}
ds^2/\alpha'=G_{tt}dt^2+G_{uu}du^2+\dots
\end{equation}
and make sure that near the horizon it is of the form:
\begin{equation}
ds^2/\alpha'^2=-(2\pi T)^2z^2dt^2+dz^2+\dots\ ,
\end{equation}
which corresponds to a going to a Rindler frame. From equation (\ref{spinning}) one can see that we have:
\begin{equation}
G_{uu}=\Delta^{1/2}f(u)^{-1},
\label{Guu}
\end{equation}
where:
\begin{eqnarray}
f(u)&=&\frac{1}{R^2u^4}\left((u^2+q_1^2)(u^2+q_2^2)(u^2+q_3^2)-u^2u_{\rm{H}} ^4\right)\label{f(u)}\\
&=&\frac{1}{R^2u^4}(u^2-u_{\rm{E}} ^2)(u^2-u_1^2)(u^2-u_2^2)\ .\nonumber
\end{eqnarray}
Here $u_1^2$ and $u_2^2$ refer to the other two roots of the polynomial in $u^2$. Note that they do not need to be real. After substituting equations (\ref{zoom-R}) and (\ref{f(u)}) into equation (\ref{Guu}) and leaving only the term leading in $z$, $y_1$ and $y_2$ we obtain :
\begin{equation}
G_{uu}du^2=\frac{2\Delta_{\rm{E}}^{1/2} R^2u_{\rm{E}} ^3a}{(u_{\rm{E}} ^2-u_1^2)(u_{\rm{E}} ^2-u_2^2)}dz^2=dz^2\ ,
\end{equation}
where $\Delta_{\rm{E}}=\Delta|_{(u_{\rm{E}} ,\theta_{\rm{E}},\psi_{\rm{E}})}$. The corresponding value of $a$ is:
\begin{equation}
a=\frac{(u_{\rm{E}} ^2-u_1^2)(u_{\rm{E}} ^2-u_2^2)}{2\Delta_{\rm{E}}^{1/2}R^2u_{\rm{E}} ^3}\ .
\label{a}
\end{equation}
Now we focus on the $G_{tt}$ term in (\ref{spinning}) but with the new gauge choice for $A_t^i$ given in equation (\ref{shifted}). The exact expression is:
\begin{equation}
G_{tt}=-\Delta^{1/2}({\cal H}_1{\cal H}_2{\cal H}_3)^{-1}f+\Delta^{-1/2}\sum_{i=1}^3{\cal H}_i\mu_i^2({A'}_t^i)^2\label{Gtt}
\end{equation}
Note that close to the horizon to the leading order in $z$ we have:
\begin{equation}
{A'}_t^i=\frac{u_{\rm{H}} ^2}{R}\left(\frac{q_i}{u^2+q_i^2}-\frac{q_i^2}{u_{\rm{E}} ^2+q_i^2}\right)=\frac{2au_{\rm{H}} u_{\rm{E}} }{R(u_{\rm{E}} ^2+q_i^2)}z^2+\dots
\end{equation}
and hence the second term in equation (\ref{Gtt}) is of order $z^4$. Thus at leading order in $z$, $y_1$ and $y_2$ we have:
\begin{equation}
G_{tt}dt^2=-\frac{(u_{\rm{E}} ^2-u_1^2)^2(u_{\rm{E}} ^2-u_2^2)^2}{u_{\rm{H}} ^4u_{\rm{E}} ^2R^4}z^2dt^2=-(2\pi T)^2dt^2\ ,
\end{equation}
where we have used equation (\ref{a}) for the value of $a$. Therefore our final expression for the temperature of the background is:
\begin{equation}
T=\frac{u_{\rm{E}} }{2\pi R^2 u_{\rm{H}} ^2}\left(2u_{\rm{E}} ^2+q_1^2+q_2^2+q_3^2-\frac{q_1^2q_2^2q_3^2}{u_{\rm{E}} ^4}\right)=\frac{1}{2\pi R^2 u_{\rm{H}} ^2 u_{\rm{E}}^{\phantom{2}}}(u_{\rm{E}}^2-u_1^2)(u_{\rm{E}}^2-u_2^2)\ ,
\label{temperatureR}
\end{equation}
which is the same as the expression obtained in ref.~\cite{Russo:1998by} by calculating the surface gravity at the event horizon.

\subsection*{6.2.3 \hspace{2pt} The extremal case and the case of three equal charges $q_i=q$}
\addcontentsline{toc}{subsection}{6.2.3 \hspace{0.15cm} The extremal case and the case of three equal charges $q_i=q$}
It is worth exploring the case when we have extremal horizon of the background (\ref{spinning}) in some more details. In particular we will consider the simplest case, when have three equal charges. The corresponding expression for the temperature is:
\begin{equation}
T=\frac{u_{\rm{E}} }{2\pi R^2u_{\rm{H}} ^2}\left(2u_{\rm{E}} ^2+3q^2-\frac{q^6}{u_{\rm{E}} ^4}\right)\ .
\label{T3q}
\end{equation}
In order to explore the extremal case let us study the function $f(u)$ determining the position of the event horizon of the geometry. It is convenient to introduce the following dimensionless parameters:
\begin{equation}
\tilde u=u/u_{\rm{H}} ;~~~\tilde q=q/u_{\rm{H}}  \ .
\end{equation}
The expression for $f(u)$ from equation (\ref{f(u)}) is then:
\begin{equation}
f(u)=\frac{u_{\rm{H}} ^6}{R^2u^4}\left(\tilde u^6+3\tilde q^2\tilde u^4+(3\tilde q^4-1)\tilde u^2+\tilde q^6\right)\ .
\label{f3q}
\end{equation}
Therefore we need to study the behavior of the zeros of the polynomial in equation (\ref{f3q}), as we change the parameter $\tilde q^2$. In particular we want to examine for which values of $\tilde q^2$ the polynomial has a double zero. Therefore we need $f(u_{\rm{E}} )=0$ and $f'(u_{\rm{E}} )=0$. The second condition amounts to:
\begin{equation}
3\tilde u_{\rm{E}} ^4+6\tilde u_{\rm{E}} ^2\tilde q^2+3\tilde q^4-1=0 \ , 
\end{equation}
which has the solution:
\begin{equation}
\tilde u_{\rm{E}} ^2=\frac{\sqrt{3}}{3}-\tilde q^2\ .
\label{uq1}
\end{equation}
Substituting $\tilde u_{\rm{E}} $ from equation (\ref{uq1}) into equation (\ref{f3q}) results to the following simple expression:
\begin{equation}
\tilde q^2-\frac{2}{3\sqrt{3}}=0
\end{equation}
and therefore we learn that for charges $q$ satisfying:
\begin{equation}
q^2=\frac{2}{3\sqrt{3}}u_{\rm{H}} ^2\ ,
\label{qext}
\end{equation}
the background has an extremal horizon at:
\begin{equation}
u_{\rm{E}} ^2=\frac{1}{3\sqrt{3}}u_{\rm{H}} ^2\ .
\label{ueext}
\end{equation}
It is easy to verify that for these values of $q$ and $u_{\rm{E}} $ the temperature of the background given by equation (\ref{T3q}) is zero (as it should be). 

An interesting question is, what is the physical meaning of the parameter $u_{\rm{H}} $ in the dual gauge theory. Recall that in Chapters~3,4 and 5 the parameter $u_{\rm{H}} $ was proportional to the temperature of the dual theory. However, here we have an extremal horizon and therefore the dual gauge theory is at zero temperature. The answer to that question becomes obvious if we calculate the R--charge chemical potential of the theory given in equation (\ref{chempot6}). Indeed, for the extremal values of $q$ and $u_{\rm{E}} $ from equations (\ref{qext}) and (\ref{ueext}) we have:
\begin{equation}
\mu_{\rm{R}}^i=\sqrt{6\sqrt{3}}\frac{u_{\rm{H}} }{R^2}\ .
\label{muext}
\end{equation}
and hence we can think of the parameter $u_{\rm{H}} $ as determining the R--charge chemical potential of the dual gauge theory. This provides a convenient framework to study of the confinement/deconfinement phase transition of the fundamental matter at zero temperature. In the next section we   
will focus on the introduction of fundamental matter to the spinning D3--branes background. 

\section*{6.3 \hspace{2pt} Introducing fundamental matter}
\addcontentsline{toc}{section}{6.3 \hspace{0.15cm} Introducing fundamental matter}

In this section we focus on the introduction of fundamental matter to the dual gauge theory. The technique that we employ is the same used in the previous chapters, namely introducing a stack of $N_f$ D7--branes in the probe limit ($N_f\ll N_c$). Before we do that, we should notice that, unlike all other cases that we have considered so far, the general spinning D3--brane background has a deformed compact part of the geometry. In particular the internal $S^5$ sphere in (\ref{spinning}) is deformed by the rotations along $\phi_1$, $\phi_2$ and $\phi_3$. Therefore, we cannot directly apply the set up that we have used so far, namely the D7--brane extendeds along the non--compact AdS-like part of the geometry and wraps an internal $S^3$ inside the $S^5$ part of the geometry. However, there is a way out of it if we consider the slightly less general case of having $q_2=q_3$. Indeed, in order to restore some of the symmetry of the metric
(\ref{spinning}), we will consider the case, when $q_2=q_3$. This
corresponds to having an $S^3$ (parameterized by $\psi,\phi_2,\phi_3$)
inside the deformed~$S^5$. 

Now if we introduce D7--branes filling the
AdS--like part of the geometry and wrapping the~$S^3$, we will add
fundamental matter to the gauge theory. Furthermore, we are free to
rotate the D7--branes along~$\phi_1$ and the corresponding angular
velocity is interpreted as a time dependent phase of the bare quark
mass. (Recall that in introducing D7--branes to the D3--brane system
we actually add flavors as chiral superfields into the ${\cal N}=2$
gauge theory).  If that phase is the same as the phase of the complex
adjoint scalar, $\mu_{\rm{R}}^1 t$, it is equivalent to an R--charge
chemical potential for both the adjoint scalar and the chiral field.

On the gravity side of the description this is equivalent to letting
the D7--branes have the same angular velocity $\mu_{\rm{R}}^1$ as the
rotating frame of the background. Moving to the frame corotating
with the D7--brane corresponds to going back to the gauge choice for
$A_t^1$ from equation~(\ref{fields}). The price that we pay is that we again have an
ergosphere in the bulk of the background. It will be also convenient to gauge shift ${A'}_t^2$ and ${A'}_t^3$ to ${A''}_t^2$ and ${A''}_t^3$, correspondingly, so that the ergosphere is at:
\begin{equation}
\Delta({\cal H}_1{\cal H}_2^2)^{-1}f-{\cal H}_1\sin^2\theta(A_t^1)^2=0
\label{ergosphere-new}. 
\end{equation}
The shifted forms, ${A''}_t^2$ and ${A''}_t^3$, vanish at the locus given by equation (\ref{ergosphere-new}). In analogy to the T--dual description of the previous chapter for the external electric field
case, the ergosphere embeddings will have to be extended along
$\phi_1$ so that they can stay non--space--like beyond the
ergosphere. The possible D7--brane embeddings then naturally split into two
classes: Minkowski embeddings that have a shrinking $S^3$ above the
ergosphere and ergosphere embeddings which reach the ergosphere. These
classes are again separated by a critical embedding which has a
conical singularity at the ergosphere. The set up is very close to the one employed in the
 T--dual description of the electric case, considered in Chapter~5. In Chapter~7, we shall shed more light on this similarity. 
 
 Now let us focus on the properties of D7--branes embedding. There are two different cases that we consider:
\begin{enumerate}
\item A single charge in the $\phi_1$ direction: $q_1 = q$ and $q_2 = q_3 =0$.  This case is a useful test case since it is the simplest problem to explore and provides valuable insight into the next, more complicated case.
\item Three equal charges: $q_1 = q_2 = q_3 = q$.  This case is of greater interest since it is the simples case involving extremal horizon and thus enables us to study the gauge theory at zero temperature and finite R--charge chemical potential.
\end{enumerate}

\subsection*{6.3.1 \hspace{2pt} Single charge: The general set up}
\addcontentsline{toc}{subsection}{6.3.1 \hspace{0.15cm} Single charge: The general set up}
%
For the single charge case, we choose $q_1 = q$ and $q_2 = q_3 =0$ in equation~(\ref{spinning}), and the metric becomes:
\begin{eqnarray} \label{spinningq1}
ds^2 / \alpha'&=&\Delta^{1/2} \left\{ -\mathcal{H}^{-1} f dt^2 + \frac{u^2}{R^2} d\vec{x}\cdot
d\vec{x}\right\} +\Delta^{1/2} \left\{f^{-1} du^2 +R^2 d\theta^2\right\} \nonumber \\
&&+ \Delta^{-1/2}\left\{\mathcal{H} \sin^2\theta\left(R d\phi_1 - A_t^1 dt\right)^2 + R^2\cos^2\theta d\Omega_3^2\right\} \ ,
\end{eqnarray}
where
\begin{eqnarray}
f = \frac{u^2}{R^2}\mathcal{H}-\frac{u_{\rm{H}} ^4}{u^2 R^2} \ , \quad 
\mathcal{H} = 1+ \frac{q^2}{u^2} \ , \quad 
A_t^1 = \frac{u_{\rm{H}} ^2 q}{R u^2 \mathcal{H}} \ , \quad 
\Delta = \sin^2\theta + \mathcal{H} \cos^2\theta \ .
\end{eqnarray}
Although there is a change of variables that allows us to rewrite this metric in a more conventional form, it will be computationally convenient to work with the metric in equation~(\ref{spinningq1}). In order to introduce dynamical quarks into our gauge theory, we probe the geometry with D7--branes~\cite{Karch:2002sh}. We choose the coordinates $\left( \theta , \phi_1 \right)$ to be the transverse coordinates to the D7--brane.  However, we will define radial coordinates in the $\IR^{6}$--subspace transverse to the D--3 branes via:
\begin{equation}
\rho=u\cos\theta \ , \quad
L=u\sin\theta \ .
\label{crd1q1}
\end{equation}
Although these radial coordinates are not transverse to each other, the non--zero $G_{L\rho}$ component of the metric vanishes at the asymptotic boundary. Therefore, by expanding:
\begin{equation} \label{eqt: L expansion}
L=m+\frac{c}{u^2}+O\left(\frac{1}{u^4}\right) \ ,
\end{equation}
we can still make sense of the coefficient $m$ as being the asymptotic separation of the D--3 and D--7 branes. This will allow us to extract the bare quark mass and the condensate value. However, it should be noted that unlike the cases studied so far, we cannot directly identify the coefficient $c$ with the quark condensate.  The correct identification can be made after an appropriate change of variables that restores the $S^5$ symmetry of the geometry at the asymptotic boundary.  We will explicitly show this later.
We choose an ansatz $\theta=\theta(u)$ and $\partial_\mu \phi_1=0$.
With these choices, the lagrangian of the D7--brane is given by:
\begin{equation} 
{\cal L}\propto u^3\cos^3\theta  \sqrt {\frac{u^2(u^2+q^2\cos^2\theta)-u_{\rm{H}} ^4}{u^2(u^2+q^2)-u_{\rm{H}} ^4}}\sqrt{1+ \frac{u^2(u^2+q^2)-u_{\rm{H}} ^4}{u^2}\theta '^2}\sin\psi\cos\psi \ .
\label{eqt: single charge worldvolume}
\end{equation}
Note that the ergosphere of the background (\ref{spinningq1}) is given by equation (\ref{ergosphere-new}) which for the case in consideration simplifies to:
\begin{equation}
u^2(u^2+q^2\cos^2\theta)-u_{\rm{H}} ^4=0\ .
\label{ergoq1}
\end{equation}
The event horizon given by the zeros of $f(u)$  from equation (\ref{f(u)}) is determined by the equation:
\begin{equation}
u_{\rm{E}} ^2(u_{\rm{E}} ^2+q^2)-u_{\rm{H}} ^4=0\ .
\label{EHq1}
\end{equation}
Now it is straightforward to see that, just like in T--dual description of the external electric field in Section~5.4,  the D7--brane probes, which reach the ergosphere, develop speed (along the $\phi_1$ direction) equal to the local speed of light and hence become superluminal beyond the ergosphere, which results in an imaginary on-shell action and hence free energy. Naively it seems that the D7--branes embeddings should become non--static beyond the ergosphere and develop angular velocity along the $\phi_1$ direction. However, there is another possibility stemming from the fact that the D7--branes are extended objects. Indeed, for extended objects the only condition for stability that we need to satisfy is that they should not be sace-like, hence if they extend along the $\phi_1$ direction, they can still describe static configurations. 

Let us make this discussion explicit and consider the ansatz $\phi_1=\phi_1(u)$ for the D7--brane action. The corresponding lagrangian is given by:
\begin{eqnarray}
{\cal L}\propto u^3\cos^3\theta\sqrt{(1-\frac{q^2}{R^2f}\sin^2\theta)(1+ R^2f\theta '^2)+R^2f\sin^2\theta\phi_1 '^2}\sin\psi\cos\psi\ .
\label{mink1}
\end{eqnarray}
The equation of motion for $\phi_1$ can be integrated to give us the constraint:
\begin{equation}
\frac{R^2u^3f\sin^2\theta\cos^3\theta\phi_1'}{\sqrt{(1-\frac{q^2}{R^2f}\sin^2\theta)(1+ R^2f\theta '^2)+R^2f\sin^2\theta\phi_1 '^2}}=K_1\ .
\label{eqphi1}
\end{equation}
We shall discuss the physical meaning of the constant of integration $K_1$ in the dual gauge theory in Subsection~6.3.7. Let us use equation (\ref{eqphi1}) to solve for $\phi_1'$ and obtain the on-shell lagrangian:
\begin{equation}
 {\cal L}\propto \sqrt{\frac{R^2f-q^2\sin^2\theta}{R^2fu^6\cos^6\theta\sin^2\theta-K_1^2}}\sqrt{1+R^2f\theta'^2}u^6\cos^6\theta\sin\theta\sin\psi\cos\psi\ .
\label{on-shellq1}
\end{equation}
If we require that the lagrangian (\ref{on-shellq1}) be regular at the ergosphere, we can fix the constant of integration:
\begin{equation}
K_1^2=q^2u_0^6\cos^6\theta_0\sin^4\theta_0\ .
\label{K1}
\end{equation}
The parameters $u_0$ and $\theta_0$ in equation (\ref{K1}) correspond to the coordinates in the $(u,\theta)$ plane for which the D7--brane enters the ergosphere. Note that $u_0$ and $\theta_0$ are related via equation (\ref{ergoq1}). One can see that for the critical embedding at $\theta_0=\pi/2$, as well as for Minkowski type of embeddings, we have $K_1=0$ and the lagrangian (\ref{on-shellq1}) reduces to the one given in equation (\ref{mink1}). 

Now let us address the question about the quark condensate of the theory and how is it encoded in the profile of the D7--brane embedding in the $(\rho,L)$ plane. As we commented above we cannot directly relate the coefficient $c$ in the expansion (\ref{eqt: L expansion}) to the condensate of the theory. The reason is that because of the deformed $S^5$ it is not so simple to calculate the variation of the action with respect to $L$. Furthermore, one can show that for the coordinates $\rho$ and $L$ defined in equation (\ref{crd1q1}) the on-shell has a logarithmic divergences which require counter term depending on the field $L$. As we saw in Chapter~4, the main advantage of using $L$ as a transverse field is that the counter terms are independent on $L$ and we can regularize the action be simply subtracting the action for the $L=0$ embedding. This is why it is convenient to perform change of coordinates that will remove the logarithmic divergences of the action.  To this end we will restore the symmetry of the $S^{5}$ as much as possible as we move away from the horizon.

By looking at the distorted $S^{5}$ part of the background in equation~(\ref{spinning}), one can see that, away from the horizon (i.e. when we see the effect of $u_{\rm{H}} $ less and less), it has the conformal form: 
\begin{equation}
{ds_{S_5}}^2=\sum_{i=1}^3{{\mathcal{H}_i} \left(d\mu_i^2+d\phi_i^2\mu_i^2\right)} \ ,
\end{equation}
where
\begin{eqnarray}
\mu_1 &=& \sin\theta \ , \quad 
\mu_2 = \cos\theta \sin \psi \ , \quad 
\mu_3 = \cos\theta \cos \psi \ .
\end{eqnarray}
Following ref.~\cite{Russo:1998mm}, the change of variables that would restore the symmetry is given by:
\begin{eqnarray}
\hspace{-1cm}&&y_{1}=\sqrt{r^2+q_{1}^2} \ \mu_1\cos\phi_1,~
y_{2}=\sqrt{r^2+q_{1}^2} \ \mu_1\sin\phi_1,~~
y_{3}=\sqrt{r^2+q_{2}^2} \ \mu_2\cos\phi_2,\quad \nonumber \\
\hspace{-1cm}&&y_{4}=\sqrt{r^2+q_{2}^2} \ \mu_2\sin\phi_2 ,~
y_{5}=\sqrt{r^2+q_{3}^2} \ \mu_3\cos\phi_3 ,~~
y_{6}=\sqrt{r^2+q_{3}^2} \ \mu_3\sin\phi_3 \ .
\end{eqnarray}
In view of this, for the single charge case, one defines the following radial coordinate:
\begin{equation}
r=\frac{u^2+\sqrt{u^2(u^2+q^2)-u_{H}^4}}{\sqrt{q^2+2u^2+2\sqrt{u^2(u^2+q^2)-u_{H}^4}}} \ .
\end{equation}
This allows us to rewrite:
\begin{equation}
f^{-1}du^2+R^2d\theta^2=R^2\left(\frac{dr^2}{r^2+q^2}+d\theta^2\right) \ .
\end{equation}
We can now define new coordinates:
\begin{eqnarray}
\widetilde{L}=\sqrt{r^2+q^2}\sin\theta\ ,\quad
 \rho=r\cos\theta\label{newL} \ ,
 \end{eqnarray}
so that we have:
 \begin{equation}
R^2(\frac{dr^2}{r^2+q^2}+d\theta^2)=\frac{R^2}{\sqrt{4q^2\rho^2+(\tilde L^2+\rho^2-q^2)^2}}(d\widetilde{L}^2+d\rho^2) \ .
\end{equation}
Then from the asymptotic behavior of $r$ for large $u$:
\begin{equation}
\sqrt{r^2+q^2}=u+\frac{q^2}{2u}+\dots \ ,
\end{equation}
we can argue that $\widetilde{L}$ has the expansion:
\begin{equation}
\widetilde{L}=m+\frac{c+m\frac{q^2}{2}}{\rho^2}+\dots \ .
\end{equation}
Using the standard argument from ref.~\cite{Kruczenski:2003uq} we
can show that:
\begin{equation}
\langle\overline{\psi}\psi\rangle\sim-c-m\frac{q^2}{2} \equiv c'\ .
\label{condq1}
\end{equation}
\subsection*{6.3.2 \hspace{2pt} Single charge: Properties of the solution}
\addcontentsline{toc}{subsection}{6.3.2 \hspace{0.15cm} Single charge: Properties of the solution}
Let us now study the properties of the D7--brane embeddings. As we already mentioned, the solutions fall into two different classes, namely Minkowski embeddings which have a shrinking $S^3$ and correspond to meson states and ergosphere embeddings which reach the ergosphere and eventually fall into the black hole. 

\begin{figure}[h] 
     \centering
     \includegraphics[width=11cm]{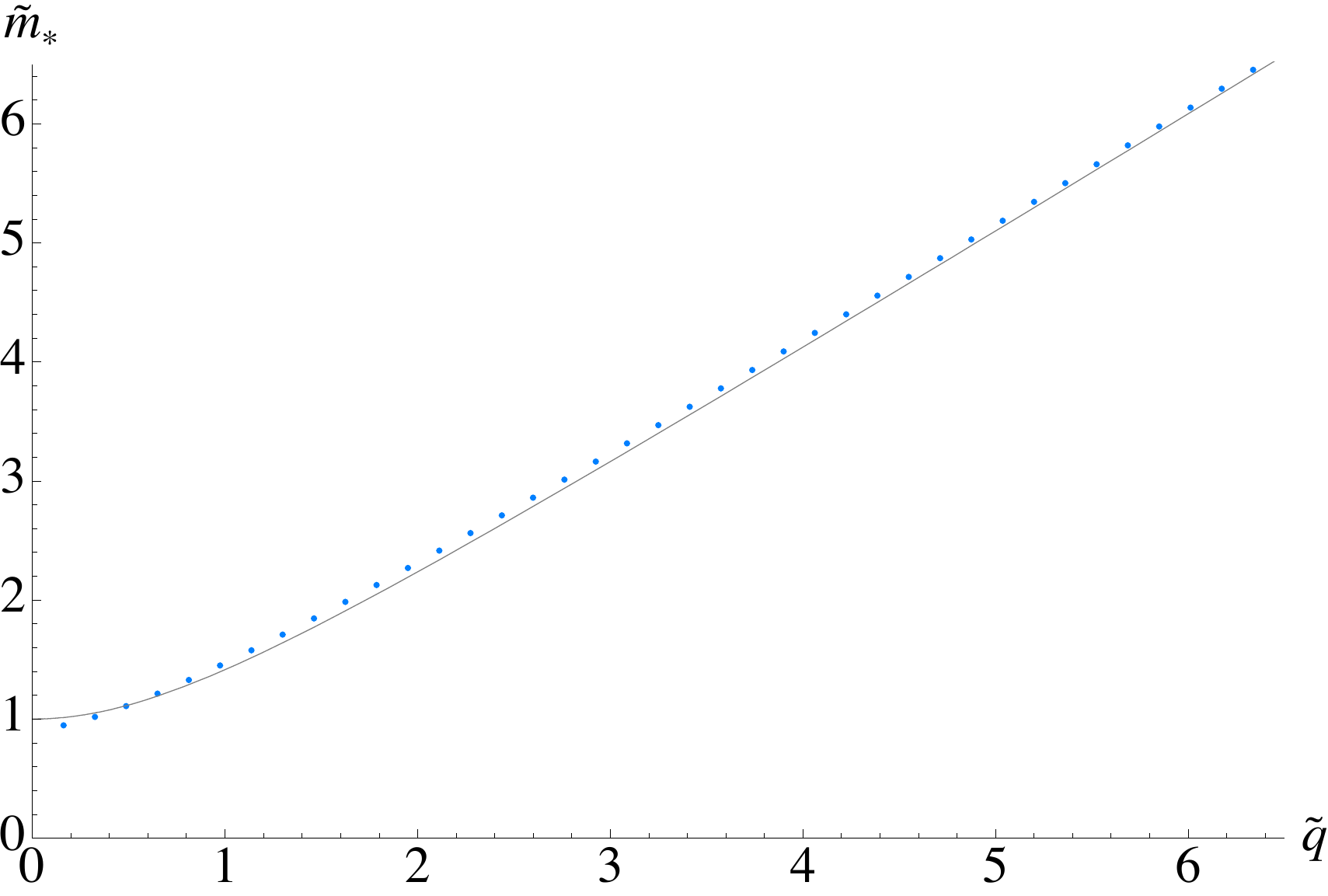}
     \caption{Plot of $\tilde m_*$ versus $\tilde q$. The solid line in the figure correspond to equation (\ref{mtldq}), one can see the good agreement even at small $\tilde q$. }
     \label{fig:massvsq1}
  \end{figure}

We will consider first the Minkowski type of embeddings. The relevant lagrangian is given in equation (\ref{eqt: single charge worldvolume}). The equation of motion for $\theta(u)$ derived from (\ref{eqt: single charge worldvolume}) is a second order non--linear differential equation, thus similar to the cases studied in the previous chapter, we will rely on numerical methods to solve it. However, It turns out that for values of the parameter $ q\gg u_{\rm{H}} $ one can derive a simple analytic solution which proves to be a very good approximation even at $\tilde q$ of the order of $u_{\rm{H}} $. Indeed, for Minkowski embeddings the D7--brane close above the ergosphere and the minimum value of the coordinate $u$ is $u_{min}\ge u_{\rm{H}} $. This suggests that for large $q$ one can ignore the $u_{\rm{H}} ^4$ terms in (\ref{eqt: single charge worldvolume}) and the lagrangian simplifies to:
\begin{equation}
{\cal L}\propto u^3\cos^3\theta\sqrt{u^2+q^2\cos^2\theta}\sqrt{\frac{1}{u^2+q^2}+\theta'^2}\ .
\end{equation}
If we change coordinates as in equation (\ref{newL}), the action becomes simply the action for a regular AdS$_5\times S^5$ space, namely:
\begin{equation}
{\cal L}\propto \rho^3\sqrt{1+\tilde L'^2} \ .
\end{equation}
The solution regular at $\rho=0$ is simply:
\begin{equation}
\tilde L=\sqrt{u^2+q^2}\sin\theta=\mathrm{const}\ .
\label{smplsol}
 \end{equation}
We can determine the constant in (\ref{smplsol}), by requiring that the D7--brane closes at $u=u_{min}$, namely $\theta(u_{min})=\pi/2$. Therefore we have that for large $q\gg u_{\rm{H}} $ the bare quark mass is related to the parameter $u_{min}$, via:
\begin{equation}
m=\sqrt{u_{min}^2+q^2}\ .
\label{mlq}
\end{equation}
Equation (\ref{mlq}) is particularly useful because it gives us an estimate of the critical mass $m_{cr}$ at which the meson melting phase transition takes place (for large $q$). More precisely, it gives us an estimate of the bare quark mass $m_*$ of the critical embedding separating the Minkowski and ergosphere classes of solutions (the one having a conical singularity at the ergosphere).  Indeed, one can show that for the critical embedding $u_{min}=u_{\rm{H}} $, then we obtain:
\begin{equation}
\tilde m_*=\sqrt{1+\tilde q^2}\ ,
\label{mtldq}
\end{equation}
where we have defined the dimensionless parameters $\tilde m=m/u_{\rm{H}} $ and $\tilde q=q/u_{\rm{H}} $. It is interesting to verify this result numerically. To this end we solve the equation of motion of the critical embedding numerically for wide range of the parameter $\tilde q$. From the asymptotic of the solution at large $\tilde u=u/u_{\rm{H}} $ we can extract the parameter $\tilde m_*$. The resulting plot is presented in Figure~\ref{fig:massvsq1}. The solid line in Figure~\ref{fig:massvsq1} corresponds to equation (\ref{mtldq}), one can see the good agreement even at small $\tilde q$. Note that the critical embedding is unstable and is by--passed by the first order phase transition of the fundamental matter. Nevertheless, the parameter $\tilde m_*$ is a good approximation of the actual critical mass $\tilde m_{cr}$ at which the phase transition happens.

\begin{figure}[h] 
     \centering
     \includegraphics[width=11cm]{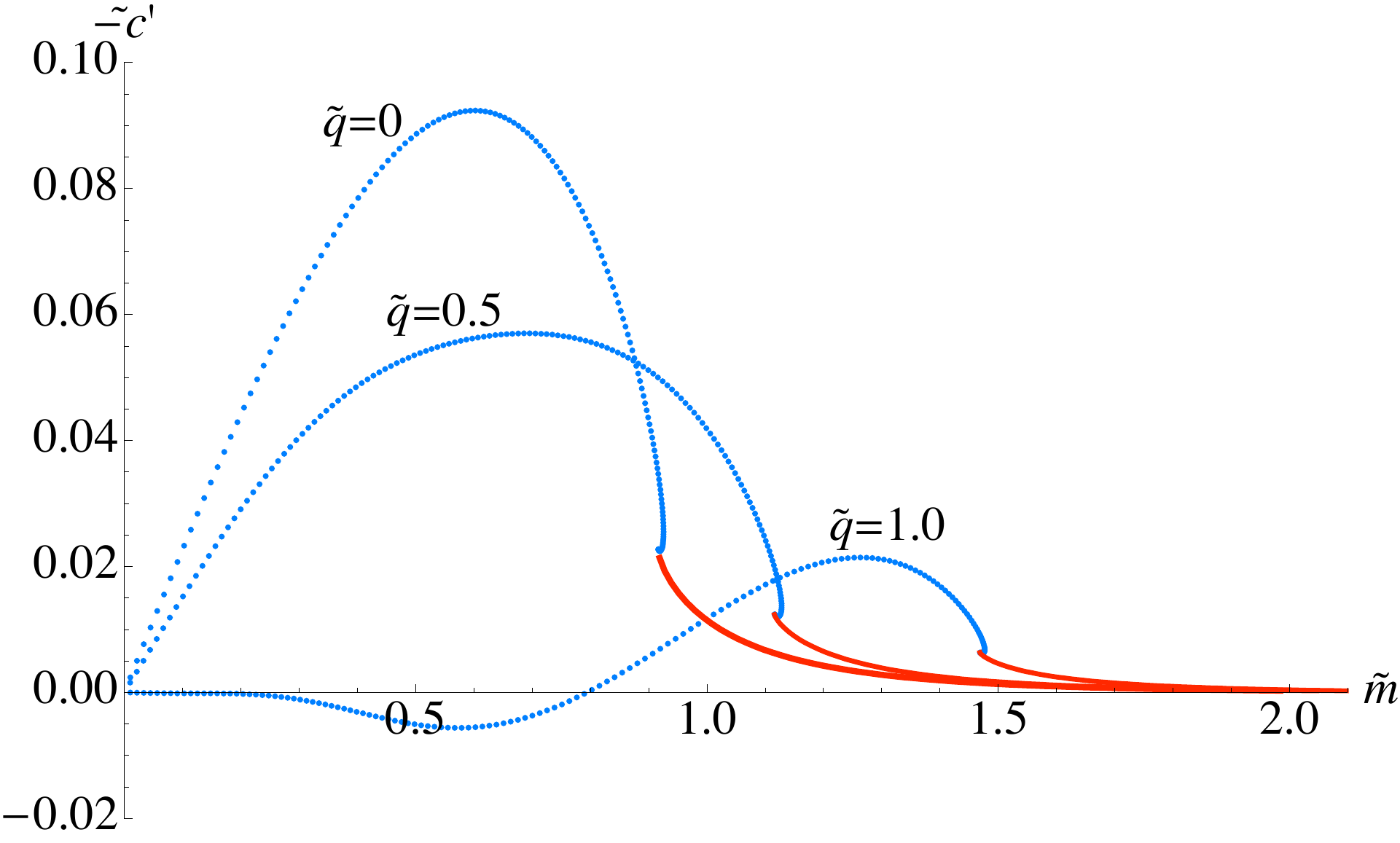}
     \caption{A plot of the equation of state for values of the parameter $\tilde q=0,0.5,1.0$. One can see that the critical mass is increasing with $\tilde q$. }
     \label{fig:q1-many}
  \end{figure}

Now let us study the effect of the R--charge chemical potential on the phase transition. To this end we derive the equations of motion for the Minkowski and ergosphere classes of embeddings, using equations (\ref{eqt: single charge worldvolume}) and (\ref{mink1}). Note that the lagrangian in equation (\ref{mink1}) gives us two equations of motion: for $\theta(u)$ and $\phi_1(u)$. The second one can be integrated and substituted into the equation of motion for $\theta(u)$ using equation~(\ref{eqphi1}). Next we define the dimensionless parameters:
\begin{equation}
\tilde u=u/u_{\rm{H}} ;~~~\tilde m=m/u_{\rm{H}} ;~~~\tilde c=c/u_{\rm{H}} ^3;~~~\tilde q=q/u_{\rm{H}}\ .
\end{equation}
 After that we solve numerically the equation of motion for $\theta(\tilde u)$ and using equations (\ref{crd1q1}), (\ref{eqt: L expansion}) and (\ref{condq1}), we calculate the  calculate the parameters $\tilde m$ and ${\tilde c}'$, corresponding to the bare quark mass and the quark condensate of the dual gauge theory.  
The resulting plot of the equation of state in $-\tilde c'$ versus $\tilde m$ coordinates, for different values of the parameter $\tilde q$, is presented in Figure~\ref{fig:q1-many}. 

\begin{figure}[h] 
     \centering
     \includegraphics[width=11cm]{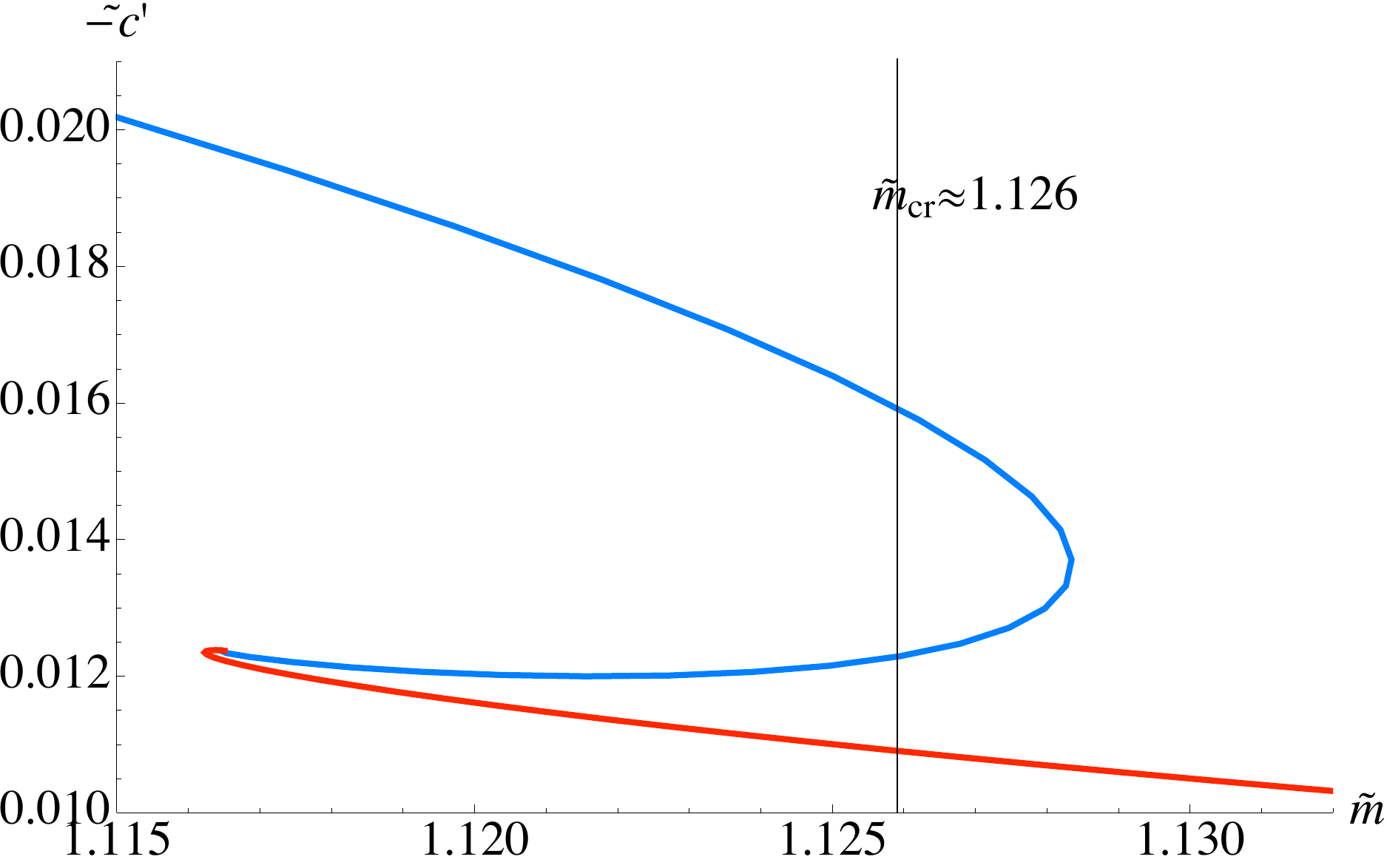}
     \caption{Magnified plot of the first order transition pattern of the equation of state for $\tilde q=0.5$. The solid vertical line corresponds to the bare quark mass of the critical embedding $\tilde m_{\rm{cr}}\approx 1.126 $. }
     \label{fig:q=.5-zoom}
  \end{figure}

One can see that the overall structure remains the same as in the $\tilde q=0$ case. However, the critical mass is increasing with $\tilde q$ as can be expected from the behavior of $\tilde m_*$ that we have studied. Furthermore, for sufficiently large $\tilde q$, the ergosphere type of embeddings develop a negative condensate $-\tilde c'$ which approaches zero from bellow in the limit $\tilde m\to 0$. The plot in Figure~\ref{fig:q=.5-zoom} presents a zoomed in version of the equation of state near the phase transition. One can see the typical first order phase transition pattern. The corresponding value of the critical mass, calculated using the equal area law, is represented by the vertical solid line. It is interesting to study the phase diagram of the theory and in particular how does the critical temperature of the theory depend on the R--charge chemical potential.

\subsection*{6.3.3 \hspace{2pt} Single charge: The phase diagram}
\addcontentsline{toc}{subsection}{6.3.3 \hspace{0.15cm} Single charge: The phase diagram}
Let us study the phase diagram of the theory. To this end let us obtain expressions for the temperature $T$ and the R--charge chemical potential $\tilde \mu_{\rm{R}}$ as functions of the parameters $u_{\rm{H}} $ and $\tilde m$. Using equations (\ref {chempot6}), (\ref{temperatureR}) and (\ref{EHq1}), we obtain the following expressions for the temperature:
\begin{equation}
T=\frac{u_{\rm{H}} }{\pi R^2}\sqrt{1+\frac{\tilde q^4}{4}}\sqrt{\sqrt{1+\frac{\tilde q^4}{4}}-\frac{\tilde q^2}{2}}
\end{equation}
and for the R--charge chemical potential:
\begin{equation}
\mu_{\rm{R}}=\frac{u_{\rm{H}} }{R^2}\tilde q\left(\sqrt{1+\frac{\tilde q^4}{4}}-\frac{\tilde q^2}{2}\right) \ .
\end{equation}
Note that in this set up the temperature $T$ and the R--charge chemical potential are not completely independent. Indeed, for the ratio $T/\mu_{\rm{R}}$ we get:
\begin{equation}  
\frac{T}{\mu_{\rm{R}}}=\frac{   \sqrt{1+\frac{\tilde q^4}{4}  }   }{\pi\tilde q\sqrt{\sqrt{1+\frac{\tilde q^4}{4}}-\frac{\tilde q^2}{2}}
}\gsim \frac{\sqrt{2}}{\pi}\ .
\end{equation}
Therefore, in the $(\mu_{\rm{R}},T)$ plane only states above the line $T=\frac{\sqrt{2}}{\pi}\mu_{\rm{R}}$ correspond to physical states. This suggests that it is more convenient to study the phase diagram in the $(\mu_{\rm{R}},T-\frac{\sqrt{2}}{\pi}\mu_{\rm{R}})$ plane. Furthermore, the change of variables from $u_H$ and $\tilde q$ to $\mu_{\rm{R}}$ and $T$ is invertible in two separate intervals of values of $\tilde q$ , namely $\tilde q^2 \leq 2/\sqrt{3}$ and $\tilde q^2 \geq 2/\sqrt{3}$. Of those, only the first interval contains the point $\tilde q=0$ which corresponds to the pure meson melting phase transition considered in Chapter~3. Therefore, we shall consider only values of $\tilde q$ satisfying $\tilde q^2 \leq 2/\sqrt{3}$. Next note that:
\begin{equation}
 R^4=4\pi g_S\alpha'^2 N=\lambda \alpha'^2;~~~ m_q=m/(2\pi\alpha')=\tilde m u_{\rm{H}} /(2\pi\alpha') \ ,
\end{equation}
where $\lambda$ is the t'Hooft coupling and $m_q$ is the bare quark mass of the dual gauge theory. This suggests introducing the following dimensionless physical parameters:
\begin{eqnarray}
\sqrt{\lambda}\frac{\mu_{\rm{R}}}{m_q}&=&\frac{2\pi\tilde q}{\tilde m}\left(\sqrt{1+\frac{\tilde q^4}{4}}-\frac{\tilde q^2}{2}\right)\\
\sqrt{\lambda}\frac{T}{m_q}&=&\frac{2}{\tilde m}\sqrt{1+\frac{\tilde q^4}{4}}\sqrt{\sqrt{1+\frac{\tilde q^4}{4}}-\frac{\tilde q^2}{2}}\label{phspar1}\nonumber
\end{eqnarray}

\begin{figure}[h] 
     \centering
     \includegraphics[width=11cm]{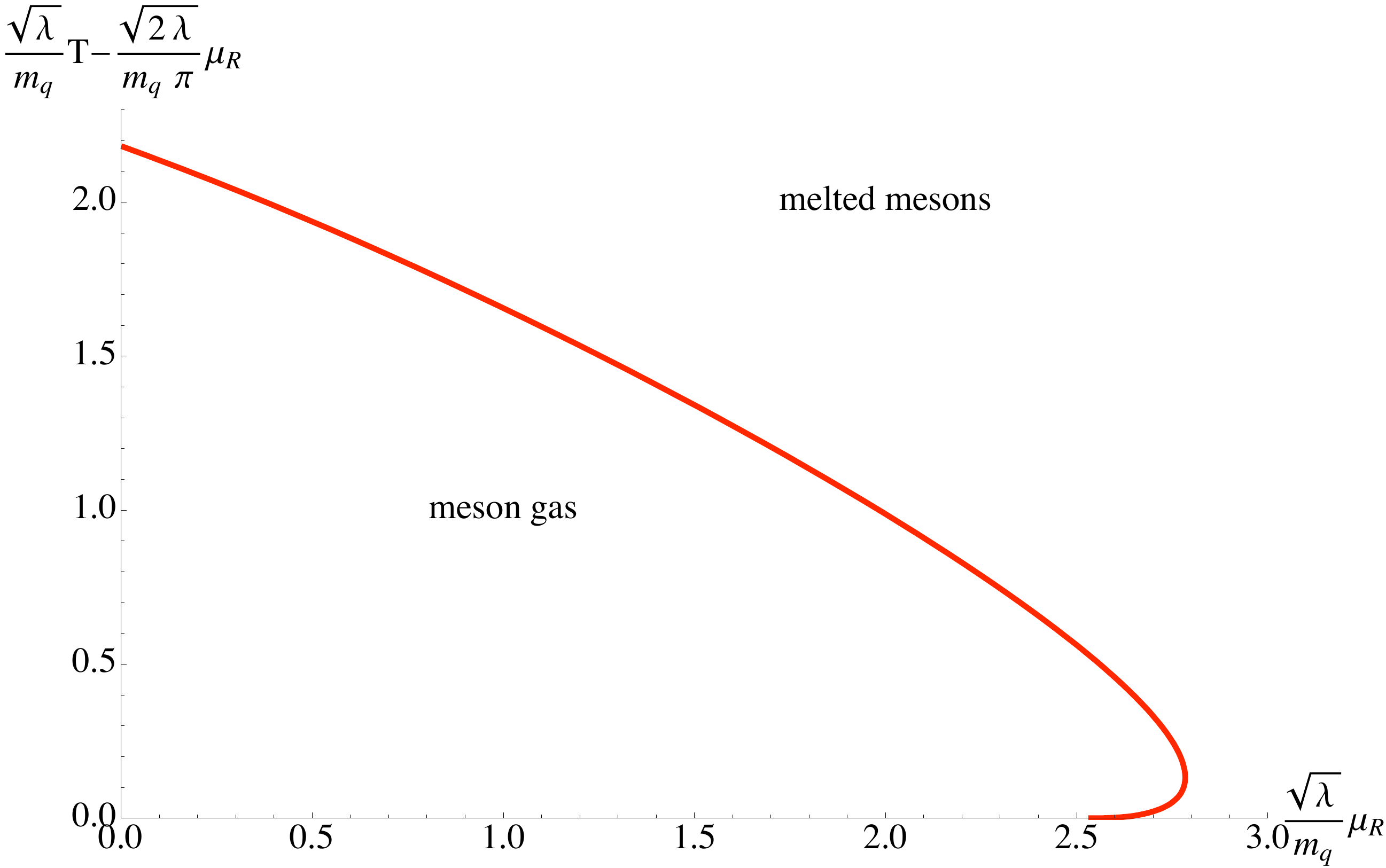}
     \caption{The area enclosed by the curve corresponds to a confined phase of the fundamental matter. The area outside corresponds to melted mesons. }
     \label{fig:phasediagq1}
  \end{figure}

The corresponding phase diagram is presented in Figure~\ref{fig:phasediagq1}. Note that The curve separates meson states from melted meson states. Note that we have used $\tilde m_*$ as an approximation for the critical mass. We have also assumed that the D7--brane embeddings which reach the ergosphere fall into the black hole. In this way we ignore effects as the one described at the end of Chapter~5, where we showed that ergosphere type of embeddings that are sufficiently close to the critical embedding have a conical singularity above the event horizon. We assume that stringy corrections which become important near the point of singularity would resolve this singularity and enable the D7--brane to connect to the black hole, via a thin tube.  

\subsection*{6.3.4 \hspace{2pt} Three equal charges: The general set up}
\addcontentsline{toc}{subsection}{6.3.4 \hspace{0.15cm} Three equal charges: The general set up}
When we set all three charges to be equal \cite{Chamblin:1999tk}, ({\it i.e.,} $q_1=q_2=q_3=q$ in equation~(\ref{spinning}), the metric is given by:
\begin{eqnarray} \label{eqt: three charge metric}
ds^2/\alpha' &=&
-\mathcal{H}^{-2} f dt^2 + \mathcal{H}f^{-1} du^2 + \mathcal{H}\frac{u^2}{R^2}d\vec{x}\cdot d\vec{x} \nonumber\\
&&+\sum_{i=1}^3\left\{R^2d\mu_i^2 +\mu_i^2\left(R d\phi_i -A_t^i dt\right)^2\right\} \ ,
\end{eqnarray}
where
\begin{eqnarray}
\mathcal{H} &=& 1+ \frac{q^2}{u^2} \ , \quad 
f =  \frac{\mathcal{H}^3 u^2}{R^2} - \frac{u_{\rm{H}} ^4}{u^2 R^2} \ , \quad 
A_t^i = \frac{u_{\rm{H}} ^2 q}{ R(u^2+  q^2)} \ . 
\label{eqaf}
\end{eqnarray}
From a computational point of view, it is again more convenient to use
\begin{equation}
L(u)=u\sin{\theta}=m+\frac{c}{u^2}+O\left(\frac{1}{u^4}\right) \ .
\label{eqn1}
\end{equation}
However, just like in the single charge case of last section, the Minkowski  solutions exhibit a positive slope of $q^2/2$ for the condensate which is again an artifact of the coordinates. To show that, we introduce a new radial coordinate $r(u)$ satisfying:
\begin{equation}
\mathcal{H} f^{-1}du^2+R^2=\frac{R^2}{r^2}(dr^2+r^2d\theta^2) \label{eqn2} \ ,
\end{equation}
and the corresponding radial coordinates are:
\begin{eqnarray}
\widetilde{L}=r\sin{\theta}\ ,\quad
\widetilde{\rho}=r\cos{\theta} \ .
\end{eqnarray}
Using that $r(u)$ from equation~(\ref{eqn2}) has the asymptotic behavior:
\begin{equation}
r(u)=u+\frac{q^2}{2u}+O\left(\frac{1}{u^3}\right) \ ,
\end{equation}
one can show that:
\begin{equation}
\widetilde{L}=m+\left(c+m\frac{q^2}{2}\right)\frac{1}{\rho^2}+O\left(\frac{1}{\rho^4}\right) \ .
\end{equation}
Note also that one can write:
\begin{equation}
 \mathcal{H} f^{-1}du^2+R^2=\frac{R^2}{L^2+\rho^2}(d\rho^2+dL^2) \ ,
\end{equation}
and following the standard argument presented in ref.~\cite{Kruczenski:2003uq} one can show that:
\begin{equation}
\langle\overline{\psi}\psi\rangle\sim-c-m\frac{q^2}{2} \ .
\end{equation}
Therefore, to explore the functional dependence of the quark condensate on the bare quark mass, we again choose $\theta$ and $\phi_1$ to be our transverse coordinates to the D7--brane. And again we work with the gauge choice ${A'}_t^i$ from equation (\ref{shifted}) for the $\phi_2$ and $\phi_3$ directions. With this choice, the world--volume of the D7--brane is given by:
\begin{equation}
{\cal L}\propto u^3\cos^3\theta H\sqrt{1-\frac{u_{\rm{H}} ^4q^2}{u^4R^2f}\sin^2\theta}\sqrt{1+\frac{R^2f}{H}\theta'^2}\sin\psi\cos\psi
\label{lagr3q}
\end{equation}
Again, just like in the single charge case, one can see that the action vanishes at the ergosphere given by equation (\ref{ergosphere-new}). For the case in consideration the equation (\ref{ergosphere-new}) reduces to:
\begin{equation}
(u^2+q^2)^3-u_{\rm{H}} ^4(u^2+q^2\sin^2\theta)=0
\label{ergosphere-3q}
\end{equation}
For Minkowski types of embeddings we can still use the lagrangian in (\ref{lagr3q}), but for ergosphere embeddings we need to generalize the ansatz and let the D7--branes extend along the $\phi_1$ direction,  thus enabling them to remain non--space like beyond the ergosphere. The resulting lagrangian is given by:
\begin{equation}
{\cal L}\propto u^3\cos^3\theta H\sqrt{(1-\frac{u_{\rm{H}} ^4q^2}{u^4R^2f}\sin^2\theta)(1+\frac{R^2f}{H}\theta'^2)+\frac{R^2f}{H}\sin^2\theta\phi_1'^2}\sin\psi\cos\psi
\label{lagr3qerg}
\end{equation}
Now, just like we did for the single charge case, we can integrate the equation of motion for $\phi_1$ derived from (\ref{lagr3qerg}). This amounts to the equation:
\begin{equation}
\frac{u^3\cos^3\theta{R^2f}\sin^2\theta\phi_1'}{\sqrt{(1-\frac{u_{\rm{H}} ^4q^2}{u^4R^2f}\sin^2\theta)(1+\frac{R^2f}{H}\theta'^2)+\frac{R^2f}{H}\sin^2\theta\phi_1'^2}}=K_1
\end{equation}
For the on-shell action we get:
\begin{equation}
{\cal L}\propto u^5\cos^6\theta\sin\theta H\sqrt{\frac{H(u^4H^3-u_{\rm{H}} ^4(1+\frac{q^2}{u^2}\sin^2\theta))\left(1+\frac{R^2f}{H}\theta'^2\right)}{((u^2+q^2)^3-u_{\rm{H}} ^4u^2)(u^2+q^2)\sin^2\theta\cos^6\theta-K_1^2}}\frac{\sin2\psi}{2}\ .
\end{equation}
Now it is straightforward to determine the constant of integration $K_1$:
\begin{equation}
K_1^2={u_{\rm{H}} ^4q^2(u_0^2+q^2)}\sin^4\theta_0\cos^6\theta_0\ ,
\label{K13q}
\end{equation}
where $u_0$ and $\theta_0$ determine the point at which the D7--brane enters the ergosphere. Note that $\theta_0$ and $u_0$ satisfy equation (\ref{ergosphere-3q}). 
\subsection*{6.3.5 \hspace{2pt} Three equal charges: Properties of the solution}
\begin{figure}[h] 
     \centering
     \includegraphics[width=11cm]{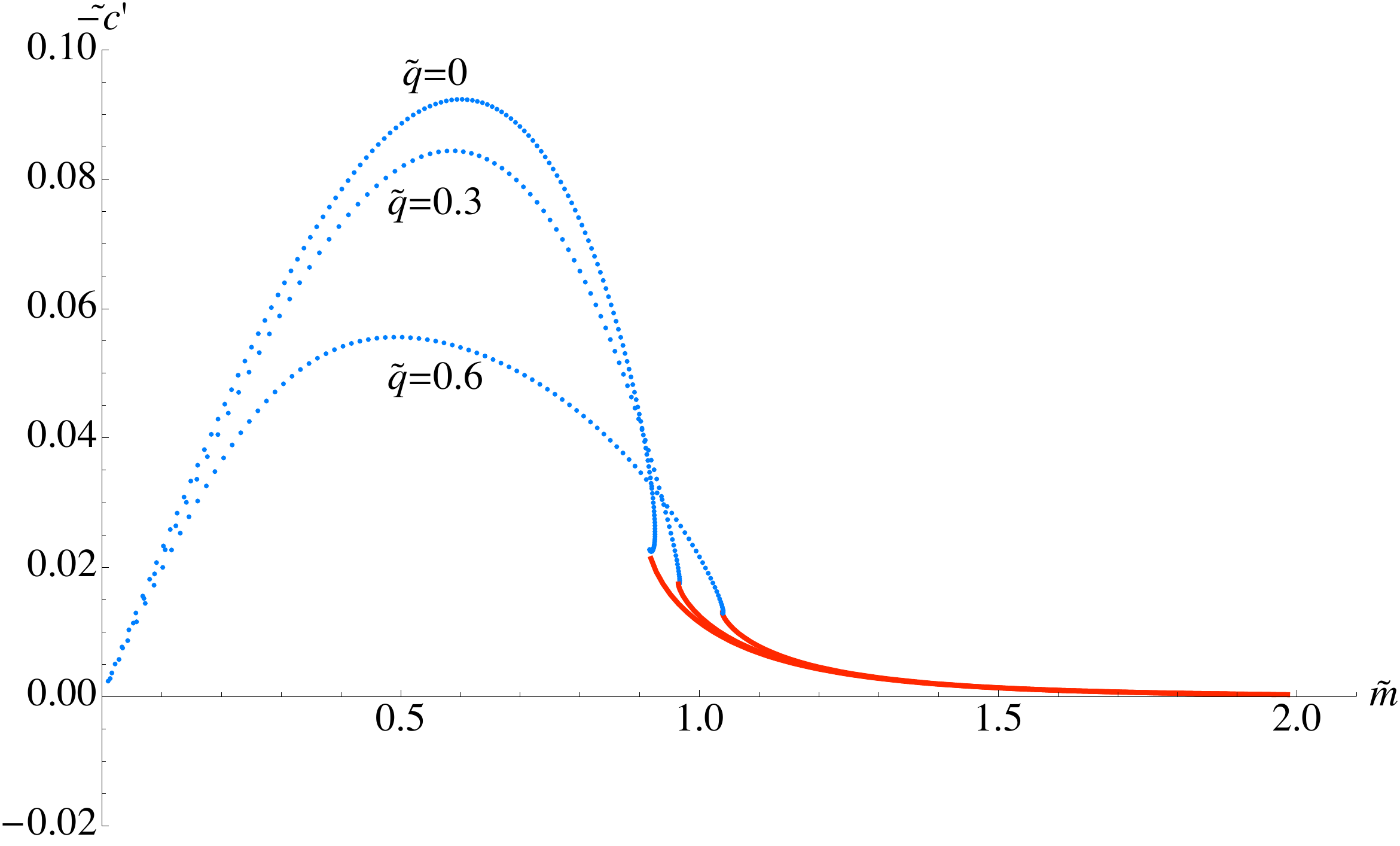}
     \caption{A plot of the equation of state for values of the parameter $\tilde q=0,0.3,0.6$. One can see that the critical mass is increasing with $\tilde q$. }
     \label{fig:3q-many}
  \end{figure}

\addcontentsline{toc}{subsection}{6.3.5 \hspace{0.15cm} Three equal charges: Properties of the solution}
The analysis of the three charge case is analogous to the on for the single charge case. A crucial difference is that there is an upper bound for the parameter $\tilde q^2\leq {2\sqrt{3}}/{3}$ at which the temperature drops to zero. Following the same procedure as in the single charge case we can derive the equation of motion for $\theta(\tilde u)$ and use it to calculate the equation of state in the $(\tilde m, -\tilde c')$-plane. The resulting plot for various values of the parameter $\tilde q$ is presented in Figure~\ref{fig:3q-many}. One can see that the critical mass is again increasing as a we increase $\tilde q$. However, since the for the three charge case we have extremal horizon at $\tilde q^2=2/(3\sqrt{3})$ and we consider only the case $\tilde q^2\leq 2/(3\sqrt{3})$ (no naked singularity), the critical mass has an upper limit (unlike the one charge case).   

\vspace{3cm}
\subsection*{6.3.6 \hspace{2pt} Three equal charges: The phase diagram}
\addcontentsline{toc}{subsection}{6.3.6 \hspace{0.15cm} Three equal charges: The phase diagram}

\begin{figure}[h] 
     \centering
     \includegraphics[width=11cm]{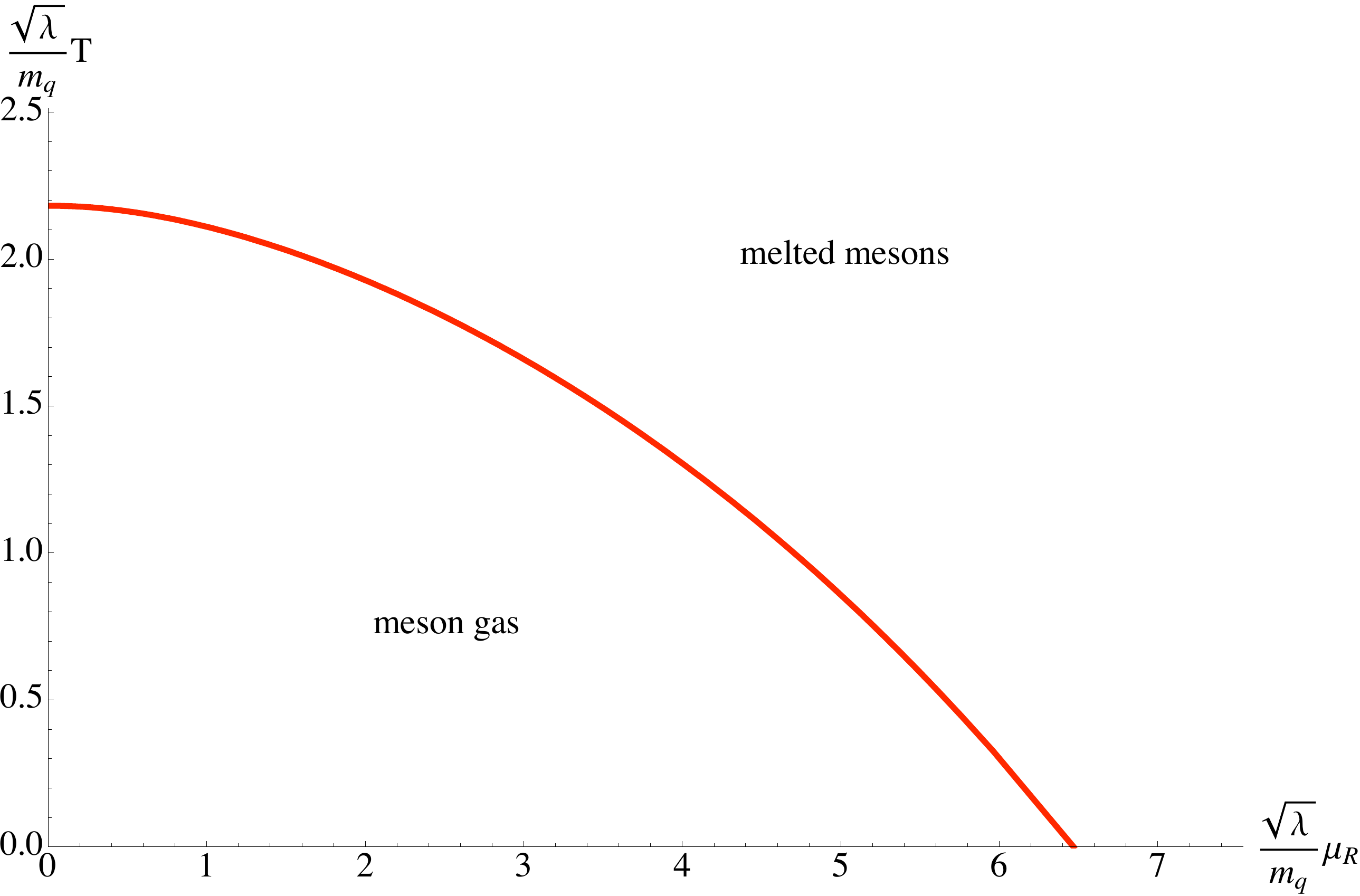}
     \caption{The area enclosed by the curve corresponds to a confined phase of the fundamental matter. The area outside corresponds to melted mesons. There are no bound meson states above the critical value of the parameter $\sqrt{\lambda}\frac{\mu_{\rm{R}}}{m_q}\approx 6.47$. }
     \label{fig: phase-diagram-3q}
  \end{figure}

Now let us focus on the study of the phase diagram. As we saw earlier in the case of three equal charges the temperature of the background is given by equation (\ref{T3q}) which we replicate below:
\begin{equation}
T=\frac{u_{\rm{E}} }{2\pi R^2u_{\rm{H}} ^2}\left(2u_{\rm{E}} ^2+3q^2-\frac{q^6}{u_{\rm{E}} ^4}\right)\ .
\label{T3q-again}
\end{equation}
The radius of the event horizon is the biggest root of $f(u)$ given in equation (\ref{eqaf}). In analogy to the one-charge case it is convenient to express that temperature as a function of $\tilde q$ and $u_{\rm{H}} $:
\begin{equation}
T=\frac{u_{\rm{H}} }{2\pi R^2}\tilde u_{\rm{E}} (\tilde q)\left(2\tilde u_{\rm{E}} (\tilde q)^2+3\tilde q^2-\frac{\tilde q^6}{\tilde u_{\rm{E}} (\tilde q)^4}\right)\ ,
\end{equation}
where
\begin{equation}
\tilde u_{\rm{E}} (\tilde q)^2=-\tilde q^2+\frac{{3}^{1/3}2+2^{1/3}(-9\tilde q^2+\sqrt{81\tilde q^4-12})^{2/3}}{6^{2/3}(-9\tilde q^2+\sqrt{81\tilde q^4-12})^{1/3}}\ .
\end{equation}
Similarly just like in the one charge case we can express the R--charge chemical potential as a function of $\tilde q$ and $u_{\rm{H}} $:
\begin{equation}
\mu_{\rm{R}}=\frac{u_{\rm{H}} }{R^2}\frac{6^{2/3}\tilde q(-9\tilde q^2+\sqrt{81\tilde q^4-12})^{1/3}}{{3}^{1/3}2+2^{1/3}(-9\tilde q^2+\sqrt{81\tilde q^4-12})^{2/3}}\ .\end{equation}
One can verify that unlike the one charge case the temperature and R--charge chemical potential are completely independent and as $u_{\rm{H}} $ and $\tilde q$ vary take all possible values in the $(T, \mu_{\rm{R}})$-pane. It is once again convenient to introduce the following dimensionless parameters: 
\begin{eqnarray}
\sqrt{\lambda}\frac{\mu_{\rm{R}}}{m_q}&=&\frac{2\pi}{\tilde m}\frac{6^{2/3}\tilde q(-9\tilde q^2+\sqrt{81\tilde q^4-12})^{1/3}}{{3}^{1/3}2+2^{1/3}(-9\tilde q^2+\sqrt{81\tilde q^4-12})^{2/3}}\\
\sqrt{\lambda}\frac{T}{m_q}&=&\frac{\tilde u_{\rm{E}} (\tilde q)}{\tilde m}\left(2\tilde u_{\rm{E}} (\tilde q)^2+3\tilde q^2-\frac{\tilde q^6}{\tilde u_{\rm{E}} (\tilde q)^4}\right)\ .\nonumber
\end{eqnarray}
Now using $\tilde m_*$ (the bare quark mass of the critical embedding) as an approximation for $\tilde m_{cr}$, we obtain the phase diagram presented in Figure~\ref{fig: phase-diagram-3q}. One can see that there is a critical value of the chemical potential above which even at zero temperature there are no bound meson states. Note that, just like in the one charge case, we assume that all of the embeddings which are type ergosphere eventually fall into the black hole and correspond to melted mesons. This is equivalent of assuming that the conical singularity of the embeddings close to the critical one is being repaired by a stringy corrections. In this way those embeddings would smoothly connect to the black hole.

\subsection*{6.3.7 \hspace{2pt} The physical meaning of $K_1$}
\addcontentsline{toc}{subsection}{6.3.7 \hspace{0.15cm} The physical meaning of $K_1$}

Let us focus on the physical interpretation of the constant of integration $K_1$ defined in equations (\ref{K1}) and (\ref{K13q}), for the single and three equal charges cases respectively. The main point is to interpret the physical meaning of the coordinate $\phi_1$ in the dual gauge theory. The correct interpretation is that $\phi_1$ is contributing to the phase of the bare quark mass. Indeed, in a T--dual description the mass separation between the D3-- and D7--branes corresponds to the expectation value of the $U(1)$ gauge field on the D7--brane \cite{Polchinski:1998rq}, and rotation in the transverse plane parameterized by $\phi_1$, rotates the components of the D7--brane's gauge field which correspond to the components of the bare quark mass. Let us make the discussion more explicit. The mass term of the theory can be written in general as:
\begin{equation}
{\cal H}=\frac{1}{2}(m_q{\cal O}_q+m_q^*{\cal O}_q^*)+{\cal H}_{0}\ ,
\label{mass-term}
 \end{equation}
Here ${\cal H}_{0}$ is the term of the hamiltonian that doesn't depend on the bare quark mass. Usually we can choose coordinates such that $m_q$ is real. In that case $\langle{\cal O}_q\rangle$ is the corresponding condensate of the theory. Here we assume that:
\begin{equation}
m_q=|m_q|e^{i\phi_m};~~~\langle {\cal O}\rangle_q=|\langle {\cal O}\rangle_q|e^{-i\phi_c}\ ,
\end{equation}
then the expectation value of the term (\ref{mass-term}) can be written as:
\begin{equation}
|m_q||\langle {\cal O}\rangle_q|\cos\Delta\phi;~~~\Delta\phi=\phi_m-\phi_c\ .
\label{conddphi}
\end{equation}
Now we identify the expectation value of the hamiltonian with the regularized action via:
\begin{equation}
\langle{\cal H}\rangle=-S_{\rm{reg}}[\tilde L,\phi_1]
\end{equation}

On gravity side, as we learned in Chapter~1, the value of the bare quark mass is controlled by the asymptotic separation of the D3-- and D7--branes $m=L_{\infty}$, via $|m_q|=m/(2\pi\alpha')$. Therefore we can relate the variations of $|m_q|$ and $\Delta\phi$ to the variations of $\tilde L_{\infty}$ and $\phi_1|_\infty$. Then using a standard variational argument one can show that:
\begin{equation}
|\langle {\cal O}\rangle_q|\cos\Delta\phi=-(2\pi\alpha')\frac{\partial{\cal L}_{\rm{reg}}}{\partial\tilde L'}|_{\infty};~~~|m_q||\langle {\cal O}\rangle_q|\sin\Delta\phi=\frac{\partial{\cal L}_{\rm{reg}}}{\partial \phi_1'}|_{\infty}
\end{equation}
Therefore we see that:
\begin{equation}
\tan\Delta\phi=-\left(\frac{\partial{\cal L}_{\rm{reg}}}{\partial \phi_1'}|_{\infty}\right)/\left(m\frac{\partial{\cal L}_{\rm{reg}}}{\partial\tilde L'}|_{\infty}\right)=\frac{K_1}{2mc'}\ .
\end{equation}
Therefore we see that the constant of integration $K_1$ is directly related to the phase difference between the bare quark mass and the condensate of the dual gauge theory $\Delta\phi$. Furthermore, using equation (\ref{conddphi}) one can write a modified version of equation (\ref{corresp1}) for the absolute value of the condensate of the theory:
\begin{equation}
|\langle {\cal O}\rangle_q|=\frac{N_f}{(2\pi\alpha')^3g_{YM}^2}\sqrt{c'^2+\frac{K_1^2}{4m^2}}\ .
\end{equation}

Note that $\Delta\phi=0$ for Minkowksi embeddings and $\Delta\phi\neq0$ for ergosphere type of embeddings. Therefore, we conclude that the phase difference $\Delta\phi$ is a natural order parameter for the first order phase transition subject of our study. Note also the striking similarity to the case of external electric field, considered in the previous chapter, when the order parameter of the phase transition is the global electric current along the direction of the electric field. In the following chapter, we shall shed more light on the discussed similarity.

%
%
%
%
%
%
%
%
%
%
%
%
%
%

\chapter*{Chapter 7:  \hspace{1pt} Universality in the large $N_c$ dynamics of flavor}
\addcontentsline{toc}{chapter}{Chapter 7:\hspace{0.15cm} Universality in the large $N_c$ dynamics of flavor}

\section*{7.1 \hspace{2pt} Thermal vs. quantum induced phase transitions }
\addcontentsline{toc}{section}{7.1 \hspace{0.15cm} Thermal vs. quantum induced phase transitions}

In this chapter we discuss common properties of the phase transitions studied in Chapters~3 ,5 and 6. In particular we focus on discrete self-similarity exhibited by the theory near criticality. To begin with, let us summarize the properties of the set up. In all cases considered so far, the background corresponded to the near horizon limit of a non--extremal D3--brane background. In addition we introduce D7--branes in the probe limit. The
D7--brane wraps a $S^3\subset S^5$ and extends in the radial direction
of the asymptotically AdS$_5$-space. The size of the $S^3$ varies as a function of the
radial coordinate.  The D7--brane embeddings then naturally form two
classes: embeddings that reach the horizon and hence fall into the
black hole, and embeddings for which the wrapped $S^3$ shrinks to zero
size at some radial position. For these, the D7--brane world--volume
simply closes smoothly before the horizon. In an Euclidean
presentation, the compact, unbounded parts of the D7--brane have the
topology $S^3\times S^1$ since the Euclidean time has a periodicity
set by the inverse temperature of the system. The classes are then
distinguished by one or the other compact space shrinking away.  The
authors of ref.~\cite{Babington:2003vm} proposed that the (topology
changing) transition of the D7--brane embeddings corresponds to a type
of  confinement/deconfinement phase transition, now in the meson
sector of the theory. This system has been extensively studied and it was shown that
it is a first order phase transition providing a holographic
description of the meson melting phase transition of the fundamental
matter.

There is also a unique critical embedding separating those two
classes. This solution reaches the horizon {\it and} has a shrinking
$S^3$. It has a conical singularity. Solutions of this type will
occupy much of our attention in this chapter. Many of these features
generalize to the general D$p$/D$q$ system.  In
ref.~\cite{Mateos:2006nu} the D$p$/D$q$ system was considered and some
universal properties, associated with this critical solution
separating the two classes of embedding, were uncovered. In particular
it was shown that for a certain temperature the theory exhibits a
discrete self--similar behavior, manifested by a double logarithmic
spiral in the solution space. This space of solutions is parameterized by
the bare quark mass and the quark condensate. (Geometrically
these correspond, respectively, to the asymptotic separation of the
D7-- and D3-- branes and the degree of bending of the D7--branes away
from the D3--branes.)

The region of solution space where the self--similar spiral is located
is unstable, in fact: There is a first order phase transition
associated with the physics of the system jumping between branches of
solutions and bypassing it entirely.  Nevertheless, it seems that
important features of the full physical story can be captured by
examining the neighborhood of this critical solution.  It is
remarkable that the critical exponents (or better ``scaling
exponents'', so as not to confuse the physics with the nomenclature of
second order phase transitions) characterizing this logarithmic
structure exhibit universal properties and depend only on the
dimension of the internal $S^n$  wrapped by the D$q$--brane.
The precise value of the critical temperature is irrelevant. The
structure is determined by focusing on the local geometry near the
conical singularity of the critical D$q$--brane embedding, and the
exponents are then naturally determined by the study of possible
embeddings in a Rindler space \cite{Frolov:2006tc,Mateos:2006nu}.

The studies described above concern a thermally driven phase
transition. As the temperature passes a certain threshold, thermal
fluctuations seek out the new global minimum that appears and the
system undergoes a transition to a new phase.  In this chapter we study
transitions of the system under the effect of two different types of
control parameters: an external electric field and an R--charge
chemical potential, revisiting work done on these systems in
 Chapter~5 and Chapter~6 of this work.  We show \cite{Filev:2008xt} corresponding scaling exponents are again universal and depend only on
the dimension of the internal sphere wrapped by the D$q$--brane.  We
find that the key properties of the critical solution can be
determined from the local properties of the geometry, and we find that
this geometry arises naturally by working in a rotating frame, arrived
at using T--duality.  The resulting physics is not controlled by
thermal dynamics, the local geometry is not Rindler, and so the
exponents are different.  The phase transition is driven by the
quantum (as opposed to thermal) fluctuations of the system, as can be
seen from the fact that they persist at zero temperature. It is
satisfying that we can cast these different types of transition into
the same classifying framework.

\section*{7.2 \hspace{2pt} Thermal phase transition }
\addcontentsline{toc}{section}{7.2 \hspace{0.15cm} Thermal phase transition}

Let us begin by reviewing the result of
refs. \cite{{Frolov:2006tc},{Mateos:2006nu}}. We will be using the
notations of ref.~\cite{Mateos:2006nu}. Consider the near--horizon
black D$p$--brane given by:
\begin{eqnarray}\label{1}
&&ds^2=H^{-\frac{1}{2}}\left(-f dt^2+\sum\limits_{i=1}^{p}dx_i^2\right)+H^{\frac{1}{2}}\left(\frac{du^2}{f}+u^2d\Omega_{8-p}^2\right)\ , \\
&&e^{\Phi}=g_s H^{(3-p)/4}\ ,~~~C_{01\dots p}=H^{-1}\ ,\nonumber
\end{eqnarray}
where $H(u)=(R/u)^{7-p}$, $f(u)=1-(u_{\rm{H}} /u)^{7-p}$ and $R$ is a length
scale (the AdS radius in the $p=3$ case). According to the
gauge/gravity correspondence, string theory on this background is dual
to a $(p+1)$--dimensional gauge theory at finite temperature. Now if
we introduce D$q$--brane probe having $d$ common space-like directions
with the D$p$--brane, wrapping an internal $S^n\subset S^{8-p}$ and
extended along the holographic coordinate $u$, we will introduce
fundamental matter to the dual gauge theory that propagates along a
$(d+1)$--dimensional defect.

If we parameterize $S^{8-p}$ by:
\begin{equation}
d\Omega_{8-p}^2=d\theta^2+\sin^2\theta d\Omega_n^2+\cos^2\theta d\Omega_{7-p-n}^2\ ,
\end{equation}
where $d\Omega_m^2$ is the metric on a round unit radius $m$--sphere,
the DBI part of the Lagrangian governing the classical embedding of
the probe is given by (We consider only systems T--dual to the D3/D7 one, which imposes the constraint $p-d+n+1=4$.):
\begin{equation}
{\cal L}\propto e^{-\Phi}\sqrt{-|g_{\alpha\beta}|}=\frac{1}{g_s}u^n\sin^n\theta\sqrt{1+f u^2\theta'^2}
\end{equation}
The embeddings split to two classes of different topologies:
`Minkowski' embeddings which have a shrinking $S^n$ above the
vanishing locus (the horizon) and yield the physics of meson states
and `black hole' embeddings that reach the vanishing locus,
corresponding to a melted/deconfined phase of the fundamental matter.
These classes are separated by a critical embedding with a conical
singularity at the vanishing locus, as depicted in Figure~\ref{fig:A1}.

\begin{figure}[h] 
    \centering \includegraphics[width=1.2in]{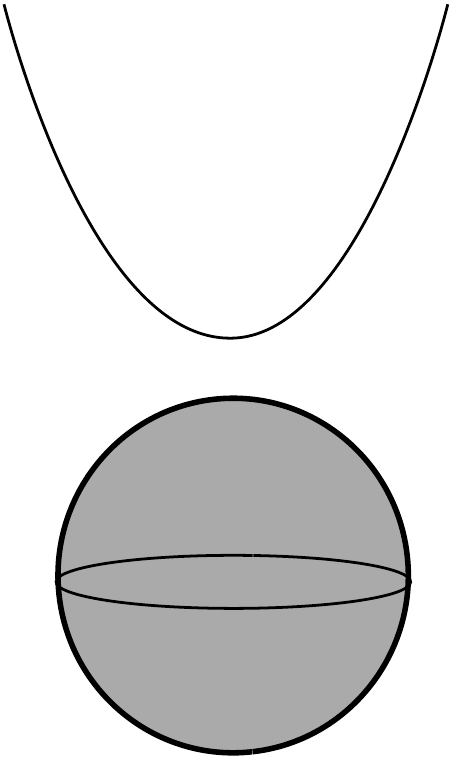}
    \includegraphics[width=1.2in]{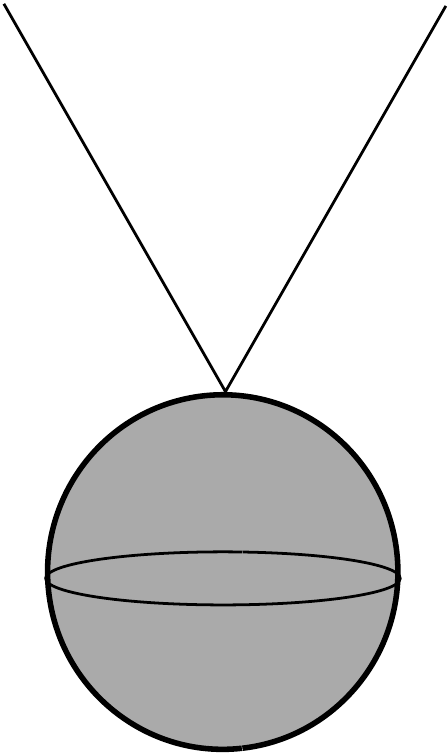}
    \includegraphics[width=1.45in]{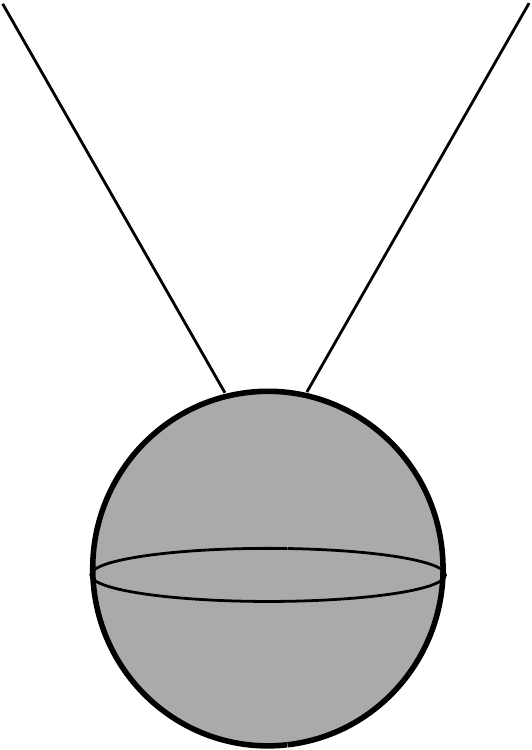}
     \caption{Schematic diagram depicting the Minkowski (left) and black hole (right) embedding solutions that are separated by a ``critical'' embedding (centre) which has a conical singularity at the event horizon. }
     \label{fig:A1}
  \end{figure}
  It is convenient to introduce the following coordinates:
\begin{eqnarray}
&& r^{\frac{7-p}{2}}=\frac{1}{2}\left(u^{\frac{7-p}{2}}+\sqrt{u^{7-p}-u_{\rm{H}} ^{7-p}}\right)\ ,\\
&& L=r\cos\theta\ ,\quad\mathrm{and}\quad \rho=r\sin\theta\ .  \nonumber
\end{eqnarray}
Then one can show \cite{Karch:2002sh,Kruczenski:2003uq} that the
asymptotic behavior of the embedding at $\rho\to\infty$ encodes the
bare quark mass $m_q={m}/{2\pi\alpha'}$ and the quark condensate
$\langle\bar\psi\psi\rangle\propto-c$ of the dual gauge theory via the
expansion:
\begin{equation}
L(\rho)=m+\frac{c}{\rho^{n-1}}+\dots
\end{equation}
After solving numerically for each embedding of the D$q$--brane, the
parameters $m$ and $c$ can be read off at infinity. From the full
family of embeddings, a plot of the equation of state of the system
$c(m)$ can be generated. The resulting plot for the D3/D7
system \cite{Albash:2006ew} is presented in Figure~\ref{fig:A2}. The
two different colors (and line types) correspond to the two different
classes of embeddings. The equation of state is a multi--valued
function, and there is a first order phase transition when the free
energies of the uppermost and lowermost branches match.

\begin{figure}[h] 
    \centering \includegraphics[width=3.1in]{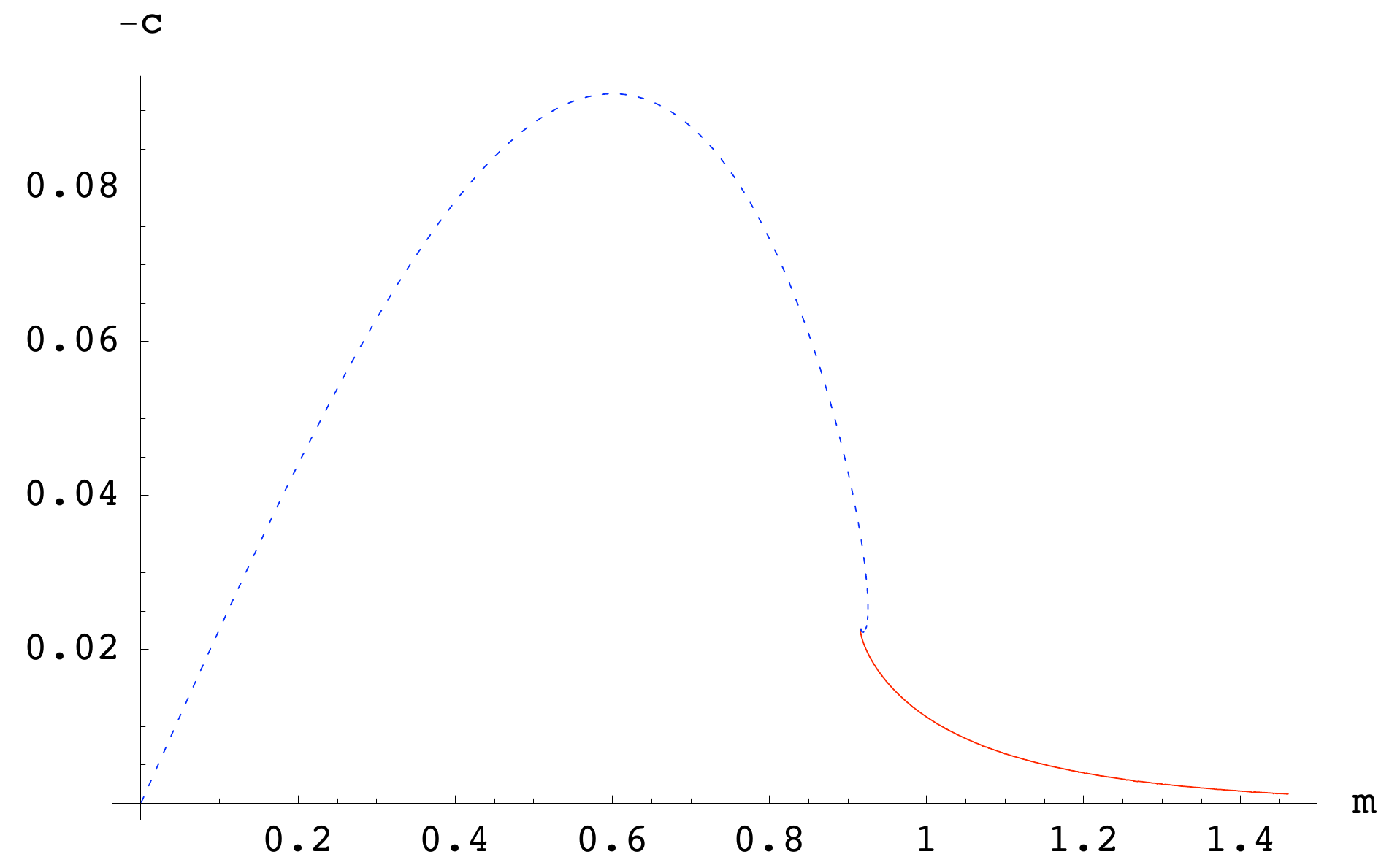}
    \includegraphics[width=3.1in]{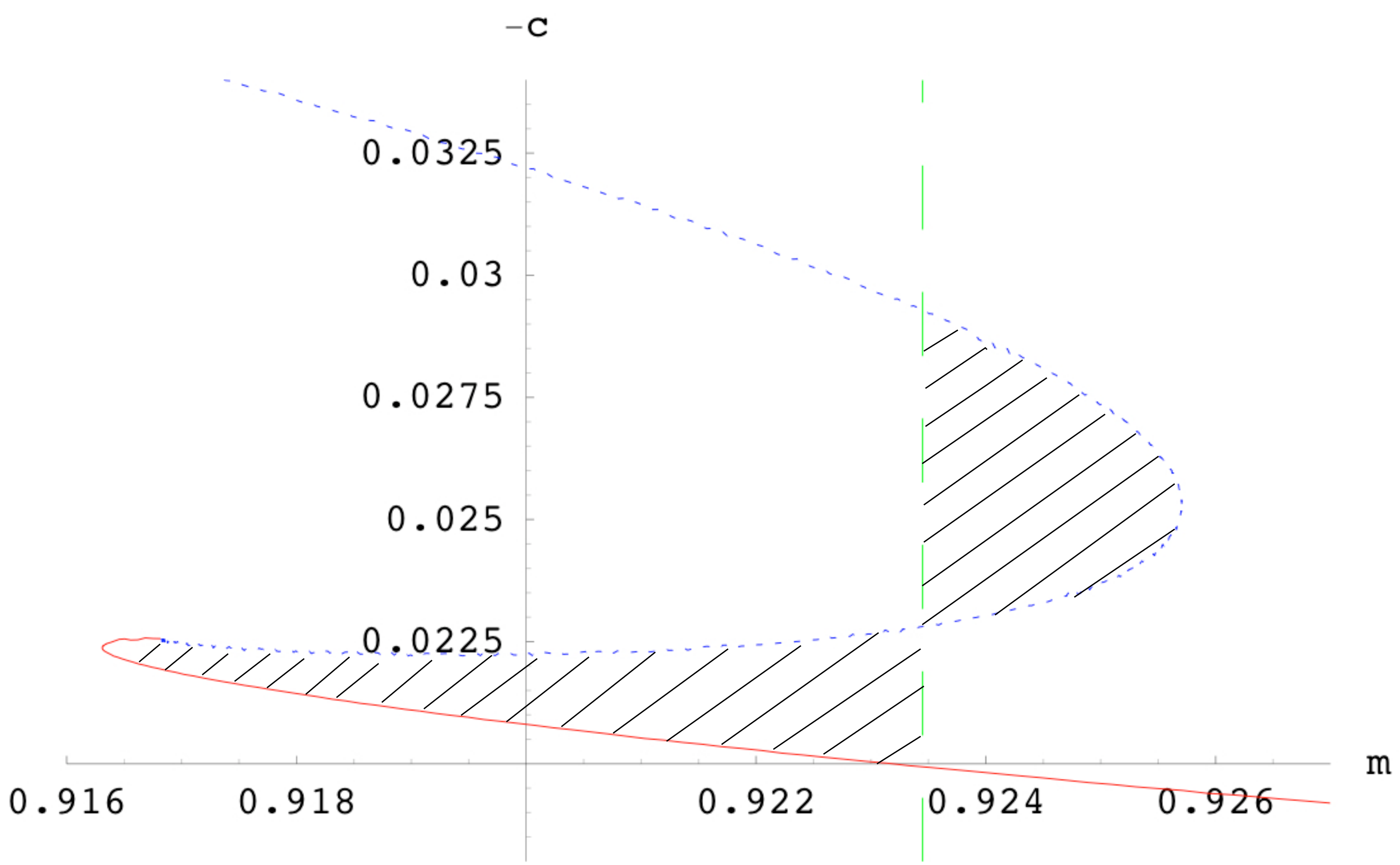}
     \caption{Plot of the equation of state $c(m)$. The zoomed region shows the location of the first order phase transition. There is a spiral structure hidden near the ``critical'' solution in the neighborhood of $m=0.9185$, $-c=0.0225$. }
     \label{fig:A2}
  \end{figure}
  
  The main subject of our discussion is the spiral structure in the
  solution space near the critical
  embedding \cite{Mateos:2006nu,Mateos:2007vn}. In the enlarged portion
  on the right in Figure~\ref{fig:A2} it is located to the lower left,
  roughly at $m=0.9185$,  $-c=0.0225$. The spiral structure that is
  hidden near this point is a signal of the discrete self--similarity
  of the theory near the critical solution.

  In order to understand the origins of the spiral, we zoom into the
  space--time region near the tip of the cone of the critical
  embedding \cite{Frolov:2006tc,Mateos:2006nu} using the change of variables:
\begin{equation}
u=u_{\rm{H}} +\pi T z^2\ ;~~~\theta=\frac{y}{R}\left(\frac{u_{\rm{H}} }{R}\right)^{\frac{3-p}{4}}\ ;~~~\hat x=x\left(\frac{u_{\rm{H}} }{R}\right)^{\frac{7-p}{4}}\ .
\end{equation}
Here $T$ is the temperature of the background given by:
\begin{equation}
{T}=\frac{7-p}{4\pi R}\left(\frac{u_{\rm{H}} }{R}\right)^{\frac{5-p}{2}}\ .
\end{equation}
 Leaving only the
leading terms in $z$ results in the following metric:
\begin{equation}
ds^2=-(2\pi T)^2z^2 dt^2+dz^2+dy^2+y^2 d\Omega_n^2+d\hat x_d^2+\dots
\label{Rindler}
\end{equation}

The metric (\ref{Rindler}) corresponds to flat space in Rindler
coordinates. The embeddings of the D$q$--branes in the background
(\ref{Rindler}) again split into two different classes: Minkowski
embeddings characterized by shrinking $S^n$ ($y=0$) at some finite
$z_0$, and black hole embeddings which reach the horizon at $z=0$
for some finite $y=y_0$ (the radius of the induced horizon). The
equation of motion is derived from the Dirac--Born--Infeld action of
the D$q$--branes which has the following Lagrangian:
\begin{equation}
{\cal L}\propto z y^n\sqrt{1+y'^2}\ .
\end{equation}
The equation of motion derived from this reads:
\begin{equation}
zyy''+(yy'-nz)(1+y'^2)=0\ .
\label{scaling}
\end{equation}
Solutions of this equation enjoy the scaling property $y(z)\to
\frac{1}{\mu}y(\mu z)$, in the sense that if $y(z)$ is a solution to
the equation (\ref{scaling}) so is $\frac{1}{\mu}y(\mu z)$. Under such
a re--scaling the initial conditions $(z_0,y_0)$ for the two classes
of embeddings scale as:
\begin{equation}
z_0\to z_0/\mu;~~~y_0\to y_0/\mu;
\label{re-scaling}
\end{equation}
This suggests the existence of a critical solution characterized by
$z_0=y_0=0$. One can check that $y=\sqrt{n}z$ is the critical
solution. It  has a conical singularity at $y=z=0$.

To analyze the parameter space of the solutions we can linearize the
equation of motion (\ref{scaling}) near the critical solution by
substituting $y(z)=\sqrt{n}z+\xi(z)$, for small $\xi(z)$. The
resulting equation of motion is:
\begin{equation}
z^2\xi''(z)+(n+1)(z\xi'(z)+\xi(z))=0\ ,
\end{equation}
which has a general solution of the form:
\begin{eqnarray}
&&\xi(z)=\frac{1}{z^{r_n}}(A\cos({\alpha_n\ln z)}+B\sin(\alpha_n\ln z))\ ,\label{linsol}\\
&&\mathrm{with}\quad r_n=\frac{n}{2}\ ;~~~\alpha_n=\frac{1}{2}\sqrt{4(n+1)-n^2}\ .\nonumber
\end{eqnarray}
Note that $\alpha_n$ are real only for $n\le4$ which are the cases
naturally realized in string theory \cite{Mateos:2007vn}. Now the
scaling property of equation (\ref{scaling}), combined with the form of
the solutions (\ref{linsol}) suggests the following transformation of
the parameters $(A,B)$ under the re-scaling of the initial conditions
given in equation~(\ref{re-scaling}):
\begin{equation} 
\begin{pmatrix} A' \\  B' \end{pmatrix}=\frac{1}{\mu^{r_n+1}}\begin{pmatrix}\cos{(\alpha_n\ln\mu)} & \sin{(\alpha_n\ln\mu)}\\-\sin{(\alpha_n\ln\mu)} & \cos(\alpha_n\ln\mu)\end{pmatrix}\begin{pmatrix} A \\  B \end{pmatrix}\ .
\label{transformation}
\end{equation}
For a fixed choice of the parameters $A$ and $B$, the parameters $(A',
B')$ describe a double spiral, whose step and periodicity are set by
the real and imaginary parts of the critical/scaling exponents $r_n$
and $\alpha_n$.  

Equation~(\ref{scaling}) has a $Z_2$ symmetry \cite{Frolov:2006tc}
relating the two classes of solutions (black hole and Minkowski
embeddings). If the parameters $(A, B)$ describe
one class of embeddings, then the parameters $(-A,-B)$ describe the
other. In this way the full parameter space near the critical solution
(given by $A=0$, $B=0$) is a double logarithmic spiral.

This self--similar structure of the embeddings near the critical
solution in our Rindler space is transferred by a linear transformation
to the structure of the solutions in the $(m, c)$ parameter space.  If
we call $(m^*, c^*)$ the parameters corresponding to the critical
embedding from Figure~\ref{fig:A1}, then sufficiently close to the
critical embedding we can expand:
\begin{equation}
\begin{pmatrix}
m-m^*\\c-c^*
\end{pmatrix}=M\begin{pmatrix}A\\B\end{pmatrix}+O(A^2)+O(B^2)+O(A,B)\ .
\end{equation}
The constant matrix $M$ cannot be determined analytically and depends
on the properties of the system. Generically it should be invertible
(numerically we have verified that it is) and therefore in the
vicinity of the parameter space close to the critical embedding
$(m^*,c^*)$ there is a discrete self--similar structure determined by
the transformation:
\begin{equation}
\begin{pmatrix} m'-m^*\\  c'-c^* \end{pmatrix}=\frac{1}{\mu^{r_n+1}}M\begin{pmatrix}\cos{(\alpha_n\ln\mu)} & \sin{(\alpha_n\ln\mu)}\\-\sin{(\alpha_n\ln\mu)} & \cos(\alpha_n\ln\mu)\end{pmatrix}M^{-1}\begin{pmatrix} m-m^*\\  c-c^* \end{pmatrix}\ . 
\label{fultrans}
\end{equation}
Let us define two solutions to be ``similar'' if:
\begin{equation}
\begin{pmatrix}|m'-m^*|\\|c'-c^*|\end{pmatrix}=\frac{1}{\mu^{r_n+1}}\begin{pmatrix} |m-m^*| \\ |c-c^*|\end{pmatrix}\ .
\end{equation}   
Then one can see from equation (\ref{fultrans}) that this is possible
only for a discrete set of $\mu$s given by:
\begin{equation}
\mu=e^{{k\pi}/{\alpha_n}};~~~k=1, 2, \dots\ .
\end{equation}
Note that in general the matrix $M$ in equation (\ref{fultrans}) will
deform the spiral structure given by the transformation
(\ref{transformation}). However, the scaling properties of the theory
remain the same as they are completely determined by the scaling
exponents: $r_n,\alpha_n$. Furthermore, one can see that the scaling
exponents depend only on the dimension of the internal sphere $S^n$
wrapped by the D$q$--brane and are thus universal, in the sense that
the detailed value of the critical temperature is irrelevant. It is
the spiral structure that ultimately seeds the multi--valuedness of the
space of solutions, twisting the $(m,-c)$ curve back on itself as in
Figure~\ref{fig:A2}. Therefore, it is the spiral --- and the
neighborhood of the critical solution from where it emanates --- that
is responsible for the presence of a first order phase transition in
the system. Whether there is a spiral or not can be read off from the
scaling parameters $(r_n,\alpha_n)$, and since \cite{Mateos:2007vn} for
all consistent D$p$/D$q$ systems the condition $n\le 4$ is satisfied
the corresponding thermal phase transition (meson melting at large
$N_c$) is a first order one.

\section*{7.3 \hspace{2pt} Quantum--induced  phase transitions}
\addcontentsline{toc}{section}{7.3 \hspace{0.15cm} Quantum--induced  phase transitions}

In this section we will consider a different class of phase
transitions. These are arise in the presence of external fields, and
can happen even at zero temperature, and so since the fluctuations
driving the transition are no longer thermal, they might be expected
to be in a different class.  Naively, the broad features of the
equation of state --- multi--valuedness and so forth --- have
similarities with the thermal case, and so it is natural to attempt to
trace the extent to which these similarities persist. We will find
that once we cast these systems in the language of the previous
section, the similarities and differences will be quite clear.

We will first concentrate on the case of an external electric field.
The flavored system, at large enough electric field, has an {\it
  insulator/conductor} phase transition, as studied in
Chapter~5. As with the thermal transition of the last
section, the mesons dissolve into their constituent quarks, but this
time it is due to the electric field overcoming their binding energy.
The transition is of first order.

As we saw in the previous section the scaling properties of the
thermally driven phase transition are naturally studied in a Rindler
frame with a temperature set by the temperature of the background.  In
Chapter~5 it was shown that in analogy to the thermally
driven phase transition there is a nice geometrical description of the
electrically driven phase transition, and the structure of the system
can be again characterized by an unstable critical embedding with a
conical singularity at an appropriate vanishing locus (analogous to
the event horizon). Here, we will generalize this description to the
case of the D$p$/D$q$ system.

Furthermore, after an appropriate T--duality transformation we will show
that the vanishing locus corresponds to an effective ``ergosphere''
due to a rotation of the coordinate frame along the compact directions
of the background. The instability near criticality is then naturally
interpreted as an instability due to the over--spinning of the
D$(q-1)$ brane probes (in the T--dual background) as they reach the
ergosphere. We then study the structure of the theory near criticality
by zooming in on the space--time region in the vicinity of the conical
singularity. Once again, we will find that the structure is entirely
controlled by the dimension of the internal sphere, $S^n$,  wrapped by
the D$(q-1)$--branes (in the T--dual background) --- details such as
the value of the electric field and the temperature of the system, are
irrelevant.

\subsection*{7.3.1 \hspace{2pt} Criticality and scaling in an external electric field}
\addcontentsline{toc}{subsection}{7.3.1 \hspace{0.15cm} Criticality and scaling in an external electric field}
Let us consider the near--horizon black D$p$--brane given by the
background in equation~(\ref{1}). Following a similar
idea \cite{Filev:2007gb} for producing a background magnetic field, if
we turn on a pure gauge $B$--field in the $(t,x_p)$
plane \cite{Albash:2007bq,Erdmenger:2007bn,Karch:2007pd}, in the dual
gauge theory this will correspond to an external electric field,
oriented along the $x_p$ direction:
\begin{equation}
B=E dt\wedge dx_p\ .
\label{beefield}
\end{equation}
The resulting Lagrangian is:
\begin{equation}
{\cal L}\propto e^{-\Phi}\sqrt{-|g_{\alpha\beta}+B_{\alpha\beta}|}=\frac{1}{g_s}\sqrt{\frac{f-E^2H}{f}}u^n\sin^n\theta\sqrt{1+f u^2\theta'^2}\ .
\label{electric}
\end{equation}
This leads to the existence of a vanishing locus at $u=u_*$ given by:
\begin{equation}
u_*^{7-p}=u_{\rm{H}} ^{7-p}+E^2R^{7-p}\ ,
\label{locus}
\end{equation}
at which the action (\ref{electric}) vanishes. Notice that this is
distinct from the horizon, and even at zero temperature will be
present. A study of the local physics near this locus will therefore
pertain to non--thermal physics.

The embeddings split into two different classes: Minkowski
embeddings which have a shrinking $S^n$ above the vanishing locus and
correspond to meson states and embeddings reaching the vanishing
locus, corresponding to a deconfined phase of the fundamental matter.
These classes are separated by a critical embedding with a conical
singularity at the vanishing locus. Our goal is to explore the
self--similar behavior of the theory near this critical embedding and
calculate the corresponding scaling exponents.

In order to make the analysis closer to the one performed in
refs. \cite{{Frolov:2006tc},{Mateos:2006nu}}, for the thermal phase
transition (described in the last section), we T--dualize along the
$x_p$ direction.  This is equivalent to a trading of the pure gauge
$B$--field for a rotating frame in the T--dual background. Indeed, the
geometry T--dual to equation~(\ref{1}), with the $B$--field given by
equation~(\ref{beefield}), is given by:
\begin{eqnarray}
&&d\tilde s^2=H^{-\frac{1}{2}}(-\tilde f dt^2+\sum\limits_{i=1}^{p-1}dx_i^2)+2H^{\frac{1}{2}}E dt d{\tilde x}_p+H^{\frac{1}{2}}\left(\frac{du^2}{f}+u^2d\Omega_{8-p}^2+d{\tilde x}_p^2\right)\ ,\label{tdual}\nonumber\\
&&e^{\tilde\Phi}=g_sH^{1-\frac{p}{4}};~~~\tilde f=1-\left(\frac{u_*}{u}\right)^{7-p}\ .
\end{eqnarray}
The background given by equation (\ref{tdual}) corresponds to the
near--horizon limit of a stack of $N_c$ D$(p-1)$--branes smeared along
the coordinate $\tilde x_p$. Now if we place a probe D$(q-1)$--brane
having $(d{-}1)$ spatial directions shared with the D$(p-1)$---branes,
filling the radial direction $u$ and wrapping an internal $S^n$ inside
the $S^{8-p}$ sphere of the background, we will recover the action
(\ref{electric}), as we should.

Note that in these coordinates we have an effective ``ergosphere''
coinciding with the vanishing locus given by equation~(\ref{locus}).
Now the critical embedding is the one touching the ergosphere and
having a conical singularity at $u=u_*$. In the $(m, c)$--plane this
embedding corresponds to the center of the spiral structure $(m_*,
c_*)$.

Despite the analogy with the analysis of the thermal phase transition,
in this case there is a crucial difference because of the necessity
(from charge conservation) for the D$(q-1)$--brane to extend beyond
the ergosphere. Indeed, since the D$(q-1)$--brane is an extended object
one can find static solutions that extend beyond the ergosphere and
are non--superluminal. To this end one should allow the
D$(q-1)$--brane to extend along the direction of rotation $\tilde
x_p$.  In the original coordinates (before T--dualization) this is
equivalent to a non--trivial profile for the $A_p$ component of the
gauge field, which corresponds to the appearance of a global electric
current along the
$x_p$--direction \cite{{Albash:2007bq},{Karch:2007pd}}. This is the
reason why we refer to the corresponding phase transition as an
insulator/conductor phase transition. After the transition, the quarks
are free to flow under the influence of the electric field, forming a
current.

Let us describe how this procedure works in the case of a general
D$(p-1)$/D$(q-1)$--intersection. Again we will work in the T--dual
background (\ref{tdual}).  Let us consider an ansatz for the
D$(q-1)$--brane embedding of the form:
\begin{equation}
\theta=\theta(u)\ ;~~~{\tilde x}_p={\tilde x}_p(u)\ ;
\label{ext}
\end{equation}
this leads to the action:
\begin{equation}
{\cal L}_*\propto\frac{1}{g_s}\sqrt{\frac{f-E^2H}{f}}u^n\sin^n\theta\sqrt{1+f u^2\theta'^2+\frac{f^2}{f-E^2H}{{\tilde x}'^2_p}}\ .
\end{equation}
Now after integrating the equation of motion for $\tilde x_p$ and
plugging the result in the original Lagrangian, we get the following
on--shell Lagrangian:
\begin{equation} 
{\cal L}_*\propto\frac{1}{g_s}\sqrt{\frac{f-E^2H}{f u^{2n}\sin^{2n}\theta-K^2}}u^{2n}\sin^{2n}\theta\sqrt{1+f u^2\theta'^2}\ .
\label{action*}
\end{equation}
It is easy to verify that if we choose the integration constant $K^2$
in equation~(\ref{action*}) to satisfy:
\begin{equation}
K^2=E^2H_{*}u_*^{2n}\sin^{2n}\theta_0\ ,
\label{inconst}
\end{equation}
then the action (\ref{action*}) is regular at the ergosphere
($u=u_*$). Note that at the critical embedding
$\theta_0=\theta_*\equiv0$ and the constant in
equation~(\ref{inconst}) is zero. As we learned in Chapter~5, this constant is proportional to the
global electric current along the $x_p$ direction of the original
D$p$/D$q$--brane system. 

We are interested in the scaling properties of the theory, near the
critical embedding solution. Despite the fact that the Lagrangians
(\ref{electric}) and (\ref{action*}) describing the Minkowski and
ergosphere classes of embeddings are different, the fact that at
the critical embedding they coincide ($K^2$=0) shows that the
corresponding equations of motion share the same critical solution.
Furthermore, as we will see, the critical exponents are the
same for both types of embedding. 

Let us introduce dimensionless coordinates by the transformation:
\begin{eqnarray}  
&&u=u_*+z\frac{Du_*}{7-p}\ ;~~\theta=\frac{y}{R}\left(\frac{u_*}{R}\right)^{\frac{3-p}{4}}\ ;~~x_i\left(\frac{u_*}{R}\right)^{\frac{7-p}{4}}\to x_i\ ;~~t\left(\frac{u_*}{R}\right)^{\frac{7-p}{4}}\to t\ ; \nonumber\label{zooming}\\
&&H_*^{\frac{3}{4}}E\tilde x_p\to\tilde x_p\ ,
\end{eqnarray}
where $D^2=(7-p)^2f_*/H_*^{\frac{1}{2}}u_*^2$, $H_*=H|_{u=u_*}$ and
$f_*=f|_{u=u_*}$. To leading order in $z$ and $y$ the metric
(\ref{tdual}) is given by:
\begin{eqnarray}
\label{rotmet}
d\tilde s^2&=&-Dz dt^2+dz^2+dy^2+y^2d\Omega_n^2+H_*^{\frac{1}{2}}u_*^2d\Omega_{7-p-n}^2\\&&+2dtd{\tilde x}_p
+\frac{1}{E^2H_*}d{\tilde x}_p^2+\sum\limits_{i=1}^{p-1}dx_i^2\ .\nonumber
\end{eqnarray}

First consider the case of Minkowski embeddings, characterized by a
distance $z_0$ above the ergosphere at which they close
($y=y(z_0)=0$). The Lagrangian describing the D$(q-1)$--brane
embedding is:
\begin{equation}
{\tilde{\cal L}_*}\propto y^n z^{1/2}\sqrt{1+y'^2}\ ,
\label{lagrscal1}
\end{equation}
The corresponding equation of motion is given by:
\begin{equation}
\partial_z\left(y^n z^{1/2}\frac{y'}{\sqrt{1+y'^2}}\right)-ny^{n-1}z^{1/2}\sqrt{1+y'^2}=0\ .
\label{eqmn}
\end{equation}
Equation (\ref{eqmn}) possesses the scaling symmetry:
\begin{equation}
y\to y/\mu\ ;~~~z\to z/\mu;
\label{scal}
\end{equation}
in the sense that if $y=y(z)$ is a solution to equation (\ref{eqmn})
so is the function $\frac{1}{\mu}y(\mu z)$. Now under the scaling
(\ref{scal}) the boundary condition for the Minkowski embedding
scales as $z_0\to z_0/\mu$. This suggests the existence of a limiting
critical embedding with $z_0=0$, and indeed:
\begin{equation}
y(z)=\sqrt{2n}z\ ,
\label{crit2}
\end{equation}
is a solution to the equation of motion in equation~(\ref{eqmn}).  The
corresponding D$(q-1)$--brane has a conical singularity at $y=z=0$.
Now before we linearize equation (\ref{eqmn}) and calculate the
critical exponents let us consider the case of the ergosphere class of
solutions characterized by the radius of the ergosphere induced on
their world--volume. Because of the possibility to extend beyond the
ergosphere we should consider the analog of the ansatz from
equation~(\ref{ext}):
\begin{equation}
y=y(z);~~~\tilde x_p=\tilde x_p(z)\ .
 \end{equation}
 The corresponding Lagrangian is:
 \begin{equation}
\tilde{\cal L_*}\propto y^n\sqrt{Dz(1+y'^2)+(Fz+1)\tilde x'^2_p}\ ,
\label{ergo-action}
 \end{equation}
 where $F=D/E^2H_*$. After integrating the equation of motion for
 $\tilde x_p$ and substituting it into the Lagrangian
 (\ref{ergo-action}), we obtain the following on--shell Lagrangian:
 \begin{equation}
\tilde{\cal L_*}\propto \frac{z^{1/2}\sqrt{Fz+1}y^{2n}}{\sqrt{(Fz+1)y^{2n}-y_0^{2n}}}\sqrt{1+y'^2}\ .
\label{on-shell-erg}
 \end{equation}
 It is easy to see that the Lagrangian (\ref{on-shell-erg}) is regular
 at $z=0, y=y_0$. The equation of motion for $y(z)$, derived from the
 Lagrangian (\ref{ergo-action}) and with the substituted solution for
 $\tilde x_p(z)$ is:
\begin{equation}
\frac{\partial}{\partial z}\left(\frac{z^{1/2}y'}{\sqrt{1+y'^2}}\sqrt{\frac{(Fz+1)y^{2n}-y_0^{2n}}{Fz+1}}\right)-ny^{2n-1}z^{1/2}\sqrt{\frac{Fz+1}{(Fz+1)y^{2n}-y_0^{2n}}}\sqrt{1+y'^2}=0\ .
\label{eom-erg}
\end{equation}
It is easy to check that equation (\ref{crit2}) is a solution to
equation (\ref{eom-erg}). Furthermore, for $z\ll 1/F$ one can see that
equation (\ref{eom-erg}) has the scaling symmetry (\ref{scal}) (note
that equation (\ref{scal}) suggests $y_0\to y_0/\mu$). Linearizing
equations (\ref{eqmn}) and (\ref{eom-erg}) near the critical solution
(\ref{crit2}) by substituting:
\begin{equation}
y(z)=\sqrt{2n}z+\xi(z)
\end{equation}
results in the same equation:
\begin{equation}
z^2\xi''(z)+(n+1/2)(z\xi'(z)+\xi(z))=0\ .
\label{lin2}
\end{equation}
The general solution of equation (\ref{lin2}) is given by:
\begin{equation}
\xi(z)=\frac{1}{z^{r_n}}(A\cos(\alpha_n\ln z)+B\sin(\alpha_n\ln z))\ ,
\end{equation}
where the scaling exponents are given by:
\begin{equation}
r_n=\frac{2n-1}{4}\ ;~~~\alpha_n=\frac{1}{4}\sqrt{7+20n-4n^2}\ .
\label{crit-exp2}
\end{equation}
Note that the scaling exponents again, while quite different from
those of the thermal case (see equation~(\ref{linsol})) depend only on
the dimension of the internal $S^n$ wrapped by the D$q$--brane and are
thus universal for all D$p$/D$q$ systems. Furthermore, the discrete
self--similarity holds for $n\leq5$. By similar reasoning to the
thermal case \cite{Mateos:2007vn}, since for all consistent systems
realized in string theory we have that $n\leq4$, for such systems we
may expect that the electrically driven confinement/deconfinement
phase transition is first order and has the described discrete
self--similar behavior near the solution that seeds the
multi--valuedness of the equation of state.

The rest of the analysis is completely analogous to the thermal case
considered in the previous section. Therefore we come to the
conclusion that close to the critical embedding (specified by~$m_*$
and $c_*$) the theory has the following scaling property:
\begin{equation}
\begin{pmatrix} m'-m^*\\  c'-c^* \end{pmatrix}=\frac{1}{\mu^{r_n+1}}M\begin{pmatrix}\cos{(\alpha_n\ln\mu)} & \sin{(\alpha_n\ln\mu)}\\-\sin{(\alpha_n\ln\mu)} & \cos(\alpha_n\ln\mu)\end{pmatrix}M^{-1}\begin{pmatrix} m-m^*\\  c-c^* \end{pmatrix}\ , 
\label{fultrans2}
\end{equation}
with $r_n$ and $\alpha_n$ given by equation (\ref{crit-exp2}).

It is interesting to compare the analytic results some numerical
studies. Let us consider the D3/D7 system. From equation~(\ref{scal})
on can see that the variation of the scaling parameter $\mu$ in
equation~(\ref{fultrans2}) can be traded for the variation of the
boundary conditions of the probe, namely $z_0$ for Minkowski and $y_0$
for ergosphere embeddings. On the other hand, close to the critical
embedding, the change of coordinates in equation (\ref{zooming})
suggests that:
\begin{equation}
\theta_0\propto y_0\quad{\rm and}\quad u_0-u_*\propto z_0\ ,
\end{equation}
where $u_0$ and $\theta_0$ are the boundary conditions for the
embeddings in the original (not zoomed in)
background. Note that the parameter $u_0$ is related to the
constituent quark mass $M_c$ \cite{Mateos:2007vn} (in the absence of
an electric field) via $M_c=(u_0-u_{\rm{H}} )/(2\pi\alpha')$. 

Close to the
critical embedding we have that:
\begin{equation}
\mu=(u_{0,{\rm in}}-u_*)/(u_0-u_*)\quad{\rm and}\quad \mu=z_{0,{\rm in}}/z_0\ ,
\end{equation}
for some fixed boundary conditions $u_{0,{\rm in}}$ and
$\theta_{0,{\rm in}}$. Now equation (\ref{fultrans2}) suggests that
for Minkowski embeddings the plot of $(m-m_*)/(u_0-u_*)^{r_n+1}$
versus $\alpha_n\ln(u_0-u_*)$ should be an harmonic function of
$\alpha_n\ln(u_0-u_*)$ with a period $2\pi$. Similarly for ergosphere
embeddings the plot of $(m-m_*)/\theta_0^{r_n+1}$ versus
$\alpha_n\ln\theta_0$ should be a harmonic function of
$\alpha_n\ln\theta_0$ with a period $2\pi$. Note that the physical
meaning of $\theta_0$ can be related to the value of the global
electric current (see equation~(\ref{inconst}) and the comment below).
 
\begin{figure}[h] 
   \centering
      \includegraphics[width=3.2in]{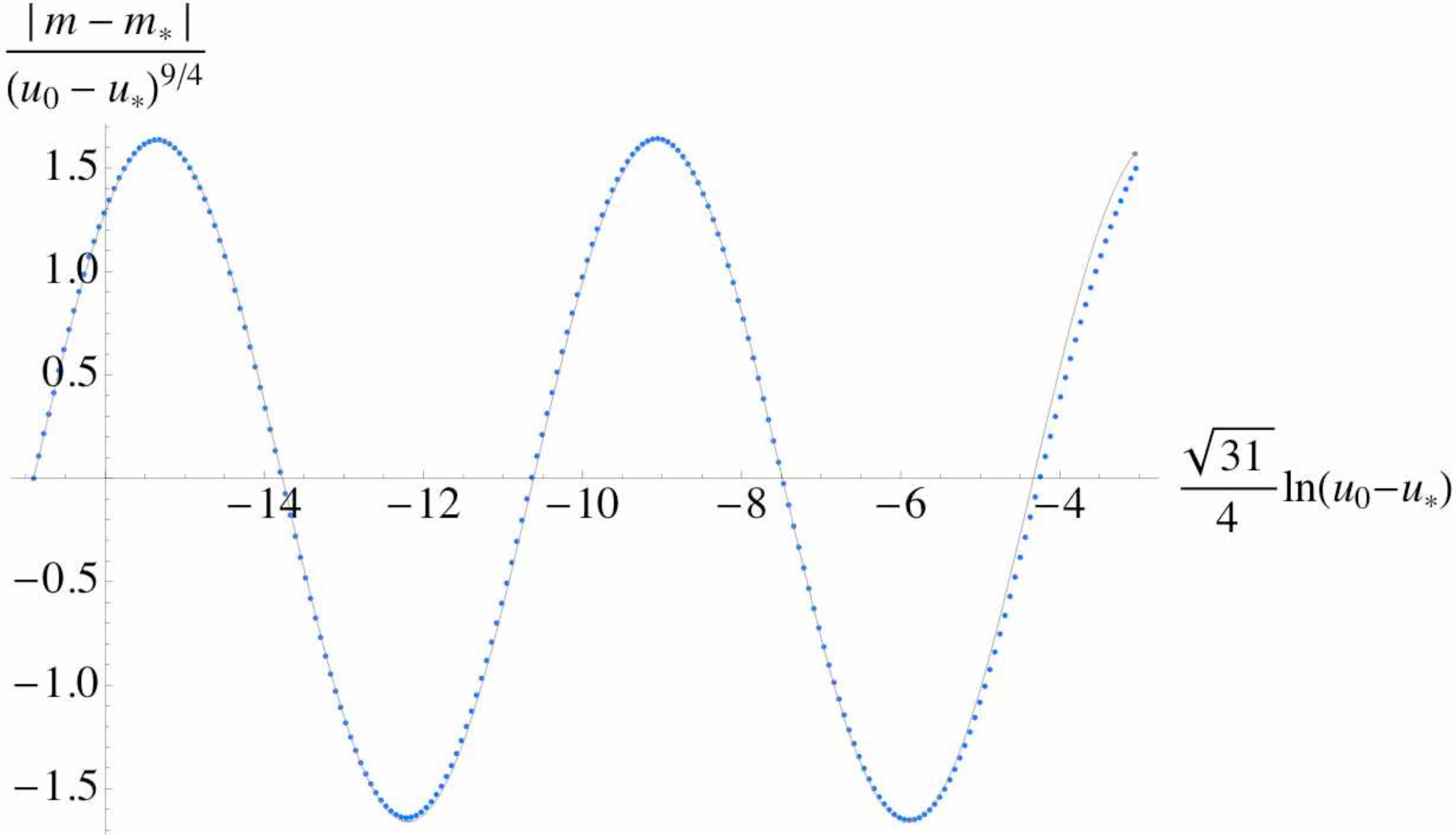} \hskip0.5cm
      \includegraphics[width=3.2in]{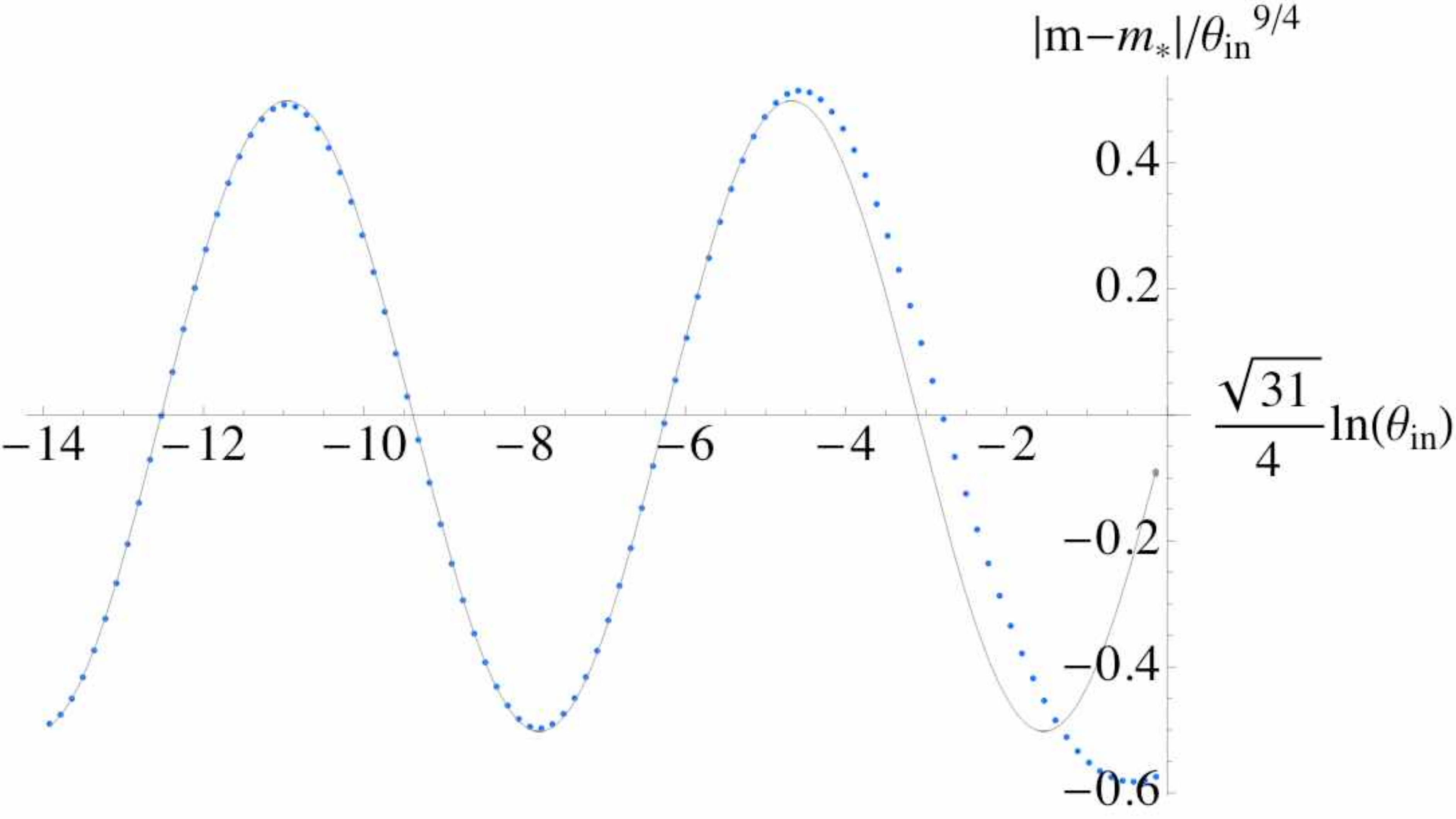} 
   \caption{The solid line is a fit with trigonometric functions of period $2\pi$. The plots confirm 
     that the scaling exponents of the theory are $r_3=5/4$ and $\alpha_3=\sqrt{31}/4$. }
   \label{fig:electric}
\end{figure}

As can be seen in Figure~\ref{fig:electric}, for both types of
embeddings the numerical results are in accord with
equation~(\ref{fultrans2}) and the analytic results improve deeper
into the spiral (large negative values on the horizontal axis).  Our
numerical results confirm that the critical exponents are indeed
$r_3=5/4$ and $\alpha_3=\sqrt{31}/4$, as the general analytic
results yield.

\subsection*{7.3.2 \hspace{2pt} Criticality and scaling with R--charge chemical potential}
\addcontentsline{toc}{subsection}{7.3.2 \hspace{0.15cm} Criticality and scaling with R--charge chemical potential}

Now we study the case when the external parameter is an R--charge
chemical potential in the dual gauge theory.  Introducing flavor to this background was the subject of our study in Chapter~6. Let us for convenience replicate some of the important equations here.

The relevant geometry is given by:
\begin{eqnarray}
&&ds^2=\Delta^{1/2}\left(-({\cal H}_1{\cal H}_2{\cal H}_3)^{-1}f dt^2+\frac{u^2}{R^2}d\vec x^2+f^{-1}du^2\right)+\label{spinning7}\\
&&\hskip3.0cm +\Delta^{-1/2}\sum_{i=1}^{3}{\cal H}_i\left(\mu_i^2(Rd\phi_i-A_t^idt)^2+R^2 d\mu_i^2\right)\ ,\nonumber
\end{eqnarray}
where
\begin{eqnarray}
&&f=\frac{u^2}{R^2}{\cal H}_1{\cal H}_2{\cal H}_3-\frac{u_{\rm{H}} ^4}{u^2R^2},~~~{\cal H}_i=1+\frac{q_i^2}{u^2},~~~A_t^i=\frac{u_{\rm{H}} ^2}{R}\frac{q_i}{u^2+q_i^2}\\ \nonumber\label{fields7} \,
&&\Delta={\cal H}_1{\cal H}_2{\cal H}_3\sum_{i=1}^{3}\frac{\mu_i^2}{{\cal H}_i},~~~{\rm with}\qquad\mu_1=\sin\theta,~~~\mu_2=\cos\theta\sin\psi,~~~\mu_3=\cos\theta\cos\psi .
\end{eqnarray}
Here the parameter $u_{\rm{H}} $ would be the radius of the event horizon if
the angular momentum of the geometry was set to zero ($q_i=0$). The
radius $u_{\rm{E}} $ of the actual event horizon is determined by the largest
root of $f(u)=0$.  The temperature of the background that we calculated in the previous chapter is given by:
\begin{equation}
T=\frac{u_{\rm{E}} }{2\pi R^2 u_{\rm{H}} ^2}\left(2u_{\rm{E}} ^2+q_1^2+q_2^2+q_3^2-\frac{q_1^2q_2^2q_3^2}{u_{\rm{E}} ^4}\right)=\frac{1}{2\pi R^2 u_{\rm{H}} ^2 u_{E}^{\phantom{2}}}(u_{E}^2-u_1^2)(u_{E}^2-u_2^2)\ ,
\label{temperatureR7}
\end{equation}
where $u_1$ and $u_2$ are the other two roots of $f(u)=0$.

The background (\ref{spinning7}) has an ergosphere determined by the
expression:
\begin{equation}
\Delta({\cal H}_1{\cal H}_2{\cal H}_3)^{-1}f-\sum_{i=1}^3{\cal H}_i\mu_i^2(A_t^i)^2=0\ .
\label{ergosphere7}
\end{equation}
Since the background (\ref{spinning7}) is asymptotically AdS$_5\times
S^5$, we can ``remove'' the ergosphere (\ref{ergosphere7}), by going to a
rotating frame. This is equivalent to gauge shifting $A_t^i$
from~(\ref{fields7}) such that ${A'_t}^i=-R\mu_{\rm{R}}^i+A_t^i$. The
parameters $\mu_{\rm{R}}^i$ are set by the condition ${A'_t}^i|_{u_{\rm{E}} }=0$ and
hence:
\begin{equation} 
\mu_{\rm{R}}^i=\frac{u_{\rm{H}} ^2}{R^2}\frac{q_i}{u_{\rm{E}} ^2+q_i^2}\ .
\label{chempot}
\end{equation}
From the behavior at infinity ($u\to\infty$), it is clear that
$\mu_{\rm{R}}^i$ correspond to the angular velocities of the frame along
$\phi_i$. In the dual gauge theory these correspond to having time
dependent phases of the adjoint complex scalars or equivalently to
R--charge chemical potentials for the corresponding
scalars \cite{Chamblin:1999tk}.

In order to restore some of the symmetry of the metric
(\ref{spinning7}), we will consider the case when $q_2=q_3$. This
corresponds to having an $S^3$ (parameterized by $\psi,\phi_2,\phi_3$)
inside the deformed~$S^5$. Now if we introduce D7--branes filling the
AdS--like part of the geometry and wrapping the~$S^3$, we will add
fundamental matter to the gauge theory. Furthermore, we are free to
rotate the D7--branes along~$\phi_1$ and the corresponding angular
velocity is interpreted as a time dependent phase of the bare quark
mass. (Recall that in introducing D7--branes to the D3--brane system
we actually add flavors as chiral superfields into the ${\cal N}=2$
gauge theory).  If that phase is the same as the phase of the complex
adjoint scalar, $\mu_{\rm{R}}^1 t$, it is equivalent to an R--charge
chemical potential for both the adjoint scalar and the chiral field.

On the gravity side of the description this is equivalent to letting
the D7--branes have the same angular velocity $\mu_{\rm{R}}^1$ as the
rotating frame of the background. Moving to the frame corotating
with the D7--brane corresponds to moving back to the gauge choice for
$A_t^1$ from equation~(\ref{fields7}).
The price that we pay is that we again have an
ergosphere in the bulk of the background. It will be convenient to gauge shift ${A'}_t^2$ and ${A'}_t^3$ to ${A''}_t^2$ and ${A''}_t^3$, correspondingly, so that the ergosphere is at:
\begin{equation}
\Delta({\cal H}_1{\cal H}_2^2)^{-1}f-{\cal H}_1\sin^2\theta(A_t^1)^2=0
\label{ergosphere-new7}. 
\end{equation}
The shifted forms, ${A''}_t^2$ and ${A''}_t^3$, vanish at the locus given by equation (\ref{ergosphere-new7}).

The possible D7--brane embeddings then naturally split into two
classes: Minkowski embeddings that have a shrinking $S^3$ above the
ergosphere and ergosphere embeddings which reach the ergosphere. These
classes are again separated by a critical embedding which has a
conical singularity at the ergosphere. In analogy to the T--dual
description of the previous subsection for the external electric field
case, the ergosphere embeddings will have to be extended along
$\phi_1$ (as we saw in Chapter~6) so that they can stay non--space--like beyond the
ergosphere.  However, in this chapter we are interested in the scaling properties of the theory
for parameters $(m, c)$ in the vicinity of the critical parameters
$(m_*,c_*)$, corresponding to the critical embedding. As we saw in the
previous section, modifying the ergosphere class of embeddings so as
to be regular at the ergosphere does not alter the properties of the
theory near the critical solution. In particular the scaling exponents
characterizing the discrete self--similar behavior of the theory
remain the same. So henceforth we will focus on the study of the
Minkowski type of embeddings. The analysis is completely analogous to
the one performed in the previous subsection.

In order to focus on the space--time region close to the conical
singularity of the critical embedding, we consider the change of
coordinates:
\begin{equation}   
u=u_{\rm erg}+\frac{u_{\rm{H}} q_1}{R u_{\rm erg}}z;~~~\theta=\frac{\pi}{2}-\frac{y}{R}\ ,
\end{equation}
where 
\begin{equation}
u_{\rm erg}^2=u_{\rm{H}} ^2-q_2^2
\end{equation}
is the radial coordinate $u$ of tip of the critical embedding or
equivalently the $\theta=\pi/2$ point of the ergosphere. It can be
shown that for the values of $q_2$ for which the geometry is
not over spun (and so has an horizon) the corresponding value of
$u_{\rm erg}$ is real.

\begin{figure}[h] 
   \centering
      \includegraphics[width=3.2in]{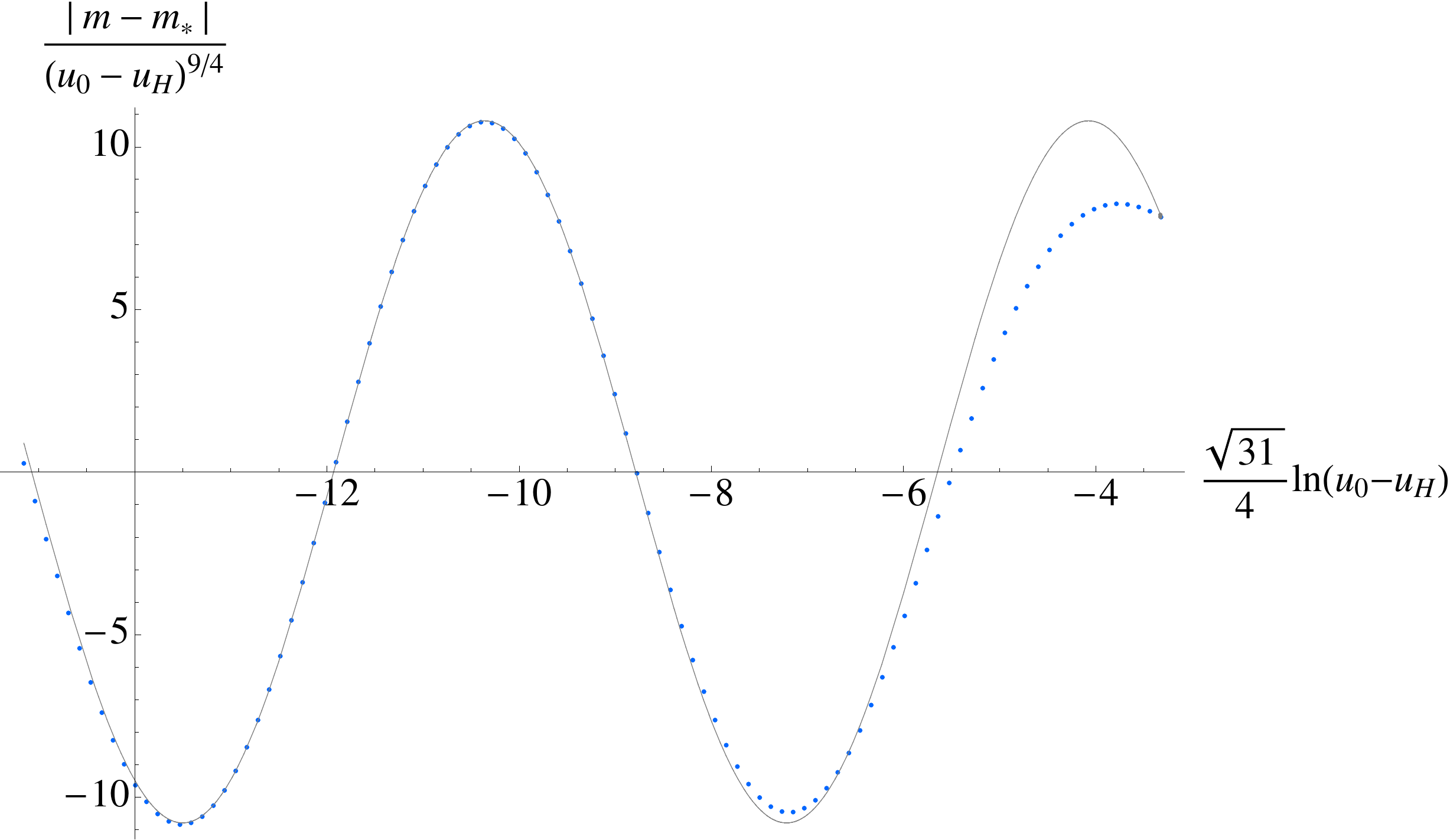} 
    \caption{Plot of the relation between the bare quark mass parameter $m$ and the distance above the ergosphere $(u_0-u_{\rm{H}} )$. The plot is for $q=0.5$ in units in which $u_{\rm{H}} =1$. The solid line is a fit with trigonometric functions of a period $2\pi$. The plot confirms that the scaling exponents of the theory are $r_3={5}/{4}$ and $\alpha_3={\sqrt{31}}/{4}$. }
   \label{fig:rcharge}
\end{figure}

After leaving only the leading terms in $z$ and $y$, we get:
\begin{eqnarray}
&&ds^2/\alpha'=-D_1zdt^2+dz^2+dy^2+y^2d\Omega_3^2-2q_1 dt d\phi_1+\frac{u_{\rm{H}} ^2}{R^2}d\vec x^2+R_1d\phi_1^2\ ,\label{rotmet2}\\
&&d\Omega^3=d\psi^2+\sin^2\psi d\phi_2^2+\cos^2\psi d\phi_3^2\ ;~~~
D_1=\frac{4q_1u_{\rm{H}} }{R^3}\ ;~~~R_1^2=\frac{u_{\rm erg}^2+q_1^2}{u_{\rm{H}} ^2}R^2\ .\nonumber
\end{eqnarray}  
The metric in equation (\ref{rotmet2}) is of the same type as that in
equation~(\ref{rotmet}), namely flat space with some compact
directions in a rotating frame. Therefore the analysis is completely
analogous to the one for the electric case and hence the scaling
exponents are again given by equation (\ref{crit-exp2}) with $n=3$,
because the D7--branes are wrapping an internal $S^3$:
\begin{equation}
r_3=5/4\ ; ~~~\alpha_n=\sqrt{31}/4\ .
\end{equation}
We can again verify this numerically. It is convenient to do this for
the single charge case, namely $q_1\neq 0$, $q_2=q_3=0$. The plot
analogous to Figure~\ref{fig:electric} for the electric case, is
presented in Figure~\ref{fig:rcharge}. The plot represents the
variation of the bare quark mass parameter $m$ as a function of the
initial boundary condition $u_0-u_{\rm{H}} $, for Minkowski--type embeddings.
The parameter $m_*$ corresponds again to the bare quark mass for the
state corresponding to the critical embedding.  The good agreement
with the result for the critical exponents in equation
(\ref{crit-exp2}) is clear, and the accuracy of the analytic
description improves as we go deeper into the spiral (to the left).

An important observation is that our result does not depend on the
values of the R--charges, nor the temperature. In fact, this physics
persists at zero temperature, such as at extremality with all three
charges equal $q_1=q_2=q_3=q$, or more generally. (Extremality is when
$u_{E}=u_1$ or $u_2$ for which $T=0$. See
equation~(\ref{temperatureR7}).) The fact that we have the same
structure at zero temperature (extremal horizon) further confirms that
the key properties of the corresponding phase transition is indeed
driven by the quantum (rather than thermal) fluctuations of the
system.

\section*{7.4 \hspace{2pt} Criticality and scaling: Some generalizations}
\addcontentsline{toc}{section}{7.4 \hspace{0.15cm} Criticality and scaling: Some generalizations}
In this section we generalize the procedure for the study of the
critical behavior employed in all three different systems of phase
transition (thermal, or in the presence of electric field  or R--charge
chemical potential). This may lay the groundwork for other
types of phase transitions that may arise in future studies, seeded by
spirals with different universal behavior.

Note that in all cases there is some vanishing locus. The different
classes of D$q$--brane's embeddings are being classified with respect
to whether they fall into that vanishing locus, or wrap an internal
$S^n$ sphere that is contracting to zero size above the vanishing
locus signaling the end of the D$q$--brane.

In all cases there is a critical embedding that separates the two
classes of embeddings. The critical embedding reaches the vanishing
locus and has a conical singularity there at some finite radius $u_*$
($u_* = u_{\rm{H}} $ or $u_{\rm erg}$ for the thermal and R--charge cases). 

The main point of
the analysis is that after zooming into the space--time region near the
conical singularity we obtain the metric:
\begin{equation}
ds^2=-Dz^k dt^2+dz^2+dy^2+y^2d\Omega_n^2+\dots\ ,
\end{equation}
where $D$ is a non--essential constant.  The Dirac--Born--Infeld
Lagrangian of the brane is then:
\begin{equation}
{\cal L}\propto z^{k/2}y^n\sqrt{1+y'^2}\ .
\label{lagrgen}
\end{equation}
Note that to extract the key behavior (that we are studying) of this
critical embedding (and its neighborhood) there is no need to modify
the embeddings which reach the vanishing locus (as we did for the
ergosphere class of embeddings). The critical solution and the
linearized equation of motion is the same for both classes. Therefore
it is sufficient to consider the Minkowski type of embeddings and
analyze the Lagrangian (\ref{lagrgen}). The resulting equation of
motion is:
\begin{equation}
\partial_z\left(\frac{z^{k/2}y^ny'}{\sqrt{1+y'^2}}\right)-ny^{n-1}z^{k/2}\sqrt{1+y'^2}=0\ .
\label{eom-gen}
\end{equation}
It is easy to check that equation (\ref{eom-gen}) has the scaling property (\ref{scal}) and the limiting critical solution is given by:
\begin{equation}
y_*(z)=\sqrt{\frac{2n}{k}}z\ .
\label{crit-gen}
\end{equation}
Now after the substitution:
\begin{equation}
y(z)=\sqrt{\frac{2n}{k}}z+\xi(z)\ ,
\end{equation}
we obtain the following linearized equation:
\begin{equation}
z^2\xi''(z)+(n+k/2)(z\xi'(z)+\xi(z))=0\ .
\label{eom-lin-gen}
\end{equation}
The general solution of equation~(\ref{eom-lin-gen}) can be written as:
\begin{equation}
\xi(z)=\frac{1}{z^{r_n^{(k)}}}(A\cos(\alpha_n^{(k)})\ln z)+B\sin(\alpha_n^{(k)}\ln z)\ ,
\label{sol-gen}
\end{equation}
where
\begin{equation}
r_n^{(k)}=(n+k/2-1)/2;~~~\alpha_n^{(k)}=\frac{1}{2}\sqrt{4(n+k/2)-(n+k/2-1)^2};
\label{crit-exp-gen}
\end{equation}
are the scaling exponents characterizing the self--similar behavior of
the theory. Both being real, they control the shape of the spiral
which emanates from the critical solution.  The oscillatory behavior
is present for $n\leq 3+2\sqrt{2}-k/2$. For these values of $n$ the
theory exhibits a discrete self--similarity and the equation of state
$c=c(m)$ is a multi--valued function suggesting that the corresponding
phase transition is a first order one.

While there is the possibility of complex scaling exponents and hence
possibly second order phase transitions (if the multi--valuedness goes
away when the spiral does), this is not realized in the examples that
we know from string theory.

Note that we have $k=2$ for a thermal induced phase transition and
$k=1$ for the quantum induced phase transitions that we studied
(external electric field and R--charge chemical potential), arising
from the two most natural types of a vanishing locus that one may
have: an horizon, and an ergosphere. Perhaps other systems will yield
different values of $k$.

\section*{7.5 \hspace{2pt} Concluding remarks}
\addcontentsline{toc}{section}{7.5 \hspace{0.15cm} Concluding remarks}

We have succeeded in casting two important types of phase transition
(in large $N_c$ gauge theory with fundamental flavors) into the same
classifying framework as the meson--melting phase transition. These
quantum fluctuation induced transitions (so--called since they persist
at zero temperature), resulting in the liberation of quarks from being
bound into mesons as a result of the application of an external
electric field, or a chemical potential for R--charge, turn out to
have the same underlying structure. It is distinct from that found for
thermal fluctuation induced transitions.  The structures are
controlled by the local geometry of the spacetime seen by a critical
D--brane embedding (it is the borderline case between two physically
distinct classes of embedding), and while it is Rindler for the
thermal case with an horizon at the origin, it is (after a T--duality
in order to geometrize the discussion as much as possible) a rotating
space with a simple ``ergosphere'' type locus.  The technique of
characterizing the physics in terms of this underlying classifying
space \cite{Frolov:2006tc,Mateos:2006nu} is rather pleasing in its
utility, and we extended our analysis to the natural generalization of
this space, extracting the scaling exponents that might pertain to
physics from future studies.

Of course, there is much interest in how much we can learn about
finite $N_c$ physics (for applications to systems such as QCD) by
studying universal features of large $N_c$. Unfortunately, it is
almost certain that much of this is far from robust against $1/N_c$
corrections. The spiral structure is rather delicate, and the stringy
corrections arising in going away from the large $N_c$ limit would
generically severely modify the classifying spacetimes we've been
studying, erasing the spiral and its self--similarity. The absence of
the spiral is necessary for there to be (at best) a second--order
transition at finite $N_c$, since it results in multi--valuedness of
the solution space, requiring the system to perform a first order
jump.

However, It is tempting to speculate that the nature by which the
spiral is destroyed by $1/N_c$ corrections might (especially since the
setting is so geometrical) be characterizable in a way that allows
universal properties of the second (or higher) order phase transitions
to be deduced from the properties of the spiral at large $N_c$. 


\chapter*{Chapter 8:  \hspace{1pt} Conclusion}
\addcontentsline{toc}{chapter}{Chapter 8:\hspace{0.15cm} Conclusion}

 In this work we employed holographic techniques to study dynamics of flavored strongly coupled large $N_c$ non--abelian gauge theories in the quenched approximation. In particular we focused on the gauge theories dual to the D$p$/D$(p+4)$ systems and most of our study was dedicated to the D3/D7 system. One of the main properties that we studied, was the phase structure of the theory under the influence of various controlling parameters such as: external magnetic field, external electric field and R--charge chemical potential. We were also interested in describing non--perturbative phenomena such as chiral symmetry breaking. Another aspect of our study was the meson spectroscopy of the dual gauge theory.

In our study of the zero temperature flavored large $N_c$ ${\cal N}=2$ Yang--Mills theory coupled to an external magnetic field we showed that, as expected, the supersymmetry is completely broken by the external field. Furthermore, in order to lower its energy, the theory develops a quark condensate that scales with the appropriate power of the energy scale which is set by the external magnetic field. Our study thus provided a holographic description of the phenomena of magnetic catalysis of chiral symmetry breaking. An intriguing property of the theory is the spiral structure at the origin of the quark condensate versus bare quark mass plane. In particular it is interesting how this pattern governs the spontaneous breaking of the chiral symmetry. The detailed analysis of this structure, performed in Chapter~2, revealed a discrete self-similar behavior of the dual gauge theory near criticality. We calculated the corresponding critical exponents for the bare quark mass, the quark condensate, and the meson spectrum. 
Our study of the meson spectrum confirmed the expectations based on thermodynamic considerations that the lowest positive branch of the spiral corresponds to a stable phase of the theory and that the inner branches are real instabilities that are characterized by a tachyonic ground state and cannot be reached by a supercooling. The lowest negative $\tilde m$ branch of the spiral is tachyon free and thus could be metastable. Further study of the meson spectrum revealed a Zeeman splitting of the energy levels and characteristic Gell-Mann-Oakes-Renner relation \cite{GellMann:1968rz} between the mass of the pion associated to the softly broken chiral symmetry and the bare quark mass. A very interesting direction of future study would be the generalization to the case of the D$p$/D$(p+4)$ system. In particular, it would be interesting to verify the universality of the pattern in which the spontaneous chiral symmetry breaking takes place.

Another important issue that we address in this work is the thermal properties of the gauge theory dual to the D3/D7 brane intersection. Following the pioneer paper ref.~\cite{Babington:2003vm}, in Chapter~3 we introduced flavor D7--brane to the near horizon limit of the black 3--brane geometry. Our numerical study revealed a first order phase transition pattern in the equation of state in the quark condensate versus bare quark mass plane. This phase transition has been extensively studied in the literature and is now believed to be describing the melting of mesons into the ${\cal N}=4$ super Yang--Mills plasma. The study of this phase transition and in particular, the generalization of the corresponding phase diagram to include the influence of additional controlling parameters is one of the main subjects of our study.

Encouraged by the rich physical picture accessible through the holographic tools employed in Chapter~2 and Chapter~3, in Chapter~4, we considered the influence of an external magnetic field on the meson melting phase transition. Inversely, one could say that we studied the effect that the temperature has on the magnetic catalysis of chiral symmetry breaking. Our study of the corresponding phase diagram supported the expectations from earlier field theory results that the temperature has a restoration effect and increase the value of the magnetic field at which spontaneous chiral symmetry breaking happens. The meson melting phase transition exists only below a critical value of the applied field. This is the critical value above which spontaneous chiral symmetry breaking is triggered (in the case of zero mass). Above this value, regardless of the quark mass (or for fixed quark mass, regardless of the temperature), the system remains in a phase with a discrete spectrum of stable masses. Evidently, for these values of the field, it is magnetically favorable for the quarks and anti--quarks to bind together, reducing the degrees of freedom of the system, as can be seen from our computation of the entropy.
Meanwhile, the magnetization is greater in this un--melted phase. Our study of the meson spectrum revealed the existence of a chiral symmetry breaking metastable phase existing bellow a critical ratio of the magnetic field and the temperature square. This metastable phase eventually wins at sufficiently strong magnetic field and the melted phase seizes to exist. We also verified that the meson spectrum of the pion, associated to the spontaneous chiral symmetry breaking, exhibit Zeeman splitting of the energy levels at weak magnetic field. Furthermore, for sufficiently strong magnetic field, when the chiral symmetry is spontaneously broken, a characteristic Gell-Mann-Oakes-Renner relation \cite{GellMann:1968rz} was observed.

 A natural extension of the phase diagram of the meson melting phase transition is the addition of an external electric field. In Chapter~5 of this work we addressed this problem. A key technical feature that makes the electric case different from the magnetic case is the existence of a vanishing locus (above the horizon) at which the D7--brane action vanishes. In order to be able to extend the D7--branes beyond that locus one needs to excite additional $U(1)$ gauge field on the D7--brane. This additional gauge field sources holographically a global electric current along the direction of the external electric field \cite{Karch:2007pd}. The D7--branes then naturally fall into two different classes, namely D7--branes which reach the vanishing locus and eventually fall into the black hole, and D7--branes which close above that locus. The scenario is quite similar to the case of finite temperature studied in Chapter~3. The transition from one type of embeddings to another type of embeddings corresponds to transition from melted to un--melted phase of the fundamental matter and correspondingly from zero global current to non--zero one. This is why we refer to this transition as to an insulator/conductor phase transition. The bound quarks dissociate under the separating electric force. It is somewhat intriguing that this process takes place also at zero temperature. We addressed this issue in more details in Chapter~7. Another interesting feature (again technical) is that even after we have regularized the D7--branes at the vanishing locus, by introducing additional $U(1)$ gauge field on the D7--brane, the embeddings, entering the vanishing locus above some critical polar angle, develop a conical singularity before falling into the black hole. Unlike the conical singularity of the critical embedding separating the two classes of embeddings, this conical singularity does not lead to a discrete self--similar behavior. We proposed that perhaps this singularity is fixed by stringy corrections. Of immediate interest is the calculation of the meson spectrum for this class of embeddings as it could shed light on the question of their stability. We leave this task for the future. It would also be interesting to study the meson spectrum at zero temperature in the dissociated phase and check for the existence of quasi-normal modes, such a study could verify the stability of that phase.  

Another controlling parameter that one can consider is an R--charge chemical potential. In order to address this problem, in Chapter~6, we introduced flavor D7--brane to the geometry of spinning D3--branes. We provided a detailed calculation of the Hawking temperature of the background and elaborated on the extremal case for three equal charges when we have vanishing temperature. We also calculated the R--charge chemical potential of the theory at zero temperature. Then we focused on the properties of the D7--brane embeddings in two special cases, namely one R--charge and three equal R--charges. We obtained the equations of state in a numerical form for both cases and demonstrated that the critical mass is increasing with the R--charge of the geometry. We also obtained the phase diagrams of the meson melting phase transition in the temperature versus R--charge chemical potential plane. While studying the properties of the probes, in both cases, we found a subtlety closely related to the one that we encountered in Chapter~5, for the study of the electric field case. This is the existence of a vanishing locus, this time due to the ergosphere of the background. Letting the probes extend along the direction of rotation of the geometry initially transverse to the D7--brane, solved the problem of extending the solutions beyond the ergosphere. Since this direction is isometry, there is a corresponding constant of integration (analogous to the global electric current from Chapter~5). It was shown that the physical interpretation of this constant of integration is related to the phase difference between the bare quark mass and the quark condensate of the dual gauge theory.  An interesting direction of future study is the analysis of the meson spectrum of theory, especially in the deconfined phase.

Finally we would like to comment on the universal properties of the gauge theories holographically dual to the D$p$/D$q$ systems T-dual to the D3/D7 intersection. One such property was described in ref.~\cite{Mateos:2006nu}, where the authors studied the meson melting phase transition for the general class of D$p$/D$q$ systems and revealed a remarkable self-similar structure of the theory. The structure is
controlled by the local geometry of the space-time seen by a critical
D--brane embedding (it is the borderline case between two physically
distinct classes of embedding) which is a Rindler space. 
This critical behavior is characterized by scaling exponents that can be calculated analytically. They are the same for all D$p$/D$q$ systems and depend only on the dimension of the internal sphere wrapped by the probe branes.  In Chapter~7, we reviewed this work and showed that similar structure exists for the quantum fluctuation induced insulator/conductor phase transition studied in Chapter~5. The theory again exhibits discrete self-similar behavior, characterized by scaling exponents which are distinct from the one for the thermal phase transition but are again universal for all D$p$/D$q$ systems. After a T--duality, in order to geometrize the discussion, we showed that the local geometry controlling the structure is a rotating space with a simple ``ergosphere'' type locus. We also considered the case of R--charge chemical potential driven phase transition, studied in Chapter~6, for the case of the D3/D7 system. We found the same structure, as for the insulator/conductor phase transition, characterized by the same scaling exponents. From technical point of view this is not surprising since in both cases the vanishing loci are of  ``ergosphere'' type. It would be interesting to extend this study to the case of spinning D$p$--brane geometries as this would provide a further check of the proposed universal behavior. Let us also clarify that these studies are in the context of the large $N_c$ limit and most likely the observed self-similar behavior is an artifact. Furthermore, at finite $N_c$ the phase transition is expected to be of second order. There is, of course, the possibility that some universal properties of the D$p$/D$q$ systems will still be present at finite $N_c$.



\end{document}